# The Surface Distributions of the Production of the Major Volatile Species, $H_2O$, $CO_2$, CO and $O_2$, from the Nucleus of Comet 67P/Churyumov-Gerasimenko throughout the Rosetta Mission as Measured by the ROSINA Double Focusing Mass Spectrometer

Michael Combi (1), Yinsi Shou (1), Nicolas Fougere (1), Valeriy Tenishev (1), Kathrin Altwegg (2), Martin Rubin (2), Dominique Bockelée-Morvan (3), Fabrizio Capaccioni (4), Yu-Chi Cheng (3), Uwe Fink (5), Tamas Gombosi (1), Kenneth C. Hansen (1), Zhenguang Huang (1), David Marshall (6), Gabor Toth (1)

(1) Department of Climate and Space Sciences and Engineering, University of Michigan, Ann Arbor, Michigan, USA
(2) Physikalisches Institut, University of Bern, Bern, Switzerland
(3) LESIA, Observatoire de Paris, PSL Research University, CNRS, Sorbonne Universites, UPMC Univ. Paris 06, Univ. Paris-Diderot, Sorbonne Paris Cité, 5 place Jules Janssen, F-92195 Meudon, France
(4) 4INAF-IAPS, Istituto di Astrofisica e Planetologia Spaziali, via del fosso del Cavaliere 100, I-00133 Rome, Italy
(5) Lunar and Planetary Laboratory, University of Arizona, Tucson, Arizona, USA
(6) Max-Planck-Institut für Sonnensystemforschung, Justus-von-Liebig-Weg 3, 37077 Göttingen, Germany

**Abstract**

The Rosetta Orbiter Spectrometer for Ion and Neutral Analysis (ROSINA) suite of instruments operated throughout the over two years of the Rosetta mission operations in the vicinity of comet 67P/Churyumov-Gerasimenko. It measured gas densities and composition throughout the comet's atmosphere, or coma. Here we present two-years' worth of measurements of the relative densities of the four major volatile species in the coma of the comet, $H_2O$, $CO_2$, $CO$ and $O_2$, by one of the ROSINA sub-systems called the Double Focusing Mass Spectrometer (DFMS). The absolute total gas densities were provided by the Comet Pressure Sensor (COPS), another ROSINA sub-system. DFMS is a very high mass resolution and high sensitivity mass spectrometer able to resolve at a tiny fraction of an atomic mass unit. We have analyzed the combined DFMS and COPS measurements using an inversion scheme based on spherical harmonics that solves for the distribution of potential surface activity of each species as the comet rotates, changing solar illumination, over short time intervals and as the comet changes distance from the sun and orientation of its spin axis over long time intervals. We also use the surface boundary conditions derived from the inversion scheme to simulate the whole coma with our fully kinetic Direct Simulation Monte Carlo model and calculate the production rates of the four major species throughout the mission. We compare the derived production rates with revised remote sensing observations by the Visible and Infrared Thermal Imaging Spectrometer (VIRTIS) as well as with published observations from the Microwave Instrument for the Rosetta Orbiter (MIRO). Finally we use the variation of the surface production of the major species to calculate the total mass loss over the mission and, for different estimates of the dust/gas ratio, calculate the variation of surface loss all over the nucleus.

**1. Introduction**

The Rosetta spacecraft orbited and maneuvered in the vicinity of the nucleus of comet 67P/Churyumov-Gerasimenko (67P hereafter) for 26 months from August 2014 through September 2016 recording various measurements from 11 instruments on the orbiter. There were 10 more instruments on the Philae Lander. The mission was able to monitor the comet carefully at heliocentric distances from 3.5 AU before perihelion, through perihelion at 1.24 AU and out to 3.5 AU after perihelion. As such the combination of temporal coverage, sensitivity and resolution of all types of measurements was unprecedented.

In previous work we (Fougere et al. 2016a, 2016b) used measurements by the Double Focused Mass Spectrometer (DFMS) and the Comet Pressure Sensor (COPS) subsystems of the Rosetta Ion and Neutral Analyzer (ROSINA) package to determine the potential surface distribution of the four major ions, $H_2O$, $CO_2$, CO and $O_2$. Those results covered the period from first close encounter in early August 2014 through the end of February 2016, two months before the comet's second equinox. We had used the determined potential surface distribution of $H_2O$ and $CO_2$ to compare with images of these two species taken with the VIRTIS-M (Migliorini et al. 2016) moderate spectral resolution imaging spectrometer and with various point spectra taken with the VIRTIS-H high-resolution spectrometer (Bockelée-Morvan et al. 2015 and 2016). The spatial distributions, abundances and derived productions of $H_2O$ and $CO_2$ from the DFMS/COPS-determined model were in quite reasonable agreement with both VIRTIS-M images of $H_2O$ and $CO_2$ taken in March 2015 and taken with VIRTIS-H spectra taken from November 2014 through January 2015. However, the DFMS/COPS-determined model generally predicted higher abundances by up to a factor of four in the weeks around perihelion of $H_2O$ than determined by VIRTIS-H and somewhat higher abundances of $CO_2$. In this paper we have improved our earlier approach and extended a similar analysis through the end of DFMS data, a month before the end of mission in September 2016.

## 2. Reduction and Calibration of the ROSINA DFMS Data

The results presented in this paper all stem from measurements by the DFMS and COPS subsystems of the ROSINA package. The ROSINA package of instruments was described by Balsiger et al. (2007). The DFMS instrument measures high resolution mass spectra with a resolution of up to $m/\Delta m = 3000$ and as such can determine molecular, elemental and isotopic abundances even for otherwise normally overlapping masses such as CO and $N_2$ (Rubin et al. 2015). The reduction procedure for the data on $H_2O$, $CO_2$, CO and $O_2$ presented here has been described in great detail in a series of previous papers (Bieler et al. 2015a; Bieler et al. 2015b; Fougere et al. 2016a; Fougere et al. 2016b). Briefly, DFMS mass spectra were obtained in several individual mass channels for the four major species and were given an absolute calibration using the total gas densities derived from the nude gauge measurements by COPS.

The full set of DFMS data analyzed here is shown in Figure 1, which shows the measured densities times the square of the comet-spacecraft distance. It is roughly broken up into 3 time periods. In the first the north pole is nearly continually exposed to the Sun and the two dominant species, $H_2O$ and $CO_2$, are generally anti-correlated with one another. The second period runs from the first equinox until the second equinox when enough or all of the south polar region is illuminated by the Sun and the two dominant species, $H_2O$ and $CO_2$, are generally correlated with one another. The third period runs to the end of mission and returns to the geometry when most of the north polar region is illuminated, the south is not, and the two dominant species, $H_2O$ and $CO_2$, are generally anti-correlated again. We will discuss the relative behavior of all four species later in this paper. Uncertainties in the COPS measurements of the total gas density are ~10%. The DFMS densities of each major species are constructed from individual DFMS spectra at each mass channel and then scaled to the total measured by COPS. The final uncertainties in the individual species densities are ~ 20%.

3. Potential Surface Activity Distributions: Methods

In an analysis of ROSINA COPS data from the first 5 months of the Rosetta mission, August through December 2014, it was shown using a simple analytic model based on assuming that a uniform potential activity from the surface when illuminated could explain much of the density variation of the gas as the comet rotated and the spacecraft position changed (Bieler et al. 2015a). One exception was that it appeared that some additional activity was required from the so-called neck region of the "duck" shape of the nucleus that is not far from the north pole. This was tested and verified in the same paper using both a hydrodynamic model for the gas flow as well as a fully kinetic Direct Simulation Monte Carlo model.

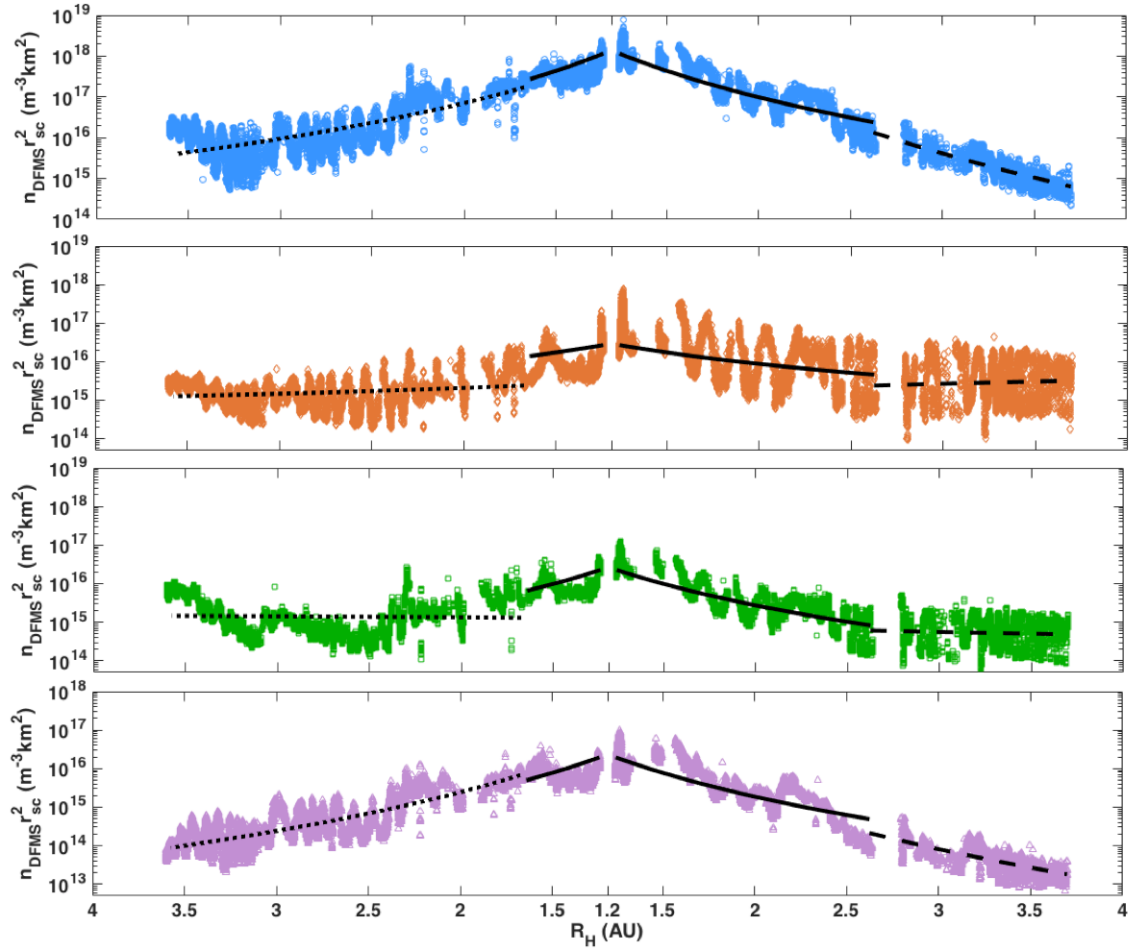

Figure 1. Heliocentric distance dependence of DFMS densities normalized by the inverse square of the spacecraft-comet distance. From the top down the plots are for the $H_2O$ (blue), $CO_2$ (orange), CO (green) and $O_2$ (purple). The lines in each give the power-law fits for the three separate time periods as given in Table 1, dotted for phase 1, solid for phase 2 and dashed for phase 3.

In order to examine this idea more fully and instead of assuming the potential surface activity was driven by illumination uniformly from the surface, Fougere et al. (2016a) analyzed the $H_2O$ and $CO_2$ gas densities measured by ROSINA DFMS for the period from August 2014 through May 2015 with a least squares inversion scheme based on expressing the potential surface activity as a spherical harmonics expansion with undetermined coefficients. This covered the time period from the near nucleus measurements soon after rendezvous until the comet reached its first equinox. The technique assumed the same analytic model connecting illumination-driven surface activity to densities at the spacecraft as introduced by Bieler et al. (2015a) and described and used by Fougere et al. (2016a, 2016b). The final potential surface activity distributions for $H_2O$ and $CO_2$ were then used as the surface boundary conditions and a physically-based coma was generated using the fully-kinetic Direct Simulation Monte Carlo (DSMC) model.

The DSMC models based on the spherical harmonics inversions were also used by Fougere et al. (2016a) to synthesize a series of $H_2O$ and $CO_2$ images obtained in April 2015 by Migliorini et al. (2016) using the Rosetta Visible and Infrared Thermal Imaging Spectrometer (VIRTIS) medium resolution mapping channel and a set of $H_2O$ and $CO_2$ column abundances obtained with the VIRTIS high resolution channel at various locations around the coma by Bockelée-Morvan et al. (2016) from November 2014 through January 2015 and found fairly consistent results. $H_2O$ and $CO_2$ production rates were calculated from both ROSINA DFMS and VIRTIS-M and VIRTIS-H measurements. Comparisons by Fougere et al. (2016a) showed consistent direct correlation of VIRTIS-H column densities from November 2014 through January 2015 with those derived from the DFMS model derived coma for both $H_2O$ and $CO_2$. Synthetic images from the DFMS-modeled coma were in excellent agreement with the VIRTIS-M images during the April 2015 period. Production rates based on the coma DFMS model distribution applied to both the VIRTIS-M and VIRTIS-H measurements were consistent on the average and showed a similar scatter (factors of 2-3) about the mean trends.

Fougere et al. (2016b) then applied the same spherical harmonics potential surface activity inversion technique to ROSINA DFMS measurements of $H_2O$, $CO_2$, CO and $O_2$ from August 2014 until the second equinox of comet 67P in March 2016. They found that approximately the same potential surface activity distributions could be used for each species for the entire apparition to that point in time. Changes in the coma were mainly driven by the changes in solar illumination from nucleus rotation, seasonal variation of the comet spin axis-solar geometry and heliocentric distance. The heliocentric distance, R, dependencies for the $H_2O$, $CO_2$, CO and $O_2$ varied as $R_{AU}^{-5.6}$, $R_{AU}^{-2.8}$, $R_{AU}^{-2.6}$ and $R_{AU}^{-4.9}$, respectively.

Comparison of the ROSINA DFMS derived coma around the time of perihelion with VIRTIS-H observations, however, were apparently not completely consistent. Although VIRTIS-H observations of $CO_2$ were reasonably consistent with column densities synthesized with the DFMS derived DSMC models for the time around perihelion, the $H_2O$ distributions inferred from the DFMS data were on the average a factor of ~4 larger than those derived from VIRTIS-H measurements. This is despite the fact that the relative spatial distributions in a set of raster-scanned images made with the VIRTIS-H aperture in late July and early August 2015 were similar to those from the surface distributions obtained from the DFMS data. Both showed an enhanced coma toward the Sun and the south in the images. The only difference in the various analysis procedures for the perihelion period measurements is that the VIRTIS-H column densities (Bockelée-Morvan et al. 2016) for $H_2O$ were calculated using some isolated hot band emissions features of $H_2O$ rather than the total fluorescence rate in the whole 2.7-μm band that was used in the earlier work (Bockelée-Morvan et al. 2015). This was done because around perihelion the 2.7-μm band is largely optically thick, while for the November 2014-January 2015 VIRTIS-H observations one can assume the whole band is optically thin. It is worth mentioning that the conversion of emission brightness to column density for the VIRTIS observations requires no assumption of the coma distribution. Finally, optical thickness would cause column densities derived from the whole 2.7-μm band to be underestimated. This is why the basic analysis of the VIRTIS-H data around perihelion uses optically thin hot bands rather than the whole 2.7-μm band.

A new approach to calculation adopted here is to account for the variation of potential surface activity over the whole apparition. This seems to resolve a portion of this inconsistency. A brief summary of the surface activity inversion method is given here for completeness. The detailed explanation was given by Fougere et al. (2016a).

The density of a species is expressed in terms of a spherical harmonics series using the analytic model of Bieler et al. (2015) that makes the connection between the geometric projection of each surface facet of the nucleus as seen by the spacecraft and the amount to which it is illuminated by the sun with the location of the spacecraft. This yields an M x N matrix, $C_{ij}$ that is written as

$$C_{ij} = \frac{1}{R_{AU}^\beta} \left( \sum_{k=1}^{N_{faces}} \frac{(g(\Theta_k))_i S_k (\cos(\alpha_k))_i}{r_k^2} Y_j(\vartheta_k, \varphi_k) \right),$$

where $N_{faces}$ is the total number of triangles in the nucleus surface mesh, $\Theta_k$ is their solar zenith angle, $g$ is defined as in Bieler et al. (2015) by $(g(\Theta_k) = max(a_{night}, cos(\Theta_k))$, $a_{night}$ is the minimum value in shadows or on the night side. The surface area of the $k$th triangle is $S_k$, $r_k$ and $\alpha_k$ are the distance and angle between the spacecraft and the outward normal of the $k$th triangle, respectively. The colatitude and longitude of the center of the $k$th triangle are $\vartheta_k$ and $\varphi_k$, respectively. Finally, the least squares problem is in the form $\min_{x \in R^N} \|Cx - d\|_2^2$, where $d$ is a vector of length $M$ elements containing the DFMS data and $x$ is a vector of length $N$ containing the undetermined coefficients of the spherical harmonics for which we are solving. There are 15000 to 25000 data points total for each species. The potential activity $f_k = \sum_{j=1}^{N} x_j Y_j(\vartheta_k, \varphi_k)$ of any surface triangular facet is constrained to be positive definite. The gas flux of any triangle is given by $F_k = \lambda \frac{G(\Theta_{SZA}) f_k}{R_{AU}^\beta}$ where the function G give the variation of the flux with solar zenith angle defined by Bieler et al. (2015) as $G(\Theta_{SZA}) = a_{night} + (1 - a_{night}) cos(\Theta_{SZA})$ which gives $G(\Theta_{SZA}) = a_{night}$ if a surface triangle is not illuminated by the Sun. The factor λ is included here to enable correction of the approximate nature of the initial power-law assumption of the production rate as a function of heliocentric distance. Most of the values of λ range from 0.5 to 4 and serve mostly to correct densities for the initial assumed power laws in heliocentric distance.

**Model Analysis**

As in the papers by Fougere et al. (2016a) and (2016b) we have employed a 5th order spherical harmonics expansion with 25 undetermined coefficients to the DFMS measurements. One change for this work is that having observed a noticeable difference in the activity between the pre-perihelion and the now extended post-perihelion variations of the four major volatiles, we have separated the data into three time groups: (1) from the beginning of the mission to the first equinox, (2) from the first to the second equinox, and (3) from the second equinox until the last data before the end of mission.

Figure 1 shows the evolution of the measured densities of the 4 species normalized by the square of the spacecraft-comet distance and fitted with power-laws for the three time periods shown as the overlain straight lines for each of the four species. Table 1 lists the values of the power-law exponents for the three time periods for the each of the four major volatile species.

Figure 2 shows the comparison of the measured DFMS number densities of $H_2O$, $CO_2$, CO and $O_2$ with those obtained by applying the potential surface activity distribution using the simple analytical model (Bieler et al. 2015a) that was used in the least squares fitting procedure.

The correlation coefficients for the model fits for the three phases (p1, p2, p3) are (0.971, 0.910, 0.895) for $H_2O$, (0.924, 0.940, 0.952) for $CO_2$, (0.832, 0.942, 0.947) for CO and (0.869, 0.559, 0.923) for $O_2$, respectively. Tables 2-4 give the spherical harmonics coefficients for the three phases, respectively.

**Table 1.** Parameters used in the analytical model to de-trend the input DFMS data for computing the surface activity distribution inversions.

| Species | β-period 1 | β-period 2 | β-period 3 | $a_{night}$ |
|---|---|---|---|---|
| $H_2O$ | 4.96 | 5.17 | 8.86 | 2 per cent |
| $CO_2$ | 0.85 | 2.37 | -0.83 | 2 per cent |
| CO | -0.15 | 4.44 | 0.69 | 10 per cent |
| $O_2$ | 5.71 | 4.94 | 7.17 | 2 per cent |

Figure 3 shows the potential surface activity distributions for the four major volatiles as a function of latitude and longitude on the nucleus of 67P. A number of interesting things can be seen by inspection. First the potential surface source distributions of $H_2O$ and $O_2$ are very similar for each of the three time periods. This further enhances the conclusions of Bieler et al. (2015b) that the $O_2$ is predominantly associated with $H_2O$ ice and despite the high volatility of frozen $O_2$, the release of $O_2$ in comets is mostly controlled by $H_2O$.

In contrast that of $CO_2$ has most of its dominant activity in the south with little in the north compared with $H_2O$ and $O_2$. This is consistent with the first reported DFMS results by Hässig et al. (2015) and with VIRTIS results first reported by Bockelée-Morvan et al. (2015). In addition the potential surface activity distributions are quite similar throughout the whole apparition, unlike those for the other species that change noticeably over the three studied periods. The distributions of CO and $CO_2$ are very similar for the second and third time periods, that is, after the first equinox, but before the first equinox the distribution of the CO surface source is quite different. CO has a major source in the north coinciding with $H_2O$ as well as activity in the south similar to but not quite as strong as $CO_2$.

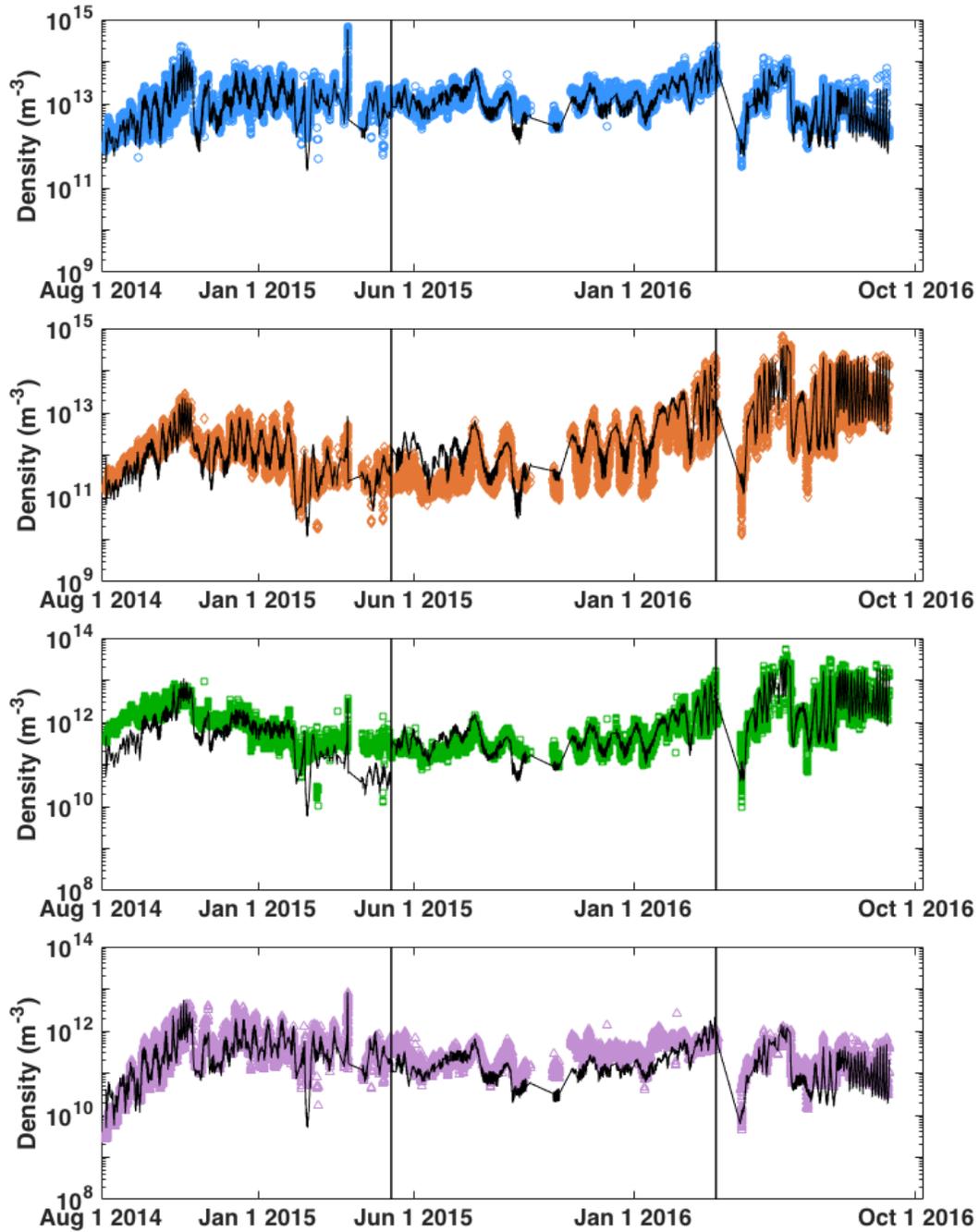

Figure 2. Best fit of potential surface activity distribution (black) compared with the ROSINA DFMS measurements of $H_2O$ (blue), $CO_2$ (orange), CO (green), and $O_2$ (purple) using just the simple analytical model of Bieler et al. (2015a). The left and right vertical lines give the times of the first and second equinox, respectively.

**Table 2.** Real spherical harmonics and the normalized coefficients for time period Phase 1.

| Term | Expression | H$_2$O coefficient | CO$_2$ coefficient | CO coefficient | O$_2$ coefficient |
|---|---|---|---|---|---|
| $Y^0_0$ | 1 | 5.561458e-01 | 1.372429e+00 | 2.447405e-01 | 6.595613e-01 |
| $Y^{-1}_1$ | $\sin(\varphi) \sin(\theta)$ | -1.368247e+00 | -9.392681e-01 | -4.207209e-01 | -8.844053e-01 |
| $Y^0_1$ | $\cos(\theta)$ | 6.153107e-01 | -2.528400e+00 | -9.513894e-03 | 4.491927e-01 |
| $Y^1_1$ | $\cos(\varphi) \sin(\theta)$ | 5.256718e-01 | -1.072431e-01 | -2.408516e-01 | 9.071971e-02 |
| $Y^{-2}_2$ | $\sin(2\varphi) \sin2(\theta)$ | -3.546880e-02 | -1.394259e-01 | -2.337115e-02 | -5.583797e-02 |
| $Y^{-1}_2$ | $\sin(\varphi) \sin(\theta) \cos(\theta)$ | -4.495567e-01 | 4.077346e-01 | 5.454155e-02 | -9.520508e-01 |
| $Y^0_2$ | $3 \cos^2(\theta) - 1$ | 3.805068e-01 | 1.211708e+00 | 1.172589e-01 | 8.753543e-01 |
| $Y^1_2$ | $\cos(\varphi) \sin(\theta) \cos(\theta)$ | -1.059403e-02 | 9.819443e-02 | 2.375706e-01 | 1.242387e-01 |
| $Y^2_2$ | $\cos(2\varphi) \sin^2(\theta)$ | -1.438441e-01 | -6.736809e-02 | -4.202433e-02 | -1.354156e-01 |
| $Y^{-3}_3$ | $\sin(3\varphi) \sin^3(\theta)$ | -1.676119e-02 | -8.508174e-02 | -1.839418e-02 | 4.155716e-02 |
| $Y^{-2}_3$ | $\sin(2\varphi) \sin^2(\theta) \cos(\theta)$ | -1.152462e-01 | 3.711349e-01 | 1.658289e-02 | -6.718495e-02 |
| $Y^{-1}_3$ | $\sin(\varphi) \sin(\theta)5(\cos2(\theta) - 1)$ | -2.968521e-01 | -1.748390e-01 | -8.638151e-02 | -1.713198e-01 |
| $Y^0_3$ | $5 \cos^3(\theta) - 3 \cos(\theta)$ | 1.413108e-01 | -5.157962e-01 | 6.093995e-02 | 1.200808e-01 |
| $Y^1_3$ | $\cos(\varphi) \sin(\theta)5(\cos^2(\theta) - 1)$ | 1.013912e-01 | -2.018093e-02 | -4.975368e-02 | 1.509013e-02 |
| $Y^2_3$ | $\cos(2\varphi) \sin^2(\theta) \cos(\theta)$ | -1.767766e-01 | 2.239736e-02 | 1.949725e-02 | -1.968264e-01 |
| $Y^3_3$ | $\cos(3\varphi) \sin^3(\theta)$ | -2.367182e-02 | 8.436323e-02 | 2.056742e-02 | -2.582434e-02 |
| $Y^{-4}_4$ | $\sin(4\varphi) \sin^4(\theta)$ | -4.993913e-02 | -1.092401e-02 | -2.965282e-03 | 2.610879e-02 |
| $Y^{-3}_4$ | $\sin(3\varphi) \sin^3(\theta) \cos(\theta)$ | -1.284025e-01 | 4.845830e-01 | 3.680527e-02 | 1.205724e-01 |
| $Y^{-2}_4$ | $\sin(2\varphi) \sin^2(\theta)(7 \cos^2(\theta) - 1)$ | -1.237713e-01 | -1.583961e-01 | -3.947024e-02 | -6.396652e-02 |
| $Y^{-1}_4$ | $\sin(\varphi) \sin(\theta)(7 \cos^3(\theta) - 3 \cos(\theta))$ | -4.875823e-01 | 7.665290e-02 | -1.547551e-02 | -2.646542e-01 |
| $Y^0_4$ | $35 \cos^4(\theta) - 30 \cos^2(\theta) + 3$ | -8.971545e-03 | -8.752650e-03 | -2.458842e-02 | 8.999660e-02 |
| $Y^1_4$ | $\cos(\varphi) \sin(\theta)(7 \cos^3(\theta) - 3 \cos(\theta))$ | -3.002277e-02 | 1.238735e-01 | 3.064848e-02 | 2.473054e-02 |
| $Y^2_4$ | $\cos(2\varphi) \sin^2(\theta)(7 \cos^2(\theta) - 1)$ | 1.622916e-02 | -7.260629e-02 | -4.074338e-02 | -1.398722e-01 |
| $Y^3_4$ | $\cos(3\varphi) \sin^3(\theta) \cos(\theta)$ | -2.928818e-01 | -2.410971e-01 | -5.985826e-02 | -7.720337e-02 |
| $Y^4_4$ | $\cos(4\varphi) \sin^4(\theta)$ | 1.826566e-02 | -1.687838e-02 | 2.094206e-02 | 1.979501e-02 |

**Table 3.** Real spherical harmonics and the normalized coefficients for time period Phase 2.

| Term | Expression | H$_2$O coefficient | CO$_2$ coefficient | CO coefficient | O$_2$ coefficient |
|---|---|---|---|---|---|
| $Y^0_0$ | 1 | 4.895834e-01 | 1.375611e+00 | 4.609846e-01 | 2.407584e-01 |
| $Y^{-1}_1$ | $\sin(\varphi) \sin(\theta)$ | -3.628282e-01 | -4.683337e-02 | 2.602201e-01 | -1.642184e-02 |
| $Y^0_1$ | $\cos(\theta)$ | -2.644014e-01 | -2.335596e+00 | -5.886954e-01 | -1.625394e-01 |
| $Y^1_1$ | $\cos(\varphi) \sin(\theta)$ | 5.399610e-01 | -4.373107e-01 | -1.177043e-01 | 5.914185e-02 |
| $Y^{-2}_2$ | $\sin(2\varphi) \sin2(\theta)$ | -8.288829e-02 | 9.081417e-02 | 6.886639e-02 | 2.291460e-02 |
| $Y^{-1}_2$ | $\sin(\varphi) \sin(\theta) \cos(\theta)$ | -1.459630e-02 | -3.300306e-01 | -2.445945e-01 | -4.651805e-03 |
| $Y^0_2$ | $3 \cos^2(\theta) - 1$ | 3.385707e-01 | 1.239087e+00 | 2.656085e-01 | 1.044153e-01 |
| $Y^1_2$ | $\cos(\varphi) \sin(\theta) \cos(\theta)$ | -3.197988e-02 | -7.304284e-02 | 3.411257e-02 | -8.059736e-04 |
| $Y^2_2$ | $\cos(2\varphi) \sin^2(\theta)$ | -1.070801e-01 | -3.457324e-01 | -9.687787e-02 | -1.179830e-01 |
| $Y^{-3}_3$ | $\sin(3\varphi) \sin^3(\theta)$ | -7.809985e-04 | -2.156722e-01 | -1.349597e-01 | 3.071106e-02 |
| $Y^{-2}_3$ | $\sin(2\varphi) \sin^2(\theta) \cos(\theta)$ | 5.920391e-02 | -5.579715e-01 | -4.069729e-01 | 2.252411e-02 |
| $Y^{-1}_3$ | $\sin(\varphi) \sin(\theta)5(\cos2(\theta) - 1)$ | -7.746035e-02 | 8.204630e-03 | 6.060292e-02 | 4.076148e-03 |
| $Y^0_3$ | $5 \cos^3(\theta) - 3 \cos(\theta)$ | -6.807962e-02 | -4.696779e-01 | -4.771903e-02 | -1.344509e-01 |
| $Y^1_3$ | $\cos(\varphi) \sin(\theta)5(\cos^2(\theta) - 1)$ | 1.293904e-01 | -1.282020e-01 | -2.840572e-02 | 1.707097e-02 |
| $Y^2_3$ | $\cos(2\varphi) \sin^2(\theta) \cos(\theta)$ | -4.824502e-02 | 1.196393e+00 | 4.563505e-01 | -4.698604e-02 |
| $Y^3_3$ | $\cos(3\varphi) \sin^3(\theta)$ | -1.677468e-02 | -1.740398e-01 | -7.938387e-02 | -9.831394e-03 |
| $Y^{-4}_4$ | $\sin(4\varphi) \sin^4(\theta)$ | 4.762068e-02 | -1.434847e-01 | -2.638165e-02 | 6.271004e-03 |
| $Y^{-3}_4$ | $\sin(3\varphi) \sin^3(\theta) \cos(\theta)$ | 7.163975e-02 | 4.958958e-01 | 2.879720e-01 | -1.438314e-02 |
| $Y^{-2}_4$ | $\sin(2\varphi) \sin^2(\theta)(7 \cos^2(\theta) - 1)$ | -1.857199e-02 | 2.723782e-01 | 1.180346e-01 | -2.667406e-02 |
| $Y^{-1}_4$ | $\sin(\varphi) \sin(\theta)(7 \cos^3(\theta) - 3 \cos(\theta))$ | -1.226328e-01 | -2.080173e-01 | -1.285519e-01 | -5.684545e-02 |
| $Y^0_4$ | $35 \cos^4(\theta) - 30 \cos^2(\theta) + 3$ | 9.768123e-02 | 8.176546e-02 | 4.075415e-02 | 6.118638e-02 |
| $Y^1_4$ | $\cos(\varphi) \sin(\theta)(7 \cos^3(\theta) - 3 \cos(\theta))$ | 7.062120e-02 | 3.730570e-02 | 1.036030e-02 | 1.607577e-02 |
| $Y^2_4$ | $\cos(2\varphi) \sin^2(\theta)(7 \cos^2(\theta) - 1)$ | 6.002072e-02 | -3.559440e-01 | -1.368811e-01 | 4.581718e-02 |
| $Y^3_4$ | $\cos(3\varphi) \sin^3(\theta) \cos(\theta)$ | 5.921993e-04 | 6.116564e-01 | 1.687368e-01 | -3.823632e-03 |
| $Y^4_4$ | $\cos(4\varphi) \sin^4(\theta)$ | 1.702889e-02 | -1.207067e-02 | 1.452672e-02 | 2.428081e-02 |

**Table 4.** Real spherical harmonics and the normalized coefficients for time period Phase 3.

| Term | Expression | $H_2O$ coefficient | $CO_2$ coefficient | CO coefficient | $O_2$ coefficient |
|---|---|---|---|---|---|
| $Y^0_0$ | 1 | 6.457323e-01 | 3.002227e+00 | 8.823522e-01 | 3.938829e-01 |
| $Y^{-1}_1$ | $\sin(\varphi)\sin(\theta)$ | -7.648887e-01 | 2.001145e-01 | 2.379021e-01 | -1.097899e+00 |
| $Y^0_1$ | $\cos(\theta)$ | -2.570970e-02 | -6.338377e+00 | -1.403207e+00 | 6.715108e-03 |
| $Y^1_1$ | $\cos(\varphi)\sin(\theta)$ | 9.965626e-01 | 2.506014e+00 | 9.738838e-01 | -1.383084e-02 |
| $Y^{-2}_2$ | $\sin(2\varphi)\sin2(\theta)$ | -1.193436e-01 | -5.466062e-02 | -5.966775e-02 | -1.717304e-02 |
| $Y^{-1}_2$ | $\sin(\varphi)\sin(\theta)\cos(\theta)$ | 1.233427e-01 | -4.396767e-02 | -1.551214e-01 | 2.092213e-01 |
| $Y^0_2$ | $3\cos^2(\theta) - 1$ | -1.605573e-02 | 3.426135e+00 | 6.769170e-01 | 1.822769e-01 |
| $Y^1_2$ | $\cos(\varphi)\sin(\theta)\cos(\theta)$ | -2.665689e-01 | -1.688630e+00 | -4.867386e-01 | 4.063866e-02 |
| $Y^2_2$ | $\cos(2\varphi)\sin^2(\theta)$ | -3.243013e-01 | -1.322311e-01 | 4.061452e-02 | -1.496820e-01 |
| $Y^{-3}_3$ | $\sin(3\varphi)\sin^3(\theta)$ | 1.734523e-01 | -2.713263e-01 | -1.446861e-01 | 1.078008e-04 |
| $Y^{-2}_3$ | $\sin(2\varphi)\sin^2(\theta)\cos(\theta)$ | 3.966970e-01 | 2.877859e-01 | -1.811304e-01 | 9.942423e-02 |
| $Y^{-1}_3$ | $\sin(\varphi)\sin(\theta)5(\cos2(\theta) - 1)$ | -1.705277e-01 | 4.682026e-02 | 5.778836e-02 | -2.493769e-01 |
| $Y^0_3$ | $5\cos^3(\theta) - 3\cos(\theta)$ | 1.261865e-01 | -2.076162e+00 | -4.891634e-01 | 1.999027e-01 |
| $Y^1_3$ | $\cos(\varphi)\sin(\theta)5(\cos^2(\theta) - 1)$ | 2.330877e-01 | 4.908625e-01 | 2.267818e-01 | -1.099441e-02 |
| $Y^2_3$ | $\cos(2\varphi)\sin^2(\theta)\cos(\theta)$ | -5.384119e-02 | 6.938674e-01 | 2.201555e-01 | 1.672629e-01 |
| $Y^3_3$ | $\cos(3\varphi)\sin^3(\theta)$ | 1.190484e-01 | -2.192789e-01 | -1.541520e-01 | -2.698796e-02 |
| $Y^{-4}_4$ | $\sin(4\varphi)\sin^4(\theta)$ | 1.453714e-01 | -7.812493e-02 | 7.363468e-02 | 2.176997e-02 |
| $Y^{-3}_4$ | $\sin(3\varphi)\sin^3(\theta)\cos(\theta)$ | 6.825370e-02 | 5.849854e-01 | 5.239300e-01 | -1.098071e-01 |
| $Y^{-2}_4$ | $\sin(2\varphi)\sin^2(\theta)(7\cos^2(\theta) - 1)$ | 3.076581e-03 | -4.271857e-02 | 7.328999e-02 | 8.599415e-03 |
| $Y^{-1}_4$ | $\sin(\varphi)\sin(\theta)(7\cos^3(\theta) - 3\cos(\theta))$ | -3.308553e-02 | -8.651249e-02 | -2.610426e-01 | 5.112065e-02 |
| $Y^0_4$ | $35\cos^4(\theta) - 30\cos^2(\theta) + 3$ | 7.553597e-02 | 2.726439e-01 | 9.291173e-02 | 2.212457e-02 |
| $Y^1_4$ | $\cos(\varphi)\sin(\theta)(7\cos^3(\theta) - 3\cos(\theta))$ | -1.062888e-01 | -2.617530e-01 | -1.241888e-01 | 1.643481e-01 |
| $Y^2_4$ | $\cos(2\varphi)\sin^2(\theta)(7\cos^2(\theta) - 1)$ | 1.755188e-01 | -1.708951e-01 | -1.398572e-01 | -1.762693e-02 |
| $Y^3_4$ | $\cos(3\varphi)\sin^3(\theta)\cos(\theta)$ | -2.466417e-02 | 3.037333e-01 | 2.470421e-01 | 2.570353e-01 |
| $Y^4_4$ | $\cos(4\varphi)\sin^4(\theta)$ | -1.072578e-01 | -9.622150e-02 | 5.915493e-02 | -1.031560e-02 |

There are likely changes in the near-surface composition throughout the mission that are responsible for the variations of the potential surface activity maps during the three mission phases. Such changes were seen in the color of the surface, which became more blue during the middle phase including perihelion (Fornasier et al. 2016) indicating the presence of more water ice near the surface caused by the removal of surface dust in combination with deposition of water ice through internal recondensation (DeSanctis et al. 2015; Ciarniello et al. 2016).

Given the new surface distributions we have run new sets of DSMC models now spanning the entire Rosetta mission. Table 5 gives the times, heliocentric distances and solar latitudes for the 15 dates for which a set of 12 DSMC models were run, each set spanning one nucleus rotation. Figure 4 shows a comparison of the DFMS densities and those determined from the surface distributions and the DSMC model that follow power-law slopes discussed above.

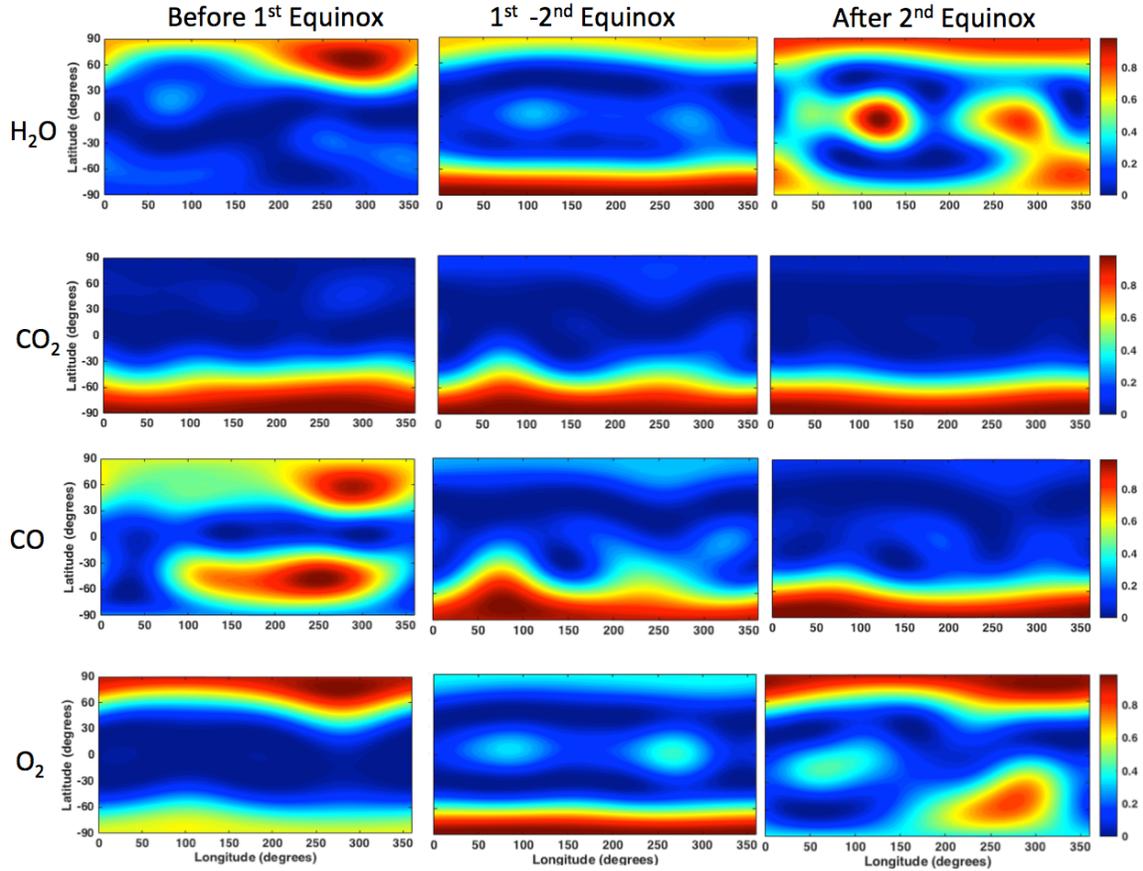

Figure 3. Potential surface activity distribution for the species $H_2O$, $CO_2$, CO, and $O_2$ for the three time periods: (1) before the first equinox, (2) from the first to the second equinox, and (3) after the second equinox. The color scale gives the relative distribution of potential surface activity in each plot on a scale from 0 to 1.

The correlation coefficients for DSMC model results to the whole data set are 0.8407, 0.7814, 0.7793, and 0.7194 for $H_2O$, $CO_2$, CO and $O_2$, respectively. It is noteworthy that the full DSMC model calculations include variations of initial temperature from the surface from the Davidsson and Gutierrez (2004, 2005, 2006) model calculations and the resulting outflow speeds and their variation with heliocentric distance are in good agreement with those found in the recent work on MIRO observations by Biver et al. (2019). The model of Davidsson and Gutierrez, first introduced in the DSMC models of Tenishev, Combi and Davidsson (2008), is a thermophysical model of the nucleus solving locally for solar insolating heating input, local temperature, and sublimation rate, where we have adopted a fast rotator for the night side average values and stationary sphere for the dayside. It accounts for sublimation from a water-dominated nucleus at a range of heliocentric distances and solves for the temperature and relative gas flux from the sublimating layer near the surface. These improve on both the initial velocity distributions and fluxes as a function of solar illumination angle and heliocentric distance (i.e., changing with rotation and heliocentric distance) compared with the simple analytical model of Bieler et al. (2015a).

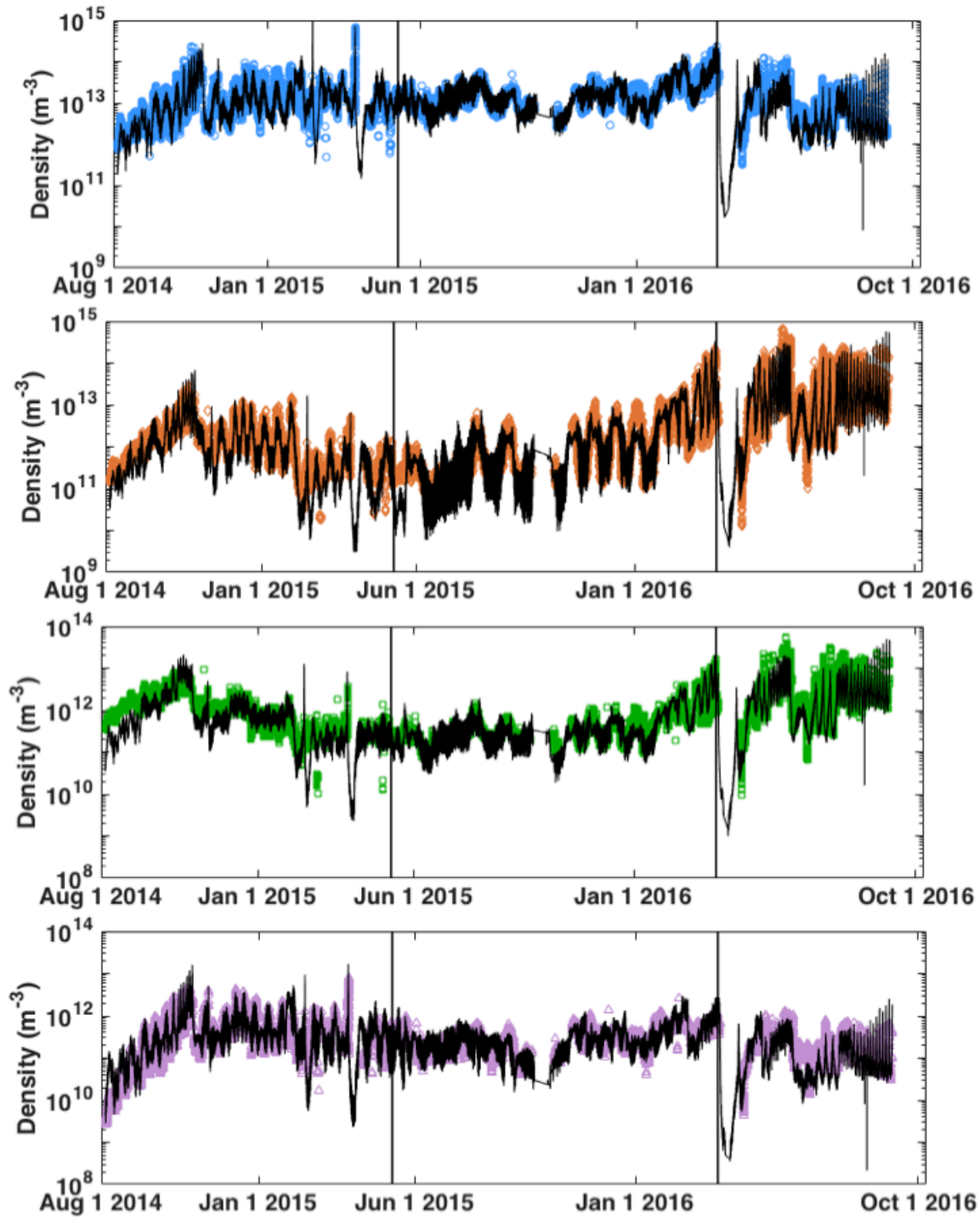

Figure 4. Densities calculated by DSMC models using the best-fit potential surface activity distribution (black) compared with the ROSINA DFMS measurements of $H_2O$ (blue), $CO_2$ (orange), CO (green), and $O_2$ (purple). The vertical lines to the left and right show the times of the two equinoxes.

**Table 5.** Geometries for the DSMC performed simulations. For each of the 15 dates in the table 12 hourly simulations were run covering one nucleus rotation each.

| Date | $R_{au}$(au) | Solar Latitude (°) |
|---|---|---|
| 2014-08-23 | 3.50 | 43.6 |
| 2014-12-23 | 2.71 | 32.6 |
| 2015-03-04 | 2.18 | 20.3 |
| 2015-05-06 | 1.70 | 1.7 |
| 2015-06-08 | 1.48 | -12.9 |
| 2015-07-01 | 1.35 | -25.3 |
| 2015-07-22 | 1.27 | -37.3 |
| 2015-08-16 | 1.24 | -49.0 |
| 2015-09-25 | 1.35 | -49.3 |
| 2015-10-28 | 1.54 | -37.8 |
| 2015-12-01 | 1.78 | -25.8 |
| 2016-01-14 | 2.12 | -13.4 |
| 2016-03-30 | 2.69 | 1.30 |
| 2016-06-06 | 3.15 | 10.3 |
| 2016-09-25 | 3.80 | 20.0 |

**Gas production rates**

Figure 5 shows a plot of the four gas production rates throughout the mission plotted as functions of heliocentric distance and time. For the first two studied periods the variations of productions rates of the four species are similar to the previous results (Fougere et al. 2016b). The most remarkable behavior of the post-second-equinox period (p3) is the leveling of the $CO_2$ and CO production rates and the precipitous drop of the $H_2O$ and $O_2$ production rates.

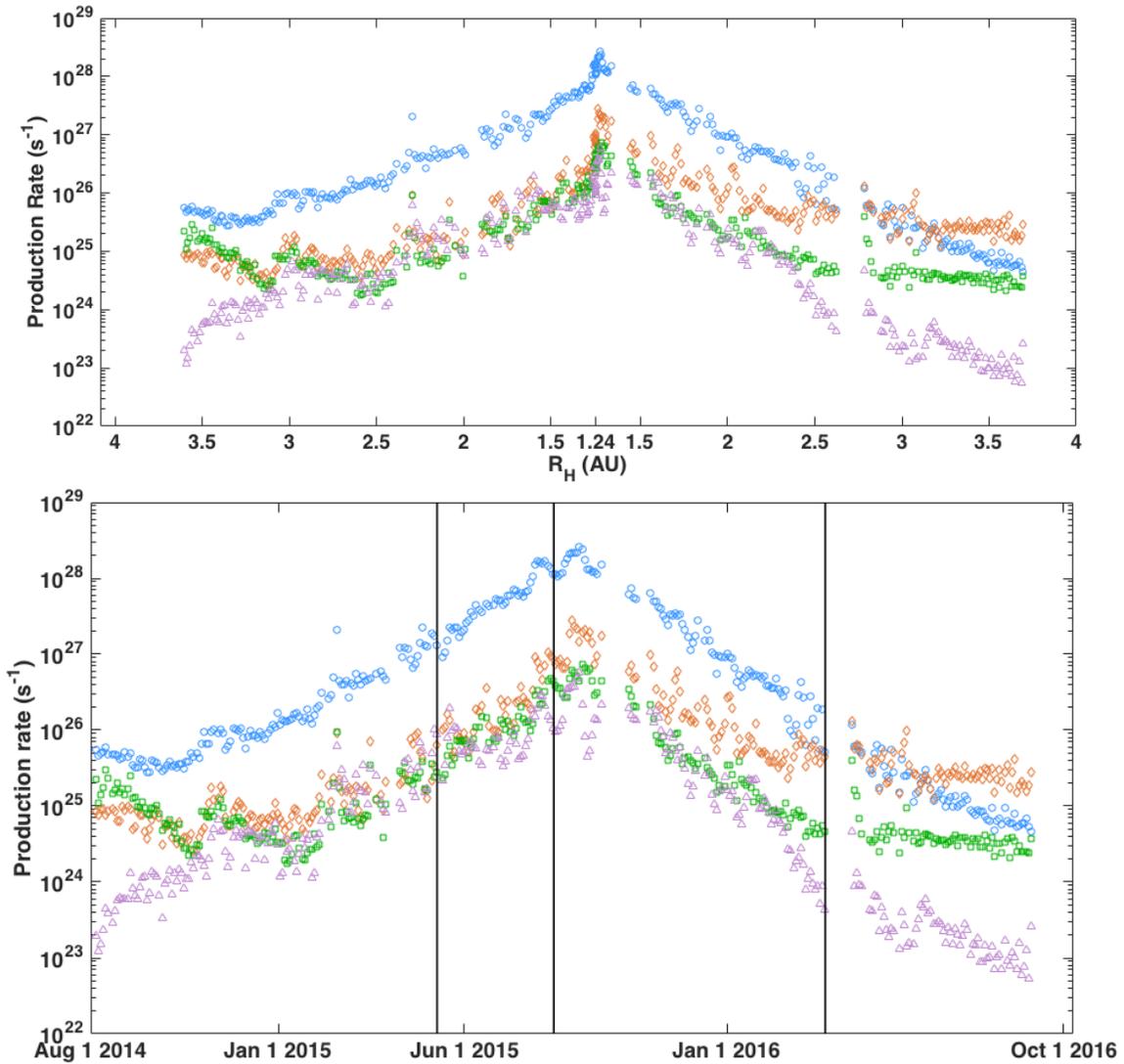

Figure 5. Production rates of $H_2O$ (blue), $CO_2$ (orange), CO (green), and $O_2$ (purple) throughout the mission as a function of heliocentric distance, $R_H$ in AU (top), and as a function of time (bottom). The vertical lines represent the first equinox, perihelion, and second equinox, respectively.

Surprisingly, although the northern hemisphere of the comet is again primarily illuminated after the second equinox as it was early in the mission before the first equinox, the water production rate drops continuously to well below the $CO_2$ production rate and nearly equaling that of CO. Throughout the early period before the first equinox when the northern hemisphere was primarily illuminated, the water production rate was factors of 2 to 3 larger than those of $CO_2$ and CO at the same large heliocentric distances (~3.5 AU) and with the subsolar latitude still in being the north but being +20° instead of +45°.

The result of calculating potential surface activity maps for three time periods compared with the previous results (Fougere et al. 2016b) which used a single set of maps for each species for all times up to approximately the second equinox is overall a lowering of the production rates by about 30% with the peak values shortly after perihelion coming down by over 35%. Fougere et al. (2016b) and Hansen et al. (2016) had a peak water production rate of 4.0 x $10^{28}$ $s^{-1}$ whereas the new map has a peak value of 2.8 x $10^{28}$ $s^{-1}$ and with values within a month or so of perihelion down to about 1 x $10^{28}$ $s^{-1}$ which is closer to previous determination of maximum water production rates in in 67P based on ground-based and space-based measurements (Bertaux et al. 2014; Schleicher 2006, Hansen et al. 2016). This narrows the gap in the previously reported discrepancy between the ROSINA DFMS water production rates (Fougere et al. 2016b; Kramer et al. 2017) and some remote sensing values, but does not completely eliminate it. We return to this comparison in the following section.

Kramer et al. (2017) performed a similar surface activity map exercise on ROSINA COPS data. Their peak value was 4.0 x $10^{28}$ $s^{-1}$ but only when using an initial temperature that resulted in an outflow velocity of 1.5 km $s^{-1}$, which is much larger than our maximum value of only 0.8 km $s^{-1}$ that is more typical of previous measurements of outflow velocities in comets (Tseng et al. 2007; Shou et al. 2017) and in reasonable agreement with those determined directly from observations of MIRO (Lee et al. 2015; Biver et al. 2019). For the Kramer et al. (2017) case of 0.8 km $s^{-1}$ they find a peak production rate of about 2 x $10^{28}$ $s^{-1}$, which is in better agreement with our new value.

Läuter et al. (2019) also used an inversion model like that of Kramer et al. (2017) in order to calculate the time-averaged surface production rates of the same major volatile species from the DFMS/COPS data set. There are several important differences between our method and theirs. They calculate essentially a rotation-averaged production fluxes from the surface, whereas we separate the effect of solar illumination explicitly and calculate the potential surface production distribution. This allows us to reproduce the coma at any given time for direct comparison with any given remote sensing line of sight and solar illumination direction. This is of critical importance for the kind of multi-instrument comparison we show in the following section of this paper. Läuter et al. (2019) calculate the surface distribution using many surface triangles providing a potentially more detailed distribution of sources on the surface whereas, as discussed earlier, we describe the surface distributions with spherical harmonics. Our final results are achieved by using our inverted potential surface distribution as the input conditions for fully-kinetic DSMC models that account explicitly for the solar illumination variation of the rotation-induced activity changes as well as the resulting time-dependent surface temperature distribution given by the thermophysical nucleus model of Davidsson and Gutierrez (2004, 2005, 2006) which yields outflow speeds that vary with local solar illumination, rotation and heliocentric distance. Läuter et al. (2019) calculated production rates over 11 day intervals whereas we can determine production rates from individual measurements following the rotationally dependent solar illumination induced effects on surface heating and then can average the results over any desired time intervals when appropriate. Between the two approached there are potentially small differences in production rates owing to the different ways that outflow speeds are determined, but these differences are rather small. In any case there is reasonable agreement with the overall variation of production rates of all four major species between the two approaches.

**Comparison of H$_2$O production rates from DFMS, VIRTIS-H, MIRO, and PROCYON/LAICA and CO$_2$ production rates from DFMS and VIRTIS-H**

In previous work (Fougere et al. 2016a & 2016b) we compared H$_2$O and CO$_2$ column densities measured by the VIRTIS-H high-resolution infrared mass spectrometer (Bockelée-Morvan et al. 2015 & 2016) with those from the model calculations resulting from the ROSINA DFMS data inversion. We also calculated production rates resulting from both sets of data using the spatial distribution predicted by the model inversion. As discussed already, because the VIRTIS remote sensing measurements were made through columns that sampled the coma much closer to the nucleus, those data can give a more detailed picture of the small scale activity distribution from the surface than do the ROSINA measurements at the spacecraft because the variable density structures seen by VIRTIS tend to smear out at larger distances from the nucleus (Combi et al. 2012) where ROSINA sampled the coma. In addition, the effective FoV of ROSINA is much larger than the apparent size of the nucleus. Uncertainties for the DFMS model-determined production rates can be estimated from the uncertainties in the densities and the correlation coefficients of the model inversions. The uncertainties in the model inversion are uncorrelated with those of the densities, so the final uncertainties in the production rates of H$_2$O, CO$_2$, CO and O$_2$ are 30%, 33%, 33% and 37%, respectively.

The VIRTIS-H H$_2$O column densities in Bockelée-Morvan et al. (2016) were derived from the analysis of faint, optically thin, hot-band fluorescence emissions falling in the 4.45--5.0 µm range (order 0 of the spectrograph) where the rotational temperature was assumed. For this paper, new determinations were performed this time using more intense hot-band transitions detected between 2.775 and 2.910 μm in order 4 (see Fig. 1 of Bockelée-Morvan et al. 2016). Synthetic fluorescence spectra were computed using the model of Crovisier (2009), assuming a Boltzmann distribution for the population of the rotational levels of the ground vibrational state, and an ortho-para ratio equal to 3. By spectral fitting, this time we derived the rotational temperature ($T_{rot}$), which best fits the relative intensities of the 2.775--2.910 μm hot-band rovibrational lines. The column density was then evaluated using the fluorescence efficiency (g-factor) in this wavelength range for a temperature equal to $T_{rot}$ (Villanueva, personal communication, and Villanueva et al. 2012). Column densities derived from the 2.9-μm hot-bands are about 35% higher than those published in Bockelée-Morvan et al. (2016). The origin of this difference is still under study. Table A1 gives the results of the column densities from the 2.775--2.910 μm hot-band, those from the previous results from the 4.45--5.0 μm hot bands and the ratios between the two.

We have calculated the average ratios of the column densities from the new 2.8 μm hot bands to the column densities derived from our surface inversion model determined from the ROSINA DFMS data. These are summarized in Table 6. As in the previous studies, the abundances determined from the model fit to the ROSINA DFMS data are generally larger than those from VIRTIS-H, but the differences vary between the major species, $H_2O$ and $CO_2$, and with the time during the mission. Although we now have the potential surface inversions performed for three separate time intervals during the mission and new VIRTIS-H column densities, the large difference for $H_2O$ between the two instruments during the perihelion period has been reduced from a factor of 4 (Fougere et al. 2016b) to a factor of ~3, so the discrepancy is smaller but remains. For $H_2O$ during the early part of the mission the average ratio is 0.69, but there is a fair amount of scatter in the individual results. For $CO_2$ in the perihelion period the results are rather consistent with a ratio of 0. 90, but for the early mission the ratio for $CO_2$ is 0.52.

**Table 6.** Ratios of $H_2O$ and $CO_2$ column densities from VIRTIS-H spectra and the DFMS fitted model

| Species and Time Period | VIRTIS-H/DFMS Model Ratio |
|---|---|
| $H_2O$ Early Mission | 0.69 |
| $CO_2$ Early Mission | 0.52 |
| $H_2O$ Perihelion Period | 0.34 |
| $CO_2$ Perihelion Period | 0.90 |

Figure 6 shows the production rates of $CO_2$ over the whole missing calculated using the potential surface activity results scaled to the DFMS data in 2-day averages (or 4 rotations of the nucleus) compared with production rates derived from the VIRTIS-H column densities of $CO_2$ using the relative spatial coma distribution taken from our new DFMS determined model integrated along all the VIRTIS-H lines of sight. As expected the differences in production rates are consistent with the ratios of column density given in Table 6.

Our peak value of 2.8 x $10^{28}$ $s^{-1}$ derived from the ROSINA DFMS data for water production is 50% lower than that found in our previous work (Fougere et al. 2016b) by using a potential surface activity distribution determined from ROSINA DFMS measurements from the perihelion phase of the comet between the two equinoxes rather than one for the whole apparition. The comparison of production rates between DFMS and VIRTIS-H in the pre-perihelion period around January 2015 shows that while the scatter overlaps, the DFMS-determined values are on the average larger than the VIRTIS-H values, as before. Similarly around perihelion the difference between DFMS and VIRTIS-H remains, but is somewhat smaller than found by Fougere et al. (2016b) as well as by Hansen et al. (2016) who used the coma distributions from Fougere et al. (2016b).

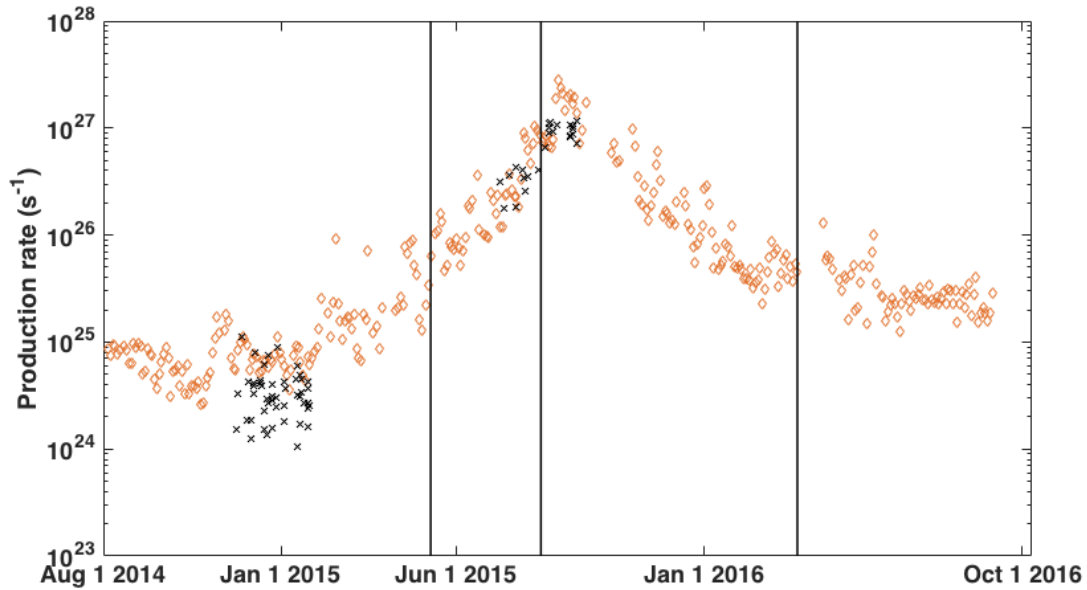

Figure 6. Comparison of production rates of $CO_2$ determined from the ROSINA DFMS data (orange circles) with those from VIRTIS-H (black x's) from the observations of Bockelée-Morvan et al. (2015 and 2016) using the spatial distributions from the DSMC model. The vertical lines left to right represent the first equinox, perihelion, and second equinox, respectively.

Marshall et al. (2017) presented the results of generally nucleus or nadir-pointed observations of the water emission in the mm-wave spectrum using the Microwave Instrument for the Rosetta Orbiter (MIRO). Here we used the water column densities obtained from the over 4000 individual observations, concentrated during the period between the two equinoxes to calculate water production rates using the coma distributions obtained from our model that was determined from the spherical harmonics inversion of the DFMS measurements. These data appear in a supplementary file associated with this paper (MIRO_Columns.txt). The MIRO column densities were derived from lookup tables, where measured MIRO spectral line area ratios for $H_2^{18}O$ and $H_2^{16}O$ were compared to simulated line area ratios. To get the simulated line area ratios, line-of-sight profiles for number density, velocity and kinetic temperature were made, assuming a Haser (1957) model comet with an expanding coma and decreasing temperature.

The profiles used in Marshall et al. (2017) to describe the density, velocity and temperature are taken from parameterizations made by Lee et al. (2015), which were inspired by DSMC simulations from Lee et al. (2011) and Davidsson et al. (2010). By varying the parameters, simulated curves for the change in the line area ratio of $H_2^{16}O$ and $H_2^{18}O$ with column density were made and thus the measured line area ratios could be used to derive an estimate for the column density. The sensitivity of the MIRO column density extends mostly from 1-50 km from the surface. There is little sensitivity at the very surface and for long columns near perihelion (~400 km nucleus/spacecraft distance) there is not much contribution close to the spacecraft. The method used here for determining production rates from the MIRO nadir-pointing column densities is the same as we have done for VIRTIS observations above and in the earlier papers (Fougere et al. 2016a and 2016b). Just as shown by the original Haser model determinations by Marshall et al. (2017), there is considerable scatter among the results for all 4000 values. This is similar to what we find if use all 25,000 individual DFMS measurements. Here we show water production rates on 2-day averages over the whole apparition for both DFMS and MIRO measurements. This corresponds to about 4 rotations of the nucleus and thus yields rotation-averaged values. For the MIRO data we take 2-day averages, which include intervals that are determined by no less than 8 points each.

Figure 7 shows a comparison of all DFMS, MIRO and VIRTIS-H water production rates over the entire mission. The agreement between the DFMS and MIRO measurements is quite good and shows a difference between both of these and VIRTIS-H, especially in the perihelion period. One caveat to the calculation of the water production rates using our coma model and the MIRO data is that the MIRO observations were made with 30-minute integrations and our relative coma model column density was extracted for the mid-point time of the integration.

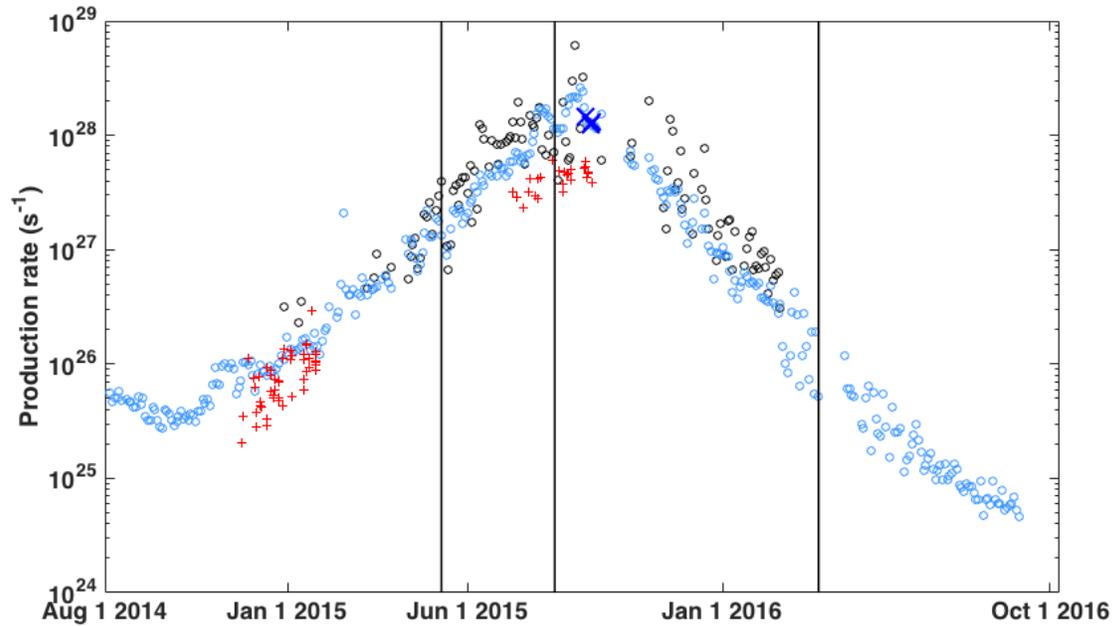

Figure 7. Comparison of Rosetta water production rates in comet 67P/Churyumov-Gerasimenko. The red +'s are the same water production rates in Figure 6 obtained by interpreting the VIRTIS-H measurements of Bockelée-Morvan et al. (2015 and 2016) with the ROSINA DFMS determined water potential surface distribution model. The blue circles are also the same two-day average values from the ROSINA DFMS measurements. The black circles are two-day averages determined from the column densities from the MIRO observations of Marshall et al. (2017) also using the coma distributions from our ROSINA DFMS determined water surface distribution. The blue X's are determined from the H Lyα observations by Shinnaka et al. (2017), the sizes of which are indicative of the given uncertainties. The vertical lines left to right represent the first equinox, perihelion, and second equinox, respectively.

Therefore, there is some comet-spacecraft distance-dependent pointing smear and associated uncertainty with the extracted value. During the time around the peak value found from the DFMS measurements ($2.8 \times 10^{28}$ s$^{-1}$) there are MIRO values both larger and smaller than the DFMS peak. Otherwise there is apparent scatter of values from both data sets that could be due to actual comet production rate variations, specific timing sampling of the 2-day rotation averaging, or real more sporadic variations of actual comet activity. There is a period around June 1, 2015 when the MIRO values are up to a factor of 2 larger than the DFMS values and even larger than the average of the scatter of the VIRTIS-H values. Keep in mind that the

VIRTIS-H values represent individual observations whereas MIRO and DFMS values are all 2-day averages.

Regarding the comparison of the MIRO and VIRTIS-H values with ROSINA DFMS near perihelion, the few lowest MIRO values are just above all the VIRTIS-H values while the average of the scattered values are generally consistent with the DFMS values. The MIRO values in the November 2014-January 2015 period are generally larger than both the DFMS and VIRTIS-H values, but the scatterings in that time period of all data sets overlap. The only non-Rosetta determinations of water production rates in 67P during the 2014-2016 apparition were derived from the atomic hydrogen Lyman-$\alpha$ observations by the PROCYON/LAICA satellite (Shinnaka et al. 2017). They found water production rates of $(1.46\pm0.47) \times 10^{28}$ s$^{-1}$, $(1.24\pm0.40) \times 10^{28}$ s$^{-1}$ and $(1.30\pm0.42) \times 10^{28}$ s$^{-1}$ on 7.40, 12.37 and 13.17 September 2015, respectively. These values are in reasonable agreement with the contemporaneous measurements from ROSINA DFMS and MIRO but larger than those from VIRTIS-H.

Marshall et al. (2017) and Shinnaka et al. (2017) show comparisons of various water production rates including those determined from ground-based and near Earth satellite observations from previous apparitions going back to 1982. These represent a wide range of types of observations including H Ly$\alpha$ and OH in the radio and UV. They cover a considerable range in values but generally spanning the range of the values shown here from the 2014-2016 apparition.

**Unusual composition variation from 1 August - 31 October 2014**

In the first 90 days of phase 1 while the water production rate drops somewhat, the CO$_2$ and CO production rates drop precipitously, and the O$_2$ production rate increases greatly. We have generated potential surface activity distributions for just the first 90 days to see if there was a change that could, in combination with solar illumination, be responsible for the unusual variations in production rates. Figure 8 shows a comparison of the potential surface activity distributions of all four species in period 1 (before the first equinox) with those for just the first 90 days. First, the water source distribution is not much different in the first 90 days compared to that determined for the whole of phase 1. The $O_2$ distribution for the first 90 days looks similar to that of water, but during the whole of phase 1 potential activity increases in the south. The distributions for $CO_2$ and CO do change.

During the first 90 days, CO has much more of its activity in the north than to the south compared with the north/south activity throughout phase 1 where the activity is more favored in the south. So there is more potential activity during the first 90 days when illumination heavily favors the north, but shifts to more to the south throughout the rest of phase 1. So it is likely there is some activation of the south polar region by the increasing solar exposure as the comet approached the first equinox. For $CO_2$ potential activity remains concentrated to the south, though the longitude distribution is more spread out in the remainder of phase 1 than in the first 90 days.

**Mass and surface loss of comet 67P throughout the apparition**

We have taken the volatile mass loss rate distribution over the surface of comet 67P using the potential surface activity and estimated the surface erosion loss as a function of location on the surface over the entire two-year mission. The results of the individual and total volatile mass loss rates at the end of the three studied time phases are given in Table 7. Not surprisingly nearly all of the volatile mass loss occurs in period 2 between the first and second equinoxes. We find a water mass loss of (4.9±1.5) x $10^9$ kg and a total mass loss for the major volatiles of (6.1±1.9) x $10^9$ kg. Consistent with our new lower production rates, this is below the total mass loss calculated by Hansen et al. (2016) using ROSINA COPS data and the previous coma models of Fougere et al. (2016b).

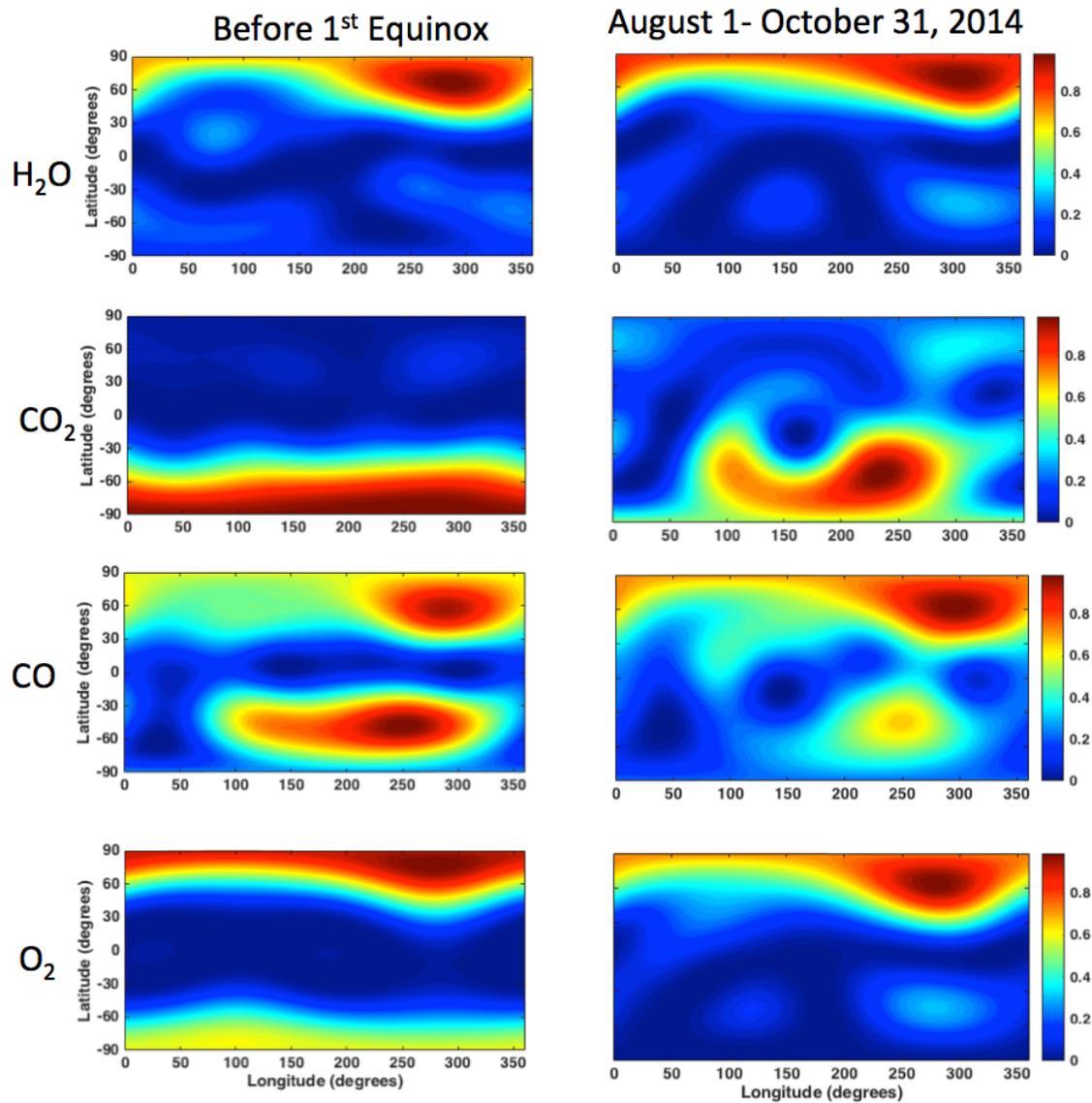

Figure 8. Comparison of potential surface activity distributions during the first 90 days with period 1, August 2014 - May 2015. The color scale gives the relative distribution in each plot on a scale from 0 to 1 of potential surface activity.

**Table 7.** Mass loss in kg from gas sublimation of $H_2O$, $CO_2$, CO and $O_2$.

| Species | Before 1st Equinox | Before 2nd Equinox | End of Mission |
|---------|--------------------|--------------------|----------------|
| $H_2O$  | $2.2 \times 10^8$  | $4.8 \times 10^9$  | $4.9 \times 10^9$ |
| $CO_2$  | $2.3 \times 10^7$  | $8.6 \times 10^8$  | $8.8 \times 10^8$ |
| CO      | $1.2 \times 10^7$  | $2.0 \times 10^8$  | $2.1 \times 10^8$ |
| $O_2$   | $1.2 \times 10^7$  | $1.3 \times 10^8$  | $1.3 \times 10^8$ |
| Total   | $2.7 \times 10^8$  | $6.0 \times 10^9$  | $6.1 \times 10^9$ |

Previously Hansen et al. (2016) found a water mass loss of $6.4 \times 10^9$ kg using ROSINA COPS data and the earlier coma model of Fougere et al. (2016b). Owing to our new modeling of the coma in three separate time intervals we find a total water mass loss of 31% lower than Hansen et al. (2016). From their Haser model analysis of the nadir pointing MIRO observations, Marshall et al. (2017) found a total water mass loss between the two equinoxes of $(2.4\pm1.1) \times 10^9$ kg. However, using a 1D spherical Haser model on many nadir-pointing columns that came from much less active regions of the nucleus would likely underestimate the production rates and the resulting total water mass loss. Using a different MIRO data set that consisted of off-limb observations and a different modeling procedure that accounted for an aspherical coma, Biver et al. (2019) nonetheless found a similar total water mass loss of $(2.4\pm0.1) \times 10^9$ kg. Our value of $(4.9\pm1.5) \times 10^9$ kg is a factor of 2 larger but is consistent with the value of $(4.8\pm1.5) \times 10^9$ kg found by Läuter et al. (2019) from the same ROSINA DFMS data set but using a different model analysis procedure.

The total mass loss of the comet, of course, includes both the loss of the nonvolatile refractory component, or dust, as well as the volatiles. Both the dust production rate from the surface and the total net dust loss from the comet remain somewhat of a controversial topic (Fulle et al. 2016a, 2016b, 2019; Levasseur-Regourd et al. 2018). While several Rosetta instruments actually detected various sizes and numbers of dust particles and remote-sensing measurements of dust in coma were made by OSIRIS, VIRTIS and MIRO, there still remain some unknown parameters and incomplete sampling of dust particle sizes and masses. There is also a suggestion regarding gas activity lifting of either icy (Keller et al. 2015) or refractory (Fulle et al., 2019) particles, or both, that then fall back to the surface contributing to the surface mantle that may decrease or limit future activity (Keller et al. 2017), with most of the fall back likely transferring mass from the south to the north and consisting of larger and more granular particles.

For our simple calculation we calculate the total surface excavation depth assuming the total nucleus mass loss of 10.5±3.4 x $10^9$ kg and total nucleus density 533 kg m$^{-3}$ from the Radio Science Investigation (RSI) of Pätzold et al. (2016, 2019). Over the whole 2-year period the volatile mass loss of the major species is (6.1±1.9) x $10^9$ kg. Accounting approximately for the mass of all the minor volatile species taken together (LeRoy et al. 2015; Rubin et al. 2019) would raise the total volatile mass loss to about (6.3±2.0) x $10^9$ kg. Figure 9 shows plots of the surface loss at the end of each of our three phases over the two-year mission, at the first equinox, at the second equinox and at the end of mission, respectively. Maximum erosion depths in the south polar region over the whole apparition of ~4 m were found for a refractory/ice loss ratio of 0.67, which is consistent with the average erosion depths in the south that were predicted by the model calculations of Keller et al. (2015). More than 95% of the mass loss occurs during the time between the first and second equinox.

Using the volatile loss from ROSINA would imply a net refractory mass loss of 4 x $10^9$ kg or a net refractory/ice mass ratio of 1:1.5. If we take the full uncertainty ranges of the nucleus mass loss and our combined model plus ROSINA uncertainties we find a rather wide range for the refractory/ice mass ratio of 0.7 (-0.7 +1.0) or basically from 0 to 1.7. Depth results can be scaled simply enough for other dust/gas (refractory/ice) ratios, though clearly this does not account for variable dust/gas mass ratios over the surface. Biver et al. (2019) has recently made a similar estimate using different assumptions for the total volatile/ice loss from the comet starting with the water ice loss calculated from observations of MIRO over the whole mission. Although their water loss is about half of what we calculate from the ROSINA/DFMS results, their total volatile mass loss rate is about 2/3. Therefore, their refractory/ice mass loss ratio is 1.5:1.

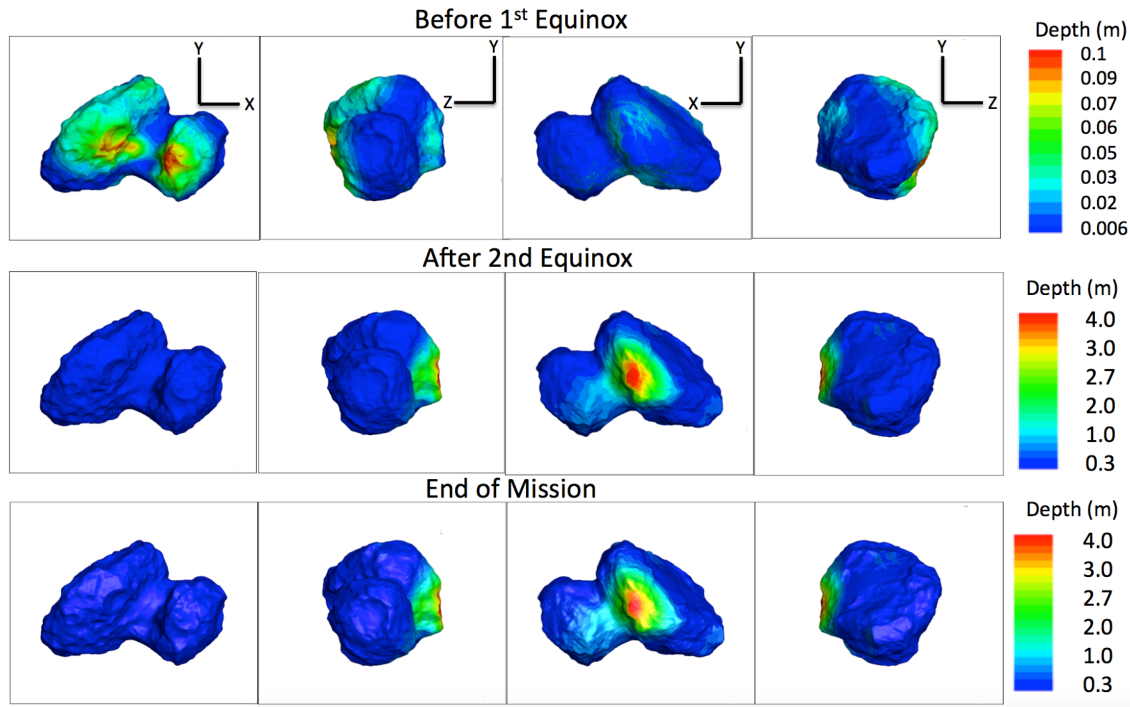

Figure 9. Cumulative Erosion Depths at the end of each of three mission phases: 1st equinox, after 2nd equinox and end of mission. Note that by necessity the color contour look-up table is different for the first phase. A dust/gas (or refractory/ice) loss ratio of 1:1.5 is assumed for a nucleus of density 533 kg m$^{-3}$. Results can be scaled for other ratios.

A similar analysis by Biver et al. (2019) that includes also the nadir-pointing MIRO observations yields total volatile mass loss rate that is about 25% larger. These estimates are in contrast to the estimate by Fulle et al. (2019) starting with the refractory mass loss from various dust measurements who find a refractory/ice mass loss ratio lower limit of 3.

Finally, these can also be compared with the dust-to-ice ratio calculated from the real part of the permittivity determined by the analysis of the propagation of the radio signal from the

Comet Nucleus Sounding Experiment by Radiowave Transmission (CONSERT) instrument (Kofman et al. 2015; Herique et al. 2016). The CONSERT measurements combined with their conclusion that the porosity of the nucleus is in the range of 75-85%, yields a dust-to-ice ratio in the nucleus in the range of 0.4 to 2.6 (Kofman et al. 2015). This range includes the separate estimates from the ROSINA/DFMS and MIRO measurements at the lower end and go nearly to the lowest limit from the dust analysis by Fulle et al. (2019). Herique et al. (2016) later assumed a priori a dust-to-ice ratio in the range of 4±2 from Rotundi et al. (2015) and from that could conclude that the dust-to-ice ratio is larger than 3. Most of the mass as measured in the coma by both mass spectrometry and remote sensing is lost during the period around southern summer, which includes perihelion. It is worth noting that Keller et al. (2017) suggest that the composition of the coma over the southern hemisphere during southern summer is then likely to be much more representative of that of the nucleus itself.

**Summary**

We present a model inversion of ROSINA DFMS density measurements of $H_2O$, $CO_2$, CO and $O_2$, in the Rosetta target comet 67P/Churyumov-Gerasimenko throughout the mission which calculates the potential surface activity of these species in response to solar illumination as the nucleus rotates and moves in its orbit over the 25 months of the near nucleus phase of the mission. The potential surface activity for the four species was calculated during three time intervals of the orbit: (1) from the start of the near nucleus phase, 1 August 2014 to the first equinox; (2) from the first equinox through perihelion to the second equinox; and (3) from the second equinox until August 2016 when the DFMS measurements ended. Using the potential surface activity distributions determined, 15 sets of DSMC models were computed covering the two-year period, with each set consisting of 12 models spanning the ~12-hour rotation period of the nucleus. Interpolation and scaling of these 180 model calculations then provides a realistic description of the full variation of the major volatile species at any time during the mission. The results of the DSMC models were then shown to compare very favorably with the ROSINA DFMS densities, consisting of more than 50,000 individual measurements.

The coma models were used to calculate the production rates of the four species throughout the mission. The coma model distributions were then used to analyze remote sensing measurements of $H_2O$ and $CO_2$ made by the VIRTIS-H infrared spectrometer and of $H_2O$ by the MIRO microwave instrument, and production rates of those species were compared with those determined from the model fit to the ROSINA DFMS densities. The discrepancy between the ROSINA DFMS production rates and those from VIRTIS-H obtained using different hot bands during the month around perihelion was improved somewhat, but remains and is still under study. However, the $H_2O$ production rates determined from nadir-pointing MIRO data and the DFMS-determined potential surface activity model were in much better agreement with those from DFMS throughout the mission, though the scatter in those results was larger.

The MIRO observations used were viewing nadir columns directly to the nucleus often from near terminator locations. ROSINA was sampling the densities from similar spacecraft locations as MIRO. PROCYON was observing the entire large-scale hydrogen coma and should provide a reasonable global estimate of the production rate independent of pointing. VIRTIS was mostly viewing the active sector nearer the sub-solar point. However, ROSINA was sampling more directly the same regions of the nucleus but at other times when the illumination conditions were different. The difference in production rates around perihelion between the ROSINA-determined model, the nadir-pointing MIRO and the PROCYON values determined from H-Ly$\alpha$ on the one hand from the lower values from our analysis using the same model on the VIRTIS columns still needs further investigation.

The variation of production rates was shown to be different during the first 90 days of the mission than during the remainder of the pre-first-equinox time period. The water production rate decreases somewhat during this period but those of $CO_2$ and CO drop precipitously, and the $O_2$ production rate increases greatly. Using the same spherical harmonics inversion technique we looked at the potential surface activities of the four species during just the first 90 days (August 1 to October 31, 2014) and compared them with those throughout phase 1 from the beginning of the mission through the first equinox. The potential surface activity distributions of $O_2$ and water look similar to one another and look similar during the first 90 days as they do throughout phase 1 with only some minor changes in the $O_2$ potential surface activity noted. This is not the case for $CO_2$ and CO. The potential surface activity distribution of CO has much more of its activity originating in the north than to the south during the first 90 days whereas throughout phase 1 the activity shifts to mostly from the south. The potential activity distribution of $CO_2$ remains concentrated to the south throughout the first 90 days and all of phase 1, but is spread out more in longitude in the remainder of phase 1 than during the first 90 days. It is also interesting to note that whereas water is the dominant species throughout the period before the second equinox, even at the beginning of the mission when the comet was at heliocentric distances larger than 3.5 AU, after the second equinox $CO_2$ becomes the dominant species in the coma and the water production rate falls to nearly as small as that of CO. Except for the unusual behavior during the first 90 days $O_2$ generally follows the variation of $H_2O$ even as the water production drops at large heliocentric distances after the second equinox.

The production rates from the ROSINA-determined model in the time interval around perihelion do still seem to vary with spacecraft latitude, although the variations are not nearly as large those seen in the density itself as the sub-spacecraft latitude moved north and south. This is worth future investigation with a more complicated model than the fifth order spherical harmonics expansion used.

Finally, we estimated the total mass loss distribution of the comet at stages throughout the mission. Not surprisingly most of the mass was lost during the few months around perihelion when the production rates were the largest. A total volatile mass loss of $(6.3\pm2.0) \times 10^9$ kg was found. Also, not surprisingly most of the surface loss throughout the apparition occurs in the southern hemisphere because of its large solar exposure during southern summer that occurs just after perihelion. We also used the total mass loss of the nucleus reported by the RSI team together with the total volatiles mass loss that implied a refractory/ice ratio somewhat less than one.

# Appendix

Table A1. Revised VIRTIS-H column densities based on the 2.775-2.910 μm hot bands

| Obs ID | Start Time | $r_h$ | ρ | P.A. | $N_{4 \mu m}$ ($H_2O$) | $N_{2 \mu m}$ ($H_2O$) | T (rot) | ratio |
|---|---|---|---|---|---|---|---|---|
| T1_00395011055 | 2015-07-08T21:11:00.3 | 1.316 | 2.82 | 259 | 9.25 (2.88) | 10.96 (0.18) | 128.7 (2.6) | 1.12 |
| T1_00395322550 | 2015-07-12T11:46:52.6 | 1.302 | 2.70 | 274 | 10.10 (0.77) | 9.80 (0.18) | 125.7 (3.2) | 0.95 |
| T1_00395742154 | 2015-07-17T08:15:55.9 | 1.286 | 3.10 | 272 | 6.33 (0.36) | 7.48 (0.13) | 120.7 (2.4) | 1.14 |
| T1_00396199623 | 2015-07-22T15:23:57.9 | 1.271 | 2.69 | 266 | 10.00 (0.41) | 12.25 (0.20) | 141.8 (2.9) | 1.19 |
| T1_00396220410 | 2015-07-22T21:10:24.7 | 1.270 | 2.98 | 266 | 9.79 (0.28) | 12.38 (0.17) | 126.4 (2.2) | 1.21 |
| T1_00396659230 | 2015-07-27T23:06:40.5 | 1.259 | 4.01 | 247 | 6.48 (0.17) | 6.14 (0.12) | 111.4 (3.1) | 0.92 |
| T1_00396826054 | 2015-07-29T21:20:59.8 | 1.255 | 3.56 | 265 | 6.62 (0.44) | 9.53 (0.16) | 125.9 (2.4) | 1.38 |
| T1_00396842044 | 2015-07-30T01:47:25.8 | 1.255 | 3.59 | 265 | 7.27 (0.48) | 9.82 (0.16) | 110.7 (2.4) | 1.31 |
| T1_00397038715 | 2015-08-01T08:25:16.8 | 1.252 | 3.08 | 270 | 7.10 (0.39) | 10.28 (0.19) | 131.5 (2.7) | 1.41 |
| T1_00397871165 | 2015-08-10T23:32:29.0 | 1.244 | 6.23 | 286 | 4.19 (0.40) | 7.35 (0.12) | 93.8 (3.3) | 1.72 |
| T1_00398347798 | 2015-08-16T11:56:25.1 | 1.244 | 3.58 | 267 | 9.21 (0.63) | 12.70 (0.22) | 121.0 (2.6) | 1.36 |
| T1_00398603680 | 2015-08-19T11:08:06.5 | 1.246 | 2.96 | 263 | 10.20 (0.50) | 12.82 (0.23) | 136.3 (2.6) | 1.25 |
| T1_00398619671 | 2015-08-19T15:34:33.4 | 1.246 | 2.97 | 262 | 11.60 (0.64) | 12.42 (0.18) | 129.3 (2.1) | 1.06 |
| T1_00398640452 | 2015-08-19T21:20:58.5 | 1.246 | 3.47 | 262 | 8.90 (0.61) | 11.27 (0.20) | 138.9 (2.6) | 1.24 |
| T1_00398729980 | 2015-08-20T22:13:06.6 | 1.247 | 4.58 | 258 | 8.32 (0.38) | 11.00 (0.20) | 113.6 (2.6) | 1.31 |
| T1_00398970868 | 2015-08-23T17:07:54.5 | 1.250 | 3.78 | 256 | 8.27 (0.33) | 11.13 (0.20) | 118.7 (2.8) | 1.31 |
| T1_00398986563 | 2015-08-23T21:29:29.5 | 1.250 | 3.82 | 255 | 7.44 (0.36) | 12.55 (0.18) | 119.9 (2.2) | 1.65 |
| T1_00399198695 | 2015-08-26T08:17:52.6 | 1.254 | 4.25 | 264 | 6.87 (0.36) | 9.47 (0.20) | 116.3 (3.1) | 1.32 |
| T1_00399208316 | 2015-08-26T10:58:13.4 | 1.254 | 3.95 | 263 | 9.30 (0.99) | 11.43 (0.17) | 112.4 (2.5) | 1.21 |
| T1_00400195914 | 2015-09-06T21:25:16.9 | 1.280 | 4.92 | 266 | 5.25 (0.49) | 7.03 (0.11) | 111.6 (2.5) | 1.31 |
| T1_00400216374 | 2015-09-07T03:06:16.9 | 1.281 | 4.87 | 268 | 5.64 (0.43) | 7.82 (0.12) | 112.1 (2.6) | 1.34 |
| T1_00400232094 | 2015-09-07T07:28:16.8 | 1.282 | 4.83 | 268 | 3.64 (0.60) | 7.94 (0.12) | 106.6 (2.5) | 2.12 |
| T1_00400252549 | 2015-09-07T13:09:16.0 | 1.282 | 4.83 | 270 | 5.35 (0.35) | 8.42 (0.11) | 107.1 (2.5) | 1.54 |
| T1_00400407713 | 2015-09-09T08:15:16.0 | 1.288 | 3.14 | 263 | 8.22 (0.49) | 12.50 (0.19) | 125.9 (2.3) | 1.46 |
| T1_00400433767 | 2015-09-09T15:29:34.0 | 1.289 | 3.72 | 262 | 7.42 (0.30) | 10.83 (0.16) | 109.8 (2.0) | 1.40 |
| T1_00400470243 | 2015-09-10T01:37:26.0 | 1.290 | 5.78 | 263 | 4.93 (0.18) | 6.99 (0.13) | 89.2 (3.1) | 1.39 |
| T1_00400748481 | 2015-09-13T06:54:44.2 | 1.301 | 3.58 | 265 | 7.13 (0.29) | 10.07 (0.12) | 119.8 (2.1) | 1.35 |
| T1_00400765976 | 2015-09-13T11:46:19.3 | 1.302 | 3.61 | 265 | 10.00 (1.00) | 12.22 (0.20) | 124.5 (2.7) | 1.20 |
| T1_00401717336 | 2015-09-24T11:55:21.0 | 1.347 | 5.02 | 275 | 3.68 (0.19) | 7.12 (0.17) | 98.5 (4.0) | 1.88 |
| T1_00401994267 | 2015-09-27T16:57:54.5 | 1.362 | 6.88 | 282 | 2.41 (0.18) | 3.31 (0.09) | 77.1 (6.9) | 1.34 |

Notes to Table A1
Column #1 Observation ID
Column #2 Start time of observation
Column #3 Heliocentric distance [au]
Column #4 LOS mean distance to the nuclei center [km]
Column #5 mean LOS position angle relative to the Sun (PA(Sun)=27°) [deg]
Column #6 $H_2O$ column density from 4.450-5.030 micron [$10^{20}$ m$^{-2}$]; uncertainty in parentheses.
Column #7 $H_2O$ column density from 2.775-2.910 micron [$10^{20}$ m$^{-2}$]; uncertainty in parentheses.
Column #8 Rotational temperature from $H_2O$ hot band [K]
Column #9 Ratio of column densities from 2.775-2.910 micron hot band to from 4.450-5.030 micron hot band
*** More detail on observing/geometric conditions please read Bockelée-Morvan et al. (2016)

**Acknowledgements**

This work was supported by contracts JPL1266313 and JPL 1266314 from the US Rosetta Project and NASA grants NNX14AG84G from the Planetary Atmospheres Program and 80NSSC18K1280 from the Rosetta Data Analysis Program. Work at UoB was funded by the State of Bern, the Swiss National Science Foundation (200021_165869, 200020_182418), and the European Space Agency PRODEX Program. The authors would like to thank ASI, Italy; CNES, France; DLR, Germany and NASA, USA for supporting this research. VIRTIS was built by a consortium, which includes Italy, France, and Germany, under the scientific responsibility of the Istituto di Astrofisica e Planetologia Spaziali of INAF, Italy, which also guides the scientific operations. The VIRTIS instrument development, led by the prime contractor Leonardo-Finmeccanica (Florence, Italy), has been funded and managed by ASI, with contributions from Observatoire de Meudon financed by CNES, and from DLR. We thank the Rosetta Science Ground Segment and the Rosetta Mission Operations Centre for their support throughout all the phases of the mission. The VIRTIS calibrated data are available through the ESA's Planetary Science Archive Web site (www.rssd.esa.int). We thank the following institutions and agencies for support of this work: Italian Space Agency (ASI, Italy) contract number I/024/12/1, Centre National d'Études Spatiales (CNES, France), DLR (Germany), NASA (USA) Rosetta Program, and Science and Technology Facilities Council (UK). We acknowledge the contributions of the entire Rosetta science and engineering teams, and especially the ROSINA, VIRTIS and MIRO science teams. For the entire Rosetta archive see https://www.cosmos.esa.int/web/psa/rosetta as well as https://pds-smallbodies.astro.umd.edu/data_sb/missions/rosetta/index.shtml.

Supplemental Data File: MIRO_Columns.txt

# Date - Time - Helio dist (AU) - Surface Area - Col den (cm^-2) - Col den error (cm^-2)

| Date | Time | Helio dist (AU) | Surface Area | Col den (cm^-2) | Col den error (cm^-2) |
|---|---|---|---|---|---|
| 2014-12-18 | 0:00:00 | 2.746258 | ['Aten'] | 1.54E+16 | -5.85002E+15 |
| 2014-12-26 | 14:30:00 | 2.68504 | ['Seth'] | 3.07E+15 | -1.10696E+15 |
| 2014-12-26 | 15:00:00 | 2.684891923 | ['Seth'] | 2.78E+15 | -1.04699E+15 |
| 2014-12-26 | 20:00:00 | 2.683404231 | ['Babi'] | 9.31E+15 | -3.22823E+15 |
| 2014-12-27 | 1:00:00 | 2.681912593 | ['Hapi'] | 5.45E+15 | -1.91377E+15 |
| 2014-12-27 | 7:00:00 | 2.680114444 | ['Ash'] | 5.38E+15 | -1.88137E+15 |
| 2014-12-27 | 7:30:00 | 2.679965385 | ['Ash'] | 6.79E+15 | -2.35199E+15 |
| 2014-12-28 | 7:30:00 | 2.672745714 | ['Hatmehit'] | 4.92E+15 | -1.86194E+15 |
| 2014-12-28 | 8:00:00 | 2.672687333 | ['Aten'] | 6.71E+15 | -2.48294E+15 |
| 2014-12-28 | 20:00:00 | 2.669066296 | ['Hapi'] | 1.10E+16 | -3.77962E+15 |
| 2014-12-28 | 20:30:00 | 2.668911481 | ['Hapi'] | 8.60E+15 | -2.9458E+15 |
| 2014-12-28 | 21:00:00 | 2.668763462 | ['Seth'] | 2.19E+15 | -8.11966E+14 |
| 2014-12-28 | 22:30:00 | 2.668316667 | ['Seth'] | 3.33E+15 | -1.20676E+15 |
| 2014-12-29 | 0:00:00 | 2.667871538 | ['Seth'] | 3.13E+15 | -1.10917E+15 |
| 2014-12-29 | 1:30:00 | 2.667414615 | ['Seth'] | 1.04E+16 | -3.58463E+15 |
| 2014-12-30 | 2:00:00 | 2.66003 | ['Seth'] | 4.83E+15 | -1.76663E+15 |
| 2014-12-30 | 14:00:00 | 2.656483704 | ['Maat'] | 5.17E+15 | -1.77344E+15 |
| 2014-12-30 | 20:00:00 | 2.654690741 | ['Babi'] | 4.55E+15 | -1.58679E+15 |
| 2014-12-30 | 20:30:00 | 2.654553182 | ['Hapi'] | 3.40E+15 | -1.22504E+15 |
| 2015-01-07 | 12:00:00 | 2.599221034 | ['Hathor'] | 5.55E+15 | -1.95317E+15 |
| 2015-01-08 | 0:00:00 | 2.595532222 | ['Maat'] | 9.22E+15 | -3.18357E+15 |
| 2015-01-08 | 2:30:00 | 2.594893333 | ['Hatmehit'] | 4.48E+15 | -1.93084E+15 |
| 2015-01-08 | 12:00:00 | 2.591924 | ['Hapi'] | 1.16E+16 | -3.98289E+15 |
| 2015-01-08 | 12:30:00 | 2.591791852 | ['Hapi'] | 1.14E+16 | -3.9025E+15 |
| 2015-01-08 | 18:00:00 | 2.590124231 | ['Serqet'] | 7.43E+15 | -2.55198E+15 |
| 2015-01-10 | 0:00:00 | 2.5810136 | ['Serqet'] | 9.75E+15 | -3.33661E+15 |
| 2015-01-10 | 0:30:00 | 2.580861154 | ['Hapi'] | 1.12E+16 | -3.85778E+15 |
| 2015-01-10 | 1:00:00 | 2.580716 | ['Hapi'] | 5.19E+15 | -1.78851E+15 |
| 2015-01-10 | 5:00:00 | 2.579498077 | ['Ash'] | 3.09E+15 | -1.08941E+15 |
| 2015-01-10 | 5:30:00 | 2.579344231 | ['Ash'] | 4.55E+15 | -1.57671E+15 |
| 2015-01-10 | 6:00:00 | 2.579193333 | ['Seth'] | 8.24E+15 | -2.829E+15 |
| 2015-01-10 | 6:30:00 | 2.57904 | ['Seth'] | 1.30E+16 | -4.43999E+15 |
| 2015-01-10 | 7:00:00 | 2.578880385 | ['Hapi'] | 7.24E+15 | -2.48193E+15 |
| 2015-01-10 | 12:30:00 | 2.577271429 | ['Hapi'] | 1.04E+16 | -3.6384E+15 |
| 2015-01-10 | 17:00:00 | 2.57584 | ['Ash'] | 3.45E+15 | -1.24047E+15 |
| 2015-01-10 | 17:30:00 | 2.5756908 | ['Ash'] | 4.46E+15 | -1.54712E+15 |
| 2015-01-10 | 18:00:00 | 2.575535769 | ['Babi'] | 8.42E+15 | -2.88268E+15 |
| 2015-01-10 | 18:30:00 | 2.575387308 | ['Hapi'] | 1.15E+16 | -3.93363E+15 |
| 2015-01-10 | 19:00:00 | 2.575239231 | ['Hapi'] | 6.66E+15 | -2.28792E+15 |
| 2015-01-10 | 23:30:00 | 2.573867037 | ['Hapi'] | 7.13E+15 | -2.45915E+15 |
| 2015-01-11 | 0:00:00 | 2.573716538 | ['Seth'] | 9.34E+15 | -3.21089E+15 |
| 2015-01-11 | 6:30:00 | 2.571732593 | ['Ash'] | 9.67E+15 | -3.32923E+15 |
| 2015-01-11 | 7:00:00 | 2.571581538 | ['Hapi'] | 6.92E+15 | -2.39648E+15 |

| Date | Time | Value | Name | Col5 | Col6 |
|---|---|---|---|---|---|
| 2015-01-11 | 12:00:00 | 2.57006 | ['Seth'] | 1.11E+16 | -3.81425E+15 |
| 2015-01-11 | 18:00:00 | 2.568233704 | ['Babi'] | 4.20E+15 | -1.47301E+15 |
| 2015-01-11 | 23:00:00 | 2.566707037 | ['Hapi'] | 4.87E+15 | -1.69641E+15 |
| 2015-01-11 | 23:30:00 | 2.566552593 | ['Seth'] | 4.54E+15 | -1.57806E+15 |
| 2015-01-12 | 0:00:00 | 2.566401538 | ['Seth'] | 7.04E+15 | -2.41074E+15 |
| 2015-01-12 | 11:30:00 | 2.562902963 | ['Anuket'] | 4.32E+15 | -1.53724E+15 |
| 2015-01-12 | 12:30:00 | 2.562591071 | ['Maat'] | 4.57E+15 | -1.63624E+15 |
| 2015-01-12 | 18:00:00 | 2.560914074 | ['Ash'] | 4.69E+15 | -1.6436E+15 |
| 2015-01-12 | 19:00:00 | 2.560611852 | ['Hapi'] | 5.12E+15 | -1.76637E+15 |
| 2015-01-13 | 0:00:00 | 2.5590852 | ['Anuket'] | 5.29E+15 | -1.81961E+15 |
| 2015-01-13 | 12:00:00 | 2.555421154 | ['Anuket'] | 4.87E+15 | -1.69345E+15 |
| 2015-01-13 | 12:30:00 | 2.55526963 | ['Hapi'] | 3.93E+15 | -1.45245E+15 |
| 2015-02-09 | 19:30:00 | 2.35207 | ['Maat'] | 1.04E+16 | -3.58419E+15 |
| 2015-02-09 | 20:00:00 | 2.3519488 | ['Maat'] | 6.46E+15 | -2.23819E+15 |
| 2015-02-09 | 20:30:00 | 2.351795926 | ['Maat'] | 3.63E+15 | -1.27119E+15 |
| 2015-02-09 | 21:30:00 | 2.351484074 | ['Seth'] | 6.59E+15 | -2.26372E+15 |
| 2015-02-15 | 20:30:00 | 2.306383429 | ['Khepry'] | 3.65E+15 | -1.41227E+15 |
| 2015-02-17 | 3:00:00 | 2.296755882 | ['Hapi'] | 7.54E+15 | -2.66869E+15 |
| 2015-02-21 | 22:30:00 | 2.260050606 | ['Hapi'] | 9.35E+15 | -3.21529E+15 |
| 2015-02-22 | 3:30:00 | 2.2584615 | ['Hapi'] | 7.48E+15 | -2.6553E+15 |
| 2015-03-01 | 3:00:00 | 2.20508725 | ['Aten'] | 1.90E+16 | -6.49739E+15 |
| 2015-03-01 | 8:30:00 | 2.203317692 | ['Aten'] | 3.07E+16 | -1.05E+16 |
| 2015-03-01 | 17:30:00 | 2.200421053 | ['Aten'] | 2.76E+16 | -9.43187E+15 |
| 2015-03-01 | 20:00:00 | 2.199704375 | ['Maat'] | 2.42E+16 | -8.35908E+15 |
| 2015-03-02 | 20:30:00 | 2.191856471 | ['Hapi'] | 2.01E+16 | -7.03512E+15 |
| 2015-03-04 | 8:00:00 | 2.18053 | ['Imhotep'] | 1.18E+16 | -4.10113E+15 |
| 2015-03-06 | 17:00:00 | 2.162282 | ['Anuket'] | 1.30E+16 | -4.5014E+15 |
| 2015-03-08 | 16:30:00 | 2.146918 | ['Anuket'] | 6.53E+15 | -2.31515E+15 |
| 2015-03-09 | 8:00:00 | 2.142014444 | ['Imhotep'] | 6.11E+15 | -2.12154E+15 |
| 2015-03-09 | 9:00:00 | 2.141698462 | ['Imhotep'] | 3.71E+15 | -1.34E+15 |
| 2015-03-09 | 15:30:00 | 2.139606667 | ['Anuket'] | 1.42E+16 | -4.87006E+15 |
| 2015-03-09 | 16:00:00 | 2.139446538 | ['Hapi'] | 1.23E+16 | -4.24494E+15 |
| 2015-03-09 | 16:30:00 | 2.139289231 | ['Seth'] | 6.39E+15 | -2.23367E+15 |
| 2015-03-09 | 20:00:00 | 2.138164615 | ['Imhotep'] | 8.01E+15 | -2.76334E+15 |
| 2015-03-09 | 20:30:00 | 2.138003846 | ['Imhotep'] | 3.64E+15 | -1.28099E+15 |
| 2015-03-11 | 8:00:00 | 2.126548333 | ['Imhotep'] | 6.06E+15 | -2.21777E+15 |
| 2015-03-11 | 9:30:00 | 2.126189 | ['Imhotep'] | 8.97E+15 | -3.12994E+15 |
| 2015-03-12 | 14:00:00 | 2.117032857 | ['Maat'] | 3.67E+15 | -1.3625E+15 |
| 2015-03-14 | 4:00:00 | 2.104797073 | ['Maat'] | 8.41E+15 | -2.88822E+15 |
| 2015-03-14 | 5:30:00 | 2.104287255 | ['Maat'] | 2.97E+16 | -1.02E+16 |
| 2015-03-14 | 6:00:00 | 2.104129423 | ['Hapi'] | 3.02E+16 | -1.03E+16 |
| 2015-03-14 | 6:30:00 | 2.103972353 | ['Hapi'] | 2.44E+16 | -8.35038E+15 |
| 2015-03-14 | 7:00:00 | 2.1038092 | ['Hapi'] | 1.89E+16 | -6.465E+15 |
| 2015-03-14 | 7:30:00 | 2.10368871 | ['Hapi'] | 1.65E+16 | -5.64059E+15 |

| Date | Time | Value | Name | Col5 | Col6 |
|---|---|---|---|---|---|
| 2015-03-14 | 8:30:00 | 2.10326 | ['Seth'] | 8.97E+15 | -3.10273E+15 |
| 2015-03-14 | 9:00:00 | 2.1031772 | ['Seth'] | 6.60E+15 | -2.25993E+15 |
| 2015-03-14 | 12:00:00 | 2.102206154 | ['Imhotep1'] | 1.11E+16 | -3.77559E+15 |
| 2015-03-15 | 6:30:00 | 2.096243421 | ['Imhotep1'] | 2.33E+16 | -7.99959E+15 |
| 2015-03-15 | 7:00:00 | 2.096165833 | ['Hapi'] | 2.90E+16 | -1.01E+16 |
| 2015-03-16 | 11:30:00 | 2.086958 | ['Hapi'] | 4.23E+16 | -1.45E+16 |
| 2015-03-16 | 23:00:00 | 2.08329125 | ['Seth'] | 2.70E+16 | -9.20912E+15 |
| 2015-03-17 | 0:00:00 | 2.082948158 | ['Hapi'] | 3.40E+16 | -1.16E+16 |
| 2015-03-17 | 0:30:00 | 2.082692857 | ['Hapi'] | 2.97E+16 | -1.02E+16 |
| 2015-03-17 | 1:00:00 | 2.082606604 | ['Hathor'] | 2.52E+16 | -8.59567E+15 |
| 2015-03-17 | 1:30:00 | 2.082440714 | ['Maat'] | 1.88E+16 | -6.41342E+15 |
| 2015-03-17 | 6:30:00 | 2.080829623 | ['Hatmehit'] | 2.00E+16 | -6.86081E+15 |
| 2015-03-17 | 7:00:00 | 2.080669623 | ['Hatmehit'] | 2.13E+16 | -7.27395E+15 |
| 2015-03-17 | 8:30:00 | 2.08012 | ['Hatmehit'] | 2.02E+16 | -7.0912E+15 |
| 2015-03-17 | 9:00:00 | 2.080032264 | ['Babi'] | 1.49E+16 | -5.10902E+15 |
| 2015-03-17 | 9:30:00 | 2.079872075 | ['Ash'] | 1.70E+16 | -5.82218E+15 |
| 2015-03-17 | 10:00:00 | 2.079707143 | ['Aker'] | 2.25E+16 | -7.69569E+15 |
| 2015-03-25 | 3:30:00 | 2.020057736 | ['Anuket'] | 4.19E+15 | -1.53587E+15 |
| 2015-03-25 | 8:00:00 | 2.018543333 | ['Imhotep'] | 5.01E+15 | -1.76452E+15 |
| 2015-03-25 | 8:30:00 | 2.018449057 | ['Imhotep'] | 3.81E+15 | -1.32639E+15 |
| 2015-03-25 | 9:00:00 | 2.018284909 | ['Imhotep'] | 2.91E+15 | -1.02895E+15 |
| 2015-03-25 | 11:00:00 | 2.017639623 | ['Khepry'] | 4.92E+15 | -1.71835E+15 |
| 2015-03-25 | 12:00:00 | 2.017321569 | ['Aker'] | 4.41E+15 | -1.51177E+15 |
| 2015-03-25 | 12:30:00 | 2.017178605 | ['Hapi'] | 4.35E+15 | -1.51856E+15 |
| 2015-03-25 | 22:30:00 | 2.013941429 | ['Khepry'] | 3.52E+15 | -1.21305E+15 |
| 2015-03-25 | 23:30:00 | 2.013619434 | ['Khepry'] | 7.58E+15 | -2.62783E+15 |
| 2015-03-26 | 0:30:00 | 2.013343077 | ['Aker'] | 3.86E+15 | -1.41265E+15 |
| 2015-03-27 | 4:00:00 | 2.0044575 | ['Sobek'] | 9.25E+15 | -3.2779E+15 |
| 2015-03-27 | 5:00:00 | 2.004160857 | ['Anubis'] | 1.07E+16 | -3.73049E+15 |
| 2015-03-27 | 10:00:00 | 2.002531887 | ['Imhotep'] | 5.62E+15 | -1.93637E+15 |
| 2015-03-27 | 10:30:00 | 2.002368545 | ['Khepry'] | 6.02E+15 | -2.07822E+15 |
| 2015-03-28 | 7:00:00 | 1.995722778 | ['Seth'] | 8.84E+15 | -3.07158E+15 |
| 2015-03-28 | 8:00:00 | 1.995462264 | ['Ash'] | 3.60E+15 | -1.25154E+15 |
| 2015-03-28 | 10:00:00 | 1.994811321 | ['Imhotep'] | 6.55E+15 | -2.25007E+15 |
| 2015-03-28 | 11:30:00 | 1.994274375 | ['Aten'] | 8.47E+15 | -2.97139E+15 |
| 2015-03-28 | 13:00:00 | 1.99385 | ['Hathor'] | 1.29E+16 | -4.40811E+15 |
| 2015-03-28 | 13:30:00 | 1.993684615 | ['Aten'] | 9.96E+15 | -3.40911E+15 |
| 2015-03-28 | 14:30:00 | 1.993338611 | ['Ash'] | 1.04E+16 | -3.57164E+15 |
| 2015-03-28 | 15:00:00 | 1.993206038 | ['Anuket'] | 1.17E+16 | -3.99772E+15 |
| 2015-03-28 | 15:30:00 | 1.993047308 | ['Aten'] | 1.31E+16 | -4.47009E+15 |
| 2015-03-28 | 16:00:00 | 1.992888868 | ['Hathor'] | 1.69E+16 | -5.77804E+15 |
| 2015-03-28 | 16:30:00 | 1.992728679 | ['Maat'] | 1.38E+16 | -4.72258E+15 |
| 2015-03-28 | 17:30:00 | 1.992401455 | ['Seth'] | 1.69E+16 | -5.79291E+15 |
| 2015-03-28 | 19:00:00 | 1.99189 | ['Hathor'] | 4.67E+15 | -1.64797E+15 |

| Date | Time | Value | Name | A | B |
|---|---|---|---|---|---|
| 2015-03-28 | 20:00:00 | 1.991600566 | ['Ash'] | 6.47E+15 | -2.29777E+15 |
| 2015-03-28 | 21:00:00 | 1.991281923 | ['Serqet'] | 2.59E+15 | -8.99495E+14 |
| 2015-03-28 | 23:00:00 | 1.990632453 | ['Hapi'] | 1.97E+16 | -6.71895E+15 |
| 2015-03-28 | 23:30:00 | 1.990472264 | ['Seth'] | 2.04E+16 | -6.96177E+15 |
| 2015-03-29 | 0:00:00 | 1.990313846 | ['Seth'] | 8.99E+15 | -3.09314E+15 |
| 2015-03-29 | 10:00:00 | 1.987078537 | ['Seth'] | 2.99E+15 | -1.04423E+15 |
| 2015-03-29 | 11:30:00 | 1.986587143 | ['Seth'] | 2.10E+16 | -7.18031E+15 |
| 2015-03-29 | 12:00:00 | 1.986504348 | ['Hapi'] | 2.62E+16 | -8.95548E+15 |
| 2015-04-10 | 10:30:00 | 1.894631569 | ['Imhotep'] | 2.47E+15 | -8.60947E+14 |
| 2015-04-10 | 11:00:00 | 1.894475283 | ['Bes'] | 2.85E+15 | -9.98391E+14 |
| 2015-04-10 | 22:30:00 | 1.890823889 | ['Imhotep'] | 3.79E+15 | -1.31564E+15 |
| 2015-04-11 | 11:00:00 | 1.88681 | ['Imhotep'] | 2.57E+15 | -8.90382E+14 |
| 2015-04-11 | 12:00:00 | 1.886492222 | ['Bes'] | 3.60E+15 | -1.26961E+15 |
| 2015-04-11 | 22:00:00 | 1.883306415 | ['Imhotep'] | 6.30E+15 | -2.15712E+15 |
| 2015-04-11 | 23:30:00 | 1.882824286 | ['Imhotep'] | 4.59E+15 | -1.58712E+15 |
| 2015-04-12 | 12:00:00 | 1.87884 | ['Imhotep'] | 8.17E+15 | -2.7989E+15 |
| 2015-04-12 | 12:30:00 | 1.878680385 | ['Bes'] | 5.57E+15 | -1.91688E+15 |
| 2015-04-12 | 13:00:00 | 1.878520392 | ['Bes'] | 4.67E+15 | -1.6678E+15 |
| 2015-04-12 | 18:00:00 | 1.8769324 | ['Anuket'] | 5.78E+15 | -2.03672E+15 |
| 2015-04-12 | 18:30:00 | 1.876771636 | ['Anuket'] | 6.94E+15 | -2.38738E+15 |
| 2015-04-12 | 23:30:00 | 1.875169623 | ['Imhotep'] | 8.79E+15 | -3.00868E+15 |
| 2015-04-13 | 0:00:00 | 1.875012308 | ['Imhotep'] | 9.25E+15 | -3.17187E+15 |
| 2015-04-13 | 0:30:00 | 1.874855283 | ['Imhotep'] | 8.44E+15 | -2.89287E+15 |
| 2015-04-13 | 7:00:00 | 1.87273625 | ['Hatmehit'] | 7.59E+15 | -2.64551E+15 |
| 2015-04-13 | 13:00:00 | 1.870873 | ['Bes'] | 7.36E+15 | -2.52267E+15 |
| 2015-04-13 | 14:00:00 | 1.870555849 | ['Bes'] | 8.69E+15 | -3.01152E+15 |
| 2015-04-13 | 14:30:00 | 1.870398269 | ['Anhur'] | 1.19E+16 | -4.14694E+15 |
| 2015-04-13 | 15:30:00 | 1.870083077 | ['Sobek'] | 8.78E+15 | -3.06914E+15 |
| 2015-04-13 | 18:30:00 | 1.869123962 | ['Anuket'] | 6.32E+15 | -2.18034E+15 |
| 2015-04-13 | 20:00:00 | 1.868651132 | ['Anubis'] | 7.91E+15 | -2.70746E+15 |
| 2015-04-14 | 0:00:00 | 1.867375283 | ['Imhotep'] | 1.30E+16 | -4.45645E+15 |
| 2015-04-14 | 0:30:00 | 1.86721902 | ['Imhotep'] | 9.78E+15 | -3.35405E+15 |
| 2015-04-14 | 1:00:00 | 1.867062115 | ['Imhotep'] | 7.54E+15 | -2.58913E+15 |
| 2015-04-14 | 1:30:00 | 1.866896909 | ['Khepry'] | 4.67E+15 | -1.60843E+15 |
| 2015-04-14 | 7:00:00 | 1.86514434 | ['Anuket'] | 8.98E+15 | -3.14738E+15 |
| 2015-04-14 | 9:00:00 | 1.864503261 | ['Anuket'] | 4.47E+15 | -1.55767E+15 |
| 2015-04-14 | 13:30:00 | 1.863080196 | ['Anuket'] | 8.06E+15 | -2.77124E+15 |
| 2015-04-14 | 14:00:00 | 1.862924151 | ['Khepry'] | 6.80E+15 | -2.33715E+15 |
| 2015-04-14 | 19:30:00 | 1.861166604 | ['Anuket'] | 1.26E+16 | -4.33742E+15 |
| 2015-04-14 | 20:00:00 | 1.861008302 | ['Hapi'] | 1.79E+16 | -6.11063E+15 |
| 2015-04-14 | 21:00:00 | 1.860692745 | ['Anubis'] | 1.03E+16 | -3.52769E+15 |
| 2015-04-15 | 3:30:00 | 1.858626429 | ['Khepry'] | 1.52E+16 | -5.19613E+15 |
| 2015-04-15 | 4:00:00 | 1.858462642 | ['Aker'] | 1.29E+16 | -4.44463E+15 |
| 2015-04-15 | 4:30:00 | 1.858321707 | ['Aker'] | 7.96E+15 | -2.73844E+15 |

| Date | Time | Value | Name | A | B |
|---|---|---|---|---|---|
| 2015-04-15 | 6:30:00 | 1.857674231 | ['Wosret'] | 5.19E+15 | -1.8301E+15 |
| 2015-04-15 | 7:30:00 | 1.857356538 | ['Wosret'] | 5.91E+15 | -2.08138E+15 |
| 2015-04-15 | 9:00:00 | 1.856879286 | ['Anuket'] | 1.84E+16 | -6.29588E+15 |
| 2015-04-15 | 9:30:00 | 1.856715283 | ['Hapi'] | 1.17E+16 | -4.00439E+15 |
| 2015-04-15 | 10:00:00 | 1.856571395 | ['Anubis'] | 5.51E+15 | -1.88587E+15 |
| 2015-04-15 | 13:30:00 | 1.85544875 | ['Imhotep'] | 1.66E+16 | -5.6833E+15 |
| 2015-04-15 | 14:00:00 | 1.855285 | ['Imhotep'] | 1.24E+16 | -4.24551E+15 |
| 2015-04-15 | 14:30:00 | 1.855128302 | ['Imhotep'] | 7.16E+15 | -2.45601E+15 |
| 2015-04-15 | 15:00:00 | 1.854969811 | ['Khepry'] | 1.09E+16 | -3.74691E+15 |
| 2015-04-15 | 15:30:00 | 1.854811887 | ['Khepry'] | 1.91E+16 | -6.53258E+15 |
| 2015-04-15 | 17:00:00 | 1.854340392 | ['Anhur'] | 6.63E+15 | -2.27767E+15 |
| 2015-04-15 | 21:00:00 | 1.853068269 | ['Anuket'] | 2.27E+16 | -7.74667E+15 |
| 2015-04-15 | 21:30:00 | 1.8529138 | ['Hapi'] | 2.03E+16 | -6.94709E+15 |
| 2015-04-15 | 22:00:00 | 1.852749423 | ['Anubis'] | 1.16E+16 | -3.96574E+15 |
| 2015-04-15 | 22:30:00 | 1.852588148 | ['Atum'] | 5.36E+15 | -1.84326E+15 |
| 2015-04-16 | 0:30:00 | 1.851955094 | ['Imhotep'] | 4.33E+15 | -1.52648E+15 |
| 2015-04-16 | 1:30:00 | 1.851639615 | ['Imhotep'] | 2.20E+16 | -7.50349E+15 |
| 2015-04-16 | 2:30:00 | 1.851320714 | ['Imhotep'] | 1.35E+16 | -4.60984E+15 |
| 2015-04-16 | 4:00:00 | 1.850840577 | ['Bes'] | 2.08E+16 | -7.12252E+15 |
| 2015-04-16 | 9:30:00 | 1.849097692 | ['Geb'] | 2.66E+16 | -9.0931E+15 |
| 2015-04-16 | 10:00:00 | 1.848940943 | ['Geb'] | 1.88E+16 | -6.41503E+15 |
| 2015-04-16 | 10:30:00 | 1.848782642 | ['Atum'] | 1.05E+16 | -3.59636E+15 |
| 2015-04-16 | 11:00:00 | 1.848625192 | ['Khonsu'] | 7.45E+15 | -2.55288E+15 |
| 2015-04-16 | 11:30:00 | 1.84846125 | ['Khonsu'] | 3.75E+15 | -1.31015E+15 |
| 2015-04-16 | 13:30:00 | 1.847826981 | ['Imhotep'] | 1.87E+16 | -6.39374E+15 |
| 2015-04-16 | 14:00:00 | 1.847668868 | ['Imhotep'] | 2.23E+16 | -7.60835E+15 |
| 2015-04-16 | 15:30:00 | 1.847196111 | ['Bes'] | 1.34E+16 | -4.58416E+15 |
| 2015-04-16 | 16:00:00 | 1.847032364 | ['Bes'] | 1.72E+16 | -5.90531E+15 |
| 2015-04-16 | 16:30:00 | 1.846871132 | ['Anhur'] | 1.86E+16 | -6.39178E+15 |
| 2015-04-16 | 17:30:00 | 1.846556346 | ['Geb'] | 7.15E+15 | -2.53016E+15 |
| 2015-04-18 | 1:00:00 | 1.836571887 | ['Ash'] | 1.00E+16 | -3.45468E+15 |
| 2015-04-18 | 1:30:00 | 1.836413774 | ['Ash'] | 1.10E+16 | -3.80336E+15 |
| 2015-04-18 | 2:00:00 | 1.836257843 | ['Imhotep'] | 2.24E+16 | -7.66217E+15 |
| 2015-04-18 | 2:30:00 | 1.836099623 | ['Imhotep'] | 2.75E+16 | -9.40056E+15 |
| 2015-04-18 | 3:00:00 | 1.835936364 | ['Imhotep'] | 2.37E+16 | -8.10345E+15 |
| 2015-04-18 | 3:30:00 | 1.835775472 | ['Ash'] | 2.38E+16 | -8.11821E+15 |
| 2015-04-18 | 4:00:00 | 1.835617736 | ['Imhotep'] | 3.91E+16 | -1.33E+16 |
| 2015-04-18 | 4:30:00 | 1.83546 | ['Ash'] | 3.85E+16 | -1.31E+16 |
| 2015-04-18 | 5:00:00 | 1.835303846 | ['Aten'] | 2.87E+16 | -9.78806E+15 |
| 2015-04-18 | 5:30:00 | 1.835147358 | ['Aten'] | 3.70E+16 | -1.26E+16 |
| 2015-04-18 | 6:00:00 | 1.834989623 | ['Babi'] | 3.11E+16 | -1.06E+16 |
| 2015-04-18 | 6:30:00 | 1.834831887 | ['Babi'] | 2.16E+16 | -7.39258E+15 |
| 2015-04-18 | 7:00:00 | 1.834670909 | ['Hapi'] | 1.34E+16 | -4.57282E+15 |
| 2015-04-18 | 7:30:00 | 1.834507778 | ['Hapi'] | 7.40E+15 | -2.53102E+15 |

| Date | Time | Value | Label | Col5 | Col6 |
|---|---|---|---|---|---|
| 2015-04-18 | 8:30:00 | 1.834193585 | ['Maat'] | 1.33E+16 | -4.53674E+15 |
| 2015-04-18 | 9:30:00 | 1.833877451 | ['Nut'] | 1.62E+16 | -5.53421E+15 |
| 2015-04-18 | 10:00:00 | 1.83372098 | ['Serqet'] | 1.76E+16 | -5.99673E+15 |
| 2015-04-18 | 10:30:00 | 1.833564808 | ['Hathor'] | 4.66E+16 | -1.59E+16 |
| 2015-04-18 | 11:00:00 | 1.833408302 | ['Serqet'] | 4.25E+16 | -1.45E+16 |
| 2015-04-18 | 11:30:00 | 1.833245536 | ['Hathor'] | 2.96E+16 | -1.01E+16 |
| 2015-04-18 | 12:30:00 | 1.832923654 | ['Hapi'] | 2.02E+16 | -6.89704E+15 |
| 2015-04-18 | 13:00:00 | 1.832768654 | ['Seth'] | 1.60E+16 | -5.46937E+15 |
| 2015-04-18 | 13:30:00 | 1.832612642 | ['Seth'] | 1.10E+16 | -3.76397E+15 |
| 2015-04-18 | 14:00:00 | 1.832454706 | ['Ash'] | 1.22E+16 | -4.15801E+15 |
| 2015-04-18 | 14:30:00 | 1.83229717 | ['Ash'] | 1.83E+16 | -6.24949E+15 |
| 2015-04-18 | 15:00:00 | 1.832140769 | ['Ash'] | 1.78E+16 | -6.07299E+15 |
| 2015-04-18 | 15:30:00 | 1.831984717 | ['Ash'] | 2.26E+16 | -7.71435E+15 |
| 2015-04-18 | 16:00:00 | 1.831821964 | ['Ash'] | 2.58E+16 | -8.82122E+15 |
| 2015-04-18 | 16:30:00 | 1.831659245 | ['Ash'] | 3.80E+16 | -1.30E+16 |
| 2015-04-18 | 17:00:00 | 1.831501923 | ['Aten'] | 3.90E+16 | -1.33E+16 |
| 2015-04-18 | 17:30:00 | 1.831342157 | ['Aten'] | 2.39E+16 | -8.15009E+15 |
| 2015-04-18 | 18:00:00 | 1.831187692 | ['Babi'] | 3.75E+16 | -1.28E+16 |
| 2015-04-18 | 18:30:00 | 1.831031509 | ['Babi'] | 3.13E+16 | -1.07E+16 |
| 2015-04-18 | 19:00:00 | 1.830873774 | ['Babi'] | 2.23E+16 | -7.62562E+15 |
| 2015-04-18 | 19:30:00 | 1.830716415 | ['Hapi'] | 1.31E+16 | -4.4782E+15 |
| 2015-04-18 | 21:00:00 | 1.830234231 | ['Maat'] | 1.31E+16 | -4.4917E+15 |
| 2015-04-18 | 21:30:00 | 1.830078113 | ['Maat'] | 1.37E+16 | -4.72049E+15 |
| 2015-04-18 | 22:30:00 | 1.82976283 | ['Maat'] | 2.60E+16 | -8.89141E+15 |
| 2015-04-18 | 23:00:00 | 1.829606731 | ['Hapi'] | 4.90E+16 | -1.67E+16 |
| 2015-04-18 | 23:30:00 | 1.829450377 | ['Serqet'] | 3.85E+16 | -1.31E+16 |
| 2015-04-19 | 0:00:00 | 1.829291731 | ['Hathor'] | 3.15E+16 | -1.08E+16 |
| 2015-04-19 | 1:00:00 | 1.828968889 | ['Hapi'] | 1.71E+16 | -5.83868E+15 |
| 2015-04-19 | 1:30:00 | 1.828808846 | ['Hapi'] | 1.40E+16 | -4.77451E+15 |
| 2015-04-19 | 2:00:00 | 1.82865451 | ['Seth'] | 9.61E+15 | -3.29048E+15 |
| 2015-04-19 | 3:00:00 | 1.828339811 | ['Ash'] | 1.33E+16 | -4.5382E+15 |
| 2015-04-19 | 3:30:00 | 1.828184038 | ['Ash'] | 1.35E+16 | -4.61238E+15 |
| 2015-04-19 | 4:30:00 | 1.827870189 | ['Ash'] | 2.39E+16 | -8.16376E+15 |
| 2015-04-19 | 5:30:00 | 1.827543137 | ['Aten'] | 3.13E+16 | -1.07E+16 |
| 2015-04-19 | 6:00:00 | 1.827387551 | ['Aten'] | 1.94E+16 | -6.63888E+15 |
| 2015-04-19 | 6:30:00 | 1.8272325 | ['Babi'] | 2.76E+16 | -9.45094E+15 |
| 2015-04-19 | 7:00:00 | 1.827074038 | ['Babi'] | 2.59E+16 | -8.84134E+15 |
| 2015-04-19 | 7:30:00 | 1.826917547 | ['Babi'] | 2.10E+16 | -7.17931E+15 |
| 2015-04-19 | 9:30:00 | 1.826285 | ['Maat'] | 9.93E+15 | -3.40219E+15 |
| 2015-04-19 | 10:00:00 | 1.826123654 | ['Maat'] | 1.31E+16 | -4.49337E+15 |
| 2015-04-19 | 11:30:00 | 1.825603636 | ['Maat'] | 5.68E+16 | -1.94E+16 |
| 2015-04-19 | 12:00:00 | 1.825493208 | ['Serqet'] | 4.17E+16 | -1.42E+16 |
| 2015-04-19 | 12:30:00 | 1.825335472 | ['Serqet'] | 3.24E+16 | -1.11E+16 |
| 2015-04-19 | 14:00:00 | 1.824866226 | ['Hapi'] | 1.49E+16 | -5.08895E+15 |

| Date | Time | Value | Name | Col5 | Col6 |
|---|---|---|---|---|---|
| 2015-04-19 | 14:30:00 | 1.824703214 | ['Seth'] | 1.15E+16 | -3.93349E+15 |
| 2015-04-19 | 15:00:00 | 1.824540943 | ['Seth'] | 9.41E+15 | -3.21758E+15 |
| 2015-04-19 | 16:00:00 | 1.8242275 | ['Ash'] | 1.01E+16 | -3.45245E+15 |
| 2015-04-19 | 17:00:00 | 1.823914151 | ['Ash'] | 1.71E+16 | -5.87049E+15 |
| 2015-04-19 | 17:30:00 | 1.823756792 | ['Ash'] | 1.80E+16 | -6.15027E+15 |
| 2015-04-19 | 18:00:00 | 1.823600769 | ['Ash'] | 2.66E+16 | -9.08169E+15 |
| 2015-04-19 | 18:30:00 | 1.823441636 | ['Ash'] | 1.71E+16 | -5.83566E+15 |
| 2015-04-19 | 19:30:00 | 1.823119811 | ['Babi'] | 2.12E+16 | -7.23396E+15 |
| 2015-04-19 | 20:00:00 | 1.822962885 | ['Babi'] | 2.04E+16 | -6.97661E+15 |
| 2015-04-19 | 21:00:00 | 1.822648654 | ['Babi'] | 9.37E+15 | -3.21575E+15 |
| 2015-04-19 | 21:30:00 | 1.822493019 | ['Hapi'] | 8.82E+15 | -3.01782E+15 |
| 2015-04-19 | 22:30:00 | 1.822178113 | ['Maat'] | 1.28E+16 | -4.38488E+15 |
| 2015-04-19 | 23:00:00 | 1.822015893 | ['Maat'] | 1.88E+16 | -6.4405E+15 |
| 2015-04-19 | 23:30:00 | 1.821853208 | ['Maat'] | 2.89E+16 | -9.88181E+15 |
| 2015-04-20 | 0:00:00 | 1.821697115 | ['Maat'] | 4.18E+16 | -1.43E+16 |
| 2015-04-20 | 0:30:00 | 1.821541321 | ['Serqet'] | 4.00E+16 | -1.37E+16 |
| 2015-04-20 | 1:00:00 | 1.821383774 | ['Maat'] | 3.48E+16 | -1.19E+16 |
| 2015-04-20 | 2:30:00 | 1.820914717 | ['Hapi'] | 1.61E+16 | -5.5119E+15 |
| 2015-04-20 | 3:00:00 | 1.820756923 | ['Hapi'] | 1.35E+16 | -4.62181E+15 |
| 2015-04-20 | 3:30:00 | 1.820595 | ['Seth'] | 9.76E+15 | -3.33832E+15 |
| 2015-04-20 | 4:00:00 | 1.820432264 | ['Seth'] | 9.51E+15 | -3.25095E+15 |
| 2015-04-20 | 4:30:00 | 1.820275283 | ['Ash'] | 1.07E+16 | -3.67307E+15 |
| 2015-04-20 | 5:30:00 | 1.819963396 | ['Ash'] | 1.18E+16 | -4.0608E+15 |
| 2015-04-20 | 7:30:00 | 1.819336792 | ['Ash'] | 1.83E+16 | -6.25764E+15 |
| 2015-04-20 | 8:00:00 | 1.819175091 | ['Ash'] | 2.00E+16 | -6.85094E+15 |
| 2015-04-20 | 8:30:00 | 1.819011698 | ['Ash'] | 2.30E+16 | -7.86763E+15 |
| 2015-04-20 | 9:00:00 | 1.818854528 | ['Babi'] | 1.87E+16 | -6.38375E+15 |
| 2015-04-20 | 9:30:00 | 1.818699412 | ['Babi'] | 2.09E+16 | -7.1533E+15 |
| 2015-04-20 | 10:00:00 | 1.818543019 | ['Hapi'] | 2.08E+16 | -7.09266E+15 |
| 2015-04-20 | 10:30:00 | 1.818385192 | ['Hapi'] | 2.19E+16 | -7.48241E+15 |
| 2015-04-20 | 11:00:00 | 1.818228302 | ['Hapi'] | 2.46E+16 | -8.39769E+15 |
| 2015-04-20 | 11:30:00 | 1.818072692 | ['Maat'] | 2.11E+16 | -7.21995E+15 |
| 2015-04-20 | 12:00:00 | 1.817913636 | ['Maat'] | 2.51E+16 | -8.57283E+15 |
| 2015-04-20 | 12:30:00 | 1.817750926 | ['Maat'] | 2.78E+16 | -9.50377E+15 |
| 2015-04-20 | 13:00:00 | 1.817591887 | ['Serqet'] | 3.13E+16 | -1.07E+16 |
| 2015-04-20 | 13:30:00 | 1.817434906 | ['Serqet'] | 2.84E+16 | -9.69276E+15 |
| 2015-04-20 | 15:30:00 | 1.816808679 | ['Hapi'] | 1.65E+16 | -5.64087E+15 |
| 2015-04-20 | 16:30:00 | 1.816489464 | ['Seth'] | 1.57E+16 | -5.37552E+15 |
| 2015-04-20 | 17:00:00 | 1.816326792 | ['Seth'] | 1.84E+16 | -6.29638E+15 |
| 2015-04-20 | 20:00:00 | 1.815389434 | ['Ash'] | 1.75E+16 | -5.99872E+15 |
| 2015-04-20 | 20:30:00 | 1.815232264 | ['Ash'] | 2.31E+16 | -7.9219E+15 |
| 2015-04-20 | 21:00:00 | 1.815069821 | ['Ash'] | 2.91E+16 | -9.93714E+15 |
| 2015-04-20 | 21:30:00 | 1.814907547 | ['Babi'] | 2.75E+16 | -9.38269E+15 |
| 2015-04-20 | 22:00:00 | 1.814750566 | ['Babi'] | 3.27E+16 | -1.12E+16 |

| Date | Time | Value | Label | E1 | E2 |
|---|---|---|---|---|---|
| 2015-04-20 | 22:30:00 | 1.814651875 | ['Hapi'] | 3.11E+16 | -1.06E+16 |
| 2015-04-20 | 23:30:00 | 1.814232727 | ['Hapi'] | 2.31E+16 | -7.91817E+15 |
| 2015-04-21 | 0:00:00 | 1.814124717 | ['Hathor'] | 2.17E+16 | -7.40159E+15 |
| 2015-04-21 | 0:30:00 | 1.813969231 | ['Maat'] | 2.31E+16 | -7.87821E+15 |
| 2015-04-21 | 1:30:00 | 1.813650893 | ['Serqet'] | 2.43E+16 | -8.30549E+15 |
| 2015-04-21 | 2:00:00 | 1.813489057 | ['Serqet'] | 2.58E+16 | -8.83073E+15 |
| 2015-04-21 | 4:00:00 | 1.812862115 | ['Anuket'] | 2.02E+16 | -6.91768E+15 |
| 2015-04-21 | 4:30:00 | 1.812706038 | ['Seth'] | 2.29E+16 | -7.80943E+15 |
| 2015-04-21 | 5:00:00 | 1.812548868 | ['Seth'] | 2.50E+16 | -8.54828E+15 |
| 2015-04-21 | 5:30:00 | 1.812390185 | ['Seth'] | 2.57E+16 | -8.79068E+15 |
| 2015-04-21 | 6:00:00 | 1.812227455 | ['Seth'] | 2.01E+16 | -6.86168E+15 |
| 2015-04-21 | 6:30:00 | 1.812068654 | ['Ash'] | 1.34E+16 | -4.57407E+15 |
| 2015-04-21 | 7:00:00 | 1.811913019 | ['Ash'] | 1.03E+16 | -3.55904E+15 |
| 2015-04-21 | 9:00:00 | 1.811221818 | ['Ash'] | 3.43E+16 | -1.18E+16 |
| 2015-04-21 | 9:30:00 | 1.811130755 | ['Ash'] | 3.14E+16 | -1.07E+16 |
| 2015-04-21 | 10:00:00 | 1.810968929 | ['Seth'] | 3.50E+16 | -1.19E+16 |
| 2015-04-21 | 10:30:00 | 1.810806415 | ['Babi'] | 3.64E+16 | -1.24E+16 |
| 2015-04-21 | 11:00:00 | 1.810649434 | ['Babi'] | 2.97E+16 | -1.02E+16 |
| 2015-04-21 | 11:30:00 | 1.810493654 | ['Hapi'] | 3.09E+16 | -1.05E+16 |
| 2015-04-21 | 12:00:00 | 1.810338113 | ['Hapi'] | 2.45E+16 | -8.37578E+15 |
| 2015-04-21 | 12:30:00 | 1.810181321 | ['Hathor'] | 2.14E+16 | -7.30253E+15 |
| 2015-04-21 | 13:00:00 | 1.810023962 | ['Maat'] | 2.19E+16 | -7.4866E+15 |
| 2015-04-21 | 14:00:00 | 1.809713208 | ['Maat'] | 2.45E+16 | -8.37345E+15 |
| 2015-04-21 | 14:30:00 | 1.809551071 | ['Serqet'] | 2.44E+16 | -8.35056E+15 |
| 2015-04-21 | 15:00:00 | 1.80947 | ['Serqet'] | 2.58E+16 | -8.94407E+15 |
| 2015-04-21 | 16:00:00 | 1.809049459 | ['Anuket'] | 2.49E+16 | -8.51008E+15 |
| 2015-04-21 | 16:30:00 | 1.808919231 | ['Anuket'] | 2.60E+16 | -8.86622E+15 |
| 2015-04-21 | 17:00:00 | 1.808763962 | ['Seth'] | 2.95E+16 | -1.01E+16 |
| 2015-04-21 | 17:30:00 | 1.808606981 | ['Seth'] | 3.03E+16 | -1.03E+16 |
| 2015-04-21 | 18:00:00 | 1.808449623 | ['Seth'] | 2.91E+16 | -9.94161E+15 |
| 2015-04-21 | 18:30:00 | 1.808294615 | ['Seth'] | 2.07E+16 | -7.06635E+15 |
| 2015-04-21 | 19:00:00 | 1.80813193 | ['Ash'] | 1.45E+16 | -4.95705E+15 |
| 2015-04-21 | 19:30:00 | 1.807969412 | ['Ash'] | 8.73E+15 | -3.02383E+15 |
| 2015-04-21 | 21:30:00 | 1.807345192 | ['Ash'] | 3.17E+16 | -1.08E+16 |
| 2015-04-21 | 22:00:00 | 1.80719 | ['Ash'] | 3.35E+16 | -1.14E+16 |
| 2015-04-21 | 22:30:00 | 1.80703283 | ['Ash'] | 3.24E+16 | -1.11E+16 |
| 2015-04-21 | 23:00:00 | 1.806874259 | ['Babi'] | 3.13E+16 | -1.07E+16 |
| 2015-04-21 | 23:30:00 | 1.806711636 | ['Babi'] | 2.57E+16 | -8.78252E+15 |
| 2015-04-22 | 4:00:00 | 1.805291509 | ['Maat'] | 2.53E+16 | -8.67061E+15 |
| 2015-04-22 | 5:30:00 | 1.804789375 | ['Hatmehit'] | 3.07E+16 | -1.05E+16 |
| 2015-04-22 | 6:00:00 | 1.8046634 | ['Hapi'] | 3.02E+16 | -1.03E+16 |
| 2015-04-22 | 6:30:00 | 1.804508868 | ['Seth'] | 2.99E+16 | -1.02E+16 |
| 2015-04-22 | 7:00:00 | 1.804352075 | ['Seth'] | 2.09E+16 | -7.13746E+15 |
| 2015-04-22 | 7:30:00 | 1.804195283 | ['Seth'] | 1.41E+16 | -4.83688E+15 |
| 2015-04-22 | 8:00:00 | 1.804039808 | ['Seth'] | 9.89E+15 | -3.39642E+15 |
| 2015-04-22 | 10:00:00 | 1.803403774 | ['Ash'] | 4.01E+16 | -1.37E+16 |
| 2015-04-22 | 12:00:00 | 1.802779623 | ['Seth'] | 2.37E+16 | -8.08937E+15 |
| 2015-04-22 | 12:30:00 | 1.802620588 | ['Babi'] | 2.26E+16 | -7.71458E+15 |
| 2015-04-22 | 13:00:00 | 1.8024675 | ['Hapi'] | 1.93E+16 | -6.58649E+15 |
| 2015-04-22 | 13:30:00 | 1.802305614 | ['Hapi'] | 1.76E+16 | -6.01649E+15 |
| 2015-04-22 | 14:00:00 | 1.802143462 | ['Hapi'] | 1.83E+16 | -6.26485E+15 |
| 2015-04-22 | 14:30:00 | 1.801988491 | ['Maat'] | 2.02E+16 | -6.91375E+15 |
| 2015-04-22 | 15:00:00 | 1.801831698 | ['Maat'] | 1.81E+16 | -6.17459E+15 |
| 2015-04-22 | 15:30:00 | 1.801675769 | ['Maat'] | 2.67E+16 | -9.13828E+15 |
| 2015-04-22 | 16:00:00 | 1.801519231 | ['Maat'] | 2.48E+16 | -8.48823E+15 |
| 2015-04-22 | 16:30:00 | 1.801365 | ['Serqet'] | 2.59E+16 | -8.84943E+15 |
| 2015-04-22 | 17:00:00 | 1.8012052 | ['Serqet'] | 3.44E+16 | -1.18E+16 |
| 2015-04-22 | 17:30:00 | 1.801051731 | ['Hathor'] | 3.24E+16 | -1.11E+16 |
| 2015-04-22 | 18:00:00 | 1.800889825 | ['Anuket'] | 3.00E+16 | -1.02E+16 |
| 2015-04-22 | 18:30:00 | 1.800728077 | ['Anuket'] | 3.08E+16 | -1.05E+16 |
| 2015-04-22 | 19:00:00 | 1.800572642 | ['Seth'] | 2.88E+16 | -9.82957E+15 |
| 2015-04-22 | 19:30:00 | 1.800416226 | ['Seth'] | 2.39E+16 | -8.15916E+15 |
| 2015-04-22 | 20:00:00 | 1.800259434 | ['Seth'] | 1.48E+16 | -5.08552E+15 |
| 2015-04-22 | 20:30:00 | 1.800104231 | ['Seth'] | 8.54E+15 | -2.92517E+15 |
| 2015-04-22 | 21:30:00 | 1.799792075 | ['Ash'] | 1.25E+16 | -4.28113E+15 |
| 2015-04-22 | 22:00:00 | 1.799633704 | ['Ash'] | 2.32E+16 | -7.93014E+15 |
| 2015-04-23 | 0:30:00 | 1.79884434 | ['Babi'] | 2.40E+16 | -8.18742E+15 |
| 2015-04-23 | 1:00:00 | 1.798689038 | ['Babi'] | 2.21E+16 | -7.5483E+15 |
| 2015-04-23 | 1:30:00 | 1.798533962 | ['Babi'] | 1.91E+16 | -6.52713E+15 |
| 2015-04-23 | 2:00:00 | 1.798377358 | ['Hapi'] | 2.23E+16 | -7.6044E+15 |
| 2015-04-23 | 2:30:00 | 1.798215714 | ['Hapi'] | 1.99E+16 | -6.78716E+15 |
| 2015-04-23 | 3:00:00 | 1.798053774 | ['Maat'] | 1.98E+16 | -6.76941E+15 |
| 2015-04-23 | 3:30:00 | 1.797896154 | ['Maat'] | 2.29E+16 | -7.81461E+15 |
| 2015-04-23 | 4:00:00 | 1.797740392 | ['Maat'] | 2.99E+16 | -1.02E+16 |
| 2015-04-23 | 4:30:00 | 1.797586792 | ['Maat'] | 3.00E+16 | -1.02E+16 |
| 2015-04-23 | 5:00:00 | 1.797430588 | ['Maat'] | 3.21E+16 | -1.10E+16 |
| 2015-04-23 | 5:30:00 | 1.797272745 | ['Hapi'] | 3.55E+16 | -1.21E+16 |
| 2015-04-23 | 7:30:00 | 1.79664 | ['Seth'] | 2.54E+16 | -8.67994E+15 |
| 2015-04-23 | 8:00:00 | 1.796483208 | ['Seth'] | 2.28E+16 | -7.78746E+15 |
| 2015-04-23 | 8:30:00 | 1.796326792 | ['Seth'] | 1.37E+16 | -4.6814E+15 |
| 2015-04-23 | 9:30:00 | 1.796016604 | ['Ash'] | 8.68E+15 | -3.00069E+15 |
| 2015-04-23 | 10:30:00 | 1.795703396 | ['Ash'] | 1.74E+16 | -5.93806E+15 |
| 2015-04-23 | 12:00:00 | 1.795224314 | ['Ash'] | 1.52E+16 | -5.19727E+15 |
| 2015-04-23 | 12:30:00 | 1.79507 | ['Aten'] | 2.10E+16 | -7.18175E+15 |
| 2015-04-23 | 13:00:00 | 1.794913585 | ['Aten'] | 2.12E+16 | -7.23618E+15 |
| 2015-04-23 | 13:30:00 | 1.794756981 | ['Babi'] | 2.17E+16 | -7.41749E+15 |
| 2015-04-23 | 14:00:00 | 1.794601923 | ['Babi'] | 2.20E+16 | -7.51153E+15 |
| 2015-04-23 | 14:30:00 | 1.794446981 | ['Babi'] | 2.73E+16 | -9.30728E+15 |

| Date | Time | Value | Region | X | Y |
|---|---|---|---|---|---|
| 2015-04-23 | 15:30:00 | 1.794132407 | ['Hathor'] | 1.74E+16 | -5.94263E+15 |
| 2015-04-23 | 16:00:00 | 1.793970364 | ['Maat'] | 2.23E+16 | -7.61741E+15 |
| 2015-04-23 | 16:30:00 | 1.793810755 | ['Maat'] | 3.46E+16 | -1.18E+16 |
| 2015-04-23 | 17:00:00 | 1.793655385 | ['Maat'] | 3.03E+16 | -1.04E+16 |
| 2015-04-23 | 17:30:00 | 1.793500755 | ['Serqet'] | 2.95E+16 | -1.01E+16 |
| 2015-04-23 | 21:00:00 | 1.792398302 | ['Seth'] | 1.29E+16 | -4.41999E+15 |
| 2015-04-23 | 21:30:00 | 1.792244118 | ['Ash'] | 7.52E+15 | -2.57905E+15 |
| 2015-04-24 | 1:30:00 | 1.790986038 | ['Ash'] | 2.77E+16 | -9.46782E+15 |
| 2015-04-24 | 2:00:00 | 1.790832157 | ['Aten'] | 2.09E+16 | -7.13011E+15 |
| 2015-04-24 | 4:30:00 | 1.790050943 | ['Maat'] | 1.95E+16 | -6.67279E+15 |
| 2015-04-24 | 5:00:00 | 1.789889286 | ['Hatmehit'] | 2.41E+16 | -8.22052E+15 |
| 2015-04-24 | 5:30:00 | 1.789727925 | ['Hatmehit'] | 2.16E+16 | -7.38283E+15 |
| 2015-04-25 | 0:30:00 | 1.783773208 | ['Hatmehit'] | 1.41E+16 | -4.8042E+15 |
| 2015-04-25 | 1:00:00 | 1.783616792 | ['Ash'] | 1.34E+16 | -4.58301E+15 |
| 2015-04-25 | 1:30:00 | 1.783460943 | ['Ash'] | 1.58E+16 | -5.39091E+15 |
| 2015-04-25 | 2:00:00 | 1.783306154 | ['Aten'] | 2.44E+16 | -8.31458E+15 |
| 2015-04-25 | 2:30:00 | 1.78315 | ['Babi'] | 2.50E+16 | -8.53624E+15 |
| 2015-04-25 | 3:00:00 | 1.782988364 | ['Babi'] | 1.97E+16 | -6.7455E+15 |
| 2015-04-25 | 3:30:00 | 1.782829057 | ['Hapi'] | 2.68E+16 | -9.13611E+15 |
| 2015-04-25 | 4:00:00 | 1.78267283 | ['Hapi'] | 2.78E+16 | -9.50523E+15 |
| 2015-04-25 | 4:30:00 | 1.782517843 | ['Maat'] | 1.45E+16 | -4.96661E+15 |
| 2015-04-25 | 5:00:00 | 1.782363774 | ['Maat'] | 1.05E+16 | -3.60067E+15 |
| 2015-04-25 | 5:30:00 | 1.782207736 | ['Maat'] | 9.44E+15 | -3.22733E+15 |
| 2015-04-25 | 6:00:00 | 1.782051321 | ['Maat'] | 9.84E+15 | -3.36478E+15 |
| 2015-04-25 | 6:30:00 | 1.781897115 | ['Serqet'] | 9.51E+15 | -3.25405E+15 |
| 2015-04-25 | 7:00:00 | 1.7817375 | ['Serqet'] | 8.62E+15 | -2.94582E+15 |
| 2015-04-25 | 7:30:00 | 1.781576038 | ['Hathor'] | 9.35E+15 | -3.19756E+15 |
| 2015-04-25 | 8:00:00 | 1.78142 | ['Anuket'] | 1.32E+16 | -4.49343E+15 |
| 2015-04-25 | 8:30:00 | 1.781263962 | ['Seth'] | 2.00E+16 | -6.82954E+15 |
| 2015-04-25 | 9:00:00 | 1.781108113 | ['Seth'] | 3.46E+16 | -1.18E+16 |
| 2015-04-25 | 9:30:00 | 1.780953654 | ['Seth'] | 3.63E+16 | -1.24E+16 |
| 2015-04-25 | 10:00:00 | 1.780798679 | ['Seth'] | 2.26E+16 | -7.72191E+15 |
| 2015-04-25 | 10:30:00 | 1.78064283 | ['Seth'] | 1.21E+16 | -4.13E+15 |
| 2015-04-25 | 11:00:00 | 1.780489608 | ['Ash'] | 8.62E+15 | -2.94602E+15 |
| 2015-04-25 | 12:00:00 | 1.78016434 | ['Ash'] | 1.22E+16 | -4.15547E+15 |
| 2015-04-25 | 12:30:00 | 1.78001 | ['Ash'] | 1.88E+16 | -6.40276E+15 |
| 2015-04-25 | 13:00:00 | 1.779855472 | ['Imhotep'] | 1.75E+16 | -5.97437E+15 |
| 2015-04-25 | 13:30:00 | 1.779699231 | ['Imhotep'] | 1.75E+16 | -5.96291E+15 |
| 2015-04-25 | 14:00:00 | 1.779543208 | ['Ash'] | 1.95E+16 | -6.66136E+15 |
| 2015-04-25 | 14:30:00 | 1.779388654 | ['Aten'] | 2.57E+16 | -8.76966E+15 |
| 2015-04-25 | 15:00:00 | 1.77923434 | ['Aten'] | 3.31E+16 | -1.13E+16 |
| 2015-04-25 | 15:30:00 | 1.779078491 | ['Babi'] | 1.75E+16 | -5.98037E+15 |
| 2015-04-25 | 16:00:00 | 1.778917143 | ['Babi'] | 1.78E+16 | -6.06493E+15 |
| 2015-04-25 | 16:30:00 | 1.778756415 | ['Hapi'] | 2.80E+16 | -9.57767E+15 |

| Date | Time | Value | Region | Val1 | Val2 |
|---|---|---|---|---|---|
| 2015-04-25 | 17:00:00 | 1.778600566 | ['Hathor'] | 1.67E+16 | -5.72069E+15 |
| 2015-04-25 | 17:30:00 | 1.778445769 | ['Maat'] | 9.38E+15 | -3.21468E+15 |
| 2015-04-25 | 18:00:00 | 1.778291321 | ['Maat'] | 7.93E+15 | -2.70942E+15 |
| 2015-04-25 | 18:30:00 | 1.77813566 | ['Hatmehit'] | 1.06E+16 | -3.60652E+15 |
| 2015-04-25 | 19:00:00 | 1.777979623 | ['Nut'] | 9.26E+15 | -3.166E+15 |
| 2015-04-25 | 19:30:00 | 1.777825192 | ['Nut'] | 7.80E+15 | -2.66745E+15 |
| 2015-04-25 | 20:00:00 | 1.777668889 | ['Serqet'] | 9.64E+15 | -3.29727E+15 |
| 2015-04-25 | 20:30:00 | 1.777507455 | ['Anuket'] | 1.39E+16 | -4.74686E+15 |
| 2015-04-25 | 21:00:00 | 1.777348491 | ['Hapi'] | 2.10E+16 | -7.18907E+15 |
| 2015-04-25 | 21:30:00 | 1.777192115 | ['Seth'] | 3.70E+16 | -1.26E+16 |
| 2015-04-25 | 22:00:00 | 1.777036604 | ['Seth'] | 3.10E+16 | -1.06E+16 |
| 2015-04-25 | 22:30:00 | 1.776882692 | ['Anubis'] | 2.20E+16 | -7.51098E+15 |
| 2015-04-25 | 23:00:00 | 1.776728302 | ['Anubis'] | 1.29E+16 | -4.42032E+15 |
| 2015-04-25 | 23:30:00 | 1.776573462 | ['Atum'] | 8.73E+15 | -2.98335E+15 |
| 2015-04-26 | 0:00:00 | 1.776416604 | ['Ash'] | 8.26E+15 | -2.82403E+15 |
| 2015-04-26 | 0:30:00 | 1.776257455 | ['Imhotep'] | 1.32E+16 | -4.51297E+15 |
| 2015-04-26 | 1:00:00 | 1.776096111 | ['Imhotep'] | 1.91E+16 | -6.53657E+15 |
| 2015-04-26 | 1:30:00 | 1.77593902 | ['Imhotep'] | 2.34E+16 | -7.98799E+15 |
| 2015-04-26 | 2:00:00 | 1.77578566 | ['Imhotep'] | 2.55E+16 | -8.70268E+15 |
| 2015-04-26 | 2:30:00 | 1.775630189 | ['Khepry'] | 2.34E+16 | -7.99393E+15 |
| 2015-04-26 | 3:00:00 | 1.77547434 | ['Khepry'] | 2.63E+16 | -8.96499E+15 |
| 2015-04-26 | 3:30:00 | 1.77532 | ['Aten'] | 3.82E+16 | -1.30E+16 |
| 2015-04-26 | 4:00:00 | 1.77516566 | ['Babi'] | 2.01E+16 | -6.87122E+15 |
| 2015-04-26 | 4:30:00 | 1.775009623 | ['Aker'] | 1.99E+16 | -6.78456E+15 |
| 2015-04-26 | 5:00:00 | 1.774848929 | ['Bastet'] | 3.21E+16 | -1.10E+16 |
| 2015-04-26 | 5:30:00 | 1.774689038 | ['Hathor'] | 1.85E+16 | -6.32215E+15 |
| 2015-04-26 | 6:00:00 | 1.774530962 | ['Bastet'] | 9.50E+15 | -3.25283E+15 |
| 2015-04-26 | 6:30:00 | 1.774377885 | ['Bastet'] | 1.00E+16 | -3.42929E+15 |
| 2015-04-26 | 7:00:00 | 1.774223585 | ['Hatmehit'] | 1.45E+16 | -4.96747E+15 |
| 2015-04-26 | 7:30:00 | 1.774067885 | ['Maftet'] | 1.13E+16 | -3.84798E+15 |
| 2015-04-26 | 8:00:00 | 1.773911887 | ['Nut'] | 7.03E+15 | -2.40497E+15 |
| 2015-04-26 | 8:30:00 | 1.773757885 | ['Anuket'] | 8.04E+15 | -2.74823E+15 |
| 2015-04-26 | 9:00:00 | 1.773603396 | ['Anuket'] | 1.23E+16 | -4.20536E+15 |
| 2015-04-26 | 9:30:00 | 1.7734425 | ['Seth'] | 2.09E+16 | -7.1266E+15 |
| 2015-04-26 | 10:00:00 | 1.773281887 | ['Seth'] | 3.58E+16 | -1.22E+16 |
| 2015-04-26 | 10:30:00 | 1.773126038 | ['Anubis'] | 3.41E+16 | -1.17E+16 |
| 2015-04-26 | 11:30:00 | 1.772769091 | ['Anubis'] | 1.14E+16 | -3.94102E+15 |
| 2015-04-26 | 12:00:00 | 1.772658654 | ['Atum'] | 6.94E+15 | -2.37411E+15 |
| 2015-04-26 | 12:30:00 | 1.772504808 | ['Imhotep1'] | 8.35E+15 | -2.8537E+15 |
| 2015-04-26 | 13:00:00 | 1.772350755 | ['Imhotep'] | 1.67E+16 | -5.69878E+15 |
| 2015-04-26 | 14:00:00 | 1.772039623 | ['Imhotep'] | 2.68E+16 | -9.14866E+15 |
| 2015-04-26 | 14:30:00 | 1.771878214 | ['Imhotep'] | 2.73E+16 | -9.32553E+15 |
| 2015-04-26 | 15:00:00 | 1.771718113 | ['Imhotep'] | 2.65E+16 | -9.04179E+15 |
| 2015-04-26 | 15:30:00 | 1.771563462 | ['Khepry'] | 2.80E+16 | -9.56781E+15 |
| 2015-04-26 | 16:00:00 | 1.771409434 | ['Khepry'] | 2.46E+16 | -8.39793E+15 |
| 2015-04-26 | 16:30:00 | 1.771253962 | ['Khepry'] | 2.34E+16 | -7.99975E+15 |
| 2015-04-26 | 17:00:00 | 1.771097925 | ['Aker'] | 2.14E+16 | -7.29907E+15 |
| 2015-04-26 | 17:30:00 | 1.770944231 | ['Bastet'] | 2.49E+16 | -8.50578E+15 |
| 2015-04-26 | 18:00:00 | 1.770789623 | ['Sobek'] | 1.96E+16 | -6.69298E+15 |
| 2015-04-26 | 18:30:00 | 1.770632778 | ['Bastet'] | 9.87E+15 | -3.38885E+15 |
| 2015-04-26 | 19:00:00 | 1.770471455 | ['Wosret'] | 9.99E+15 | -3.41297E+15 |
| 2015-04-26 | 19:30:00 | 1.770315306 | ['Wosret'] | 1.40E+16 | -4.77081E+15 |
| 2015-04-26 | 20:00:00 | 1.770156981 | ['Wosret'] | 1.00E+16 | -3.43121E+15 |
| 2015-04-26 | 20:30:00 | 1.770002885 | ['Maftet'] | 7.65E+15 | -2.62174E+15 |
| 2015-04-26 | 21:00:00 | 1.769848679 | ['Anuket'] | 7.31E+15 | -2.50189E+15 |
| 2015-04-26 | 21:30:00 | 1.769693208 | ['Hapi'] | 1.07E+16 | -3.65389E+15 |
| 2015-04-26 | 22:00:00 | 1.769537547 | ['Hapi'] | 2.23E+16 | -7.60442E+15 |
| 2015-04-26 | 22:30:00 | 1.769383077 | ['Anubis'] | 3.57E+16 | -1.22E+16 |
| 2015-04-26 | 23:00:00 | 1.769224464 | ['Anubis'] | 3.35E+16 | -1.15E+16 |
| 2015-04-26 | 23:30:00 | 1.769063585 | ['Atum'] | 1.52E+16 | -5.17896E+15 |
| 2015-04-27 | 0:00:00 | 1.768907736 | ['Atum'] | 8.77E+15 | -3.00322E+15 |
| 2015-04-27 | 0:30:00 | 1.768752264 | ['Khonsu'] | 6.05E+15 | -2.0731E+15 |
| 2015-04-27 | 1:00:00 | 1.768598824 | ['Imhotep1'] | 7.71E+15 | -2.63422E+15 |
| 2015-04-27 | 1:30:00 | 1.768441373 | ['Imhotep'] | 1.67E+16 | -5.71E+15 |
| 2015-04-27 | 2:00:00 | 1.768289231 | ['Imhotep'] | 3.02E+16 | -1.03E+16 |
| 2015-04-27 | 2:30:00 | 1.768133019 | ['Imhotep'] | 4.08E+16 | -1.39E+16 |
| 2015-04-27 | 3:00:00 | 1.767977547 | ['Imhotep'] | 3.16E+16 | -1.08E+16 |
| 2015-04-27 | 3:30:00 | 1.767817143 | ['Imhotep'] | 2.78E+16 | -9.51259E+15 |
| 2015-04-27 | 4:00:00 | 1.767660426 | ['Khepry'] | 2.30E+16 | -7.84311E+15 |
| 2015-04-27 | 4:30:00 | 1.767497755 | ['Khepry'] | 1.62E+16 | -5.53728E+15 |
| 2015-04-27 | 5:00:00 | 1.767348302 | ['Khepry'] | 1.91E+16 | -6.52671E+15 |
| 2015-04-27 | 5:30:00 | 1.767192692 | ['Aker'] | 2.09E+16 | -7.13044E+15 |
| 2015-04-27 | 7:30:00 | 1.766573962 | ['Wosret'] | 9.93E+15 | -3.39872E+15 |
| 2015-04-27 | 8:00:00 | 1.766417308 | ['Wosret'] | 9.18E+15 | -3.15217E+15 |
| 2015-04-27 | 8:30:00 | 1.766250962 | ['Wosret'] | 8.85E+15 | -3.03034E+15 |
| 2015-04-27 | 9:00:00 | 1.766096981 | ['Wosret'] | 7.61E+15 | -2.6057E+15 |
| 2015-04-27 | 9:30:00 | 1.765943077 | ['Hapi'] | 8.92E+15 | -3.04814E+15 |
| 2015-04-27 | 10:00:00 | 1.765789245 | ['Seth'] | 1.30E+16 | -4.45601E+15 |
| 2015-04-27 | 10:30:00 | 1.765633774 | ['Seth'] | 2.37E+16 | -8.0812E+15 |
| 2015-04-27 | 11:00:00 | 1.765478302 | ['Anubis'] | 3.57E+16 | -1.22E+16 |
| 2015-04-27 | 11:30:00 | 1.765324902 | ['Atum'] | 3.09E+16 | -1.05E+16 |
| 2015-04-27 | 12:30:00 | 1.765006415 | ['Khonsu'] | 7.35E+15 | -2.5254E+15 |
| 2015-04-27 | 13:00:00 | 1.764853673 | ['Khonsu'] | 6.64E+15 | -2.27301E+15 |
| 2015-04-27 | 13:30:00 | 1.764694151 | ['Imhotep'] | 9.18E+15 | -3.14077E+15 |
| 2015-04-27 | 14:00:00 | 1.764538868 | ['Imhotep'] | 1.56E+16 | -5.32042E+15 |
| 2015-04-27 | 14:30:00 | 1.764384808 | ['Imhotep'] | 2.56E+16 | -8.74579E+15 |
| 2015-04-27 | 15:00:00 | 1.764231346 | ['Imhotep'] | 4.57E+16 | -1.56E+16 |
| 2015-04-27 | 15:30:00 | 1.764075472 | ['Imhotep'] | 3.75E+16 | -1.28E+16 |
| 2015-04-27 | 16:00:00 | 1.763919811 | ['Imhotep'] | 2.37E+16 | -8.10939E+15 |
| 2015-04-27 | 17:00:00 | 1.763600741 | ['Bes'] | 1.17E+16 | -4.01369E+15 |
| 2015-04-27 | 17:30:00 | 1.763444118 | ['Anhur'] | 1.59E+16 | -5.42629E+15 |
| 2015-04-27 | 20:00:00 | 1.762606 | ['Wosret'] | 1.27E+16 | -4.35942E+15 |
| 2015-04-27 | 20:30:00 | 1.762517736 | ['Wosret'] | 9.35E+15 | -3.19845E+15 |
| 2015-04-27 | 21:00:00 | 1.7623575 | ['Sobek'] | 6.94E+15 | -2.37475E+15 |
| 2015-04-27 | 21:30:00 | 1.762196981 | ['Hapi'] | 8.30E+15 | -2.83733E+15 |
| 2015-04-27 | 22:00:00 | 1.762042745 | ['Seth'] | 9.80E+15 | -3.35099E+15 |
| 2015-04-27 | 22:30:00 | 1.761889245 | ['Seth'] | 1.52E+16 | -5.18759E+15 |
| 2015-04-27 | 23:00:00 | 1.761734423 | ['Anubis'] | 2.30E+16 | -7.85108E+15 |
| 2015-04-27 | 23:30:00 | 1.761578868 | ['Atum'] | 3.25E+16 | -1.11E+16 |
| 2015-04-28 | 0:00:00 | 1.761425192 | ['Khonsu'] | 2.35E+16 | -8.0226E+15 |
| 2015-04-28 | 0:30:00 | 1.761271132 | ['Khonsu'] | 1.13E+16 | -3.89362E+15 |
| 2015-04-28 | 1:00:00 | 1.761116038 | ['Khonsu'] | 5.70E+15 | -1.97311E+15 |
| 2015-04-28 | 1:30:00 | 1.760955536 | ['Khonsu'] | 6.74E+15 | -2.31763E+15 |
| 2015-04-28 | 2:00:00 | 1.760795472 | ['Imhotep1'] | 9.71E+15 | -3.31896E+15 |
| 2015-04-28 | 2:30:00 | 1.760640189 | ['Imhotep'] | 1.44E+16 | -4.91781E+15 |
| 2015-04-28 | 3:00:00 | 1.760486346 | ['Imhotep'] | 2.45E+16 | -8.35891E+15 |
| 2015-04-28 | 3:30:00 | 1.760333846 | ['Imhotep'] | 4.29E+16 | -1.46E+16 |
| 2015-04-28 | 4:00:00 | 1.760177547 | ['Bes'] | 3.67E+16 | -1.25E+16 |
| 2015-04-28 | 4:30:00 | 1.760022264 | ['Bes'] | 2.28E+16 | -7.78695E+15 |
| 2015-04-28 | 5:00:00 | 1.759868462 | ['Bes'] | 1.86E+16 | -6.36501E+15 |
| 2015-04-28 | 5:30:00 | 1.759713333 | ['Anhur'] | 1.10E+16 | -3.75206E+15 |
| 2015-04-28 | 8:00:00 | 1.758930392 | ['Sobek'] | 2.82E+16 | -9.64016E+15 |
| 2015-04-28 | 8:30:00 | 1.758776604 | ['Geb'] | 2.18E+16 | -7.43774E+15 |
| 2015-04-28 | 9:00:00 | 1.758621321 | ['Geb'] | 1.03E+16 | -3.52047E+15 |
| 2015-04-28 | 9:30:00 | 1.758466415 | ['Geb'] | 9.24E+15 | -3.15648E+15 |
| 2015-04-28 | 10:00:00 | 1.758308 | ['Geb'] | 1.04E+16 | -3.55386E+15 |
| 2015-04-28 | 10:30:00 | 1.758147593 | ['Geb'] | 1.32E+16 | -4.5025E+15 |
| 2015-04-28 | 11:00:00 | 1.757992308 | ['Atum'] | 1.66E+16 | -5.67331E+15 |
| 2015-04-28 | 11:30:00 | 1.757838868 | ['Atum'] | 2.52E+16 | -8.60331E+15 |
| 2015-04-28 | 12:00:00 | 1.757684615 | ['Khonsu'] | 3.07E+16 | -1.05E+16 |
| 2015-04-28 | 13:30:00 | 1.757221731 | ['Khonsu'] | 8.01E+15 | -2.73866E+15 |
| 2015-04-28 | 14:00:00 | 1.757066415 | ['Imhotep'] | 8.14E+15 | -2.78475E+15 |
| 2015-04-28 | 14:30:00 | 1.75690625 | ['Imhotep'] | 1.02E+16 | -3.47654E+15 |
| 2015-04-28 | 15:00:00 | 1.756745849 | ['Imhotep'] | 1.55E+16 | -5.27762E+15 |
| 2015-04-28 | 15:30:00 | 1.756591132 | ['Bes'] | 2.52E+16 | -8.61459E+15 |
| 2015-04-28 | 16:00:00 | 1.756437308 | ['Bes'] | 4.32E+16 | -1.47E+16 |
| 2015-04-28 | 16:30:00 | 1.75633087 | ['Bes'] | 3.89E+16 | -1.33E+16 |
| 2015-04-28 | 17:30:00 | 1.75591625 | ['Bes'] | 1.26E+16 | -4.34125E+15 |
| 2015-04-28 | 18:00:00 | 1.755820192 | ['Sobek'] | 1.23E+16 | -4.21101E+15 |
| 2015-04-28 | 21:00:00 | 1.754883077 | ['Hapi'] | 2.62E+16 | -8.94458E+15 |
| 2015-04-28 | 21:30:00 | 1.754729434 | ['Geb'] | 1.36E+16 | -4.66276E+15 |
| 2015-04-28 | 22:00:00 | 1.754574528 | ['Geb'] | 1.23E+16 | -4.21518E+15 |
| 2015-04-28 | 22:30:00 | 1.754419245 | ['Geb'] | 1.34E+16 | -4.58241E+15 |
| 2015-04-28 | 23:00:00 | 1.754264528 | ['Atum'] | 1.34E+16 | -4.59533E+15 |
| 2015-04-28 | 23:30:00 | 1.754102909 | ['Atum'] | 1.70E+16 | -5.79921E+15 |
| 2015-04-29 | 2:30:00 | 1.753176471 | ['Atum'] | 8.69E+15 | -2.97601E+15 |
| 2015-04-29 | 3:00:00 | 1.753018723 | ['Imhotep'] | 1.28E+16 | -4.38845E+15 |
| 2015-04-29 | 3:30:00 | 1.752861429 | ['Imhotep'] | 2.20E+16 | -7.50018E+15 |
| 2015-04-29 | 4:00:00 | 1.752703542 | ['Imhotep'] | 3.42E+16 | -1.17E+16 |
| 2015-04-29 | 4:30:00 | 1.75256125 | ['Imhotep'] | 4.49E+16 | -1.53E+16 |
| 2015-04-29 | 5:30:00 | 1.752206333 | ['Hatmehit'] | 1.64E+16 | -5.60827E+15 |
| 2015-04-29 | 6:00:00 | 1.75208125 | ['Khepry'] | 1.28E+16 | -4.36506E+15 |
| 2015-04-29 | 6:30:00 | 1.751928302 | ['Khepry'] | 9.69E+15 | -3.31197E+15 |
| 2015-04-29 | 7:00:00 | 1.75177449 | ['Aker'] | 1.31E+16 | -4.48864E+15 |
| 2015-04-29 | 7:30:00 | 1.751615111 | ['Aker'] | 1.65E+16 | -5.6476E+15 |
| 2015-04-29 | 8:00:00 | 1.751462273 | ['Sobek'] | 2.13E+16 | -7.28915E+15 |
| 2015-04-29 | 8:30:00 | 1.751312444 | ['Neith'] | 2.87E+16 | -9.81294E+15 |
| 2015-04-29 | 9:00:00 | 1.751150816 | ['Wosret'] | 3.41E+16 | -1.16E+16 |
| 2015-04-29 | 9:30:00 | 1.750991702 | ['Wosret'] | 2.22E+16 | -7.59822E+15 |
| 2015-04-29 | 10:00:00 | 1.750832979 | ['Wosret'] | 1.17E+16 | -3.99691E+15 |
| 2015-04-29 | 10:30:00 | 1.750681373 | ['Wosret'] | 8.77E+15 | -3.00664E+15 |
| 2015-04-29 | 11:00:00 | 1.750528113 | ['Neith'] | 7.48E+15 | -2.55677E+15 |
| 2015-04-29 | 11:30:00 | 1.750374118 | ['Anuket'] | 1.18E+16 | -4.04441E+15 |
| 2015-04-29 | 12:00:00 | 1.750222917 | ['Seth'] | 2.03E+16 | -6.94186E+15 |
| 2015-04-29 | 12:30:00 | 1.750064898 | ['Seth'] | 3.09E+16 | -1.05E+16 |
| 2015-04-29 | 13:00:00 | 1.749911633 | ['Anubis'] | 2.90E+16 | -9.89459E+15 |
| 2015-04-29 | 14:00:00 | 1.749593333 | ['Atum'] | 8.32E+15 | -2.86494E+15 |
| 2015-04-29 | 14:30:00 | 1.749437843 | ['Khonsu'] | 9.65E+15 | -3.2997E+15 |
| 2015-04-29 | 15:00:00 | 1.749282 | ['Imhotep1'] | 1.02E+16 | -3.49273E+15 |
| 2015-04-29 | 15:30:00 | 1.74913 | ['Imhotep'] | 1.56E+16 | -5.32914E+15 |
| 2015-04-29 | 16:00:00 | 1.7489792 | ['Imhotep'] | 2.42E+16 | -8.26885E+15 |
| 2015-04-29 | 17:00:00 | 1.748670192 | ['Imhotep'] | 3.63E+16 | -1.24E+16 |
| 2015-04-29 | 17:30:00 | 1.748516 | ['Imhotep'] | 2.44E+16 | -8.33778E+15 |
| 2015-04-29 | 18:00:00 | 1.748354821 | ['Khepry'] | 1.63E+16 | -5.55238E+15 |
| 2015-04-29 | 18:30:00 | 1.74819566 | ['Khepry'] | 1.32E+16 | -4.5182E+15 |
| 2015-04-29 | 19:00:00 | 1.748040755 | ['Khepry'] | 1.13E+16 | -3.85957E+15 |
| 2015-04-29 | 19:30:00 | 1.74789 | ['Aker'] | 1.38E+16 | -4.69936E+15 |
| 2015-04-29 | 20:00:00 | 1.747734314 | ['Sobek'] | 2.20E+16 | -7.5241E+15 |
| 2015-04-29 | 20:30:00 | 1.74758 | ['Neith'] | 3.04E+16 | -1.04E+16 |
| 2015-04-29 | 21:00:00 | 1.747427708 | ['Neith'] | 3.00E+16 | -1.03E+16 |
| 2015-04-29 | 21:30:00 | 1.747272353 | ['Wosret'] | 3.69E+16 | -1.26E+16 |
| 2015-04-29 | 22:30:00 | 1.746956909 | ['Wosret'] | 9.43E+15 | -3.2207E+15 |
| 2015-04-29 | 23:00:00 | 1.746799245 | ['Wosret'] | 8.69E+15 | -2.96846E+15 |
| 2015-04-29 | 23:30:00 | 1.746643462 | ['Anuket'] | 1.01E+16 | -3.43987E+15 |
| 2015-04-30 | 0:00:00 | 1.746491538 | ['Hapi'] | 1.59E+16 | -5.43652E+15 |
| 2015-04-30 | 0:30:00 | 1.746338302 | ['Seth'] | 2.63E+16 | -8.98582E+15 |

| Date | Time | Value | Label | Col5 | Col6 |
|---|---|---|---|---|---|
| 2015-04-30 | 1:00:00 | 1.746183774 | ['Anubis'] | 3.13E+16 | -1.07E+16 |
| 2015-04-30 | 1:30:00 | 1.746028868 | ['Atum'] | 2.23E+16 | -7.62354E+15 |
| 2015-04-30 | 2:00:00 | 1.745876154 | ['Atum'] | 1.22E+16 | -4.17502E+15 |
| 2015-04-30 | 2:30:00 | 1.745717857 | ['Khonsu'] | 1.04E+16 | -3.57192E+15 |
| 2015-04-30 | 3:00:00 | 1.74555902 | ['Khonsu'] | 9.57E+15 | -3.27195E+15 |
| 2015-04-30 | 3:30:00 | 1.745403846 | ['Imhotep'] | 1.14E+16 | -3.90668E+15 |
| 2015-04-30 | 4:00:00 | 1.745248868 | ['Imhotep'] | 1.65E+16 | -5.62418E+15 |
| 2015-04-30 | 4:30:00 | 1.74509434 | ['Imhotep'] | 2.64E+16 | -9.01768E+15 |
| 2015-04-30 | 5:00:00 | 1.744941538 | ['Imhotep'] | 3.29E+16 | -1.12E+16 |
| 2015-04-30 | 5:30:00 | 1.744788113 | ['Imhotep'] | 2.95E+16 | -1.01E+16 |
| 2015-04-30 | 6:00:00 | 1.744633774 | ['Imhotep'] | 2.05E+16 | -7.02012E+15 |
| 2015-04-30 | 6:30:00 | 1.744479057 | ['Khepry'] | 2.01E+16 | -6.86279E+15 |
| 2015-04-30 | 7:00:00 | 1.74432 | ['Bes'] | 1.28E+16 | -4.38783E+15 |
| 2015-04-30 | 7:30:00 | 1.744160189 | ['Aker'] | 1.41E+16 | -4.81815E+15 |
| 2015-04-30 | 8:00:00 | 1.744007115 | ['Aker'] | 2.29E+16 | -7.81322E+15 |
| 2015-04-30 | 8:30:00 | 1.743853962 | ['Sobek'] | 3.37E+16 | -1.15E+16 |
| 2015-04-30 | 9:00:00 | 1.743699623 | ['Neith'] | 3.21E+16 | -1.10E+16 |
| 2015-04-30 | 10:30:00 | 1.743238868 | ['Wosret'] | 1.32E+16 | -4.51949E+15 |
| 2015-04-30 | 11:00:00 | 1.743084717 | ['Wosret'] | 1.00E+16 | -3.42737E+15 |
| 2015-04-30 | 11:30:00 | 1.742924821 | ['Anuket'] | 1.10E+16 | -3.76708E+15 |
| 2015-04-30 | 12:00:00 | 1.74276566 | ['Hapi'] | 1.55E+16 | -5.29712E+15 |
| 2015-04-30 | 12:30:00 | 1.742611321 | ['Seth'] | 2.40E+16 | -8.19288E+15 |
| 2015-04-30 | 13:00:00 | 1.742458077 | ['Anubis'] | 3.02E+16 | -1.03E+16 |
| 2015-04-30 | 13:30:00 | 1.742305283 | ['Anubis'] | 2.45E+16 | -8.36004E+15 |
| 2015-04-30 | 14:00:00 | 1.742150943 | ['Atum'] | 1.71E+16 | -5.83524E+15 |
| 2015-04-30 | 14:30:00 | 1.741996604 | ['Khonsu'] | 1.21E+16 | -4.1388E+15 |
| 2015-04-30 | 15:00:00 | 1.741843654 | ['Khonsu'] | 9.82E+15 | -3.35963E+15 |
| 2015-04-30 | 15:30:00 | 1.741689259 | ['Imhotep1'] | 9.34E+15 | -3.19015E+15 |
| 2015-04-30 | 16:00:00 | 1.741529273 | ['Imhotep'] | 1.23E+16 | -4.19337E+15 |
| 2015-04-30 | 16:30:00 | 1.741371698 | ['Imhotep'] | 1.85E+16 | -6.32092E+15 |
| 2015-04-30 | 17:00:00 | 1.741217692 | ['Imhotep'] | 2.49E+16 | -8.50529E+15 |
| 2015-04-30 | 17:30:00 | 1.741064314 | ['Imhotep'] | 2.49E+16 | -8.51424E+15 |
| 2015-04-30 | 18:00:00 | 1.740909216 | ['Imhotep'] | 2.29E+16 | -7.82706E+15 |
| 2015-04-30 | 18:30:00 | 1.740757885 | ['Khepry'] | 2.20E+16 | -7.524E+15 |
| 2015-04-30 | 19:00:00 | 1.740621707 | ['Bes'] | 2.03E+16 | -6.95205E+15 |
| 2015-04-30 | 20:30:00 | 1.740131481 | ['Sobek'] | 2.72E+16 | -9.29189E+15 |
| 2015-04-30 | 21:00:00 | 1.739977115 | ['Neith'] | 3.02E+16 | -1.03E+16 |
| 2015-04-30 | 23:30:00 | 1.739211538 | ['Anuket'] | 1.30E+16 | -4.43847E+15 |
| 2015-05-01 | 0:00:00 | 1.739056038 | ['Anuket'] | 1.74E+16 | -5.95051E+15 |
| 2015-05-01 | 0:30:00 | 1.738896607 | ['Seth'] | 2.36E+16 | -8.05624E+15 |
| 2015-05-01 | 1:00:00 | 1.738737358 | ['Anubis'] | 2.64E+16 | -9.02235E+15 |
| 2015-05-01 | 1:30:00 | 1.738583019 | ['Anubis'] | 2.67E+16 | -9.1267E+15 |
| 2015-05-01 | 2:00:00 | 1.738430192 | ['Atum'] | 2.01E+16 | -6.85134E+15 |
| 2015-05-01 | 2:30:00 | 1.738276538 | ['Khonsu'] | 1.49E+16 | -5.09827E+15 |

| Date | Time | Value | Name | Col5 | Col6 |
|---|---|---|---|---|---|
| 2015-05-01 | 3:00:00 | 1.738123396 | ['Khonsu'] | 1.11E+16 | -3.79341E+15 |
| 2015-05-01 | 3:30:00 | 1.737968868 | ['Khonsu'] | 8.24E+15 | -2.81715E+15 |
| 2015-05-01 | 4:00:00 | 1.737816346 | ['Imhotep'] | 1.03E+16 | -3.52457E+15 |
| 2015-05-01 | 4:30:00 | 1.737663396 | ['Imhotep'] | 1.44E+16 | -4.91787E+15 |
| 2015-05-01 | 5:00:00 | 1.737504107 | ['Imhotep'] | 1.75E+16 | -5.96221E+15 |
| 2015-05-01 | 5:30:00 | 1.737345283 | ['Imhotep'] | 2.05E+16 | -6.98646E+15 |
| 2015-05-01 | 6:00:00 | 1.737190943 | ['Imhotep'] | 2.15E+16 | -7.35816E+15 |
| 2015-05-01 | 6:30:00 | 1.737036604 | ['Khepry'] | 2.20E+16 | -7.49656E+15 |
| 2015-05-01 | 7:00:00 | 1.736884038 | ['Bes'] | 2.04E+16 | -6.96099E+15 |
| 2015-05-01 | 7:30:00 | 1.736731509 | ['Bes'] | 1.75E+16 | -5.96515E+15 |
| 2015-05-01 | 8:00:00 | 1.73657717 | ['Anhur'] | 1.81E+16 | -6.19552E+15 |
| 2015-05-01 | 8:30:00 | 1.736423019 | ['Sobek'] | 2.11E+16 | -7.20283E+15 |
| 2015-05-01 | 9:00:00 | 1.736269808 | ['Sobek'] | 2.68E+16 | -9.16633E+15 |
| 2015-05-01 | 9:30:00 | 1.736105 | ['Neith'] | 2.79E+16 | -9.53501E+15 |
| 2015-05-01 | 12:00:00 | 1.735338462 | ['Anuket'] | 1.98E+16 | -6.75893E+15 |
| 2015-05-01 | 12:30:00 | 1.735185849 | ['Seth'] | 2.56E+16 | -8.74203E+15 |
| 2015-05-01 | 13:00:00 | 1.735031509 | ['Seth'] | 2.37E+16 | -8.07768E+15 |
| 2015-05-01 | 13:30:00 | 1.734874364 | ['Anubis'] | 2.25E+16 | -7.67479E+15 |
| 2015-05-01 | 14:00:00 | 1.73471463 | ['Atum'] | 1.99E+16 | -6.79459E+15 |
| 2015-05-01 | 14:30:00 | 1.734560577 | ['Khonsu'] | 1.69E+16 | -5.78734E+15 |
| 2015-05-01 | 15:00:00 | 1.734407925 | ['Khonsu'] | 1.25E+16 | -4.26946E+15 |
| 2015-05-01 | 15:30:00 | 1.734253962 | ['Khonsu'] | 7.77E+15 | -2.65663E+15 |
| 2015-05-01 | 16:00:00 | 1.734099038 | ['Imhotep'] | 8.53E+15 | -2.91544E+15 |
| 2015-05-01 | 16:30:00 | 1.733948039 | ['Imhotep'] | 1.22E+16 | -4.18307E+15 |
| 2015-05-01 | 17:00:00 | 1.733794906 | ['Imhotep'] | 1.50E+16 | -5.13528E+15 |
| 2015-05-01 | 17:30:00 | 1.733640755 | ['Imhotep'] | 1.67E+16 | -5.71752E+15 |
| 2015-05-01 | 18:00:00 | 1.733481964 | ['Imhotep'] | 1.86E+16 | -6.3519E+15 |
| 2015-05-01 | 18:30:00 | 1.733322642 | ['Khepry'] | 2.16E+16 | -7.38171E+15 |
| 2015-05-01 | 19:30:00 | 1.733015962 | ['Bes'] | 1.75E+16 | -5.99686E+15 |
| 2015-05-01 | 20:00:00 | 1.732864038 | ['Anhur'] | 1.76E+16 | -6.00608E+15 |
| 2015-05-01 | 20:30:00 | 1.732709623 | ['Sobek'] | 2.10E+16 | -7.18923E+15 |
| 2015-05-01 | 21:00:00 | 1.732556731 | ['Neith'] | 2.09E+16 | -7.15373E+15 |
| 2015-05-01 | 21:30:00 | 1.732403462 | ['Neith'] | 2.70E+16 | -9.22524E+15 |
| 2015-05-01 | 22:30:00 | 1.732092545 | ['Wosret'] | 2.36E+16 | -8.07278E+15 |
| 2015-05-02 | 0:30:00 | 1.7314725 | ['Bastet'] | 1.55E+16 | -5.28978E+15 |
| 2015-05-02 | 1:00:00 | 1.731320377 | ['Wosret'] | 2.04E+16 | -6.96889E+15 |
| 2015-05-02 | 1:30:00 | 1.731166415 | ['Neith'] | 2.07E+16 | -7.08516E+15 |
| 2015-05-02 | 2:00:00 | 1.731012453 | ['Geb'] | 1.53E+16 | -5.22151E+15 |
| 2015-05-02 | 4:30:00 | 1.730234808 | ['Imhotep'] | 1.06E+16 | -3.63672E+15 |
| 2015-05-02 | 5:00:00 | 1.730082264 | ['Imhotep'] | 1.27E+16 | -4.32334E+15 |
| 2015-05-02 | 5:30:00 | 1.729928824 | ['Imhotep'] | 1.56E+16 | -5.33685E+15 |
| 2015-05-02 | 6:00:00 | 1.729777736 | ['Imhotep'] | 1.68E+16 | -5.72531E+15 |
| 2015-05-02 | 6:30:00 | 1.729622308 | ['Bes'] | 1.71E+16 | -5.85506E+15 |
| 2015-05-02 | 7:00:00 | 1.729466909 | ['Bes'] | 1.67E+16 | -5.69018E+15 |

| Date | Time | Value | Label | E1 | E2 |
|---|---|---|---|---|---|
| 2015-05-02 | 7:30:00 | 1.729307593 | ['Bes'] | 1.60E+16 | -5.47677E+15 |
| 2015-05-02 | 8:00:00 | 1.729152264 | ['Anhur'] | 1.20E+16 | -4.08989E+15 |
| 2015-05-02 | 8:30:00 | 1.729 | ['Anhur'] | 1.36E+16 | -4.64321E+15 |
| 2015-05-02 | 9:00:00 | 1.728847547 | ['Anhur'] | 1.93E+16 | -6.58034E+15 |
| 2015-05-02 | 9:30:00 | 1.728693585 | ['Sobek'] | 2.49E+16 | -8.50673E+15 |
| 2015-05-02 | 12:00:00 | 1.727918462 | ['Sobek'] | 1.69E+16 | -5.78051E+15 |
| 2015-05-02 | 12:30:00 | 1.727763962 | ['Anuket'] | 1.90E+16 | -6.4736E+15 |
| 2015-05-02 | 13:00:00 | 1.727610377 | ['Hapi'] | 1.77E+16 | -6.03727E+15 |
| 2015-05-02 | 13:30:00 | 1.727458235 | ['Seth'] | 1.46E+16 | -4.98217E+15 |
| 2015-05-02 | 14:30:00 | 1.727150769 | ['Atum'] | 1.08E+16 | -3.69724E+15 |
| 2015-05-02 | 15:00:00 | 1.726998302 | ['Khonsu'] | 1.03E+16 | -3.53115E+15 |
| 2015-05-02 | 16:00:00 | 1.726687368 | ['Imhotep1'] | 9.57E+15 | -3.2715E+15 |
| 2015-05-02 | 16:30:00 | 1.726528846 | ['Imhotep'] | 1.13E+16 | -3.85185E+15 |
| 2015-05-02 | 17:00:00 | 1.726376604 | ['Imhotep'] | 1.29E+16 | -4.41822E+15 |
| 2015-05-02 | 17:30:00 | 1.726223208 | ['Imhotep'] | 1.62E+16 | -5.5283E+15 |
| 2015-05-02 | 18:00:00 | 1.726069808 | ['Imhotep'] | 1.43E+16 | -4.87024E+15 |
| 2015-05-02 | 18:30:00 | 1.725918039 | ['Bes'] | 1.50E+16 | -5.1387E+15 |
| 2015-05-02 | 19:00:00 | 1.725764615 | ['Bes'] | 1.56E+16 | -5.342E+15 |
| 2015-05-02 | 19:30:00 | 1.725611321 | ['Bes'] | 1.40E+16 | -4.77479E+15 |
| 2015-05-02 | 20:00:00 | 1.725458113 | ['Bes'] | 1.19E+16 | -4.06994E+15 |
| 2015-05-02 | 20:30:00 | 1.72529875 | ['Anhur'] | 1.46E+16 | -5.00364E+15 |
| 2015-05-02 | 21:00:00 | 1.725139216 | ['Anhur'] | 2.01E+16 | -6.87001E+15 |
| 2015-05-02 | 21:30:00 | 1.724988846 | ['Sobek'] | 2.49E+16 | -8.52376E+15 |
| 2015-05-02 | 22:00:00 | 1.724836226 | ['Sobek'] | 2.45E+16 | -8.38723E+15 |
| 2015-05-02 | 23:00:00 | 1.724529434 | ['Wosret'] | 1.46E+16 | -5.01075E+15 |
| 2015-05-03 | 0:00:00 | 1.724225472 | ['Sobek'] | 1.65E+16 | -5.66447E+15 |
| 2015-05-03 | 0:30:00 | 1.724068545 | ['Anuket'] | 1.72E+16 | -5.86243E+15 |
| 2015-05-03 | 1:00:00 | 1.72390963 | ['Hapi'] | 1.50E+16 | -5.11231E+15 |
| 2015-05-03 | 1:30:00 | 1.723754717 | ['Hapi'] | 1.21E+16 | -4.15212E+15 |
| 2015-05-03 | 2:00:00 | 1.723601132 | ['Anubis'] | 1.05E+16 | -3.59845E+15 |
| 2015-05-03 | 3:00:00 | 1.723296981 | ['Khonsu'] | 1.03E+16 | -3.53652E+15 |
| 2015-05-03 | 3:30:00 | 1.723143774 | ['Khonsu'] | 8.84E+15 | -3.03434E+15 |
| 2015-05-03 | 4:30:00 | 1.722838077 | ['Imhotep'] | 1.33E+16 | -4.54963E+15 |
| 2015-05-03 | 5:00:00 | 1.72268 | ['Imhotep'] | 1.47E+16 | -5.01822E+15 |
| 2015-05-03 | 5:30:00 | 1.722521346 | ['Imhotep'] | 1.39E+16 | -4.76264E+15 |
| 2015-05-03 | 6:00:00 | 1.722369245 | ['Imhotep'] | 1.38E+16 | -4.72254E+15 |
| 2015-05-03 | 6:30:00 | 1.722215472 | ['Imhotep'] | 1.38E+16 | -4.71733E+15 |
| 2015-05-03 | 7:00:00 | 1.722062453 | ['Bes'] | 1.48E+16 | -5.05725E+15 |
| 2015-05-03 | 7:30:00 | 1.721946207 | ['Bes'] | 1.43E+16 | -4.90206E+15 |
| 2015-05-03 | 9:00:00 | 1.721452885 | ['Anhur'] | 2.27E+16 | -7.76322E+15 |
| 2015-05-03 | 9:30:00 | 1.721294561 | ['Anhur'] | 3.16E+16 | -1.08E+16 |
| 2015-05-03 | 11:30:00 | 1.720630952 | ['Anhur'] | 1.29E+16 | -4.47122E+15 |
| 2015-05-03 | 12:30:00 | 1.720371321 | ['Sobek'] | 1.77E+16 | -6.06133E+15 |
| 2015-05-03 | 13:00:00 | 1.720218113 | ['Anuket'] | 1.46E+16 | -4.97769E+15 |

| Date | Time | Value | Region | Col5 | Col6 |
|---|---|---|---|---|---|
| 2015-05-03 | 14:30:00 | 1.719755893 | ['Anubis'] | 9.34E+15 | -3.19462E+15 |
| 2015-05-03 | 15:00:00 | 1.719598113 | ['Khonsu'] | 8.74E+15 | -2.9916E+15 |
| 2015-05-03 | 15:30:00 | 1.719444528 | ['Khonsu'] | 6.33E+15 | -2.1659E+15 |
| 2015-05-03 | 16:00:00 | 1.719290192 | ['Khonsu'] | 1.23E+16 | -4.1893E+15 |
| 2015-05-03 | 16:30:00 | 1.719139038 | ['Imhotep1'] | 1.59E+16 | -5.42744E+15 |
| 2015-05-03 | 17:00:00 | 1.71898717 | ['Imhotep'] | 1.66E+16 | -5.68518E+15 |
| 2015-05-03 | 17:30:00 | 1.718833962 | ['Imhotep'] | 1.59E+16 | -5.43062E+15 |
| 2015-05-03 | 18:00:00 | 1.718680566 | ['Imhotep'] | 1.44E+16 | -4.90567E+15 |
| 2015-05-03 | 18:30:00 | 1.718525926 | ['Imhotep'] | 1.27E+16 | -4.33908E+15 |
| 2015-05-03 | 19:00:00 | 1.718367593 | ['Imhotep'] | 1.51E+16 | -5.15709E+15 |
| 2015-05-03 | 19:30:00 | 1.718211569 | ['Bes'] | 1.31E+16 | -4.48679E+15 |
| 2015-05-03 | 20:00:00 | 1.718060755 | ['Bes'] | 1.50E+16 | -5.13861E+15 |
| 2015-05-03 | 20:30:00 | 1.717907358 | ['Anhur'] | 2.16E+16 | -7.37729E+15 |
| 2015-05-03 | 21:00:00 | 1.717755686 | ['Anhur'] | 2.88E+16 | -9.82621E+15 |
| 2015-05-03 | 21:30:00 | 1.717603962 | ['Anhur'] | 4.15E+16 | -1.42E+16 |
| 2015-05-03 | 22:00:00 | 1.717450566 | ['Sobek'] | 3.92E+16 | -1.34E+16 |
| 2015-05-03 | 23:00:00 | 1.717139821 | ['Wosret'] | 1.52E+16 | -5.23086E+15 |
| 2015-05-03 | 23:30:00 | 1.716981132 | ['Wosret'] | 1.04E+16 | -3.59234E+15 |
| 2015-05-04 | 0:00:00 | 1.716828491 | ['Wosret'] | 1.62E+16 | -5.55489E+15 |
| 2015-05-04 | 0:30:00 | 1.716676078 | ['Sobek'] | 2.00E+16 | -6.82556E+15 |
| 2015-05-04 | 1:00:00 | 1.716525385 | ['Anuket'] | 1.72E+16 | -5.86616E+15 |
| 2015-05-04 | 3:00:00 | 1.715915094 | ['Atum'] | 9.92E+15 | -3.40257E+15 |
| 2015-05-04 | 3:30:00 | 1.715756182 | ['Khonsu'] | 7.20E+15 | -2.46806E+15 |
| 2015-05-04 | 4:30:00 | 1.715447115 | ['Imhotep1'] | 1.92E+16 | -6.55309E+15 |
| 2015-05-04 | 5:00:00 | 1.715292692 | ['Imhotep'] | 1.92E+16 | -6.55993E+15 |
| 2015-05-04 | 6:00:00 | 1.714990392 | ['Imhotep'] | 1.55E+16 | -5.28255E+15 |
| 2015-05-04 | 6:30:00 | 1.714836415 | ['Imhotep'] | 1.52E+16 | -5.18938E+15 |
| 2015-05-04 | 7:00:00 | 1.714685577 | ['Imhotep'] | 1.87E+16 | -6.36961E+15 |
| 2015-05-04 | 7:30:00 | 1.714533585 | ['Bes'] | 1.57E+16 | -5.35714E+15 |
| 2015-05-04 | 8:00:00 | 1.714373684 | ['Bes'] | 1.45E+16 | -4.95467E+15 |
| 2015-05-04 | 8:30:00 | 1.714216154 | ['Bes'] | 2.48E+16 | -8.46675E+15 |
| 2015-05-04 | 9:00:00 | 1.714064528 | ['Anhur'] | 3.57E+16 | -1.22E+16 |
| 2015-05-04 | 9:30:00 | 1.713911698 | ['Anhur'] | 4.73E+16 | -1.62E+16 |
| 2015-05-04 | 10:00:00 | 1.713758302 | ['Sobek'] | 5.72E+16 | -1.95E+16 |
| 2015-05-04 | 10:30:00 | 1.713606923 | ['Neith'] | 4.04E+16 | -1.38E+16 |
| 2015-05-04 | 11:30:00 | 1.7133048 | ['Wosret'] | 1.04E+16 | -3.57035E+15 |
| 2015-05-04 | 12:00:00 | 1.713147308 | ['Wosret'] | 1.08E+16 | -3.71971E+15 |
| 2015-05-04 | 12:30:00 | 1.71299 | ['Neith'] | 1.69E+16 | -5.78287E+15 |
| 2015-05-04 | 13:00:00 | 1.712833585 | ['Anuket'] | 1.91E+16 | -6.53195E+15 |
| 2015-05-04 | 15:00:00 | 1.712224906 | ['Anubis'] | 1.06E+16 | -3.61786E+15 |
| 2015-05-04 | 15:30:00 | 1.712073846 | ['Atum'] | 8.27E+15 | -2.82958E+15 |
| 2015-05-04 | 17:00:00 | 1.711606415 | ['Imhotep'] | 2.46E+16 | -8.41501E+15 |
| 2015-05-04 | 17:30:00 | 1.711453019 | ['Imhotep'] | 2.35E+16 | -8.03422E+15 |
| 2015-05-04 | 18:00:00 | 1.711301731 | ['Imhotep'] | 2.05E+16 | -7.01391E+15 |

| Date | Time | Value | Label | Col5 | Col6 |
|---|---|---|---|---|---|
| 2015-05-04 | 18:30:00 | 1.711149038 | ['Imhotep'] | 1.67E+16 | -5.69753E+15 |
| 2015-05-04 | 19:00:00 | 1.710997547 | ['Imhotep'] | 1.83E+16 | -6.23654E+15 |
| 2015-05-04 | 19:30:00 | 1.710845094 | ['Imhotep'] | 2.00E+16 | -6.83092E+15 |
| 2015-05-04 | 20:00:00 | 1.710691887 | ['Bes'] | 1.43E+16 | -4.90282E+15 |
| 2015-05-04 | 20:30:00 | 1.710540962 | ['Bes'] | 1.98E+16 | -6.77329E+15 |
| 2015-05-04 | 21:00:00 | 1.710382982 | ['Khepry'] | 3.39E+16 | -1.16E+16 |
| 2015-05-04 | 21:30:00 | 1.710225192 | ['Aker'] | 5.17E+16 | -1.77E+16 |
| 2015-05-04 | 22:00:00 | 1.710074151 | ['Bastet'] | 6.48E+16 | -2.21E+16 |
| 2015-05-04 | 22:30:00 | 1.709921321 | ['Neith'] | 5.61E+16 | -1.92E+16 |
| 2015-05-05 | 0:30:00 | 1.709313019 | ['Wosret'] | 1.24E+16 | -4.23517E+15 |
| 2015-05-05 | 1:00:00 | 1.709161765 | ['Neith'] | 2.26E+16 | -7.72663E+15 |
| 2015-05-05 | 3:30:00 | 1.708389811 | ['Atum'] | 1.20E+16 | -4.08717E+15 |
| 2015-05-05 | 4:00:00 | 1.7082375 | ['Khonsu'] | 8.57E+15 | -2.95648E+15 |
| 2015-05-05 | 5:00:00 | 1.707934528 | ['Imhotep1'] | 1.94E+16 | -6.65122E+15 |
| 2015-05-05 | 5:30:00 | 1.70778037 | ['Imhotep'] | 2.93E+16 | -9.99676E+15 |
| 2015-05-05 | 6:00:00 | 1.707622182 | ['Imhotep'] | 2.65E+16 | -9.07012E+15 |
| 2015-05-05 | 6:30:00 | 1.707466078 | ['Imhotep'] | 2.04E+16 | -6.97902E+15 |
| 2015-05-05 | 7:00:00 | 1.707313774 | ['Imhotep'] | 1.89E+16 | -6.44478E+15 |
| 2015-05-05 | 8:00:00 | 1.707011698 | ['Khepry'] | 1.56E+16 | -5.3423E+15 |
| 2015-05-05 | 8:30:00 | 1.706859057 | ['Khepry'] | 1.80E+16 | -6.16794E+15 |
| 2015-05-05 | 9:00:00 | 1.706706226 | ['Khepry'] | 2.54E+16 | -8.6846E+15 |
| 2015-05-05 | 9:30:00 | 1.706555385 | ['Aker'] | 3.52E+16 | -1.20E+16 |
| 2015-05-05 | 10:30:00 | 1.706241698 | ['Bastet'] | 7.51E+16 | -2.56E+16 |
| 2015-05-05 | 11:00:00 | 1.706089434 | ['Neith'] | 4.90E+16 | -1.67E+16 |
| 2015-05-05 | 12:00:00 | 1.705783846 | ['Wosret'] | 1.22E+16 | -4.19223E+15 |
| 2015-05-05 | 12:30:00 | 1.705632692 | ['Wosret'] | 1.21E+16 | -4.12987E+15 |
| 2015-05-05 | 13:00:00 | 1.705507778 | ['Wosret'] | 1.91E+16 | -6.53353E+15 |
| 2015-05-05 | 16:00:00 | 1.70456 | ['Atum'] | 1.13E+16 | -3.87999E+15 |
| 2015-05-05 | 16:30:00 | 1.7044075 | ['Khonsu'] | 8.92E+15 | -3.05144E+15 |
| 2015-05-05 | 17:00:00 | 1.704256078 | ['Khonsu'] | 9.56E+15 | -3.28551E+15 |
| 2015-05-05 | 17:30:00 | 1.704105094 | ['Imhotep1'] | 3.15E+16 | -1.08E+16 |
| 2015-05-05 | 18:00:00 | 1.703952264 | ['Imhotep'] | 3.80E+16 | -1.30E+16 |
| 2015-05-05 | 18:30:00 | 1.703799811 | ['Imhotep'] | 2.93E+16 | -1.00E+16 |
| 2015-05-05 | 19:00:00 | 1.703642321 | ['Imhotep'] | 2.08E+16 | -7.10947E+15 |
| 2015-05-05 | 19:30:00 | 1.703484906 | ['Imhotep'] | 2.07E+16 | -7.07031E+15 |
| 2015-05-05 | 20:30:00 | 1.703183208 | ['Khepry'] | 1.74E+16 | -5.95052E+15 |
| 2015-05-05 | 21:00:00 | 1.703032115 | ['Khepry'] | 2.33E+16 | -7.9703E+15 |
| 2015-05-05 | 21:30:00 | 1.702878302 | ['Khepry'] | 2.59E+16 | -8.86294E+15 |
| 2015-05-05 | 23:00:00 | 1.702424038 | ['Sobek'] | 7.10E+16 | -2.42E+16 |
| 2015-05-06 | 10:00:00 | 1.699061154 | ['Anhur'] | 3.02E+16 | -1.03E+16 |
| 2015-05-06 | 10:30:00 | 1.698919348 | ['Anhur'] | 3.17E+16 | -1.08E+16 |
| 2015-05-06 | 14:30:00 | 1.697686415 | ['Maftet'] | 2.84E+16 | -9.69673E+15 |
| 2015-05-06 | 15:00:00 | 1.697553947 | ['Anuket'] | 2.53E+16 | -8.65778E+15 |
| 2015-05-06 | 19:30:00 | 1.69612 | ['Anuket'] | 2.61E+16 | -8.92227E+15 |

| Date | Time | Value | Name | Col5 | Col6 |
|---|---|---|---|---|---|
| 2015-05-06 | 20:00:00 | 1.695999811 | ['Imhotep'] | 2.36E+16 | -8.04373E+15 |
| 2015-05-06 | 20:30:00 | 1.695848113 | ['Imhotep'] | 2.56E+16 | -8.75249E+15 |
| 2015-05-07 | 6:00:00 | 1.692953208 | ['Atum'] | 9.55E+15 | -3.27614E+15 |
| 2015-05-07 | 11:30:00 | 1.691280962 | ['Anhur'] | 2.56E+16 | -8.76097E+15 |
| 2015-05-07 | 12:00:00 | 1.691130566 | ['Anhur'] | 4.78E+16 | -1.63E+16 |
| 2015-05-07 | 16:30:00 | 1.689756981 | ['Seth'] | 1.78E+16 | -6.07214E+15 |
| 2015-05-07 | 17:00:00 | 1.689605472 | ['Anubis'] | 1.47E+16 | -5.03072E+15 |
| 2015-05-07 | 20:30:00 | 1.688540385 | ['Anubis'] | 4.43E+16 | -1.52E+16 |
| 2015-05-07 | 21:00:00 | 1.688390377 | ['Hathor'] | 2.10E+16 | -7.15909E+15 |
| 2015-05-07 | 21:30:00 | 1.688238679 | ['Imhotep'] | 1.67E+16 | -5.72081E+15 |
| 2015-05-08 | 1:00:00 | 1.687172642 | ['Geb'] | 1.99E+16 | -6.82322E+15 |
| 2015-05-08 | 3:30:00 | 1.686406538 | ['Sobek'] | 5.01E+16 | -1.71E+16 |
| 2015-05-08 | 4:00:00 | 1.686256604 | ['Sobek'] | 4.19E+16 | -1.43E+16 |
| 2015-05-08 | 4:30:00 | 1.686105472 | ['Hapi'] | 3.11E+16 | -1.06E+16 |
| 2015-05-08 | 5:00:00 | 1.685954151 | ['Anubis'] | 1.83E+16 | -6.26097E+15 |
| 2015-05-08 | 5:30:00 | 1.685836452 | ['Anubis'] | 1.45E+16 | -4.95772E+15 |
| 2015-05-08 | 7:00:00 | 1.685351321 | ['Khonsu'] | 1.10E+16 | -3.77233E+15 |
| 2015-05-08 | 7:30:00 | 1.685194821 | ['Imhotep'] | 4.74E+16 | -1.62E+16 |
| 2015-05-08 | 8:00:00 | 1.685040392 | ['Imhotep'] | 5.42E+16 | -1.85E+16 |
| 2015-05-08 | 8:30:00 | 1.684890566 | ['Imhotep'] | 3.99E+16 | -1.36E+16 |
| 2015-05-08 | 9:00:00 | 1.684739245 | ['Imhotep'] | 3.67E+16 | -1.25E+16 |
| 2015-05-08 | 9:30:00 | 1.684587925 | ['Imhotep'] | 2.48E+16 | -8.48013E+15 |
| 2015-05-08 | 10:00:00 | 1.684436154 | ['Imhotep'] | 2.09E+16 | -7.14868E+15 |
| 2015-05-08 | 10:30:00 | 1.684286923 | ['Bes'] | 4.91E+16 | -1.68E+16 |
| 2015-05-08 | 11:00:00 | 1.684136981 | ['Bes'] | 4.87E+16 | -1.66E+16 |
| 2015-05-08 | 11:30:00 | 1.683982727 | ['Anhur'] | 4.23E+16 | -1.44E+16 |
| 2015-05-08 | 12:00:00 | 1.683826111 | ['Anhur'] | 1.94E+16 | -6.62025E+15 |
| 2015-05-08 | 12:30:00 | 1.683673396 | ['Anhur'] | 1.28E+16 | -4.36545E+15 |
| 2015-05-08 | 17:30:00 | 1.682126207 | ['Anhur'] | 1.86E+16 | -6.36593E+15 |
| 2015-05-08 | 18:00:00 | 1.6820175 | ['Atum'] | 1.60E+16 | -5.4784E+15 |
| 2015-05-08 | 20:30:00 | 1.68118 | ['Atum'] | 6.45E+16 | -2.21E+16 |
| 2015-05-08 | 23:00:00 | 1.680490385 | ['Bes'] | 5.31E+16 | -1.81E+16 |
| 2015-05-08 | 23:30:00 | 1.680340769 | ['Khepry'] | 4.57E+16 | -1.56E+16 |
| 2015-05-09 | 0:00:00 | 1.680189623 | ['Aker'] | 2.70E+16 | -9.22081E+15 |
| 2015-05-09 | 0:30:00 | 1.680038868 | ['Hapi'] | 1.94E+16 | -6.61152E+15 |
| 2015-05-09 | 1:00:00 | 1.679889231 | ['Bastet'] | 1.32E+16 | -4.49966E+15 |
| 2015-05-09 | 1:30:00 | 1.679734643 | ['Wosret'] | 2.50E+16 | -8.52574E+15 |
| 2015-05-09 | 3:30:00 | 1.679126154 | ['Anuket'] | 1.63E+16 | -5.5647E+15 |
| 2015-05-09 | 4:00:00 | 1.678976731 | ['Anuket'] | 4.82E+16 | -1.65E+16 |
| 2015-05-09 | 4:30:00 | 1.67882717 | ['Hapi'] | 5.28E+16 | -1.80E+16 |
| 2015-05-09 | 5:00:00 | 1.678676038 | ['Hapi'] | 4.21E+16 | -1.44E+16 |
| 2015-05-09 | 5:30:00 | 1.678526923 | ['Seth'] | 3.29E+16 | -1.12E+16 |
| 2015-05-09 | 6:00:00 | 1.678370702 | ['Anubis'] | 2.76E+16 | -9.42475E+15 |
| 2015-05-09 | 6:30:00 | 1.678213585 | ['Atum'] | 2.09E+16 | -7.12456E+15 |

| Date | Time | Value | Label | Col5 | Col6 |
|---|---|---|---|---|---|
| 2015-05-09 | 8:00:00 | 1.677765882 | ['Imhotep1'] | 2.41E+16 | -8.2226E+15 |
| 2015-05-09 | 8:30:00 | 1.6776104 | ['Imhotep'] | 5.45E+16 | -1.86E+16 |
| 2015-05-09 | 9:00:00 | 1.677463654 | ['Imhotep'] | 6.22E+16 | -2.12E+16 |
| 2015-05-09 | 9:30:00 | 1.677314151 | ['Imhotep'] | 5.40E+16 | -1.84E+16 |
| 2015-05-09 | 10:00:00 | 1.677161852 | ['Imhotep'] | 5.72E+16 | -1.95E+16 |
| 2015-05-09 | 10:30:00 | 1.677041034 | ['Imhotep'] | 4.77E+16 | -1.63E+16 |
| 2015-05-09 | 14:30:00 | 1.675797679 | ['Hatmehit'] | 2.74E+16 | -9.37008E+15 |
| 2015-05-09 | 19:30:00 | 1.674281731 | ['Atum'] | 1.37E+16 | -4.69554E+15 |
| 2015-05-09 | 23:30:00 | 1.67307 | ['Hathor'] | 5.56E+16 | -1.90E+16 |
| 2015-05-10 | 0:30:00 | 1.672771731 | ['Babi'] | 3.25E+16 | -1.11E+16 |
| 2015-05-10 | 1:00:00 | 1.672620943 | ['Aker'] | 2.32E+16 | -7.94136E+15 |
| 2015-05-10 | 4:30:00 | 1.671563846 | ['Seth'] | 3.08E+16 | -1.05E+16 |
| 2015-05-10 | 5:00:00 | 1.671414717 | ['Seth'] | 5.49E+16 | -1.88E+16 |
| 2015-05-10 | 5:30:00 | 1.67126434 | ['Anuket'] | 5.93E+16 | -2.02E+16 |
| 2015-05-10 | 6:00:00 | 1.671113962 | ['Anuket'] | 4.75E+16 | -1.62E+16 |
| 2015-05-10 | 6:30:00 | 1.670963396 | ['Hapi'] | 2.97E+16 | -1.02E+16 |
| 2015-05-10 | 7:00:00 | 1.670807857 | ['Anubis'] | 2.47E+16 | -8.4449E+15 |
| 2015-05-10 | 7:30:00 | 1.670654423 | ['Atum'] | 1.88E+16 | -6.41622E+15 |
| 2015-05-10 | 8:00:00 | 1.670505283 | ['Atum'] | 1.92E+16 | -6.56893E+15 |
| 2015-05-10 | 8:30:00 | 1.670354808 | ['Ash'] | 3.04E+16 | -1.04E+16 |
| 2015-05-10 | 9:00:00 | 1.670204717 | ['Imhotep1'] | 4.31E+16 | -1.48E+16 |
| 2015-05-10 | 9:30:00 | 1.670055769 | ['Imhotep'] | 5.70E+16 | -1.95E+16 |
| 2015-05-10 | 10:00:00 | 1.669906981 | ['Imhotep'] | 5.32E+16 | -1.82E+16 |
| 2015-05-10 | 10:30:00 | 1.669755962 | ['Imhotep'] | 6.02E+16 | -2.06E+16 |
| 2015-05-10 | 11:00:00 | 1.669660833 | ['Imhotep'] | 8.31E+16 | -2.84E+16 |
| 2015-05-10 | 11:30:00 | 1.669410385 | ['Imhotep'] | 3.70E+16 | -1.26E+16 |
| 2015-05-10 | 12:00:00 | 1.669300943 | ['Aten'] | 4.33E+16 | -1.48E+16 |
| 2015-05-10 | 12:30:00 | 1.669152115 | ['Babi'] | 4.96E+16 | -1.69E+16 |
| 2015-05-10 | 13:00:00 | 1.669003396 | ['Babi'] | 3.24E+16 | -1.11E+16 |
| 2015-05-10 | 13:30:00 | 1.66885 | ['Babi'] | 2.49E+16 | -8.48795E+15 |
| 2015-05-10 | 14:00:00 | 1.668694259 | ['Aker'] | 1.30E+16 | -4.43168E+15 |
| 2015-05-10 | 14:30:00 | 1.66854283 | ['Bastet'] | 2.06E+16 | -7.04813E+15 |
| 2015-05-10 | 15:00:00 | 1.668392453 | ['Hatmehit'] | 2.31E+16 | -7.90084E+15 |
| 2015-05-10 | 15:30:00 | 1.668243462 | ['Hatmehit'] | 1.72E+16 | -5.86659E+15 |
| 2015-05-10 | 16:00:00 | 1.668094906 | ['Maat'] | 1.94E+16 | -6.62156E+15 |
| 2015-05-10 | 16:30:00 | 1.667944528 | ['Nut'] | 2.32E+16 | -7.94034E+15 |
| 2015-05-10 | 17:00:00 | 1.667794528 | ['Serqet'] | 3.09E+16 | -1.06E+16 |
| 2015-05-10 | 17:30:00 | 1.667645577 | ['Anuket'] | 5.44E+16 | -1.86E+16 |
| 2015-05-10 | 18:00:00 | 1.667490877 | ['Anuket'] | 5.46E+16 | -1.86E+16 |
| 2015-05-10 | 18:30:00 | 1.667334706 | ['Anuket'] | 4.22E+16 | -1.44E+16 |
| 2015-05-10 | 19:00:00 | 1.667185962 | ['Hapi'] | 2.82E+16 | -9.61984E+15 |
| 2015-05-10 | 19:30:00 | 1.667036604 | ['Seth'] | 2.55E+16 | -8.72536E+15 |
| 2015-05-10 | 20:00:00 | 1.666888824 | ['Anubis'] | 2.00E+16 | -6.82521E+15 |
| 2015-05-10 | 20:30:00 | 1.666735417 | ['Atum'] | 1.73E+16 | -5.92353E+15 |

| Date | Time | Value | Label | E1 | E2 |
|---|---|---|---|---|---|
| 2015-05-10 | 21:00:00 | 1.666588868 | ['Ash'] | 2.94E+16 | -1.00E+16 |
| 2015-05-10 | 21:30:00 | 1.666438491 | ['Imhotep'] | 4.02E+16 | -1.37E+16 |
| 2015-05-10 | 22:00:00 | 1.666290192 | ['Imhotep'] | 4.71E+16 | -1.61E+16 |
| 2015-05-10 | 22:30:00 | 1.666134727 | ['Imhotep'] | 4.47E+16 | -1.53E+16 |
| 2015-05-10 | 23:00:00 | 1.665978491 | ['Imhotep'] | 5.45E+16 | -1.86E+16 |
| 2015-05-10 | 23:30:00 | 1.665829808 | ['Imhotep'] | 6.32E+16 | -2.16E+16 |
| 2015-05-11 | 0:00:00 | 1.665681765 | ['Khepry'] | 4.27E+16 | -1.46E+16 |
| 2015-05-11 | 0:30:00 | 1.665531321 | ['Aten'] | 4.28E+16 | -1.46E+16 |
| 2015-05-11 | 1:00:00 | 1.6653825 | ['Babi'] | 4.66E+16 | -1.59E+16 |
| 2015-05-11 | 1:30:00 | 1.665234151 | ['Babi'] | 2.65E+16 | -9.05129E+15 |
| 2015-05-11 | 2:00:00 | 1.665083962 | ['Babi'] | 1.93E+16 | -6.58757E+15 |
| 2015-05-11 | 2:30:00 | 1.664934151 | ['Aker'] | 1.12E+16 | -3.81803E+15 |
| 2015-05-11 | 3:00:00 | 1.664779455 | ['Maat'] | 1.81E+16 | -6.19726E+15 |
| 2015-05-11 | 3:30:00 | 1.664624151 | ['Maat'] | 2.13E+16 | -7.2659E+15 |
| 2015-05-11 | 4:00:00 | 1.664474528 | ['Hatmehit'] | 1.83E+16 | -6.25845E+15 |
| 2015-05-11 | 5:00:00 | 1.664177308 | ['Nut'] | 2.18E+16 | -7.44495E+15 |
| 2015-05-11 | 5:30:00 | 1.664027308 | ['Serqet'] | 3.17E+16 | -1.08E+16 |
| 2015-05-11 | 6:00:00 | 1.663879038 | ['Anuket'] | 5.51E+16 | -1.88E+16 |
| 2015-05-11 | 6:30:00 | 1.663730377 | ['Anuket'] | 5.40E+16 | -1.84E+16 |
| 2015-05-11 | 7:00:00 | 1.663577455 | ['Anuket'] | 3.74E+16 | -1.28E+16 |
| 2015-05-11 | 11:30:00 | 1.662212 | ['Anuket'] | 5.13E+16 | -1.75E+16 |
| 2015-05-11 | 12:00:00 | 1.662067647 | ['Imhotep'] | 5.95E+16 | -2.03E+16 |
| 2015-05-11 | 12:30:00 | 1.661918113 | ['Khepry'] | 3.77E+16 | -1.29E+16 |
| 2015-05-11 | 13:00:00 | 1.661768302 | ['Aten'] | 4.14E+16 | -1.41E+16 |
| 2015-05-11 | 13:30:00 | 1.661619216 | ['Babi'] | 3.52E+16 | -1.20E+16 |
| 2015-05-11 | 14:00:00 | 1.661470192 | ['Babi'] | 2.49E+16 | -8.49805E+15 |
| 2015-05-11 | 14:30:00 | 1.661321887 | ['Babi'] | 1.41E+16 | -4.82624E+15 |
| 2015-05-11 | 15:00:00 | 1.661172075 | ['Aker'] | 9.55E+15 | -3.26993E+15 |
| 2015-05-11 | 15:30:00 | 1.661022264 | ['Maat'] | 1.58E+16 | -5.40882E+15 |
| 2015-05-11 | 16:00:00 | 1.660867857 | ['Maat'] | 2.00E+16 | -6.81516E+15 |
| 2015-05-11 | 16:30:00 | 1.660711923 | ['Maat'] | 1.85E+16 | -6.33464E+15 |
| 2015-05-11 | 17:00:00 | 1.660565745 | ['Nut'] | 2.20E+16 | -7.52841E+15 |
| 2015-05-11 | 17:30:00 | 1.6604136 | ['Serqet'] | 1.81E+16 | -6.2001E+15 |
| 2015-05-11 | 18:00:00 | 1.660266981 | ['Serqet'] | 3.91E+16 | -1.34E+16 |
| 2015-05-11 | 18:30:00 | 1.660118269 | ['Anuket'] | 5.45E+16 | -1.86E+16 |
| 2015-05-11 | 19:00:00 | 1.65996902 | ['Anuket'] | 4.48E+16 | -1.53E+16 |
| 2015-05-11 | 21:00:00 | 1.659360962 | ['Seth'] | 1.48E+16 | -5.06296E+15 |
| 2015-05-11 | 21:30:00 | 1.659276364 | ['Ash'] | 1.28E+16 | -4.38519E+15 |
| 2015-05-11 | 22:30:00 | 1.658876667 | ['Imhotep'] | 2.77E+16 | -9.47337E+15 |
| 2015-05-11 | 23:00:00 | 1.658766415 | ['Imhotep'] | 3.78E+16 | -1.29E+16 |
| 2015-05-11 | 23:30:00 | 1.658618269 | ['Imhotep'] | 4.01E+16 | -1.37E+16 |
| 2015-05-12 | 0:00:00 | 1.658469623 | ['Imhotep'] | 4.57E+16 | -1.56E+16 |
| 2015-05-12 | 1:00:00 | 1.658162593 | ['Khepry'] | 3.18E+16 | -1.08E+16 |
| 2015-05-12 | 1:30:00 | 1.658013077 | ['Aten'] | 3.85E+16 | -1.31E+16 |

| Date | Time | Value | Label | Col5 | Col6 |
|---|---|---|---|---|---|
| 2015-05-12 | 2:00:00 | 1.657861887 | ['Babi'] | 3.19E+16 | -1.09E+16 |
| 2015-05-12 | 2:30:00 | 1.6577136 | ['Babi'] | 2.40E+16 | -8.18743E+15 |
| 2015-05-12 | 3:00:00 | 1.65756566 | ['Babi'] | 1.32E+16 | -4.49501E+15 |
| 2015-05-12 | 3:30:00 | 1.657416226 | ['Aker'] | 9.20E+15 | -3.14468E+15 |
| 2015-05-12 | 4:30:00 | 1.657118654 | ['Maat'] | 2.10E+16 | -7.18267E+15 |
| 2015-05-12 | 5:00:00 | 1.656964737 | ['Maat'] | 2.11E+16 | -7.21779E+15 |
| 2015-05-12 | 5:30:00 | 1.65681 | ['Nut'] | 2.13E+16 | -7.27138E+15 |
| 2015-05-12 | 6:00:00 | 1.656662075 | ['Serqet'] | 1.33E+16 | -4.54003E+15 |
| 2015-05-12 | 6:30:00 | 1.656512642 | ['Serqet'] | 4.27E+16 | -1.46E+16 |
| 2015-05-12 | 7:00:00 | 1.656363208 | ['Hathor'] | 4.75E+16 | -1.62E+16 |
| 2015-05-12 | 9:30:00 | 1.655615439 | ['Seth'] | 1.25E+16 | -4.27133E+15 |
| 2015-05-12 | 10:00:00 | 1.655460377 | ['Ash'] | 1.41E+16 | -4.81225E+15 |
| 2015-05-12 | 10:30:00 | 1.655312308 | ['Ash'] | 1.56E+16 | -5.34827E+15 |
| 2015-05-12 | 11:00:00 | 1.65516434 | ['Ash'] | 2.06E+16 | -7.04004E+15 |
| 2015-05-12 | 11:30:00 | 1.655014906 | ['Imhotep'] | 3.42E+16 | -1.17E+16 |
| 2015-05-12 | 12:00:00 | 1.654865472 | ['Imhotep'] | 3.59E+16 | -1.22E+16 |
| 2015-05-12 | 12:30:00 | 1.654718269 | ['Imhotep'] | 3.98E+16 | -1.36E+16 |
| 2015-05-12 | 13:00:00 | 1.654580652 | ['Imhotep'] | 3.91E+16 | -1.34E+16 |
| 2015-05-12 | 18:00:00 | 1.653069423 | ['Nut'] | 1.79E+16 | -6.12899E+15 |
| 2015-05-12 | 18:30:00 | 1.652921698 | ['Serqet'] | 1.25E+16 | -4.2723E+15 |
| 2015-05-12 | 22:30:00 | 1.651678333 | ['Serqet'] | 1.20E+16 | -4.11432E+15 |
| 2015-05-12 | 23:00:00 | 1.6515694 | ['Ash'] | 1.28E+16 | -4.38288E+15 |
| 2015-05-12 | 23:30:00 | 1.651420377 | ['Ash'] | 1.36E+16 | -4.64872E+15 |
| 2015-05-13 | 7:30:00 | 1.649038409 | ['Serqet'] | 3.87E+16 | -1.32E+16 |
| 2015-05-13 | 10:00:00 | 1.648287736 | ['Seth'] | 1.29E+16 | -4.40189E+15 |
| 2015-05-13 | 20:00:00 | 1.645294615 | ['Serqet'] | 4.89E+16 | -1.67E+16 |
| 2015-05-13 | 20:30:00 | 1.645147547 | ['Hathor'] | 5.93E+16 | -2.02E+16 |
| 2015-05-14 | 7:00:00 | 1.642028125 | ['Maat'] | 1.66E+16 | -5.67694E+15 |
| 2015-05-14 | 7:30:00 | 1.641871636 | ['Seth'] | 1.41E+16 | -4.82933E+15 |
| 2015-05-14 | 8:30:00 | 1.641567708 | ['Seth'] | 5.05E+16 | -1.73E+16 |
| 2015-05-14 | 9:00:00 | 1.641425472 | ['Seth'] | 4.71E+16 | -1.61E+16 |
| 2015-05-14 | 9:30:00 | 1.641278936 | ['Ash'] | 3.36E+16 | -1.15E+16 |
| 2015-05-14 | 10:00:00 | 1.641130377 | ['Seth'] | 2.63E+16 | -8.99884E+15 |
| 2015-05-14 | 10:30:00 | 1.640977679 | ['Seth'] | 1.85E+16 | -6.32626E+15 |
| 2015-05-14 | 11:00:00 | 1.640824151 | ['Seth'] | 1.51E+16 | -5.16796E+15 |
| 2015-05-14 | 12:30:00 | 1.640382727 | ['Babi'] | 1.24E+16 | -4.26699E+15 |
| 2015-05-14 | 16:30:00 | 1.639194286 | ['Babi'] | 2.82E+16 | -9.63949E+15 |
| 2015-05-14 | 17:00:00 | 1.639041321 | ['Hapi'] | 1.96E+16 | -6.69926E+15 |
| 2015-05-14 | 21:30:00 | 1.637703962 | ['Anuket'] | 6.06E+16 | -2.07E+16 |
| 2015-05-14 | 22:00:00 | 1.637558235 | ['Anuket'] | 3.77E+16 | -1.29E+16 |
| 2015-05-14 | 22:30:00 | 1.637410943 | ['Seth'] | 2.95E+16 | -1.01E+16 |
| 2015-05-14 | 23:00:00 | 1.637263019 | ['Seth'] | 2.50E+16 | -8.54025E+15 |
| 2015-05-14 | 23:30:00 | 1.637115094 | ['Seth'] | 1.98E+16 | -6.7633E+15 |
| 2015-05-15 | 0:00:00 | 1.636968846 | ['Seth'] | 1.50E+16 | -5.11016E+15 |

| Date | Time | Value | Label | Col5 | Col6 |
|---|---|---|---|---|---|
| 2015-05-15 | 0:30:00 | 1.636819091 | ['Ash'] | 1.24E+16 | -4.23582E+15 |
| 2015-05-15 | 1:00:00 | 1.636666111 | ['Ash'] | 1.43E+16 | -4.8933E+15 |
| 2015-05-15 | 1:30:00 | 1.636516792 | ['Ash'] | 1.26E+16 | -4.32801E+15 |
| 2015-05-15 | 2:30:00 | 1.636220943 | ['Ash'] | 1.54E+16 | -5.26571E+15 |
| 2015-05-15 | 3:30:00 | 1.635927925 | ['Aten'] | 2.77E+16 | -9.47824E+15 |
| 2015-05-15 | 4:00:00 | 1.635780192 | ['Ash'] | 3.45E+16 | -1.18E+16 |
| 2015-05-15 | 5:00:00 | 1.635479821 | ['Hapi'] | 3.71E+16 | -1.27E+16 |
| 2015-05-15 | 5:30:00 | 1.635326792 | ['Hapi'] | 2.54E+16 | -8.68655E+15 |
| 2015-05-15 | 7:30:00 | 1.634742041 | ['Maat'] | 1.73E+16 | -5.91212E+15 |
| 2015-05-15 | 9:30:00 | 1.634145536 | ['Anuket'] | 4.63E+16 | -1.58E+16 |
| 2015-05-15 | 10:00:00 | 1.633993019 | ['Hapi'] | 5.53E+16 | -1.89E+16 |
| 2015-05-15 | 10:30:00 | 1.633846731 | ['Seth'] | 3.60E+16 | -1.23E+16 |
| 2015-05-15 | 11:00:00 | 1.633698846 | ['Seth'] | 2.92E+16 | -9.98402E+15 |
| 2015-05-15 | 11:30:00 | 1.63355283 | ['Seth'] | 2.80E+16 | -9.56004E+15 |
| 2015-05-15 | 12:00:00 | 1.633406346 | ['Seth'] | 2.43E+16 | -8.31295E+15 |
| 2015-05-15 | 12:30:00 | 1.633257547 | ['Seth'] | 1.94E+16 | -6.64288E+15 |
| 2015-05-15 | 17:30:00 | 1.631750571 | ['Seth'] | 3.98E+16 | -1.36E+16 |
| 2015-05-15 | 23:00:00 | 1.630142453 | ['Seth'] | 3.26E+16 | -1.11E+16 |
| 2015-05-15 | 23:30:00 | 1.6299976 | ['Seth'] | 3.05E+16 | -1.04E+16 |
| 2015-05-16 | 0:00:00 | 1.629852115 | ['Hapi'] | 3.54E+16 | -1.21E+16 |
| 2015-05-16 | 0:30:00 | 1.629703585 | ['Anuket'] | 3.38E+16 | -1.15E+16 |
| 2015-05-16 | 1:00:00 | 1.629556226 | ['Hapi'] | 2.83E+16 | -9.66081E+15 |
| 2015-05-16 | 1:30:00 | 1.62941098 | ['Hapi'] | 2.47E+16 | -8.42639E+15 |
| 2015-05-16 | 2:00:00 | 1.62926434 | ['Hapi'] | 2.40E+16 | -8.21118E+15 |
| 2015-05-16 | 3:00:00 | 1.628963091 | ['Seth'] | 1.36E+16 | -4.66488E+15 |
| 2015-05-16 | 3:30:00 | 1.628812453 | ['Seth'] | 7.53E+15 | -2.61062E+15 |
| 2015-05-16 | 4:00:00 | 1.628665283 | ['Seth'] | 6.17E+15 | -2.16969E+15 |
| 2015-05-16 | 7:00:00 | 1.627782909 | ['Babi'] | 3.46E+16 | -1.18E+16 |
| 2015-05-16 | 7:30:00 | 1.627630755 | ['Babi'] | 3.55E+16 | -1.21E+16 |
| 2015-05-16 | 8:00:00 | 1.627483585 | ['Hapi'] | 3.74E+16 | -1.28E+16 |
| 2015-05-16 | 8:30:00 | 1.627336604 | ['Hapi'] | 3.17E+16 | -1.08E+16 |
| 2015-05-16 | 9:30:00 | 1.6270454 | ['Maat'] | 1.21E+16 | -4.1567E+15 |
| 2015-05-16 | 10:00:00 | 1.626897925 | ['Maat'] | 1.08E+16 | -3.68198E+15 |
| 2015-05-16 | 10:30:00 | 1.626750943 | ['Maat'] | 2.86E+16 | -9.78322E+15 |
| 2015-05-16 | 11:00:00 | 1.626603585 | ['Maat'] | 4.54E+16 | -1.55E+16 |
| 2015-05-16 | 11:30:00 | 1.626450727 | ['Hathor'] | 4.32E+16 | -1.48E+16 |
| 2015-05-16 | 12:30:00 | 1.626154423 | ['Hapi'] | 4.07E+16 | -1.39E+16 |
| 2015-05-16 | 13:00:00 | 1.626008491 | ['Hapi'] | 3.80E+16 | -1.30E+16 |
| 2015-05-16 | 13:30:00 | 1.625862692 | ['Seth'] | 3.49E+16 | -1.19E+16 |
| 2015-05-16 | 14:00:00 | 1.625714528 | ['Seth'] | 2.97E+16 | -1.01E+16 |
| 2015-05-16 | 14:30:00 | 1.625569038 | ['Seth'] | 2.84E+16 | -9.70264E+15 |
| 2015-05-16 | 15:00:00 | 1.625423585 | ['Seth'] | 2.18E+16 | -7.45017E+15 |
| 2015-05-16 | 17:30:00 | 1.62468 | ['Ash'] | 1.10E+16 | -3.75908E+15 |
| 2015-05-16 | 18:00:00 | 1.624534528 | ['Ash'] | 1.74E+16 | -5.94634E+15 |

| Date | Time | Value | Region | Col5 | Col6 |
|---|---|---|---|---|---|
| 2015-05-16 | 18:30:00 | 1.62439 | ['Ash'] | 2.41E+16 | -8.22105E+15 |
| 2015-05-16 | 19:00:00 | 1.624242115 | ['Ash'] | 2.74E+16 | -9.35626E+15 |
| 2015-05-16 | 20:00:00 | 1.623948333 | ['Babi'] | 3.28E+16 | -1.12E+16 |
| 2015-05-16 | 20:30:00 | 1.623793846 | ['Hapi'] | 3.59E+16 | -1.22E+16 |
| 2015-05-16 | 21:00:00 | 1.623645577 | ['Hapi'] | 3.49E+16 | -1.19E+16 |
| 2015-05-16 | 21:30:00 | 1.6234988 | ['Maat'] | 2.45E+16 | -8.38364E+15 |
| 2015-05-16 | 22:30:00 | 1.6232075 | ['Maat'] | 1.07E+16 | -3.68985E+15 |
| 2015-05-16 | 23:30:00 | 1.622919 | ['Maat'] | 4.51E+16 | -1.54E+16 |
| 2015-05-17 | 10:30:00 | 1.619669184 | ['Maat'] | 1.50E+16 | -5.12014E+15 |
| 2015-05-17 | 11:30:00 | 1.619331176 | ['Maat'] | 1.82E+16 | -6.226E+15 |
| 2015-05-17 | 13:00:00 | 1.618935769 | ['Ash'] | 3.04E+16 | -1.04E+16 |
| 2015-05-17 | 14:00:00 | 1.618646667 | ['Ash'] | 3.38E+16 | -1.16E+16 |
| 2015-05-17 | 14:30:00 | 1.6185008 | ['Maat'] | 2.87E+16 | -9.79094E+15 |
| 2015-05-17 | 15:30:00 | 1.618205536 | ['Serqet'] | 4.27E+16 | -1.46E+16 |
| 2015-05-17 | 16:00:00 | 1.618054151 | ['Seth'] | 2.93E+16 | -1.00E+16 |
| 2015-05-17 | 16:30:00 | 1.617909038 | ['Seth'] | 1.54E+16 | -5.24579E+15 |
| 2015-05-17 | 17:00:00 | 1.617759615 | ['Maat'] | 1.23E+16 | -4.2178E+15 |
| 2015-05-17 | 17:30:00 | 1.617613269 | ['Maat'] | 9.06E+15 | -3.11316E+15 |
| 2015-05-17 | 19:00:00 | 1.617182708 | ['Maat'] | 1.43E+16 | -4.9095E+15 |
| 2015-05-17 | 19:30:00 | 1.617034255 | ['Ash'] | 1.84E+16 | -6.28344E+15 |
| 2015-05-17 | 20:00:00 | 1.616883929 | ['Ash'] | 1.99E+16 | -6.81673E+15 |
| 2015-05-17 | 20:30:00 | 1.616733019 | ['Ash'] | 3.10E+16 | -1.06E+16 |
| 2015-05-17 | 21:00:00 | 1.616588958 | ['Babi'] | 3.29E+16 | -1.12E+16 |
| 2015-05-17 | 23:00:00 | 1.616003208 | ['Seth'] | 2.72E+16 | -9.31963E+15 |
| 2015-05-17 | 23:30:00 | 1.615858039 | ['Seth'] | 1.49E+16 | -5.11405E+15 |
| 2015-05-18 | 0:00:00 | 1.615707925 | ['Ash'] | 1.45E+16 | -4.95361E+15 |
| 2015-05-18 | 0:30:00 | 1.615561321 | ['Babi'] | 1.29E+16 | -4.39607E+15 |
| 2015-05-18 | 1:00:00 | 1.615416731 | ['Babi'] | 1.10E+16 | -3.77393E+15 |
| 2015-05-18 | 2:00:00 | 1.615126731 | ['Hapi'] | 2.04E+16 | -6.97367E+15 |
| 2015-05-18 | 2:30:00 | 1.614976545 | ['Hapi'] | 2.92E+16 | -9.97687E+15 |
| 2015-05-18 | 3:00:00 | 1.614825556 | ['Seth'] | 3.46E+16 | -1.18E+16 |
| 2015-05-18 | 3:30:00 | 1.614677925 | ['Seth'] | 4.01E+16 | -1.37E+16 |
| 2015-05-18 | 4:00:00 | 1.614533137 | ['Seth'] | 4.73E+16 | -1.62E+16 |
| 2015-05-18 | 4:30:00 | 1.6143858 | ['Seth'] | 3.18E+16 | -1.09E+16 |
| 2015-05-18 | 5:00:00 | 1.614242075 | ['Seth'] | 1.61E+16 | -5.51107E+15 |
| 2015-05-18 | 5:30:00 | 1.6140975 | ['Seth'] | 8.94E+15 | -3.06516E+15 |
| 2015-05-18 | 6:00:00 | 1.613953019 | ['Ash'] | 8.09E+15 | -2.76766E+15 |
| 2015-05-18 | 6:30:00 | 1.613806981 | ['Ash'] | 1.23E+16 | -4.19401E+15 |
| 2015-05-18 | 11:30:00 | 1.612338571 | ['Ash'] | 2.22E+16 | -7.57989E+15 |
| 2015-05-18 | 12:00:00 | 1.612188462 | ['Maat'] | 1.41E+16 | -4.82929E+15 |
| 2015-05-18 | 12:30:00 | 1.612051364 | ['Maat'] | 1.16E+16 | -3.98788E+15 |
| 2015-05-18 | 16:30:00 | 1.610842581 | ['Hatmehit'] | 4.60E+16 | -1.57E+16 |
| 2015-05-18 | 17:00:00 | 1.610758 | ['Seth'] | 3.56E+16 | -1.22E+16 |
| 2015-05-18 | 18:30:00 | 1.61022 | ['Hatmehit'] | 1.31E+16 | -4.52387E+15 |

| Date | Time | Value | Region | X | Y |
|---|---|---|---|---|---|
| 2015-05-18 | 19:00:00 | 1.610146415 | ['Ash'] | 1.30E+16 | -4.43587E+15 |
| 2015-05-18 | 19:30:00 | 1.610001731 | ['Ash'] | 1.91E+16 | -6.52626E+15 |
| 2015-05-18 | 20:00:00 | 1.609852 | ['Ash'] | 2.18E+16 | -7.42905E+15 |
| 2015-05-18 | 20:30:00 | 1.609701481 | ['Ash'] | 2.44E+16 | -8.3402E+15 |
| 2015-05-18 | 21:00:00 | 1.60955434 | ['Ash'] | 2.63E+16 | -8.99155E+15 |
| 2015-05-18 | 21:30:00 | 1.609409808 | ['Aten'] | 3.36E+16 | -1.15E+16 |
| 2015-05-18 | 22:00:00 | 1.609266038 | ['Babi'] | 3.97E+16 | -1.35E+16 |
| 2015-05-18 | 22:30:00 | 1.609120377 | ['Babi'] | 2.73E+16 | -9.32489E+15 |
| 2015-05-18 | 23:00:00 | 1.608976154 | ['Hapi'] | 2.89E+16 | -9.8765E+15 |
| 2015-05-18 | 23:30:00 | 1.608832075 | ['Hapi'] | 4.52E+16 | -1.55E+16 |
| 2015-05-19 | 0:00:00 | 1.608686604 | ['Maat'] | 3.13E+16 | -1.07E+16 |
| 2015-05-19 | 0:30:00 | 1.608536429 | ['Hathor'] | 1.64E+16 | -5.61785E+15 |
| 2015-05-19 | 1:00:00 | 1.608384314 | ['Maat'] | 1.05E+16 | -3.57417E+15 |
| 2015-05-19 | 1:30:00 | 1.6082396 | ['Maat'] | 1.24E+16 | -4.24798E+15 |
| 2015-05-19 | 2:00:00 | 1.6080954 | ['Maat'] | 1.24E+16 | -4.23876E+15 |
| 2015-05-19 | 2:30:00 | 1.6079484 | ['Serqet'] | 1.27E+16 | -4.34367E+15 |
| 2015-05-19 | 3:00:00 | 1.607806792 | ['Hathor'] | 1.36E+16 | -4.6325E+15 |
| 2015-05-19 | 3:30:00 | 1.6076625 | ['Anuket'] | 1.92E+16 | -6.5621E+15 |
| 2015-05-19 | 4:00:00 | 1.607517674 | ['Hapi'] | 2.60E+16 | -8.86513E+15 |
| 2015-05-19 | 4:30:00 | 1.607371923 | ['Seth'] | 3.81E+16 | -1.30E+16 |
| 2015-05-19 | 5:00:00 | 1.607220536 | ['Seth'] | 4.93E+16 | -1.68E+16 |
| 2015-05-19 | 5:30:00 | 1.607069796 | ['Seth'] | 3.52E+16 | -1.20E+16 |
| 2015-05-19 | 6:00:00 | 1.606928269 | ['Seth'] | 1.97E+16 | -6.73675E+15 |
| 2015-05-19 | 6:30:00 | 1.606780196 | ['Ash'] | 1.20E+16 | -4.10263E+15 |
| 2015-05-19 | 7:00:00 | 1.606636731 | ['Ash'] | 1.07E+16 | -3.65149E+15 |
| 2015-05-19 | 7:30:00 | 1.606492885 | ['Ash'] | 1.55E+16 | -5.2919E+15 |
| 2015-05-19 | 8:00:00 | 1.606349245 | ['Ash'] | 2.27E+16 | -7.76052E+15 |
| 2015-05-19 | 8:30:00 | 1.606203878 | ['Ash'] | 2.64E+16 | -9.01837E+15 |
| 2015-05-19 | 9:00:00 | 1.606060196 | ['Imhotep'] | 2.82E+16 | -9.63377E+15 |
| 2015-05-19 | 9:30:00 | 1.605907091 | ['Ash'] | 2.88E+16 | -9.83688E+15 |
| 2015-05-19 | 10:00:00 | 1.605759216 | ['Aten'] | 3.57E+16 | -1.22E+16 |
| 2015-05-19 | 10:30:00 | 1.605613137 | ['Aten'] | 4.47E+16 | -1.53E+16 |
| 2015-05-19 | 11:00:00 | 1.605470755 | ['Babi'] | 3.22E+16 | -1.10E+16 |
| 2015-05-19 | 11:30:00 | 1.605325472 | ['Babi'] | 2.66E+16 | -9.069E+15 |
| 2015-05-19 | 12:00:00 | 1.605181961 | ['Hapi'] | 4.50E+16 | -1.54E+16 |
| 2015-05-19 | 12:30:00 | 1.605037547 | ['Hathor'] | 3.34E+16 | -1.14E+16 |
| 2015-05-19 | 13:00:00 | 1.604901064 | ['Maat'] | 1.87E+16 | -6.39963E+15 |
| 2015-05-19 | 17:30:00 | 1.603532667 | ['Hatmehit'] | 5.02E+16 | -1.72E+16 |
| 2015-05-19 | 18:00:00 | 1.603430909 | ['Seth'] | 3.80E+16 | -1.30E+16 |
| 2015-05-19 | 18:30:00 | 1.603281538 | ['Seth'] | 1.99E+16 | -6.80321E+15 |
| 2015-05-19 | 22:30:00 | 1.60207125 | ['Hatmehit'] | 3.52E+16 | -1.20E+16 |
| 2015-05-19 | 23:00:00 | 1.601974898 | ['Aten'] | 4.62E+16 | -1.58E+16 |
| 2015-05-19 | 23:30:00 | 1.601828077 | ['Babi'] | 4.10E+16 | -1.40E+16 |
| 2015-05-20 | 5:30:00 | 1.600057097 | ['Babi'] | 4.74E+16 | -1.62E+16 |

| Date | Time | Value | Region | Col5 | Col6 |
|---|---|---|---|---|---|
| 2015-05-20 | 6:00:00 | 1.599958205 | ['Seth'] | 5.77E+16 | -1.97E+16 |
| 2015-05-20 | 9:30:00 | 1.598917 | ['Seth'] | 3.44E+16 | -1.18E+16 |
| 2015-05-20 | 10:00:00 | 1.59877566 | ['Imhotep'] | 3.72E+16 | -1.27E+16 |
| 2015-05-20 | 10:30:00 | 1.598651053 | ['Imhotep'] | 3.57E+16 | -1.22E+16 |
| 2015-05-20 | 13:00:00 | 1.59789619 | ['Imhotep'] | 4.46E+16 | -1.52E+16 |
| 2015-05-20 | 13:30:00 | 1.597766792 | ['Hapi'] | 4.23E+16 | -1.45E+16 |
| 2015-05-20 | 14:00:00 | 1.597623654 | ['Hathor'] | 2.27E+16 | -7.74559E+15 |
| 2015-05-20 | 14:30:00 | 1.597473455 | ['Maat'] | 1.34E+16 | -4.58526E+15 |
| 2015-05-20 | 15:00:00 | 1.597324808 | ['Maat'] | 1.38E+16 | -4.70812E+15 |
| 2015-05-20 | 15:30:00 | 1.597180577 | ['Maat'] | 1.51E+16 | -5.17115E+15 |
| 2015-05-20 | 16:00:00 | 1.597037925 | ['Nut'] | 1.16E+16 | -3.97687E+15 |
| 2015-05-20 | 16:30:00 | 1.596893208 | ['Serqet'] | 1.16E+16 | -3.96479E+15 |
| 2015-05-20 | 17:00:00 | 1.596749 | ['Hathor'] | 1.53E+16 | -5.21714E+15 |
| 2015-05-20 | 17:30:00 | 1.596605769 | ['Anuket'] | 2.18E+16 | -7.45654E+15 |
| 2015-05-20 | 18:00:00 | 1.59646098 | ['Anuket'] | 3.92E+16 | -1.34E+16 |
| 2015-05-20 | 18:30:00 | 1.596317451 | ['Seth'] | 5.82E+16 | -1.99E+16 |
| 2015-05-20 | 19:00:00 | 1.596167593 | ['Seth'] | 4.05E+16 | -1.38E+16 |
| 2015-05-20 | 19:30:00 | 1.596020377 | ['Seth'] | 2.13E+16 | -7.28826E+15 |
| 2015-05-20 | 20:00:00 | 1.59587566 | ['Atum'] | 1.46E+16 | -4.97359E+15 |
| 2015-05-20 | 20:30:00 | 1.595732885 | ['Ash'] | 1.41E+16 | -4.80356E+15 |
| 2015-05-20 | 21:00:00 | 1.595590377 | ['Imhotep1'] | 2.05E+16 | -6.98757E+15 |
| 2015-05-20 | 21:30:00 | 1.595445192 | ['Imhotep'] | 2.99E+16 | -1.02E+16 |
| 2015-05-20 | 22:00:00 | 1.595301698 | ['Imhotep'] | 4.20E+16 | -1.43E+16 |
| 2015-05-20 | 22:30:00 | 1.595159608 | ['Imhotep'] | 4.36E+16 | -1.49E+16 |
| 2015-05-20 | 23:00:00 | 1.595014528 | ['Imhotep'] | 4.21E+16 | -1.44E+16 |
| 2015-05-20 | 23:30:00 | 1.594863846 | ['Ash'] | 3.57E+16 | -1.22E+16 |
| 2015-05-21 | 0:00:00 | 1.594718846 | ['Aten'] | 4.90E+16 | -1.67E+16 |
| 2015-05-21 | 0:30:00 | 1.594573396 | ['Aten'] | 3.46E+16 | -1.18E+16 |
| 2015-05-21 | 1:00:00 | 1.594427451 | ['Babi'] | 2.83E+16 | -9.67997E+15 |
| 2015-05-21 | 6:00:00 | 1.59298434 | ['Atum'] | 3.10E+16 | -1.06E+16 |
| 2015-05-21 | 6:30:00 | 1.592841176 | ['Seth'] | 4.89E+16 | -1.67E+16 |
| 2015-05-21 | 7:00:00 | 1.592719737 | ['Seth'] | 5.10E+16 | -1.74E+16 |
| 2015-05-21 | 11:00:00 | 1.591532955 | ['Seth'] | 5.16E+16 | -1.76E+16 |
| 2015-05-21 | 11:30:00 | 1.59142871 | ['Imhotep'] | 5.57E+16 | -1.90E+16 |
| 2015-05-21 | 13:00:00 | 1.590958431 | ['Imhotep'] | 3.47E+16 | -1.18E+16 |
| 2015-05-21 | 13:30:00 | 1.590813654 | ['Aker'] | 3.02E+16 | -1.03E+16 |
| 2015-05-21 | 14:00:00 | 1.590672353 | ['Bastet'] | 3.63E+16 | -1.24E+16 |
| 2015-05-21 | 14:30:00 | 1.590528462 | ['Bastet'] | 3.70E+16 | -1.26E+16 |
| 2015-05-21 | 15:00:00 | 1.5903846 | ['Bastet'] | 2.13E+16 | -7.28642E+15 |
| 2015-05-21 | 15:30:00 | 1.590241538 | ['Hatmehit'] | 1.73E+16 | -5.89869E+15 |
| 2015-05-21 | 16:00:00 | 1.5901 | ['Hatmehit'] | 2.17E+16 | -7.41492E+15 |
| 2015-05-21 | 16:30:00 | 1.589954545 | ['Maftet'] | 2.00E+16 | -6.81716E+15 |
| 2015-05-21 | 17:00:00 | 1.589805741 | ['Maftet'] | 1.36E+16 | -4.64348E+15 |
| 2015-05-21 | 17:30:00 | 1.589658367 | ['Serqet'] | 1.35E+16 | -4.60873E+15 |

| Date | Time | Value | Region |  | Col6 |
|---|---|---|---|---|---|
| 2015-05-21 | 18:00:00 | 1.589560833 | ['Hatmehit'] | 1.31E+16 | -4.4817E+15 |
| 2015-05-21 | 21:30:00 | 1.588509423 | ['Imhotep1'] | 1.64E+16 | -5.61131E+15 |
| 2015-05-22 | 0:00:00 | 1.587786087 | ['Ash'] | 5.62E+16 | -1.92E+16 |
| 2015-05-22 | 0:30:00 | 1.587646957 | ['Ash'] | 4.68E+16 | -1.60E+16 |
| 2015-05-22 | 2:30:00 | 1.587071321 | ['Sobek'] | 3.14E+16 | -1.07E+16 |
| 2015-05-22 | 3:00:00 | 1.586929038 | ['Neith'] | 3.36E+16 | -1.15E+16 |
| 2015-05-22 | 3:30:00 | 1.586803846 | ['Wosret'] | 2.97E+16 | -1.01E+16 |
| 2015-05-22 | 7:00:00 | 1.585777692 | ['Hatmehit'] | 3.17E+16 | -1.08E+16 |
| 2015-05-22 | 8:00:00 | 1.585506061 | ['Atum'] | 5.96E+16 | -2.03E+16 |
| 2015-05-22 | 12:30:00 | 1.584187547 | ['Imhotep'] | 5.20E+16 | -1.78E+16 |
| 2015-05-22 | 13:00:00 | 1.584044717 | ['Khepry'] | 4.47E+16 | -1.53E+16 |
| 2015-05-22 | 17:30:00 | 1.582751538 | ['Wosret'] | 2.40E+16 | -8.17931E+15 |
| 2015-05-22 | 18:00:00 | 1.582609792 | ['Geb'] | 2.17E+16 | -7.40803E+15 |
| 2015-05-22 | 22:30:00 | 1.581319623 | ['Geb'] | 2.07E+16 | -7.05931E+15 |
| 2015-05-22 | 23:00:00 | 1.581176226 | ['Bes'] | 3.31E+16 | -1.13E+16 |
| 2015-05-23 | 5:30:00 | 1.579266111 | ['Hatmehit'] | 1.97E+16 | -6.74509E+15 |
| 2015-05-23 | 6:00:00 | 1.579165577 | ['Wosret'] | 2.39E+16 | -8.15918E+15 |
| 2015-05-23 | 6:30:00 | 1.579024528 | ['Wosret'] | 2.41E+16 | -8.23855E+15 |
| 2015-05-23 | 7:00:00 | 1.578882 | ['Neith'] | 2.05E+16 | -7.01623E+15 |
| 2015-05-23 | 7:30:00 | 1.578741373 | ['Anuket'] | 2.76E+16 | -9.41247E+15 |
| 2015-05-23 | 8:00:00 | 1.578599038 | ['Hapi'] | 4.12E+16 | -1.41E+16 |
| 2015-05-23 | 8:30:00 | 1.578456538 | ['Seth'] | 5.81E+16 | -1.99E+16 |
| 2015-05-23 | 9:00:00 | 1.578314151 | ['Anubis'] | 5.79E+16 | -1.98E+16 |
| 2015-05-23 | 9:30:00 | 1.5781702 | ['Atum'] | 2.96E+16 | -1.01E+16 |
| 2015-05-23 | 10:00:00 | 1.578027308 | ['Atum'] | 2.25E+16 | -7.69221E+15 |
| 2015-05-23 | 10:30:00 | 1.577878039 | ['Khonsu'] | 2.15E+16 | -7.34916E+15 |
| 2015-05-23 | 11:00:00 | 1.577737547 | ['Imhotep1'] | 2.16E+16 | -7.37589E+15 |
| 2015-05-23 | 11:30:00 | 1.5775928 | ['Imhotep'] | 3.34E+16 | -1.14E+16 |
| 2015-05-23 | 12:00:00 | 1.5774528 | ['Imhotep'] | 5.38E+16 | -1.84E+16 |
| 2015-05-23 | 12:30:00 | 1.577310769 | ['Imhotep'] | 9.01E+16 | -3.07E+16 |
| 2015-05-23 | 13:00:00 | 1.577170566 | ['Imhotep'] | 7.56E+16 | -2.58E+16 |
| 2015-05-23 | 13:30:00 | 1.577027358 | ['Imhotep'] | 5.28E+16 | -1.80E+16 |
| 2015-05-23 | 14:00:00 | 1.576885283 | ['Khepry'] | 3.71E+16 | -1.27E+16 |
| 2015-05-23 | 14:30:00 | 1.576738113 | ['Khepry'] | 2.28E+16 | -7.78804E+15 |
| 2015-05-23 | 15:00:00 | 1.576591731 | ['Khepry'] | 2.45E+16 | -8.37837E+15 |
| 2015-05-23 | 15:30:00 | 1.576449615 | ['Aker'] | 2.66E+16 | -9.08581E+15 |
| 2015-05-23 | 16:00:00 | 1.576309412 | ['Aker'] | 2.34E+16 | -8.00597E+15 |
| 2015-05-23 | 16:30:00 | 1.576165385 | ['Sobek'] | 3.53E+16 | -1.21E+16 |
| 2015-05-23 | 17:00:00 | 1.576024717 | ['Neith'] | 3.63E+16 | -1.24E+16 |
| 2015-05-23 | 17:30:00 | 1.575882885 | ['Wosret'] | 3.18E+16 | -1.09E+16 |
| 2015-05-23 | 18:00:00 | 1.575742642 | ['Wosret'] | 2.53E+16 | -8.62894E+15 |
| 2015-05-23 | 18:30:00 | 1.575598704 | ['Wosret'] | 2.87E+16 | -9.80386E+15 |
| 2015-05-23 | 19:00:00 | 1.575451455 | ['Wosret'] | 2.90E+16 | -9.88508E+15 |
| 2015-05-23 | 19:30:00 | 1.575306604 | ['Anuket'] | 2.22E+16 | -7.57364E+15 |

| Date | Time | Value | Region | Col5 | Col6 |
|---|---|---|---|---|---|
| 2015-05-23 | 20:00:00 | 1.575163774 | ['Hapi'] | 2.96E+16 | -1.01E+16 |
| 2015-05-23 | 20:30:00 | 1.575023654 | ['Hapi'] | 4.95E+16 | -1.69E+16 |
| 2015-05-23 | 21:00:00 | 1.574881887 | ['Anubis'] | 5.99E+16 | -2.05E+16 |
| 2015-05-23 | 21:30:00 | 1.574740189 | ['Anubis'] | 4.79E+16 | -1.64E+16 |
| 2015-05-23 | 22:00:00 | 1.574598302 | ['Atum'] | 2.34E+16 | -7.99226E+15 |
| 2015-05-23 | 23:00:00 | 1.574312321 | ['Khonsu'] | 2.54E+16 | -8.68057E+15 |
| 2015-05-23 | 23:30:00 | 1.574163889 | ['Imhotep1'] | 2.86E+16 | -9.76976E+15 |
| 2015-05-24 | 0:00:00 | 1.574021154 | ['Imhotep'] | 3.66E+16 | -1.25E+16 |
| 2015-05-24 | 0:30:00 | 1.573882308 | ['Imhotep'] | 5.53E+16 | -1.89E+16 |
| 2015-05-24 | 1:00:00 | 1.573738302 | ['Imhotep'] | 8.98E+16 | -3.07E+16 |
| 2015-05-24 | 1:30:00 | 1.573597885 | ['Imhotep'] | 7.85E+16 | -2.68E+16 |
| 2015-05-24 | 2:00:00 | 1.573457925 | ['Imhotep'] | 5.23E+16 | -1.79E+16 |
| 2015-05-24 | 2:30:00 | 1.573314717 | ['Khepry'] | 3.30E+16 | -1.13E+16 |
| 2015-05-24 | 3:00:00 | 1.573173654 | ['Khepry'] | 2.27E+16 | -7.74994E+15 |
| 2015-05-24 | 3:30:00 | 1.573027818 | ['Khepry'] | 2.38E+16 | -8.12136E+15 |
| 2015-05-24 | 4:00:00 | 1.572879811 | ['Anhur'] | 2.77E+16 | -9.47107E+15 |
| 2015-05-24 | 4:30:00 | 1.572738235 | ['Anhur'] | 3.30E+16 | -1.13E+16 |
| 2015-05-24 | 5:00:00 | 1.572599245 | ['Sobek'] | 4.43E+16 | -1.51E+16 |
| 2015-05-24 | 5:30:00 | 1.572456415 | ['Neith'] | 5.17E+16 | -1.77E+16 |
| 2015-05-24 | 6:00:00 | 1.57231434 | ['Wosret'] | 4.53E+16 | -1.55E+16 |
| 2015-05-24 | 6:30:00 | 1.572174615 | ['Wosret'] | 3.51E+16 | -1.20E+16 |
| 2015-05-24 | 7:00:00 | 1.572034615 | ['Wosret'] | 2.87E+16 | -9.81208E+15 |
| 2015-05-24 | 7:30:00 | 1.5718916 | ['Sobek'] | 2.67E+16 | -9.10381E+15 |
| 2015-05-24 | 8:00:00 | 1.571800909 | ['Anuket'] | 1.99E+16 | -6.79788E+15 |
| 2015-05-24 | 9:30:00 | 1.571311957 | ['Anuket'] | 5.91E+16 | -2.02E+16 |
| 2015-05-24 | 10:00:00 | 1.571218261 | ['Atum'] | 4.54E+16 | -1.55E+16 |
| 2015-05-24 | 14:00:00 | 1.569977778 | ['Atum'] | 5.21E+16 | -1.78E+16 |
| 2015-05-24 | 14:30:00 | 1.569893673 | ['Imhotep'] | 4.68E+16 | -1.60E+16 |
| 2015-05-24 | 15:00:00 | 1.569755094 | ['Bes'] | 3.52E+16 | -1.20E+16 |
| 2015-05-24 | 15:30:00 | 1.569645 | ['Bes'] | 2.67E+16 | -9.10533E+15 |
| 2015-05-24 | 19:30:00 | 1.568475472 | ['Wosret'] | 3.17E+16 | -1.08E+16 |
| 2015-05-25 | 0:00:00 | 1.567172258 | ['Wosret'] | 3.16E+16 | -1.08E+16 |
| 2015-05-25 | 0:30:00 | 1.567057925 | ['Imhotep'] | 3.36E+16 | -1.15E+16 |
| 2015-05-25 | 1:00:00 | 1.566916038 | ['Imhotep'] | 3.76E+16 | -1.28E+16 |
| 2015-05-25 | 5:00:00 | 1.56577186 | ['Imhotep'] | 4.02E+16 | -1.37E+16 |
| 2015-05-25 | 5:30:00 | 1.565683684 | ['Sobek'] | 4.44E+16 | -1.52E+16 |
| 2015-05-25 | 9:30:00 | 1.5644375 | ['Sobek'] | 5.49E+16 | -1.88E+16 |
| 2015-05-25 | 10:00:00 | 1.564361509 | ['Atum'] | 6.43E+16 | -2.20E+16 |
| 2015-05-25 | 10:30:00 | 1.564220377 | ['Anubis'] | 5.92E+16 | -2.02E+16 |
| 2015-05-25 | 11:00:00 | 1.564103714 | ['Atum'] | 3.66E+16 | -1.25E+16 |
| 2015-05-25 | 14:30:00 | 1.563038824 | ['Atum'] | 6.16E+16 | -2.11E+16 |
| 2015-05-25 | 15:00:00 | 1.562948302 | ['Imhotep'] | 5.63E+16 | -1.92E+16 |
| 2015-05-25 | 20:00:00 | 1.561536415 | ['Wosret'] | 4.34E+16 | -1.48E+16 |
| 2015-05-25 | 20:30:00 | 1.561396154 | ['Sobek'] | 3.64E+16 | -1.24E+16 |

| Date | Time | Value | Region | E1 | E2 |
|---|---|---|---|---|---|
| 2015-05-25 | 21:00:00 | 1.56131 | ['Hapi'] | 2.15E+16 | -7.39717E+15 |
| 2015-05-26 | 1:00:00 | 1.56015 | ['Imhotep1'] | 3.43E+16 | -1.17E+16 |
| 2015-05-26 | 5:30:00 | 1.558849231 | ['Imhotep1'] | 3.64E+16 | -1.24E+16 |
| 2015-05-26 | 6:00:00 | 1.5587075 | ['Sobek'] | 5.04E+16 | -1.72E+16 |
| 2015-05-26 | 6:30:00 | 1.558589722 | ['Sobek'] | 5.63E+16 | -1.92E+16 |
| 2015-05-26 | 10:00:00 | 1.557526364 | ['Sobek'] | 3.91E+16 | -1.34E+16 |
| 2015-05-26 | 10:30:00 | 1.557438654 | ['Anubis'] | 4.90E+16 | -1.67E+16 |
| 2015-05-26 | 11:00:00 | 1.557299231 | ['Anubis'] | 6.28E+16 | -2.14E+16 |
| 2015-05-26 | 11:30:00 | 1.557159434 | ['Atum'] | 5.28E+16 | -1.80E+16 |
| 2015-05-26 | 12:00:00 | 1.557019623 | ['Khonsu'] | 3.95E+16 | -1.35E+16 |
| 2015-05-26 | 12:30:00 | 1.556880962 | ['Khonsu'] | 3.17E+16 | -1.08E+16 |
| 2015-05-26 | 13:00:00 | 1.556741698 | ['Khonsu'] | 3.29E+16 | -1.12E+16 |
| 2015-05-26 | 13:30:00 | 1.556602453 | ['Khonsu'] | 3.16E+16 | -1.08E+16 |
| 2015-05-26 | 14:00:00 | 1.556461321 | ['Imhotep1'] | 2.88E+16 | -9.8231E+15 |
| 2015-05-26 | 14:30:00 | 1.5563175 | ['Imhotep'] | 3.94E+16 | -1.34E+16 |
| 2015-05-26 | 15:00:00 | 1.556172075 | ['Imhotep'] | 5.19E+16 | -1.77E+16 |
| 2015-05-26 | 15:30:00 | 1.556034038 | ['Imhotep'] | 5.59E+16 | -1.91E+16 |
| 2015-05-26 | 16:00:00 | 1.555894906 | ['Imhotep'] | 5.55E+16 | -1.90E+16 |
| 2015-05-26 | 16:30:00 | 1.555756154 | ['Bes'] | 4.93E+16 | -1.68E+16 |
| 2015-05-26 | 17:00:00 | 1.555614906 | ['Bes'] | 4.47E+16 | -1.53E+16 |
| 2015-05-26 | 17:30:00 | 1.555475769 | ['Anhur'] | 3.34E+16 | -1.14E+16 |
| 2015-05-26 | 18:00:00 | 1.555337736 | ['Anhur'] | 3.97E+16 | -1.36E+16 |
| 2015-05-26 | 18:30:00 | 1.555193818 | ['Sobek'] | 5.58E+16 | -1.91E+16 |
| 2015-05-26 | 19:00:00 | 1.55505 | ['Neith'] | 5.95E+16 | -2.03E+16 |
| 2015-05-26 | 19:30:00 | 1.554907358 | ['Wosret'] | 6.18E+16 | -2.11E+16 |
| 2015-05-26 | 20:00:00 | 1.554768113 | ['Wosret'] | 6.05E+16 | -2.07E+16 |
| 2015-05-27 | 1:00:00 | 1.553352955 | ['Wosret'] | 2.84E+16 | -9.70795E+15 |
| 2015-05-27 | 1:30:00 | 1.553227255 | ['Imhotep1'] | 3.14E+16 | -1.07E+16 |
| 2015-05-27 | 2:00:00 | 1.553107 | ['Imhotep'] | 2.84E+16 | -9.69264E+15 |
| 2015-05-27 | 6:00:00 | 1.551963774 | ['Sobek'] | 3.44E+16 | -1.18E+16 |
| 2015-05-27 | 6:30:00 | 1.55189 | ['Sobek'] | 2.97E+16 | -1.02E+16 |
| 2015-05-27 | 10:30:00 | 1.550656471 | ['Sobek'] | 3.52E+16 | -1.20E+16 |
| 2015-05-27 | 11:00:00 | 1.550564717 | ['Anubis'] | 4.08E+16 | -1.39E+16 |
| 2015-05-27 | 17:30:00 | 1.54871 | ['Anubis'] | 4.85E+16 | -1.66E+16 |
| 2015-05-27 | 18:00:00 | 1.548605192 | ['Bes'] | 4.42E+16 | -1.51E+16 |
| 2015-05-27 | 21:00:00 | 1.54773 | ['Bes'] | 5.31E+16 | -1.81E+16 |
| 2015-05-27 | 21:30:00 | 1.547626038 | ['Geb'] | 4.71E+16 | -1.61E+16 |
| 2015-05-27 | 22:00:00 | 1.547488269 | ['Geb'] | 4.82E+16 | -1.65E+16 |
| 2015-05-27 | 22:30:00 | 1.547408 | ['Geb'] | 4.99E+16 | -1.71E+16 |
| 2015-05-28 | 2:00:00 | 1.54634129 | ['Geb'] | 2.79E+16 | -9.52055E+15 |
| 2015-05-28 | 2:30:00 | 1.546253243 | ['Imhotep1'] | 2.75E+16 | -9.37528E+15 |
| 2015-05-28 | 7:00:00 | 1.544953226 | ['Hatmehit'] | 3.12E+16 | -1.07E+16 |
| 2015-05-28 | 7:30:00 | 1.544843019 | ['Anhur'] | 4.11E+16 | -1.40E+16 |
| 2015-05-28 | 8:00:00 | 1.5447076 | ['Anhur'] | 5.45E+16 | -1.86E+16 |

| Date | Time | Value | Label | Col5 | Col6 |
|---|---|---|---|---|---|
| 2015-05-28 | 12:00:00 | 1.543566129 | ['Anhur'] | 4.24E+16 | -1.45E+16 |
| 2015-05-28 | 12:30:00 | 1.5434775 | ['Atum'] | 4.56E+16 | -1.56E+16 |
| 2015-05-28 | 17:00:00 | 1.542182609 | ['Atum'] | 5.13E+16 | -1.75E+16 |
| 2015-05-28 | 17:30:00 | 1.542052264 | ['Bes'] | 5.13E+16 | -1.75E+16 |
| 2015-05-28 | 18:00:00 | 1.541934595 | ['Bes'] | 4.49E+16 | -1.53E+16 |
| 2015-05-28 | 21:30:00 | 1.540945893 | ['Geb'] | 4.33E+16 | -1.48E+16 |
| 2015-05-28 | 22:00:00 | 1.540803774 | ['Geb'] | 4.41E+16 | -1.51E+16 |
| 2015-05-28 | 22:30:00 | 1.540667647 | ['Geb'] | 4.47E+16 | -1.53E+16 |
| 2015-05-28 | 23:00:00 | 1.540528868 | ['Geb'] | 4.71E+16 | -1.61E+16 |
| 2015-05-28 | 23:30:00 | 1.540391698 | ['Geb'] | 4.33E+16 | -1.48E+16 |
| 2015-05-29 | 0:00:00 | 1.540253396 | ['Atum'] | 3.89E+16 | -1.33E+16 |
| 2015-05-29 | 0:30:00 | 1.540115962 | ['Atum'] | 3.93E+16 | -1.34E+16 |
| 2015-05-29 | 1:00:00 | 1.539979434 | ['Khonsu'] | 3.95E+16 | -1.35E+16 |
| 2015-05-29 | 1:30:00 | 1.539838727 | ['Khonsu'] | 4.04E+16 | -1.38E+16 |
| 2015-05-29 | 2:00:00 | 1.53969537 | ['Imhotep1'] | 3.20E+16 | -1.09E+16 |
| 2015-05-29 | 2:30:00 | 1.539556415 | ['Imhotep'] | 3.05E+16 | -1.04E+16 |
| 2015-05-29 | 3:00:00 | 1.539418113 | ['Imhotep'] | 2.77E+16 | -9.45877E+15 |
| 2015-05-29 | 3:30:00 | 1.539280962 | ['Imhotep'] | 3.05E+16 | -1.04E+16 |
| 2015-05-29 | 4:00:00 | 1.53916275 | ['Bes'] | 3.53E+16 | -1.21E+16 |
| 2015-05-29 | 8:00:00 | 1.538036429 | ['Geb'] | 2.92E+16 | -9.97573E+15 |
| 2015-05-29 | 13:00:00 | 1.536650769 | ['Khonsu'] | 3.36E+16 | -1.15E+16 |
| 2015-05-29 | 17:30:00 | 1.5353475 | ['Khonsu'] | 5.19E+16 | -1.77E+16 |
| 2015-05-29 | 18:00:00 | 1.535273333 | ['Bes'] | 5.28E+16 | -1.80E+16 |
| 2015-05-29 | 23:00:00 | 1.533894151 | ['Geb'] | 3.90E+16 | -1.33E+16 |
| 2015-05-29 | 23:30:00 | 1.533756604 | ['Atum'] | 3.67E+16 | -1.25E+16 |
| 2015-05-30 | 1:00:00 | 1.533335472 | ['Bes'] | 4.43E+16 | -1.51E+16 |
| 2015-05-30 | 1:30:00 | 1.533200392 | ['Geb'] | 4.12E+16 | -1.41E+16 |
| 2015-05-30 | 2:00:00 | 1.533063962 | ['Geb'] | 3.90E+16 | -1.33E+16 |
| 2015-05-30 | 2:30:00 | 1.53292549 | ['Geb'] | 3.34E+16 | -1.14E+16 |
| 2015-05-30 | 3:00:00 | 1.532788491 | ['Geb'] | 2.49E+16 | -8.51873E+15 |
| 2015-05-30 | 3:30:00 | 1.532653269 | ['Atum'] | 2.19E+16 | -7.47858E+15 |
| 2015-05-30 | 4:00:00 | 1.53251717 | ['Atum'] | 2.36E+16 | -8.04933E+15 |
| 2015-05-30 | 4:30:00 | 1.532375273 | ['Khonsu'] | 2.86E+16 | -9.75397E+15 |
| 2015-05-30 | 5:00:00 | 1.532233774 | ['Imhotep1'] | 3.78E+16 | -1.29E+16 |
| 2015-05-30 | 5:30:00 | 1.532096792 | ['Imhotep'] | 4.68E+16 | -1.60E+16 |
| 2015-05-30 | 6:00:00 | 1.531961765 | ['Imhotep'] | 5.60E+16 | -1.91E+16 |
| 2015-05-30 | 9:00:00 | 1.531080909 | ['Hatmehit'] | 2.91E+16 | -9.93749E+15 |
| 2015-05-30 | 9:30:00 | 1.530999434 | ['Anhur'] | 3.94E+16 | -1.34E+16 |
| 2015-05-30 | 10:00:00 | 1.530862642 | ['Anhur'] | 4.07E+16 | -1.39E+16 |
| 2015-05-30 | 10:30:00 | 1.530763913 | ['Anhur'] | 4.10E+16 | -1.40E+16 |
| 2015-05-30 | 14:30:00 | 1.529628491 | ['Geb'] | 2.28E+16 | -7.80259E+15 |
| 2015-05-30 | 15:00:00 | 1.529545 | ['Atum'] | 2.72E+16 | -9.30032E+15 |
| 2015-05-30 | 19:30:00 | 1.528257547 | ['Bes'] | 3.42E+16 | -1.17E+16 |
| 2015-05-31 | 0:30:00 | 1.526880926 | ['Anuket'] | 3.94E+16 | -1.35E+16 |

| Date | Time | Value | Region | Col5 | Col6 |
|---|---|---|---|---|---|
| 2015-05-31 | 5:30:00 | 1.525510769 | ['Anuket'] | 2.94E+16 | -1.00E+16 |
| 2015-05-31 | 6:00:00 | 1.525376038 | ['Imhotep'] | 3.30E+16 | -1.13E+16 |
| 2015-05-31 | 6:30:00 | 1.525265 | ['Imhotep'] | 3.64E+16 | -1.24E+16 |
| 2015-05-31 | 10:30:00 | 1.524144118 | ['Imhotep'] | 2.83E+16 | -9.66563E+15 |
| 2015-05-31 | 11:00:00 | 1.524065 | ['Anhur'] | 2.75E+16 | -9.40473E+15 |
| 2015-05-31 | 16:00:00 | 1.522645472 | ['Khonsu'] | 2.32E+16 | -7.9249E+15 |
| 2015-05-31 | 16:30:00 | 1.52252725 | ['Khonsu'] | 2.30E+16 | -7.86556E+15 |
| 2015-06-01 | 6:30:00 | 1.518684615 | ['Hatmehit'] | 1.81E+16 | -6.18245E+15 |
| 2015-06-01 | 7:00:00 | 1.518590606 | ['Imhotep'] | 1.88E+16 | -6.43815E+15 |
| 2015-06-01 | 11:30:00 | 1.5173225 | ['Hatmehit'] | 1.78E+16 | -6.10559E+15 |
| 2015-06-01 | 12:00:00 | 1.517209057 | ['Sobek'] | 1.59E+16 | -5.45114E+15 |
| 2015-06-01 | 12:30:00 | 1.517075 | ['Neith'] | 1.50E+16 | -5.14032E+15 |
| 2015-06-01 | 16:30:00 | 1.515959091 | ['Neith'] | 4.08E+16 | -1.39E+16 |
| 2015-06-01 | 17:00:00 | 1.515857111 | ['Khonsu'] | 3.06E+16 | -1.05E+16 |
| 2015-06-01 | 21:30:00 | 1.514615366 | ['Khonsu'] | 5.44E+16 | -1.86E+16 |
| 2015-06-01 | 22:00:00 | 1.514496981 | ['Anhur'] | 5.39E+16 | -1.84E+16 |
| 2015-06-02 | 12:30:00 | 1.5105435 | ['Anhur'] | 1.96E+16 | -6.69232E+15 |
| 2015-06-02 | 13:00:00 | 1.510460638 | ['Wosret'] | 2.01E+16 | -6.87496E+15 |
| 2015-06-02 | 18:00:00 | 1.509104464 | ['Khonsu'] | 3.54E+16 | -1.21E+16 |
| 2015-06-02 | 18:30:00 | 1.508965849 | ['Imhotep1'] | 2.27E+16 | -7.75569E+15 |
| 2015-06-03 | 6:00:00 | 1.505893404 | ['Imhotep'] | 4.42E+16 | -1.51E+16 |
| 2015-06-03 | 6:30:00 | 1.50568 | ['Imhotep'] | 4.01E+16 | -1.39E+16 |
| 2015-06-03 | 7:00:00 | 1.505610769 | ['Bes'] | 2.26E+16 | -7.72141E+15 |
| 2015-06-03 | 9:30:00 | 1.504935814 | ['Bes'] | 3.33E+16 | -1.14E+16 |
| 2015-06-03 | 10:00:00 | 1.504814717 | ['Anhur'] | 4.38E+16 | -1.50E+16 |
| 2015-06-03 | 10:30:00 | 1.504697436 | ['Anhur'] | 5.29E+16 | -1.81E+16 |
| 2015-06-03 | 14:30:00 | 1.503585806 | ['Anhur'] | 3.12E+16 | -1.07E+16 |
| 2015-06-03 | 15:00:00 | 1.503503438 | ['Geb'] | 2.74E+16 | -9.35155E+15 |
| 2015-06-03 | 19:30:00 | 1.502232609 | ['Geb'] | 2.14E+16 | -7.3201E+15 |
| 2015-06-03 | 20:00:00 | 1.502135192 | ['Bes'] | 1.78E+16 | -6.09208E+15 |
| 2015-06-03 | 20:30:00 | 1.502002885 | ['Bes'] | 2.04E+16 | -6.96717E+15 |
| 2015-06-04 | 1:00:00 | 1.500804423 | ['Sobek'] | 2.30E+16 | -7.86772E+15 |
| 2015-06-04 | 6:00:00 | 1.499467174 | ['Sobek'] | 4.97E+16 | -1.70E+16 |
| 2015-06-04 | 6:30:00 | 1.49934283 | ['Imhotep'] | 5.75E+16 | -1.96E+16 |
| 2015-06-04 | 7:00:00 | 1.499238125 | ['Imhotep'] | 5.91E+16 | -2.02E+16 |
| 2015-06-04 | 11:00:00 | 1.498136364 | ['Imhotep'] | 3.38E+16 | -1.15E+16 |
| 2015-06-04 | 16:30:00 | 1.496691569 | ['Khonsu'] | 2.67E+16 | -9.10395E+15 |
| 2015-06-04 | 21:00:00 | 1.495449231 | ['Hatmehit'] | 1.87E+16 | -6.40753E+15 |
| 2015-06-04 | 21:30:00 | 1.49536283 | ['Bes'] | 2.59E+16 | -8.83899E+15 |
| 2015-06-04 | 22:00:00 | 1.495227358 | ['Bes'] | 3.45E+16 | -1.18E+16 |
| 2015-06-04 | 22:30:00 | 1.49509434 | ['Anhur'] | 2.98E+16 | -1.02E+16 |
| 2015-06-04 | 23:00:00 | 1.494962308 | ['Anhur'] | 3.15E+16 | -1.08E+16 |
| 2015-06-04 | 23:30:00 | 1.4948325 | ['Anhur'] | 3.30E+16 | -1.13E+16 |
| 2015-06-05 | 0:00:00 | 1.494702075 | ['Anhur'] | 4.79E+16 | -1.64E+16 |

| Date | Time | Value | Region | A | B |
|---|---|---|---|---|---|
| 2015-06-05 | 0:30:00 | 1.494570577 | ['Anhur'] | 7.54E+16 | -2.57E+16 |
| 2015-06-05 | 1:00:00 | 1.494438113 | ['Anhur'] | 6.08E+16 | -2.08E+16 |
| 2015-06-05 | 1:30:00 | 1.49430902 | ['Anhur'] | 4.42E+16 | -1.51E+16 |
| 2015-06-05 | 2:00:00 | 1.494172281 | ['Sobek'] | 4.04E+16 | -1.38E+16 |
| 2015-06-05 | 2:30:00 | 1.494035962 | ['Sobek'] | 4.22E+16 | -1.44E+16 |
| 2015-06-05 | 3:00:00 | 1.493904423 | ['Sobek'] | 5.11E+16 | -1.75E+16 |
| 2015-06-05 | 3:30:00 | 1.493775192 | ['Sobek'] | 5.19E+16 | -1.77E+16 |
| 2015-06-05 | 4:00:00 | 1.493643725 | ['Sobek'] | 4.09E+16 | -1.40E+16 |
| 2015-06-05 | 4:30:00 | 1.493512115 | ['Sobek'] | 3.88E+16 | -1.32E+16 |
| 2015-06-05 | 5:00:00 | 1.493381887 | ['Atum'] | 3.11E+16 | -1.06E+16 |
| 2015-06-05 | 5:30:00 | 1.493250189 | ['Khonsu'] | 2.05E+16 | -6.98943E+15 |
| 2015-06-05 | 6:00:00 | 1.493118491 | ['Khonsu'] | 2.92E+16 | -9.97691E+15 |
| 2015-06-05 | 6:30:00 | 1.492987 | ['Imhotep1'] | 5.49E+16 | -1.87E+16 |
| 2015-06-05 | 7:00:00 | 1.492845577 | ['Imhotep1'] | 6.76E+16 | -2.31E+16 |
| 2015-06-05 | 8:00:00 | 1.492627778 | ['Imhotep'] | 5.88E+16 | -2.01E+16 |
| 2015-06-05 | 12:30:00 | 1.491401132 | ['Imhotep'] | 6.00E+16 | -2.05E+16 |
| 2015-06-05 | 13:00:00 | 1.491270784 | ['Anhur'] | 6.52E+16 | -2.23E+16 |
| 2015-06-05 | 13:30:00 | 1.491146667 | ['Anhur'] | 5.46E+16 | -1.87E+16 |
| 2015-06-05 | 22:30:00 | 1.488727368 | ['Anhur'] | 2.96E+16 | -1.01E+16 |
| 2015-06-05 | 23:00:00 | 1.488641154 | ['Anhur'] | 3.21E+16 | -1.10E+16 |
| 2015-06-05 | 23:30:00 | 1.488508085 | ['Anhur'] | 2.22E+16 | -7.58454E+15 |
| 2015-06-06 | 2:30:00 | 1.487722041 | ['Atum'] | 5.23E+16 | -1.79E+16 |
| 2015-06-06 | 5:30:00 | 1.4869016 | ['Atum'] | 3.10E+16 | -1.06E+16 |
| 2015-06-06 | 9:00:00 | 1.485967 | ['Atum'] | 3.64E+16 | -1.24E+16 |
| 2015-06-06 | 9:30:00 | 1.485887547 | ['Bes'] | 3.00E+16 | -1.02E+16 |
| 2015-06-06 | 10:00:00 | 1.485757358 | ['Bes'] | 2.36E+16 | -8.0514E+15 |
| 2015-06-06 | 10:30:00 | 1.485652353 | ['Bes'] | 2.05E+16 | -7.01731E+15 |
| 2015-06-06 | 14:30:00 | 1.4845834 | ['Sobek'] | 6.24E+16 | -2.13E+16 |
| 2015-06-06 | 19:30:00 | 1.483280192 | ['Khonsu'] | 3.12E+16 | -1.07E+16 |
| 2015-06-06 | 20:00:00 | 1.483145789 | ['Imhotep'] | 8.51E+16 | -2.90E+16 |
| 2015-06-06 | 20:30:00 | 1.483044583 | ['Imhotep'] | 8.83E+16 | -3.02E+16 |
| 2015-06-07 | 0:30:00 | 1.481971455 | ['Imhotep'] | 1.80E+16 | -6.16012E+15 |
| 2015-06-07 | 5:30:00 | 1.480671176 | ['Imhotep'] | 3.69E+16 | -1.26E+16 |
| 2015-06-07 | 6:00:00 | 1.480542157 | ['Atum'] | 3.94E+16 | -1.35E+16 |
| 2015-06-07 | 6:30:00 | 1.480419565 | ['Khonsu'] | 3.05E+16 | -1.04E+16 |
| 2015-06-07 | 10:30:00 | 1.479376909 | ['Khepry'] | 2.94E+16 | -1.00E+16 |
| 2015-06-07 | 15:30:00 | 1.478082745 | ['Anuket'] | 6.54E+16 | -2.23E+16 |
| 2015-06-07 | 16:00:00 | 1.477951887 | ['Anhur'] | 6.41E+16 | -2.19E+16 |
| 2015-06-07 | 19:00:00 | 1.477168824 | ['Khonsu'] | 2.39E+16 | -8.17639E+15 |
| 2015-06-07 | 19:30:00 | 1.477042264 | ['Imhotep'] | 2.29E+16 | -7.82061E+15 |
| 2015-06-07 | 20:00:00 | 1.476913654 | ['Imhotep'] | 2.35E+16 | -8.02778E+15 |
| 2015-06-07 | 20:30:00 | 1.47678898 | ['Imhotep'] | 4.01E+16 | -1.37E+16 |
| 2015-06-07 | 21:00:00 | 1.47665766 | ['Imhotep'] | 8.52E+16 | -2.91E+16 |
| 2015-06-07 | 21:30:00 | 1.4765292 | ['Bes'] | 9.72E+16 | -3.32E+16 |

| Date | Time | Value | Label | Col5 | Col6 |
|---|---|---|---|---|---|
| 2015-06-07 | 22:00:00 | 1.476402128 | ['Bes'] | 7.17E+16 | -2.45E+16 |
| 2015-06-07 | 22:30:00 | 1.476268036 | ['Bes'] | 4.61E+16 | -1.57E+16 |
| 2015-06-07 | 23:00:00 | 1.476137551 | ['Bes'] | 2.42E+16 | -8.25172E+15 |
| 2015-06-07 | 23:30:00 | 1.476005294 | ['Anhur'] | 1.99E+16 | -6.81526E+15 |
| 2015-06-08 | 0:00:00 | 1.47588 | ['Anhur'] | 3.07E+16 | -1.05E+16 |
| 2015-06-08 | 0:30:00 | 1.475754118 | ['Anhur'] | 3.27E+16 | -1.12E+16 |
| 2015-06-08 | 1:00:00 | 1.475621569 | ['Anhur'] | 2.68E+16 | -9.15932E+15 |
| 2015-06-08 | 1:30:00 | 1.475494808 | ['Anhur'] | 2.22E+16 | -7.57211E+15 |
| 2015-06-08 | 2:00:00 | 1.475366471 | ['Anhur'] | 2.18E+16 | -7.46003E+15 |
| 2015-06-08 | 2:30:00 | 1.475239811 | ['Anhur'] | 3.46E+16 | -1.18E+16 |
| 2015-06-08 | 3:00:00 | 1.475106786 | ['Geb'] | 3.43E+16 | -1.17E+16 |
| 2015-06-08 | 3:30:00 | 1.474974151 | ['Geb'] | 5.02E+16 | -1.71E+16 |
| 2015-06-08 | 4:00:00 | 1.474846038 | ['Geb'] | 6.35E+16 | -2.17E+16 |
| 2015-06-08 | 4:30:00 | 1.474717358 | ['Geb'] | 6.35E+16 | -2.17E+16 |
| 2015-06-08 | 5:00:00 | 1.474590192 | ['Geb'] | 6.94E+16 | -2.37E+16 |
| 2015-06-08 | 5:30:00 | 1.474463208 | ['Atum'] | 5.71E+16 | -1.95E+16 |
| 2015-06-10 | 6:00:00 | 1.462082791 | ['Serqet'] | 3.59E+16 | -1.23E+16 |
| 2015-06-10 | 7:00:00 | 1.461812453 | ['Serqet'] | 6.63E+16 | -2.27E+16 |
| 2015-06-10 | 7:30:00 | 1.461690204 | ['Anuket'] | 6.13E+16 | -2.09E+16 |
| 2015-06-10 | 14:30:00 | 1.459903235 | ['Anuket'] | 4.66E+16 | -1.59E+16 |
| 2015-06-10 | 20:00:00 | 1.458530192 | ['Hathor'] | 5.55E+16 | -1.89E+16 |
| 2015-06-10 | 20:30:00 | 1.458406226 | ['Anuket'] | 2.84E+16 | -9.69361E+15 |
| 2015-06-11 | 0:30:00 | 1.457347778 | ['Anuket'] | 8.52E+16 | -2.91E+16 |
| 2015-06-11 | 1:00:00 | 1.457272885 | ['Imhotep'] | 8.87E+16 | -3.03E+16 |
| 2015-06-12 | 2:30:00 | 1.450892642 | ['Aten'] | 4.92E+16 | -1.68E+16 |
| 2015-06-12 | 8:30:00 | 1.449403396 | ['Seth'] | 5.04E+16 | -1.72E+16 |
| 2015-06-12 | 9:00:00 | 1.449280769 | ['Seth'] | 5.11E+16 | -1.74E+16 |
| 2015-06-12 | 13:00:00 | 1.448287547 | ['Ash'] | 5.96E+16 | -2.03E+16 |
| 2015-06-12 | 13:30:00 | 1.448164423 | ['Ash'] | 7.49E+16 | -2.56E+16 |
| 2015-06-13 | 2:00:00 | 1.445072453 | ['Khepry'] | 8.48E+16 | -2.90E+16 |
| 2015-06-13 | 2:30:00 | 1.444950962 | ['Aten'] | 7.30E+16 | -2.49E+16 |
| 2015-06-13 | 9:30:00 | 1.443223673 | ['Hathor'] | 5.65E+16 | -1.93E+16 |
| 2015-06-13 | 10:00:00 | 1.443101373 | ['Anuket'] | 5.17E+16 | -1.77E+16 |
| 2015-06-13 | 14:30:00 | 1.44199717 | ['Ash'] | 6.39E+16 | -2.18E+16 |
| 2015-06-13 | 19:30:00 | 1.440773333 | ['Maat'] | 2.14E+16 | -7.30973E+15 |
| 2015-06-13 | 20:00:00 | 1.440653396 | ['Maat'] | 2.07E+16 | -7.10287E+15 |
| 2015-06-14 | 3:00:00 | 1.438937115 | ['Maat'] | 5.49E+16 | -1.88E+16 |
| 2015-06-14 | 3:30:00 | 1.4388725 | ['Aten'] | 3.88E+16 | -1.33E+16 |
| 2015-06-14 | 5:30:00 | 1.438330204 | ['Aten'] | 4.39E+16 | -1.50E+16 |
| 2015-06-14 | 6:00:00 | 1.438212222 | ['Babi'] | 2.28E+16 | -7.79562E+15 |
| 2015-06-14 | 10:30:00 | 1.437106957 | ['Babi'] | 4.85E+16 | -1.66E+16 |
| 2015-06-14 | 11:00:00 | 1.437040833 | ['Hapi'] | 4.49E+16 | -1.54E+16 |
| 2015-06-14 | 15:30:00 | 1.43589119 | ['Hapi'] | 5.55E+16 | -1.90E+16 |
| 2015-06-14 | 16:00:00 | 1.435777857 | ['Ash'] | 4.12E+16 | -1.41E+16 |

| Date | Time | Value | Label | E+ | E- |
|---|---|---|---|---|---|
| 2015-06-15 | 11:30:00 | 1.431048 | ['Ash'] | 4.84E+16 | -1.65E+16 |
| 2015-06-15 | 12:00:00 | 1.43094383 | ['Hapi'] | 4.75E+16 | -1.62E+16 |
| 2015-06-15 | 23:00:00 | 1.42830717 | ['Maat'] | 1.69E+16 | -5.76669E+15 |
| 2015-06-15 | 23:30:00 | 1.428189231 | ['Hatmehit'] | 4.50E+16 | -1.54E+16 |
| 2015-06-16 | 4:00:00 | 1.427113019 | ['Ash'] | 1.52E+16 | -5.17978E+15 |
| 2015-06-16 | 4:30:00 | 1.426995385 | ['Ash'] | 1.76E+16 | -5.99922E+15 |
| 2015-06-16 | 5:00:00 | 1.426871607 | ['Ash'] | 2.53E+16 | -8.65131E+15 |
| 2015-06-16 | 13:00:00 | 1.424975217 | ['Hathor'] | 4.49E+16 | -1.54E+16 |
| 2015-06-16 | 18:00:00 | 1.423778393 | ['Ash'] | 1.61E+16 | -5.51966E+15 |
| 2015-06-17 | 7:30:00 | 1.4205812 | ['Ash'] | 3.01E+16 | -1.03E+16 |
| 2015-06-17 | 12:30:00 | 1.419404 | ['Maat'] | 3.05E+16 | -1.04E+16 |
| 2015-06-17 | 13:00:00 | 1.419289808 | ['Maat'] | 4.20E+16 | -1.43E+16 |
| 2015-06-17 | 13:30:00 | 1.419172453 | ['Serqet'] | 5.36E+16 | -1.83E+16 |
| 2015-06-17 | 16:30:00 | 1.418464694 | ['Seth'] | 7.63E+15 | -2.63865E+15 |
| 2015-06-17 | 19:30:00 | 1.417763704 | ['Aten'] | 3.20E+16 | -1.09E+16 |
| 2015-06-18 | 1:00:00 | 1.416474314 | ['Hathor'] | 4.36E+16 | -1.49E+16 |
| 2015-06-18 | 1:30:00 | 1.416353269 | ['Hathor'] | 5.20E+16 | -1.78E+16 |
| 2015-06-18 | 2:00:00 | 1.416240417 | ['Hapi'] | 4.53E+16 | -1.55E+16 |
| 2015-06-18 | 2:30:00 | 1.416123396 | ['Anuket'] | 2.66E+16 | -9.07553E+15 |
| 2015-06-18 | 5:00:00 | 1.415543654 | ['Ash'] | 9.72E+15 | -3.34139E+15 |
| 2015-06-18 | 6:00:00 | 1.415303396 | ['Ash'] | 3.32E+16 | -1.13E+16 |
| 2015-06-18 | 7:00:00 | 1.415074894 | ['Ash'] | 3.80E+16 | -1.30E+16 |
| 2015-06-18 | 7:30:00 | 1.414960588 | ['Ash'] | 4.31E+16 | -1.47E+16 |
| 2015-06-18 | 8:00:00 | 1.414840727 | ['Aten'] | 3.31E+16 | -1.13E+16 |
| 2015-06-18 | 8:30:00 | 1.414719804 | ['Aten'] | 2.74E+16 | -9.35992E+15 |
| 2015-06-18 | 9:00:00 | 1.414605577 | ['Babi'] | 4.01E+16 | -1.37E+16 |
| 2015-06-18 | 9:30:00 | 1.414488269 | ['Babi'] | 2.08E+16 | -7.12467E+15 |
| 2015-06-18 | 11:30:00 | 1.414025192 | ['Maat'] | 1.95E+16 | -6.66722E+15 |
| 2015-06-18 | 13:30:00 | 1.413563654 | ['Hathor'] | 4.82E+16 | -1.65E+16 |
| 2015-06-18 | 14:00:00 | 1.413444286 | ['Anuket'] | 5.30E+16 | -1.81E+16 |
| 2015-06-18 | 16:30:00 | 1.412864231 | ['Ash'] | 7.78E+15 | -2.67075E+15 |
| 2015-06-18 | 17:30:00 | 1.412634151 | ['Ash'] | 1.82E+16 | -6.22903E+15 |
| 2015-06-18 | 18:00:00 | 1.412519423 | ['Imhotep'] | 3.92E+16 | -1.34E+16 |
| 2015-06-18 | 18:30:00 | 1.412399455 | ['Imhotep'] | 6.06E+16 | -2.07E+16 |
| 2015-06-18 | 19:00:00 | 1.412280196 | ['Imhotep'] | 6.50E+16 | -2.22E+16 |
| 2015-06-18 | 19:30:00 | 1.412165577 | ['Imhotep'] | 6.62E+16 | -2.26E+16 |
| 2015-06-18 | 20:00:00 | 1.412052308 | ['Khepry'] | 5.88E+16 | -2.01E+16 |
| 2015-06-18 | 20:30:00 | 1.4119336 | ['Aten'] | 2.83E+16 | -9.67455E+15 |
| 2015-06-18 | 21:00:00 | 1.411820377 | ['Aten'] | 4.16E+16 | -1.42E+16 |
| 2015-06-18 | 21:30:00 | 1.411706154 | ['Babi'] | 4.16E+16 | -1.42E+16 |
| 2015-06-18 | 22:00:00 | 1.411591887 | ['Babi'] | 2.27E+16 | -7.74028E+15 |
| 2015-06-18 | 22:30:00 | 1.411474364 | ['Aker'] | 1.66E+16 | -5.68939E+15 |
| 2015-06-18 | 23:00:00 | 1.411355 | ['Hathor'] | 6.93E+15 | -2.37163E+15 |
| 2015-06-18 | 23:30:00 | 1.411238846 | ['Maat'] | 2.26E+16 | -7.71425E+15 |
| 2015-06-19 | 0:00:00 | 1.411124423 | ['Hatmehit'] | 2.25E+16 | -7.69403E+15 |
| 2015-06-19 | 0:30:00 | 1.411010566 | ['Nut'] | 2.82E+16 | -9.64609E+15 |
| 2015-06-19 | 4:30:00 | 1.41008434 | ['Atum'] | 1.64E+16 | -5.58715E+15 |
| 2015-06-19 | 5:30:00 | 1.409856604 | ['Imhotep1'] | 1.48E+16 | -5.07896E+15 |
| 2015-06-19 | 6:00:00 | 1.409741538 | ['Imhotep'] | 3.66E+16 | -1.25E+16 |
| 2015-06-19 | 6:30:00 | 1.409628077 | ['Imhotep'] | 7.90E+16 | -2.70E+16 |
| 2015-06-19 | 7:00:00 | 1.409513462 | ['Imhotep'] | 9.74E+16 | -3.32E+16 |
| 2015-06-19 | 7:30:00 | 1.409394912 | ['Imhotep'] | 8.53E+16 | -2.91E+16 |
| 2015-06-19 | 8:00:00 | 1.409276346 | ['Imhotep'] | 8.53E+16 | -2.91E+16 |
| 2015-06-19 | 8:30:00 | 1.409163269 | ['Imhotep'] | 4.97E+16 | -1.70E+16 |
| 2015-06-19 | 9:00:00 | 1.409047308 | ['Aten'] | 2.86E+16 | -9.75074E+15 |
| 2015-06-19 | 9:30:00 | 1.408933585 | ['Khepry'] | 4.67E+16 | -1.59E+16 |
| 2015-06-19 | 10:00:00 | 1.408819808 | ['Aker'] | 3.85E+16 | -1.32E+16 |
| 2015-06-19 | 10:30:00 | 1.408706415 | ['Aker'] | 3.39E+16 | -1.16E+16 |
| 2015-06-19 | 11:00:00 | 1.40859 | ['Aker'] | 1.75E+16 | -5.97108E+15 |
| 2015-06-19 | 12:00:00 | 1.408357636 | ['Hatmehit'] | 3.91E+16 | -1.33E+16 |
| 2015-06-19 | 12:30:00 | 1.40824 | ['Hatmehit'] | 2.44E+16 | -8.34001E+15 |
| 2015-06-19 | 13:00:00 | 1.408127692 | ['Maftet'] | 2.83E+16 | -9.65565E+15 |
| 2015-06-19 | 17:00:00 | 1.407209423 | ['Atum'] | 1.94E+16 | -6.62152E+15 |
| 2015-06-19 | 17:30:00 | 1.407096667 | ['Apis'] | 1.62E+16 | -5.54474E+15 |
| 2015-06-19 | 18:00:00 | 1.406983846 | ['Imhotep1'] | 2.10E+16 | -7.18798E+15 |
| 2015-06-19 | 18:30:00 | 1.406868113 | ['Imhotep'] | 5.67E+16 | -1.94E+16 |
| 2015-06-19 | 19:00:00 | 1.406753962 | ['Imhotep'] | 1.14E+17 | -3.89E+16 |
| 2015-06-19 | 19:30:00 | 1.406641346 | ['Imhotep'] | 1.17E+17 | -3.98E+16 |
| 2015-06-19 | 20:00:00 | 1.406528113 | ['Imhotep'] | 9.07E+16 | -3.10E+16 |
| 2015-06-19 | 21:30:00 | 1.406178868 | ['Khepry'] | 3.24E+16 | -1.11E+16 |
| 2015-06-19 | 22:00:00 | 1.40606434 | ['Khepry'] | 4.02E+16 | -1.37E+16 |
| 2015-06-19 | 22:30:00 | 1.4059525 | ['Anhur'] | 4.93E+16 | -1.68E+16 |
| 2015-06-19 | 23:00:00 | 1.405838868 | ['Anhur'] | 3.10E+16 | -1.06E+16 |
| 2015-06-20 | 6:00:00 | 1.404237358 | ['Ash'] | 1.72E+16 | -5.88357E+15 |
| 2015-06-20 | 6:30:00 | 1.404124615 | ['Imhotep'] | 2.07E+16 | -7.08455E+15 |
| 2015-06-20 | 7:00:00 | 1.40401283 | ['Imhotep'] | 5.03E+16 | -1.72E+16 |
| 2015-06-20 | 7:30:00 | 1.4039 | ['Imhotep'] | 1.05E+17 | -3.58E+16 |
| 2015-06-20 | 8:00:00 | 1.403786415 | ['Ash'] | 1.02E+17 | -3.50E+16 |
| 2015-06-20 | 8:30:00 | 1.403673654 | ['Aten'] | 9.32E+16 | -3.18E+16 |
| 2015-06-20 | 9:00:00 | 1.403561132 | ['Aten'] | 5.86E+16 | -2.00E+16 |
| 2015-06-20 | 9:30:00 | 1.403446 | ['Khepry'] | 3.81E+16 | -1.30E+16 |
| 2015-06-20 | 10:00:00 | 1.403329074 | ['Aten'] | 3.30E+16 | -1.13E+16 |
| 2015-06-20 | 10:30:00 | 1.403215769 | ['Babi'] | 4.85E+16 | -1.66E+16 |
| 2015-06-20 | 11:00:00 | 1.403103333 | ['Hapi'] | 3.18E+16 | -1.09E+16 |
| 2015-06-20 | 13:00:00 | 1.402651731 | ['Maat'] | 3.02E+16 | -1.03E+16 |
| 2015-06-20 | 14:00:00 | 1.402423036 | ['Serqet'] | 2.61E+16 | -8.90976E+15 |
| 2015-06-20 | 14:30:00 | 1.402306226 | ['Serqet'] | 4.15E+16 | -1.42E+16 |
| 2015-06-20 | 15:00:00 | 1.4021925 | ['Anuket'] | 6.24E+16 | -2.13E+16 |

| Date | Time | Value | Label | Col5 | Col6 |
|---|---|---|---|---|---|
| 2015-06-20 | 15:30:00 | 1.40208 | ['Anuket'] | 7.59E+16 | -2.59E+16 |
| 2015-06-20 | 16:00:00 | 1.401968269 | ['Hapi'] | 6.75E+16 | -2.30E+16 |
| 2015-06-20 | 16:30:00 | 1.401856415 | ['Seth'] | 4.20E+16 | -1.43E+16 |
| 2015-06-20 | 17:00:00 | 1.40174434 | ['Seth'] | 2.44E+16 | -8.35066E+15 |
| 2015-06-20 | 18:00:00 | 1.401519808 | ['Ash'] | 8.64E+15 | -2.95645E+15 |
| 2015-06-20 | 19:00:00 | 1.401288302 | ['Imhotep'] | 2.10E+16 | -7.16495E+15 |
| 2015-06-20 | 19:30:00 | 1.401175283 | ['Imhotep'] | 6.58E+16 | -2.25E+16 |
| 2015-06-20 | 20:00:00 | 1.401062642 | ['Imhotep'] | 9.99E+16 | -3.41E+16 |
| 2015-06-20 | 20:30:00 | 1.4009816 | ['Imhotep'] | 1.09E+17 | -3.71E+16 |
| 2015-06-21 | 0:30:00 | 1.400045769 | ['Imhotep'] | 9.74E+15 | -3.32695E+15 |
| 2015-06-21 | 6:00:00 | 1.39881 | ['Hapi'] | 1.71E+16 | -5.84894E+15 |
| 2015-06-21 | 10:30:00 | 1.39775875 | ['Hatmehit'] | 2.84E+16 | -9.70405E+15 |
| 2015-06-21 | 16:00:00 | 1.39656717 | ['Serqet'] | 5.86E+16 | -2.00E+16 |
| 2015-06-21 | 16:30:00 | 1.396473514 | ['Hathor'] | 7.22E+16 | -2.47E+16 |
| 2015-06-21 | 20:30:00 | 1.395552093 | ['Hathor'] | 5.32E+16 | -1.82E+16 |
| 2015-06-21 | 21:00:00 | 1.39548619 | ['Imhotep'] | 7.75E+16 | -2.65E+16 |
| 2015-06-22 | 2:30:00 | 1.394226792 | ['Maat'] | 2.42E+16 | -8.27303E+15 |
| 2015-06-22 | 3:00:00 | 1.394116731 | ['Maat'] | 1.73E+16 | -5.92089E+15 |
| 2015-06-22 | 3:30:00 | 1.394001071 | ['Seth'] | 1.67E+16 | -5.71341E+15 |
| 2015-06-22 | 4:00:00 | 1.3938875 | ['Seth'] | 2.71E+16 | -9.26085E+15 |
| 2015-06-22 | 5:30:00 | 1.393556604 | ['Seth'] | 3.93E+16 | -1.34E+16 |
| 2015-06-22 | 11:30:00 | 1.392208 | ['Seth'] | 2.93E+16 | -1.00E+16 |
| 2015-06-22 | 12:00:00 | 1.392119038 | ['Aten'] | 3.79E+16 | -1.29E+16 |
| 2015-06-22 | 12:30:00 | 1.392015957 | ['Aten'] | 3.60E+16 | -1.23E+16 |
| 2015-06-22 | 16:30:00 | 1.391124423 | ['Serqet'] | 2.82E+16 | -9.62005E+15 |
| 2015-06-22 | 17:00:00 | 1.391033235 | ['Serqet'] | 5.95E+16 | -2.03E+16 |
| 2015-06-22 | 22:00:00 | 1.389912264 | ['Ash'] | 4.42E+16 | -1.51E+16 |
| 2015-06-23 | 5:30:00 | 1.388263019 | ['Maat'] | 6.56E+16 | -2.24E+16 |
| 2015-06-23 | 6:00:00 | 1.388160213 | ['Hathor'] | 5.37E+16 | -1.83E+16 |
| 2015-06-23 | 16:00:00 | 1.385973077 | ['Aker'] | 1.18E+16 | -4.02788E+15 |
| 2015-06-23 | 18:00:00 | 1.385533585 | ['Serqet'] | 6.24E+16 | -2.13E+16 |
| 2015-06-23 | 18:30:00 | 1.385425962 | ['Serqet'] | 5.43E+16 | -1.85E+16 |
| 2015-06-23 | 23:00:00 | 1.384448679 | ['Ash'] | 2.62E+16 | -8.93857E+15 |
| 2015-06-23 | 23:30:00 | 1.384341346 | ['Ash'] | 2.99E+16 | -1.02E+16 |
| 2015-06-24 | 6:30:00 | 1.382776667 | ['Ash'] | 4.35E+16 | -1.51E+16 |
| 2015-06-24 | 10:30:00 | 1.381982703 | ['Ash'] | 1.19E+16 | -4.07465E+15 |
| 2015-06-24 | 14:30:00 | 1.381082667 | ['Ash'] | 2.48E+16 | -8.4573E+15 |
| 2015-06-24 | 15:00:00 | 1.380997308 | ['Babi'] | 1.69E+16 | -5.79125E+15 |
| 2015-06-24 | 15:30:00 | 1.380891132 | ['Babi'] | 8.58E+15 | -2.93528E+15 |
| 2015-06-24 | 17:00:00 | 1.380564717 | ['Maat'] | 2.45E+16 | -8.39384E+15 |
| 2015-06-24 | 18:30:00 | 1.380244528 | ['Serqet'] | 5.52E+16 | -1.88E+16 |
| 2015-06-24 | 19:00:00 | 1.380138846 | ['Anuket'] | 5.52E+16 | -1.89E+16 |
| 2015-06-25 | 0:30:00 | 1.378915556 | ['Anuket'] | 7.00E+16 | -2.39E+16 |
| 2015-06-25 | 1:00:00 | 1.378854706 | ['Imhotep'] | 6.29E+16 | -2.15E+16 |
| 2015-06-25 | 4:30:00 | 1.378105849 | ['Imhotep'] | 2.75E+16 | -9.37769E+15 |
| 2015-06-25 | 5:00:00 | 1.377999623 | ['Hatmehit'] | 2.70E+16 | -9.23391E+15 |
| 2015-06-25 | 5:30:00 | 1.377893654 | ['Hatmehit'] | 2.80E+16 | -9.5565E+15 |
| 2015-06-25 | 6:00:00 | 1.377787547 | ['Nut'] | 2.90E+16 | -9.90965E+15 |
| 2015-06-27 | 6:00:00 | 1.367721154 | ['Wosret'] | 3.49E+16 | -1.19E+16 |
| 2015-06-27 | 6:30:00 | 1.367618491 | ['Wosret'] | 4.03E+16 | -1.37E+16 |
| 2015-06-27 | 7:00:00 | 1.367517255 | ['Wosret'] | 3.82E+16 | -1.31E+16 |
| 2015-06-27 | 7:30:00 | 1.3674134 | ['Wosret'] | 5.37E+16 | -1.83E+16 |
| 2015-06-27 | 8:00:00 | 1.367312075 | ['Neith'] | 8.77E+16 | -2.99E+16 |
| 2015-06-27 | 8:30:00 | 1.367207091 | ['Neith'] | 1.00E+17 | -3.42E+16 |
| 2015-06-27 | 9:00:00 | 1.367100741 | ['Anuket'] | 8.54E+16 | -2.92E+16 |
| 2015-06-27 | 9:30:00 | 1.366997115 | ['Anuket'] | 5.58E+16 | -1.90E+16 |
| 2015-06-27 | 10:00:00 | 1.366895385 | ['Hapi'] | 5.20E+16 | -1.78E+16 |
| 2015-06-27 | 10:30:00 | 1.366793774 | ['Anubis'] | 4.35E+16 | -1.49E+16 |
| 2015-06-27 | 11:00:00 | 1.366691132 | ['Atum'] | 2.86E+16 | -9.77568E+15 |
| 2015-06-27 | 11:30:00 | 1.3665906 | ['Khonsu'] | 3.13E+16 | -1.07E+16 |
| 2015-06-27 | 12:00:00 | 1.366487925 | ['Imhotep1'] | 4.77E+16 | -1.63E+16 |
| 2015-06-27 | 12:30:00 | 1.366385283 | ['Imhotep1'] | 5.59E+16 | -1.91E+16 |
| 2015-06-27 | 13:00:00 | 1.366280185 | ['Imhotep'] | 1.25E+17 | -4.25E+16 |
| 2015-06-27 | 13:30:00 | 1.366173077 | ['Imhotep'] | 1.39E+17 | -4.73E+16 |
| 2015-06-27 | 14:00:00 | 1.366071509 | ['Imhotep'] | 9.57E+16 | -3.27E+16 |
| 2015-06-27 | 15:00:00 | 1.365869057 | ['Imhotep'] | 4.69E+16 | -1.60E+16 |
| 2015-06-27 | 15:30:00 | 1.365766415 | ['Khepry'] | 4.03E+16 | -1.37E+16 |
| 2015-06-27 | 16:00:00 | 1.365664694 | ['Khepry'] | 5.92E+16 | -2.02E+16 |
| 2015-06-27 | 16:30:00 | 1.365563111 | ['Khepry'] | 4.83E+16 | -1.65E+16 |
| 2015-06-27 | 17:00:00 | 1.365463043 | ['Aker'] | 4.85E+16 | -1.65E+16 |
| 2015-06-27 | 17:30:00 | 1.365356604 | ['Aker'] | 4.59E+16 | -1.57E+16 |
| 2015-06-27 | 18:00:00 | 1.365250385 | ['Sobek'] | 2.94E+16 | -1.00E+16 |
| 2015-06-27 | 18:30:00 | 1.3651502 | ['Wosret'] | 3.60E+16 | -1.23E+16 |
| 2015-06-27 | 19:00:00 | 1.365047 | ['Wosret'] | 3.59E+16 | -1.22E+16 |
| 2015-06-27 | 19:30:00 | 1.36494549 | ['Wosret'] | 3.67E+16 | -1.25E+16 |
| 2015-06-27 | 20:00:00 | 1.364844717 | ['Maftet'] | 5.61E+16 | -1.92E+16 |
| 2015-06-27 | 20:30:00 | 1.364742642 | ['Anuket'] | 8.61E+16 | -2.94E+16 |
| 2015-06-27 | 21:00:00 | 1.364641132 | ['Neith'] | 9.67E+16 | -3.30E+16 |
| 2015-06-27 | 21:30:00 | 1.36454 | ['Anuket'] | 6.79E+16 | -2.32E+16 |
| 2015-06-27 | 22:00:00 | 1.364432453 | ['Anuket'] | 5.17E+16 | -1.77E+16 |
| 2015-06-27 | 22:30:00 | 1.364330392 | ['Anuket'] | 4.54E+16 | -1.55E+16 |
| 2015-06-27 | 23:00:00 | 1.364229423 | ['Anubis'] | 4.22E+16 | -1.44E+16 |
| 2015-06-27 | 23:30:00 | 1.364129804 | ['Atum'] | 2.68E+16 | -9.15425E+15 |
| 2015-06-28 | 0:00:00 | 1.3640264 | ['Khonsu'] | 2.63E+16 | -8.98912E+15 |
| 2015-06-28 | 0:30:00 | 1.363924 | ['Imhotep1'] | 4.47E+16 | -1.52E+16 |
| 2015-06-28 | 1:00:00 | 1.363865 | ['Imhotep1'] | 5.35E+16 | -1.84E+16 |
| 2015-06-28 | 5:30:00 | 1.362900213 | ['Hatmehit'] | 4.96E+16 | -1.69E+16 |
| 2015-06-28 | 6:00:00 | 1.362805094 | ['Aker'] | 3.96E+16 | -1.35E+16 |

| Date | Time | Value | Label | Col5 | Col6 |
|---|---|---|---|---|---|
| 2015-06-28 | 6:30:00 | 1.362706531 | ['Wosret'] | 2.60E+16 | -8.88057E+15 |
| 2015-06-28 | 8:00:00 | 1.362394808 | ['Seth'] | 3.42E+16 | -1.17E+16 |
| 2015-06-28 | 8:30:00 | 1.362295435 | ['Seth'] | 6.26E+16 | -2.14E+16 |
| 2015-06-28 | 9:00:00 | 1.3621925 | ['Seth'] | 8.67E+16 | -2.96E+16 |
| 2015-06-28 | 9:30:00 | 1.362096226 | ['Seth'] | 8.01E+16 | -2.74E+16 |
| 2015-06-28 | 10:00:00 | 1.361995472 | ['Anubis'] | 7.36E+16 | -2.51E+16 |
| 2015-06-28 | 15:30:00 | 1.36088 | ['Anubis'] | 6.08E+16 | -2.08E+16 |
| 2015-06-28 | 16:00:00 | 1.360779057 | ['Imhotep'] | 5.02E+16 | -1.71E+16 |
| 2015-06-28 | 16:30:00 | 1.360694595 | ['Khepry'] | 4.21E+16 | -1.44E+16 |
| 2015-06-28 | 20:30:00 | 1.3598726 | ['Wosret'] | 4.44E+16 | -1.52E+16 |
| 2015-06-28 | 21:00:00 | 1.359772308 | ['Maftet'] | 5.42E+16 | -1.85E+16 |
| 2015-07-01 | 4:30:00 | 1.348823585 | ['Ash'] | 7.12E+16 | -2.43E+16 |
| 2015-07-01 | 20:00:00 | 1.345844528 | ['Babi'] | 3.42E+16 | -1.17E+16 |
| 2015-07-02 | 0:00:00 | 1.345080769 | ['Seth'] | 3.74E+16 | -1.28E+16 |
| 2015-07-02 | 0:30:00 | 1.344986604 | ['Seth'] | 5.65E+16 | -1.93E+16 |
| 2015-07-02 | 1:00:00 | 1.344891509 | ['Hapi'] | 5.01E+16 | -1.71E+16 |
| 2015-07-02 | 6:00:00 | 1.343940909 | ['Seth'] | 6.71E+16 | -2.29E+16 |
| 2015-07-02 | 6:30:00 | 1.343843019 | ['Ash'] | 4.13E+16 | -1.41E+16 |
| 2015-07-02 | 12:30:00 | 1.342709423 | ['Seth'] | 3.80E+16 | -1.30E+16 |
| 2015-07-03 | 0:00:00 | 1.340548302 | ['Hapi'] | 1.45E+16 | -4.95575E+15 |
| 2015-07-03 | 12:00:00 | 1.338320566 | ['Babi'] | 9.15E+15 | -3.14043E+15 |
| 2015-07-03 | 12:30:00 | 1.338228868 | ['Babi'] | 1.21E+16 | -4.13409E+15 |
| 2015-07-04 | 7:30:00 | 1.334734706 | ['Ash'] | 2.64E+16 | -9.0547E+15 |
| 2015-07-04 | 8:00:00 | 1.334667778 | ['Ash'] | 3.20E+16 | -1.09E+16 |
| 2015-07-04 | 9:00:00 | 1.334463962 | ['Ash'] | 3.29E+16 | -1.12E+16 |
| 2015-07-04 | 9:30:00 | 1.334374423 | ['Aten'] | 4.11E+16 | -1.40E+16 |
| 2015-07-04 | 10:00:00 | 1.334284528 | ['Aten'] | 3.90E+16 | -1.33E+16 |
| 2015-07-04 | 10:30:00 | 1.334191818 | ['Babi'] | 3.02E+16 | -1.03E+16 |
| 2015-07-04 | 15:00:00 | 1.333374561 | ['Maat'] | 5.79E+16 | -1.98E+16 |
| 2015-07-04 | 15:30:00 | 1.333281154 | ['Hathor'] | 3.43E+16 | -1.17E+16 |
| 2015-07-04 | 16:00:00 | 1.333191923 | ['Hapi'] | 3.39E+16 | -1.16E+16 |
| 2015-07-04 | 16:30:00 | 1.333101887 | ['Hapi'] | 2.75E+16 | -9.40761E+15 |
| 2015-07-05 | 15:30:00 | 1.328984423 | ['Maat'] | 7.85E+16 | -2.70E+16 |
| 2015-07-05 | 16:00:00 | 1.328896981 | ['Maat'] | 4.17E+16 | -1.43E+16 |
| 2015-07-05 | 16:30:00 | 1.328810196 | ['Hathor'] | 3.45E+16 | -1.18E+16 |
| 2015-07-05 | 22:30:00 | 1.3277475 | ['Khepry'] | 5.03E+16 | -1.72E+16 |
| 2015-07-06 | 4:00:00 | 1.326779808 | ['Serqet'] | 7.24E+16 | -2.48E+16 |
| 2015-07-06 | 4:30:00 | 1.326693462 | ['Serqet'] | 4.20E+16 | -1.44E+16 |
| 2015-07-06 | 5:00:00 | 1.326605962 | ['Hathor'] | 3.43E+16 | -1.17E+16 |
| 2015-07-06 | 5:30:00 | 1.326536364 | ['Hathor'] | 3.50E+16 | -1.19E+16 |
| 2015-07-06 | 16:00:00 | 1.324689298 | ['Maat'] | 6.10E+16 | -2.08E+16 |
| 2015-07-06 | 16:30:00 | 1.324599423 | ['Maat'] | 6.72E+16 | -2.29E+16 |
| 2015-07-06 | 21:30:00 | 1.323717241 | ['Maat'] | 5.82E+16 | -1.99E+16 |
| 2015-07-06 | 22:00:00 | 1.323650566 | ['Ash'] | 6.25E+16 | -2.13E+16 |

| Date | Time | Value | Region | Col5 | Col6 |
|---|---|---|---|---|---|
| 2015-07-06 | 22:30:00 | 1.323564906 | ['Ash'] | 5.31E+16 | -1.81E+16 |
| 2015-07-07 | 4:30:00 | 1.32253549 | ['Maat'] | 6.89E+16 | -2.35E+16 |
| 2015-07-07 | 5:00:00 | 1.322448393 | ['Maat'] | 5.90E+16 | -2.02E+16 |
| 2015-07-07 | 9:30:00 | 1.321677818 | ['Ash'] | 8.21E+16 | -2.81E+16 |
| 2015-07-07 | 10:00:00 | 1.32159 | ['Ash'] | 7.67E+16 | -2.62E+16 |
| 2015-07-07 | 10:30:00 | 1.321507843 | ['Ash'] | 6.58E+16 | -2.25E+16 |
| 2015-07-07 | 11:00:00 | 1.32142234 | ['Ash'] | 5.38E+16 | -1.84E+16 |
| 2015-07-07 | 11:30:00 | 1.321335833 | ['Aten'] | 4.03E+16 | -1.38E+16 |
| 2015-07-07 | 12:00:00 | 1.321253061 | ['Ash'] | 4.93E+16 | -1.68E+16 |
| 2015-07-07 | 16:30:00 | 1.320487551 | ['Hatmehit'] | 4.07E+16 | -1.39E+16 |
| 2015-07-07 | 17:00:00 | 1.320404792 | ['Hatmehit'] | 6.43E+16 | -2.19E+16 |
| 2015-07-07 | 17:30:00 | 1.3203208 | ['Maat'] | 7.50E+16 | -2.56E+16 |
| 2015-07-07 | 18:00:00 | 1.320234545 | ['Hathor'] | 5.81E+16 | -1.98E+16 |
| 2015-07-07 | 18:30:00 | 1.320147547 | ['Hathor'] | 4.58E+16 | -1.56E+16 |
| 2015-07-07 | 22:30:00 | 1.319441333 | ['Hatmehit'] | 7.15E+16 | -2.44E+16 |
| 2015-07-07 | 23:00:00 | 1.319385686 | ['Imhotep'] | 7.87E+16 | -2.69E+16 |
| 2015-07-07 | 23:30:00 | 1.319303585 | ['Imhotep'] | 6.82E+16 | -2.33E+16 |
| 2015-07-08 | 4:30:00 | 1.318462885 | ['Serqet'] | 4.19E+16 | -1.43E+16 |
| 2015-07-08 | 9:30:00 | 1.317627358 | ['Ash'] | 9.35E+16 | -3.19E+16 |
| 2015-07-08 | 10:00:00 | 1.317544528 | ['Imhotep'] | 1.16E+17 | -3.97E+16 |
| 2015-07-08 | 10:30:00 | 1.317475714 | ['Imhotep'] | 1.17E+17 | -4.00E+16 |
| 2015-07-08 | 15:00:00 | 1.316714314 | ['Hatmehit'] | 2.51E+16 | -8.55903E+15 |
| 2015-07-08 | 16:30:00 | 1.316465714 | ['Serqet'] | 2.88E+16 | -9.83784E+15 |
| 2015-07-08 | 17:00:00 | 1.316380189 | ['Anuket'] | 5.38E+16 | -1.84E+16 |
| 2015-07-08 | 17:30:00 | 1.316297736 | ['Hapi'] | 7.48E+16 | -2.56E+16 |
| 2015-07-08 | 18:00:00 | 1.316215686 | ['Seth'] | 1.03E+17 | -3.51E+16 |
| 2015-07-08 | 18:30:00 | 1.316134423 | ['Seth'] | 7.88E+16 | -2.69E+16 |
| 2015-07-08 | 19:00:00 | 1.316052075 | ['Anubis'] | 4.23E+16 | -1.45E+16 |
| 2015-07-08 | 19:30:00 | 1.315970189 | ['Atum'] | 3.12E+16 | -1.07E+16 |
| 2015-07-08 | 20:00:00 | 1.315888462 | ['Atum'] | 3.09E+16 | -1.05E+16 |
| 2015-07-08 | 20:30:00 | 1.315806981 | ['Apis'] | 1.81E+16 | -6.20668E+15 |
| 2015-07-09 | 1:00:00 | 1.3150638 | ['Khepry'] | 5.93E+16 | -2.03E+16 |
| 2015-07-09 | 4:30:00 | 1.314491111 | ['Seth'] | 2.38E+16 | -8.12776E+15 |
| 2015-07-09 | 5:00:00 | 1.314406538 | ['Seth'] | 4.73E+16 | -1.62E+16 |
| 2015-07-09 | 5:30:00 | 1.314322941 | ['Seth'] | 8.05E+16 | -2.75E+16 |
| 2015-07-09 | 6:00:00 | 1.314242642 | ['Seth'] | 8.52E+16 | -2.91E+16 |
| 2015-07-09 | 11:00:00 | 1.31341725 | ['Seth'] | 1.40E+17 | -4.77E+16 |
| 2015-07-09 | 11:30:00 | 1.313347556 | ['Imhotep'] | 8.38E+16 | -2.87E+16 |
| 2015-07-09 | 16:30:00 | 1.312538491 | ['Wosret'] | 3.42E+16 | -1.17E+16 |
| 2015-07-09 | 17:30:00 | 1.312377692 | ['Maftet'] | 6.29E+16 | -2.15E+16 |
| 2015-07-09 | 18:00:00 | 1.312294211 | ['Geb'] | 1.12E+17 | -3.82E+16 |
| 2015-07-09 | 18:30:00 | 1.312238462 | ['Geb'] | 8.34E+16 | -2.85E+16 |
| 2015-07-09 | 20:00:00 | 1.311971321 | ['Khonsu'] | 4.06E+16 | -1.39E+16 |
| 2015-07-09 | 20:30:00 | 1.311891538 | ['Anubis'] | 2.89E+16 | -9.88644E+15 |

| Date | Time | Value | Label | Col5 | Col6 |
|---|---|---|---|---|---|
| 2015-07-09 | 21:00:00 | 1.311811887 | ['Seth'] | 2.64E+16 | -9.00807E+15 |
| 2015-07-09 | 21:30:00 | 1.311737234 | ['Imhotep1'] | 2.77E+16 | -9.47446E+15 |
| 2015-07-09 | 23:30:00 | 1.311439167 | ['Bes'] | 1.52E+17 | -5.20E+16 |
| 2015-07-10 | 0:30:00 | 1.311226154 | ['Ash'] | 8.77E+16 | -3.00E+16 |
| 2015-07-10 | 1:00:00 | 1.311168302 | ['Ash'] | 5.18E+16 | -1.77E+16 |
| 2015-07-10 | 1:30:00 | 1.311089423 | ['Aten'] | 7.08E+16 | -2.42E+16 |
| 2015-07-10 | 2:00:00 | 1.311009623 | ['Aker'] | 6.92E+16 | -2.36E+16 |
| 2015-07-10 | 2:30:00 | 1.310927857 | ['Anhur'] | 7.09E+16 | -2.42E+16 |
| 2015-07-10 | 3:00:00 | 1.31084566 | ['Anhur'] | 5.70E+16 | -1.94E+16 |
| 2015-07-10 | 6:00:00 | 1.31037 | ['Serqet'] | 7.45E+16 | -2.54E+16 |
| 2015-07-10 | 6:30:00 | 1.310291321 | ['Geb'] | 1.21E+17 | -4.14E+16 |
| 2015-07-10 | 10:00:00 | 1.309733019 | ['Imhotep1'] | 3.29E+16 | -1.12E+16 |
| 2015-07-10 | 11:30:00 | 1.30949386 | ['Imhotep'] | 1.64E+17 | -5.60E+16 |
| 2015-07-10 | 12:00:00 | 1.309412308 | ['Imhotep'] | 1.26E+17 | -4.31E+16 |
| 2015-07-10 | 12:30:00 | 1.309333774 | ['Bes'] | 9.43E+16 | -3.22E+16 |
| 2015-07-10 | 13:00:00 | 1.30925566 | ['Bes'] | 5.87E+16 | -2.00E+16 |
| 2015-07-10 | 13:30:00 | 1.30918 | ['Bes'] | 5.37E+16 | -1.83E+16 |
| 2015-07-10 | 17:30:00 | 1.308541837 | ['Hatmehit'] | 3.96E+16 | -1.35E+16 |
| 2015-07-10 | 18:00:00 | 1.308478649 | ['Geb'] | 5.53E+16 | -1.89E+16 |
| 2015-07-10 | 20:00:00 | 1.3081525 | ['Khonsu'] | 6.94E+16 | -2.37E+16 |
| 2015-07-10 | 20:30:00 | 1.308071509 | ['Anuket'] | 4.16E+16 | -1.42E+16 |
| 2015-07-10 | 21:30:00 | 1.307915577 | ['Hatmehit'] | 3.05E+16 | -1.04E+16 |
| 2015-07-10 | 22:00:00 | 1.307838302 | ['Apis'] | 2.73E+16 | -9.32663E+15 |
| 2015-07-10 | 22:30:00 | 1.307760385 | ['Imhotep'] | 5.07E+16 | -1.73E+16 |
| 2015-07-10 | 23:00:00 | 1.307682075 | ['Imhotep'] | 9.12E+16 | -3.12E+16 |
| 2015-07-11 | 6:30:00 | 1.306510962 | ['Anuket'] | 5.87E+16 | -2.01E+16 |
| 2015-07-11 | 7:00:00 | 1.306433585 | ['Hapi'] | 1.10E+17 | -3.75E+16 |
| 2015-07-11 | 7:30:00 | 1.306356731 | ['Seth'] | 1.19E+17 | -4.08E+16 |
| 2015-07-11 | 8:00:00 | 1.30628 | ['Seth'] | 9.48E+16 | -3.24E+16 |
| 2015-07-11 | 8:30:00 | 1.306203396 | ['Anubis'] | 6.75E+16 | -2.30E+16 |
| 2015-07-11 | 9:00:00 | 1.306125185 | ['Atum'] | 4.48E+16 | -1.53E+16 |
| 2015-07-11 | 9:30:00 | 1.306045273 | ['Anubis'] | 4.45E+16 | -1.52E+16 |
| 2015-07-11 | 10:00:00 | 1.305966415 | ['Atum'] | 2.44E+16 | -8.33667E+15 |
| 2015-07-11 | 10:30:00 | 1.305890385 | ['Khonsu'] | 2.65E+16 | -9.05314E+15 |
| 2015-07-11 | 11:00:00 | 1.305813962 | ['Imhotep'] | 4.82E+16 | -1.65E+16 |
| 2015-07-11 | 11:30:00 | 1.305736792 | ['Imhotep'] | 7.48E+16 | -2.56E+16 |
| 2015-07-11 | 12:00:00 | 1.305659623 | ['Imhotep'] | 1.34E+17 | -4.58E+16 |
| 2015-07-11 | 12:30:00 | 1.305583725 | ['Imhotep'] | 1.56E+17 | -5.34E+16 |
| 2015-07-11 | 13:00:00 | 1.305507547 | ['Imhotep'] | 1.12E+17 | -3.84E+16 |
| 2015-07-11 | 13:30:00 | 1.305427692 | ['Khepry'] | 7.44E+16 | -2.54E+16 |
| 2015-07-11 | 14:00:00 | 1.305348679 | ['Aten'] | 4.47E+16 | -1.53E+16 |
| 2015-07-11 | 14:30:00 | 1.305272264 | ['Khepry'] | 6.68E+16 | -2.28E+16 |
| 2015-07-11 | 15:00:00 | 1.305195769 | ['Khepry'] | 6.12E+16 | -2.09E+16 |
| 2015-07-11 | 15:30:00 | 1.30512 | ['Hapi'] | 5.70E+16 | -1.95E+16 |
| 2015-07-11 | 16:00:00 | 1.305043585 | ['Hathor'] | 5.09E+16 | -1.74E+16 |
| 2015-07-11 | 16:30:00 | 1.304967115 | ['Bastet'] | 3.29E+16 | -1.13E+16 |
| 2015-07-11 | 17:30:00 | 1.304815472 | ['Hatmehit'] | 3.11E+16 | -1.06E+16 |
| 2015-07-11 | 18:30:00 | 1.304657547 | ['Nut'] | 3.88E+16 | -1.33E+16 |
| 2015-07-11 | 19:00:00 | 1.304581132 | ['Anuket'] | 5.34E+16 | -1.82E+16 |
| 2015-07-11 | 19:30:00 | 1.304504717 | ['Anuket'] | 9.31E+16 | -3.18E+16 |
| 2015-07-11 | 20:00:00 | 1.304429608 | ['Anuket'] | 1.16E+17 | -3.96E+16 |
| 2015-07-11 | 20:30:00 | 1.304353774 | ['Anuket'] | 1.01E+17 | -3.44E+16 |
| 2015-07-11 | 21:00:00 | 1.304277547 | ['Seth'] | 6.91E+16 | -2.36E+16 |
| 2015-07-11 | 21:30:00 | 1.304201321 | ['Anubis'] | 3.72E+16 | -1.27E+16 |
| 2015-07-11 | 22:00:00 | 1.304126154 | ['Atum'] | 3.51E+16 | -1.20E+16 |
| 2015-07-11 | 23:00:00 | 1.303969231 | ['Imhotep1'] | 2.65E+16 | -9.06764E+15 |
| 2015-07-11 | 23:30:00 | 1.303893396 | ['Imhotep'] | 4.62E+16 | -1.58E+16 |
| 2015-07-12 | 0:00:00 | 1.303817547 | ['Imhotep'] | 8.47E+16 | -2.89E+16 |
| 2015-07-12 | 0:30:00 | 1.303742075 | ['Imhotep'] | 1.51E+17 | -5.14E+16 |
| 2015-07-12 | 1:00:00 | 1.303696 | ['Imhotep'] | 1.00E+17 | -3.43E+16 |
| 2015-07-12 | 5:30:00 | 1.3029826 | ['Imhotep'] | 1.65E+16 | -5.63049E+15 |
| 2015-07-12 | 6:00:00 | 1.302908679 | ['Hatmehit'] | 4.79E+16 | -1.63E+16 |
| 2015-07-12 | 6:30:00 | 1.302831887 | ['Hatmehit'] | 3.52E+16 | -1.20E+16 |
| 2015-07-12 | 7:30:00 | 1.302682264 | ['Seth'] | 5.38E+16 | -1.84E+16 |
| 2015-07-12 | 8:00:00 | 1.302602885 | ['Seth'] | 1.00E+17 | -3.42E+16 |
| 2015-07-12 | 8:30:00 | 1.302528113 | ['Seth'] | 1.20E+17 | -4.09E+16 |
| 2015-07-12 | 10:30:00 | 1.302229273 | ['Seth'] | 3.50E+16 | -1.19E+16 |
| 2015-07-12 | 16:00:00 | 1.301401887 | ['Aten'] | 6.30E+16 | -2.15E+16 |
| 2015-07-12 | 16:30:00 | 1.301326863 | ['Khepry'] | 5.29E+16 | -1.81E+16 |
| 2015-07-12 | 20:30:00 | 1.300730189 | ['Hathor'] | 8.28E+16 | -2.83E+16 |
| 2015-07-12 | 23:00:00 | 1.300356731 | ['Seth'] | 2.33E+16 | -7.97045E+15 |
| 2015-07-12 | 23:30:00 | 1.300282453 | ['Ash'] | 1.62E+16 | -5.56321E+15 |
| 2015-07-13 | 0:30:00 | 1.300136226 | ['Ash'] | 3.91E+16 | -1.34E+16 |
| 2015-07-13 | 1:00:00 | 1.300062453 | ['Imhotep'] | 9.03E+16 | -3.08E+16 |
| 2015-07-13 | 1:30:00 | 1.299988679 | ['Ash'] | 1.06E+17 | -3.63E+16 |
| 2015-07-13 | 2:00:00 | 1.299915577 | ['Imhotep'] | 1.28E+17 | -4.37E+16 |
| 2015-07-13 | 2:30:00 | 1.299849762 | ['Imhotep'] | 9.95E+16 | -3.40E+16 |
| 2015-07-13 | 7:00:00 | 1.299176346 | ['Maat'] | 4.18E+16 | -1.43E+16 |
| 2015-07-13 | 7:30:00 | 1.299103774 | ['Maat'] | 2.06E+16 | -7.09024E+15 |
| 2015-07-13 | 9:00:00 | 1.29888283 | ['Hapi'] | 7.91E+16 | -2.70E+16 |
| 2015-07-13 | 9:30:00 | 1.29881098 | ['Seth'] | 9.79E+16 | -3.34E+16 |
| 2015-07-13 | 11:30:00 | 1.298518113 | ['Hatmehit'] | 2.17E+16 | -7.40478E+15 |
| 2015-07-13 | 12:00:00 | 1.298445283 | ['Anuket'] | 1.83E+16 | -6.24015E+15 |
| 2015-07-13 | 14:30:00 | 1.298066667 | ['Anuket'] | 9.92E+16 | -3.39E+16 |
| 2015-07-13 | 17:30:00 | 1.29764283 | ['Seth'] | 2.75E+16 | -9.41789E+15 |
| 2015-07-13 | 18:00:00 | 1.297570566 | ['Seth'] | 2.24E+16 | -7.68751E+15 |
| 2015-07-13 | 18:30:00 | 1.297496538 | ['Maat'] | 1.03E+16 | -3.58239E+15 |
| 2015-07-13 | 21:00:00 | 1.297136604 | ['Babi'] | 3.70E+16 | -1.27E+16 |

| Date | Time | Value | Label | Col5 | Col6 |
|---|---|---|---|---|---|
| 2015-07-13 | 21:30:00 | 1.297064717 | ['Babi'] | 7.77E+16 | -2.65E+16 |
| 2015-07-13 | 22:00:00 | 1.29699 | ['Serqet'] | 6.75E+16 | -2.30E+16 |
| 2015-07-13 | 22:30:00 | 1.29691566 | ['Anuket'] | 7.70E+16 | -2.63E+16 |
| 2015-07-13 | 23:00:00 | 1.296843846 | ['Anuket'] | 5.80E+16 | -1.98E+16 |
| 2015-07-14 | 8:00:00 | 1.295553208 | ['Aker'] | 3.16E+16 | -1.08E+16 |
| 2015-07-14 | 8:30:00 | 1.295482642 | ['Maat'] | 2.74E+16 | -9.3425E+15 |
| 2015-07-14 | 9:30:00 | 1.295340769 | ['Nut'] | 3.67E+16 | -1.25E+16 |
| 2015-07-14 | 10:00:00 | 1.295270755 | ['Nut'] | 6.54E+16 | -2.23E+16 |
| 2015-07-14 | 10:30:00 | 1.295199623 | ['Serqet'] | 6.95E+16 | -2.37E+16 |
| 2015-07-14 | 11:00:00 | 1.295126786 | ['Anuket'] | 7.69E+16 | -2.63E+16 |
| 2015-07-14 | 11:30:00 | 1.295053208 | ['Anuket'] | 6.06E+16 | -2.07E+16 |
| 2015-07-14 | 15:00:00 | 1.294561509 | ['Ash'] | 7.91E+16 | -2.70E+16 |
| 2015-07-14 | 15:30:00 | 1.294488333 | ['Ash'] | 7.16E+16 | -2.44E+16 |
| 2015-07-14 | 16:00:00 | 1.294416226 | ['Aten'] | 8.26E+16 | -2.82E+16 |
| 2015-07-14 | 23:00:00 | 1.293434314 | ['Serqet'] | 6.72E+16 | -2.29E+16 |
| 2015-07-14 | 23:30:00 | 1.293364906 | ['Serqet'] | 6.21E+16 | -2.12E+16 |
| 2015-07-15 | 4:00:00 | 1.292736604 | ['Ash'] | 6.26E+16 | -2.14E+16 |
| 2015-07-15 | 4:30:00 | 1.292665714 | ['Imhotep'] | 7.32E+16 | -2.50E+16 |
| 2015-07-15 | 10:00:00 | 1.291903019 | ['Nut'] | 2.55E+16 | -8.7263E+15 |
| 2015-07-15 | 10:30:00 | 1.291833774 | ['Serqet'] | 4.35E+16 | -1.49E+16 |
| 2015-07-15 | 11:00:00 | 1.291765962 | ['Hapi'] | 5.73E+16 | -1.96E+16 |
| 2015-07-15 | 15:30:00 | 1.291146792 | ['Ash'] | 5.44E+16 | -1.86E+16 |
| 2015-07-15 | 16:00:00 | 1.291079231 | ['Ash'] | 6.02E+16 | -2.06E+16 |
| 2015-07-15 | 16:30:00 | 1.291011887 | ['Ash'] | 5.69E+16 | -1.94E+16 |
| 2015-07-15 | 17:00:00 | 1.290941071 | ['Aten'] | 5.18E+16 | -1.77E+16 |
| 2015-07-15 | 18:00:00 | 1.290802308 | ['Babi'] | 5.19E+16 | -1.78E+16 |
| 2015-07-15 | 18:30:00 | 1.290734528 | ['Babi'] | 4.11E+16 | -1.40E+16 |
| 2015-07-15 | 19:00:00 | 1.2906675 | ['Babi'] | 3.32E+16 | -1.14E+16 |
| 2015-07-15 | 23:00:00 | 1.290122264 | ['Maat'] | 6.47E+16 | -2.21E+16 |
| 2015-07-15 | 23:30:00 | 1.290054528 | ['Hapi'] | 5.95E+16 | -2.04E+16 |
| 2015-07-16 | 6:00:00 | 1.289178 | ['Ash'] | 4.87E+16 | -1.66E+16 |
| 2015-07-16 | 6:30:00 | 1.289108519 | ['Ash'] | 5.34E+16 | -1.82E+16 |
| 2015-07-16 | 11:30:00 | 1.288438491 | ['Serqet'] | 5.42E+16 | -1.85E+16 |
| 2015-07-16 | 12:00:00 | 1.288371731 | ['Hapi'] | 5.35E+16 | -1.83E+16 |
| 2015-07-16 | 16:30:00 | 1.287773077 | ['Ash'] | 3.38E+16 | -1.16E+16 |
| 2015-07-16 | 17:00:00 | 1.28770902 | ['Ash'] | 4.36E+16 | -1.49E+16 |
| 2015-07-17 | 0:00:00 | 1.286784717 | ['Maat'] | 7.91E+16 | -2.71E+16 |
| 2015-07-17 | 6:00:00 | 1.286000727 | ['Ash'] | 3.99E+16 | -1.36E+16 |
| 2015-07-17 | 7:30:00 | 1.285804231 | ['Ash'] | 5.47E+16 | -1.87E+16 |
| 2015-07-17 | 12:30:00 | 1.285158679 | ['Maat'] | 6.61E+16 | -2.26E+16 |
| 2015-07-17 | 13:00:00 | 1.285093962 | ['Serqet'] | 5.81E+16 | -1.98E+16 |
| 2015-07-17 | 13:30:00 | 1.285030962 | ['Hathor'] | 5.37E+16 | -1.83E+16 |
| 2015-07-17 | 18:00:00 | 1.284453542 | ['Ash'] | 2.91E+16 | -9.9289E+15 |
| 2015-07-17 | 23:00:00 | 1.283814906 | ['Hapi'] | 6.87E+15 | -2.35668E+15 |

| Date | Time | Value | Region | Col5 | Col6 |
|---|---|---|---|---|---|
| 2015-07-17 | 23:30:00 | 1.283749464 | ['Maat'] | 9.49E+15 | -3.24639E+15 |
| 2015-07-18 | 13:30:00 | 1.281988491 | ['Maat'] | 6.43E+16 | -2.20E+16 |
| 2015-07-18 | 14:30:00 | 1.281866038 | ['Hathor'] | 6.61E+16 | -2.26E+16 |
| 2015-07-18 | 19:00:00 | 1.28128 | ['Hathor'] | 4.33E+16 | -1.49E+16 |
| 2015-07-18 | 19:30:00 | 1.281247736 | ['Ash'] | 3.68E+16 | -1.26E+16 |
| 2015-07-18 | 20:00:00 | 1.281186226 | ['Imhotep'] | 3.46E+16 | -1.18E+16 |
| 2015-07-18 | 20:30:00 | 1.281125094 | ['Ash'] | 4.52E+16 | -1.54E+16 |
| 2015-07-18 | 21:00:00 | 1.281062037 | ['Aten'] | 6.81E+16 | -2.32E+16 |
| 2015-07-18 | 21:30:00 | 1.280999074 | ['Babi'] | 3.78E+16 | -1.29E+16 |
| 2015-07-18 | 22:30:00 | 1.280877547 | ['Babi'] | 1.04E+16 | -3.57703E+15 |
| 2015-07-19 | 0:00:00 | 1.280694808 | ['Maat'] | 1.61E+16 | -5.50832E+15 |
| 2015-07-19 | 0:30:00 | 1.280633962 | ['Maat'] | 1.58E+16 | -5.39274E+15 |
| 2015-07-19 | 1:30:00 | 1.280510357 | ['Serqet'] | 4.54E+16 | -1.55E+16 |
| 2015-07-19 | 2:00:00 | 1.280447358 | ['Serqet'] | 7.34E+16 | -2.51E+16 |
| 2015-07-19 | 2:30:00 | 1.280386154 | ['Anuket'] | 4.97E+16 | -1.70E+16 |
| 2015-07-19 | 6:00:00 | 1.279962679 | ['Ash'] | 3.25E+16 | -1.11E+16 |
| 2015-07-19 | 6:30:00 | 1.279900189 | ['Imhotep'] | 4.47E+16 | -1.52E+16 |
| 2015-07-19 | 7:00:00 | 1.279839811 | ['Imhotep'] | 4.84E+16 | -1.65E+16 |
| 2015-07-19 | 7:30:00 | 1.279779623 | ['Imhotep'] | 4.69E+16 | -1.60E+16 |
| 2015-07-19 | 8:00:00 | 1.279720192 | ['Imhotep'] | 3.86E+16 | -1.32E+16 |
| 2015-07-19 | 8:30:00 | 1.279660566 | ['Khepry'] | 5.17E+16 | -1.76E+16 |
| 2015-07-19 | 9:00:00 | 1.2796 | ['Aten'] | 6.50E+16 | -2.22E+16 |
| 2015-07-19 | 10:00:00 | 1.279479811 | ['Babi'] | 3.02E+16 | -1.03E+16 |
| 2015-07-19 | 10:30:00 | 1.279417857 | ['Babi'] | 1.79E+16 | -6.11409E+15 |
| 2015-07-19 | 20:30:00 | 1.278205417 | ['Ash'] | 7.30E+16 | -2.49E+16 |
| 2015-07-20 | 1:30:00 | 1.277632075 | ['Nut'] | 2.44E+16 | -8.34362E+15 |
| 2015-07-20 | 13:30:00 | 1.276230769 | ['Seth'] | 1.87E+16 | -6.40398E+15 |
| 2015-07-20 | 15:00:00 | 1.276061837 | ['Atum'] | 6.38E+16 | -2.18E+16 |
| 2015-07-20 | 15:30:00 | 1.276001273 | ['Seth'] | 4.31E+16 | -1.47E+16 |
| 2015-07-20 | 16:00:00 | 1.275944681 | ['Seth'] | 3.25E+16 | -1.11E+16 |
| 2015-07-21 | 16:30:00 | 1.273186471 | ['Atum'] | 6.17E+16 | -2.11E+16 |
| 2015-07-21 | 17:00:00 | 1.273129091 | ['Atum'] | 3.70E+16 | -1.26E+16 |
| 2015-07-22 | 4:00:00 | 1.271926667 | ['Seth'] | 6.21E+16 | -2.12E+16 |
| 2015-07-22 | 4:30:00 | 1.271874906 | ['Anuket'] | 6.42E+16 | -2.19E+16 |
| 2015-07-22 | 9:30:00 | 1.271335472 | ['Ash'] | 8.02E+16 | -2.74E+16 |
| 2015-07-22 | 10:00:00 | 1.271282264 | ['Imhotep'] | 8.39E+16 | -2.86E+16 |
| 2015-07-22 | 10:30:00 | 1.2712375 | ['Khepry'] | 5.21E+16 | -1.78E+16 |
| 2015-07-22 | 20:00:00 | 1.270223208 | ['Imhotep1'] | 6.36E+16 | -2.17E+16 |
| 2015-07-22 | 20:30:00 | 1.270171346 | ['Imhotep'] | 1.21E+17 | -4.13E+16 |
| 2015-07-23 | 1:00:00 | 1.269700962 | ['Hapi'] | 2.61E+16 | -8.90517E+15 |
| 2015-07-23 | 3:30:00 | 1.269440755 | ['Seth'] | 4.70E+16 | -1.61E+16 |
| 2015-07-23 | 4:30:00 | 1.269337273 | ['Seth'] | 7.81E+16 | -2.67E+16 |
| 2015-07-23 | 5:00:00 | 1.269283962 | ['Seth'] | 1.00E+17 | -3.44E+16 |
| 2015-07-23 | 6:00:00 | 1.269181132 | ['Seth'] | 3.25E+16 | -1.11E+16 |

| Date | Time | Value | Region | Col5 | Col6 |
|---|---|---|---|---|---|
| 2015-07-23 | 6:30:00 | 1.269129434 | ['Anubis'] | 2.98E+16 | -1.02E+16 |
| 2015-07-23 | 7:00:00 | 1.269091481 | ['Atum'] | 2.68E+16 | -9.16533E+15 |
| 2015-07-23 | 11:00:00 | 1.268661389 | ['Atum'] | 5.58E+16 | -1.91E+16 |
| 2015-07-23 | 11:30:00 | 1.268633 | ['Aten'] | 5.92E+16 | -2.02E+16 |
| 2015-07-23 | 15:30:00 | 1.268201765 | ['Aten'] | 3.10E+16 | -1.06E+16 |
| 2015-07-23 | 16:00:00 | 1.26816 | ['Seth'] | 5.51E+16 | -1.88E+16 |
| 2015-07-23 | 16:30:00 | 1.268110192 | ['Anuket'] | 9.59E+16 | -3.27E+16 |
| 2015-07-23 | 17:00:00 | 1.2680616 | ['Anuket'] | 1.03E+17 | -3.53E+16 |
| 2015-07-23 | 21:00:00 | 1.267639231 | ['Anuket'] | 1.34E+17 | -4.56E+16 |
| 2015-07-23 | 21:30:00 | 1.267609592 | ['Imhotep'] | 1.65E+17 | -5.63E+16 |
| 2015-07-24 | 3:00:00 | 1.267057255 | ['Wosret'] | 3.61E+16 | -1.23E+16 |
| 2015-07-24 | 13:00:00 | 1.266078113 | ['Aker'] | 5.34E+16 | -1.82E+16 |
| 2015-07-24 | 13:30:00 | 1.266035238 | ['Aker'] | 4.96E+16 | -1.69E+16 |
| 2015-07-24 | 22:30:00 | 1.265149412 | ['Aker'] | 1.32E+17 | -4.51E+16 |
| 2015-07-24 | 23:00:00 | 1.265117925 | ['Imhotep'] | 1.16E+17 | -3.95E+16 |
| 2015-07-24 | 23:30:00 | 1.26507 | ['Imhotep'] | 7.62E+16 | -2.60E+16 |
| 2015-07-25 | 5:30:00 | 1.2644896 | ['Imhotep'] | 7.85E+16 | -2.68E+16 |
| 2015-07-25 | 9:30:00 | 1.264128113 | ['Imhotep1'] | 8.08E+16 | -2.76E+16 |
| 2015-07-25 | 10:00:00 | 1.264081132 | ['Imhotep'] | 1.48E+17 | -5.04E+16 |
| 2015-07-25 | 10:30:00 | 1.264050526 | ['Imhotep'] | 1.82E+17 | -6.21E+16 |
| 2015-07-25 | 14:30:00 | 1.263666981 | ['Anhur'] | 5.97E+16 | -2.04E+16 |
| 2015-07-25 | 18:00:00 | 1.263343654 | ['Seth'] | 1.11E+17 | -3.80E+16 |
| 2015-07-25 | 18:30:00 | 1.263298868 | ['Anubis'] | 8.30E+16 | -2.83E+16 |
| 2015-07-25 | 19:00:00 | 1.263253396 | ['Atum'] | 5.78E+16 | -1.97E+16 |
| 2015-07-25 | 20:00:00 | 1.263163396 | ['Atum'] | 4.06E+16 | -1.39E+16 |
| 2015-07-25 | 20:30:00 | 1.263129231 | ['Khonsu'] | 3.60E+16 | -1.23E+16 |
| 2015-07-26 | 0:30:00 | 1.262755294 | ['Khonsu'] | 6.86E+16 | -2.34E+16 |
| 2015-07-26 | 1:00:00 | 1.262721667 | ['Bes'] | 1.00E+17 | -3.43E+16 |
| 2015-07-26 | 5:30:00 | 1.262305833 | ['Bes'] | 1.18E+17 | -4.05E+16 |
| 2015-07-26 | 6:00:00 | 1.262263208 | ['Anhur'] | 1.31E+17 | -4.46E+16 |
| 2015-07-26 | 6:30:00 | 1.262219074 | ['Anhur'] | 8.49E+16 | -2.90E+16 |
| 2015-07-26 | 7:00:00 | 1.262172727 | ['Anubis'] | 9.91E+16 | -3.38E+16 |
| 2015-07-26 | 7:30:00 | 1.262127925 | ['Atum'] | 9.25E+16 | -3.16E+16 |
| 2015-07-26 | 8:00:00 | 1.262083585 | ['Khonsu'] | 5.22E+16 | -1.78E+16 |
| 2015-07-26 | 8:30:00 | 1.262039808 | ['Khonsu'] | 4.87E+16 | -1.66E+16 |
| 2015-07-26 | 9:30:00 | 1.261952075 | ['Imhotep'] | 8.03E+16 | -2.74E+16 |
| 2015-07-26 | 10:30:00 | 1.261865098 | ['Imhotep'] | 1.58E+17 | -5.41E+16 |
| 2015-07-26 | 11:00:00 | 1.2618375 | ['Bes'] | 1.34E+17 | -4.58E+16 |
| 2015-07-26 | 11:30:00 | 1.261767273 | ['Bes'] | 1.10E+17 | -3.77E+16 |
| 2015-07-26 | 12:00:00 | 1.261732264 | ['Bes'] | 9.15E+16 | -3.12E+16 |
| 2015-07-26 | 12:30:00 | 1.261689038 | ['Bes'] | 5.70E+16 | -1.95E+16 |
| 2015-07-26 | 13:00:00 | 1.261644717 | ['Anhur'] | 9.42E+16 | -3.22E+16 |
| 2015-07-26 | 13:30:00 | 1.261601111 | ['Anhur'] | 9.64E+16 | -3.29E+16 |
| 2015-07-26 | 14:00:00 | 1.261556182 | ['Anhur'] | 7.55E+16 | -2.58E+16 |
| 2015-07-26 | 14:30:00 | 1.261511731 | ['Anhur'] | 6.69E+16 | -2.28E+16 |
| 2015-07-26 | 15:00:00 | 1.261469057 | ['Geb'] | 6.11E+16 | -2.09E+16 |
| 2015-07-26 | 15:30:00 | 1.26142566 | ['Geb'] | 6.42E+16 | -2.19E+16 |
| 2015-07-26 | 16:00:00 | 1.261381887 | ['Geb'] | 6.49E+16 | -2.22E+16 |
| 2015-07-26 | 16:30:00 | 1.261339615 | ['Geb'] | 5.92E+16 | -2.02E+16 |
| 2015-07-26 | 17:00:00 | 1.261296604 | ['Geb'] | 6.16E+16 | -2.10E+16 |
| 2015-07-26 | 17:30:00 | 1.26125283 | ['Geb'] | 1.04E+17 | -3.56E+16 |
| 2015-07-26 | 18:00:00 | 1.261208727 | ['Geb'] | 1.22E+17 | -4.15E+16 |
| 2015-07-26 | 18:30:00 | 1.261163774 | ['Atum'] | 1.24E+17 | -4.25E+16 |
| 2015-07-26 | 19:00:00 | 1.261120755 | ['Atum'] | 1.12E+17 | -3.83E+16 |
| 2015-07-26 | 19:30:00 | 1.261078077 | ['Khonsu'] | 1.12E+17 | -3.84E+16 |
| 2015-07-26 | 21:00:00 | 1.2609494 | ['Bes'] | 4.60E+16 | -1.57E+16 |
| 2015-07-26 | 21:30:00 | 1.260908077 | ['Imhotep'] | 9.21E+16 | -3.14E+16 |
| 2015-07-26 | 22:00:00 | 1.260864906 | ['Imhotep'] | 1.13E+17 | -3.85E+16 |
| 2015-07-26 | 22:30:00 | 1.260820385 | ['Bes'] | 1.35E+17 | -4.60E+16 |
| 2015-07-26 | 23:00:00 | 1.260776792 | ['Bes'] | 1.44E+17 | -4.93E+16 |
| 2015-07-26 | 23:30:00 | 1.260734151 | ['Bes'] | 1.22E+17 | -4.16E+16 |
| 2015-07-27 | 0:00:00 | 1.260691321 | ['Bes'] | 6.93E+16 | -2.37E+16 |
| 2015-07-27 | 1:30:00 | 1.260564717 | ['Bes'] | 1.01E+17 | -3.45E+16 |
| 2015-07-27 | 2:00:00 | 1.260522453 | ['Bes'] | 8.60E+16 | -2.94E+16 |
| 2015-07-27 | 2:30:00 | 1.26048 | ['Bes'] | 4.77E+16 | -1.63E+16 |
| 2015-07-27 | 3:00:00 | 1.260438113 | ['Wosret'] | 7.38E+16 | -2.52E+16 |
| 2015-07-27 | 3:30:00 | 1.260394286 | ['Seth'] | 9.70E+16 | -3.31E+16 |
| 2015-07-27 | 4:30:00 | 1.26031 | ['Khonsu'] | 4.70E+16 | -1.61E+16 |
| 2015-07-27 | 5:00:00 | 1.26027 | ['Khonsu'] | 5.99E+16 | -2.05E+16 |
| 2015-07-27 | 5:30:00 | 1.260226792 | ['Hapi'] | 9.20E+16 | -3.14E+16 |
| 2015-07-27 | 6:00:00 | 1.260183393 | ['Seth'] | 1.31E+17 | -4.47E+16 |
| 2015-07-27 | 7:30:00 | 1.2600575 | ['Khonsu'] | 8.27E+16 | -2.83E+16 |
| 2015-07-27 | 8:00:00 | 1.260016863 | ['Imhotep'] | 9.55E+16 | -3.26E+16 |
| 2015-07-27 | 8:30:00 | 1.259973208 | ['Imhotep'] | 5.51E+16 | -1.88E+16 |
| 2015-07-27 | 10:30:00 | 1.259807115 | ['Imhotep'] | 1.03E+17 | -3.53E+16 |
| 2015-07-27 | 11:00:00 | 1.259765283 | ['Imhotep'] | 1.35E+17 | -4.62E+16 |
| 2015-07-27 | 13:00:00 | 1.259598929 | ['Wosret'] | 1.48E+17 | -5.07E+16 |
| 2015-07-27 | 15:30:00 | 1.259392105 | ['Sobek'] | 9.49E+16 | -3.24E+16 |
| 2015-07-27 | 16:00:00 | 1.259349808 | ['Geb'] | 1.05E+17 | -3.59E+16 |
| 2015-07-27 | 17:30:00 | 1.259228302 | ['Atum'] | 1.12E+17 | -3.83E+16 |
| 2015-07-27 | 18:00:00 | 1.259188269 | ['Khonsu'] | 1.37E+17 | -4.69E+16 |
| 2015-07-27 | 20:30:00 | 1.258983529 | ['Imhotep1'] | 1.05E+17 | -3.59E+16 |
| 2015-07-27 | 21:00:00 | 1.258943077 | ['Imhotep'] | 9.32E+16 | -3.18E+16 |
| 2015-07-29 | 6:00:00 | 1.256374651 | ['Geb'] | 9.36E+16 | -3.20E+16 |
| 2015-07-29 | 9:30:00 | 1.256112642 | ['Imhotep'] | 1.34E+17 | -4.58E+16 |
| 2015-07-29 | 10:00:00 | 1.256075472 | ['Bes'] | 1.39E+17 | -4.75E+16 |
| 2015-07-29 | 10:30:00 | 1.256039423 | ['Bes'] | 1.26E+17 | -4.31E+16 |
| 2015-07-29 | 14:00:00 | 1.255782453 | ['Anhur'] | 1.56E+17 | -5.40E+16 |

| Date | Time | Value | Region | Col5 | Col6 |
|---|---|---|---|---|---|
| 2015-07-29 | 14:30:00 | 1.255746038 | ['Khonsu'] | 2.09E+17 | -7.15E+16 |
| 2015-07-29 | 15:00:00 | 1.255710385 | ['Bes'] | 2.26E+17 | -7.71E+16 |
| 2015-07-29 | 15:30:00 | 1.255685556 | ['Bes'] | 1.23E+17 | -4.27E+16 |
| 2015-07-29 | 18:00:00 | 1.25550875 | ['Bes'] | 6.92E+16 | -2.37E+16 |
| 2015-07-29 | 19:00:00 | 1.255431111 | ['Bes'] | 9.53E+16 | -3.27E+16 |
| 2015-07-29 | 19:30:00 | 1.255384717 | ['Bes'] | 1.04E+17 | -3.56E+16 |
| 2015-07-29 | 20:00:00 | 1.255347321 | ['Bes'] | 1.13E+17 | -3.84E+16 |
| 2015-07-29 | 20:30:00 | 1.255310577 | ['Bes'] | 1.11E+17 | -3.77E+16 |
| 2015-07-30 | 1:00:00 | 1.254991132 | ['Bes'] | 1.02E+17 | -3.50E+16 |
| 2015-07-30 | 6:00:00 | 1.254640189 | ['Bes'] | 1.26E+17 | -4.29E+16 |
| 2015-07-30 | 6:30:00 | 1.254605577 | ['Bes'] | 1.29E+17 | -4.41E+16 |
| 2015-07-30 | 7:00:00 | 1.25458 | ['Bes'] | 1.17E+17 | -3.98E+16 |
| 2015-07-30 | 11:00:00 | 1.25429 | ['Bes'] | 1.34E+17 | -4.56E+16 |
| 2015-07-30 | 11:30:00 | 1.254271364 | ['Sobek'] | 1.13E+17 | -3.87E+16 |
| 2015-07-30 | 16:30:00 | 1.253920189 | ['Bes'] | 1.25E+17 | -4.26E+16 |
| 2015-07-30 | 17:00:00 | 1.253886531 | ['Bes'] | 1.12E+17 | -3.82E+16 |
| 2015-07-30 | 21:00:00 | 1.253607059 | ['Bes'] | 1.43E+17 | -4.90E+16 |
| 2015-07-30 | 21:30:00 | 1.253584615 | ['Bes'] | 1.52E+17 | -5.20E+16 |
| 2015-07-30 | 22:00:00 | 1.253552075 | ['Sobek'] | 1.67E+17 | -5.70E+16 |
| 2015-07-30 | 22:30:00 | 1.253518654 | ['Neith'] | 1.62E+17 | -5.53E+16 |
| 2015-07-30 | 23:00:00 | 1.253486154 | ['Wosret'] | 1.26E+17 | -4.30E+16 |
| 2015-07-30 | 23:30:00 | 1.253453585 | ['Neith'] | 9.73E+16 | -3.32E+16 |
| 2015-07-31 | 1:30:00 | 1.253320385 | ['Geb'] | 1.38E+17 | -4.77E+16 |
| 2015-07-31 | 2:00:00 | 1.253287308 | ['Geb'] | 2.01E+17 | -6.87E+16 |
| 2015-07-31 | 3:00:00 | 1.253223019 | ['Geb'] | 1.36E+17 | -4.63E+16 |
| 2015-07-31 | 3:30:00 | 1.253191556 | ['Geb'] | 1.34E+17 | -4.56E+16 |
| 2015-07-31 | 4:00:00 | 1.2531578 | ['Bes'] | 1.29E+17 | -4.40E+16 |
| 2015-07-31 | 4:30:00 | 1.253126038 | ['Bes'] | 1.05E+17 | -3.59E+16 |
| 2015-07-31 | 5:00:00 | 1.253092909 | ['Bes'] | 8.49E+16 | -2.93E+16 |
| 2015-07-31 | 5:30:00 | 1.253059608 | ['Bes'] | 1.11E+17 | -3.85E+16 |
| 2015-07-31 | 6:00:00 | 1.253026346 | ['Bes'] | 1.31E+17 | -4.49E+16 |
| 2015-07-31 | 6:30:00 | 1.252994717 | ['Bes'] | 1.41E+17 | -4.83E+16 |
| 2015-07-31 | 8:00:00 | 1.252887222 | ['Bes'] | 1.51E+17 | -5.15E+16 |
| 2015-07-31 | 9:30:00 | 1.252802281 | ['Sobek'] | 1.61E+17 | -5.50E+16 |
| 2015-07-31 | 10:00:00 | 1.252769231 | ['Neith'] | 1.81E+17 | -6.17E+16 |
| 2015-07-31 | 10:30:00 | 1.252737358 | ['Neith'] | 1.60E+17 | -5.45E+16 |
| 2015-07-31 | 11:00:00 | 1.252706038 | ['Wosret'] | 1.44E+17 | -4.91E+16 |
| 2015-07-31 | 11:30:00 | 1.252673774 | ['Neith'] | 9.21E+16 | -3.15E+16 |
| 2015-07-31 | 14:00:00 | 1.252516154 | ['Geb'] | 1.01E+17 | -3.49E+16 |
| 2015-07-31 | 17:00:00 | 1.25232 | ['Hatmehit'] | 1.35E+17 | -4.75E+16 |
| 2015-07-31 | 17:30:00 | 1.252286667 | ['Hatmehit'] | 1.18E+17 | -4.08E+16 |
| 2015-07-31 | 22:30:00 | 1.2519775 | ['Hatmehit'] | 1.51E+17 | -5.17E+16 |
| 2015-07-31 | 23:30:00 | 1.251925385 | ['Neith'] | 1.26E+17 | -4.30E+16 |
| 2015-08-01 | 5:30:00 | 1.251561961 | ['Neith'] | 1.15E+17 | -3.92E+16 |

| Date | Time | Value | Region | Col5 | Col6 |
|---|---|---|---|---|---|
| 2015-08-01 | 7:30:00 | 1.251441698 | ['Imhotep1'] | 1.47E+17 | -5.01E+16 |
| 2015-08-01 | 8:00:00 | 1.251418667 | ['Imhotep1'] | 8.01E+16 | -2.76E+16 |
| 2015-08-01 | 9:30:00 | 1.251324528 | ['Imhotep'] | 1.29E+17 | -4.42E+16 |
| 2015-08-01 | 10:00:00 | 1.251295 | ['Bes'] | 1.18E+17 | -4.04E+16 |
| 2015-08-01 | 10:30:00 | 1.251266226 | ['Bes'] | 1.31E+17 | -4.48E+16 |
| 2015-08-01 | 11:00:00 | 1.251236667 | ['Bes'] | 1.42E+17 | -4.84E+16 |
| 2015-08-01 | 11:30:00 | 1.251206182 | ['Bes'] | 1.04E+17 | -3.54E+16 |
| 2015-08-01 | 12:00:00 | 1.251176415 | ['Bes'] | 9.52E+16 | -3.25E+16 |
| 2015-08-01 | 12:30:00 | 1.25114717 | ['Anhur'] | 1.06E+17 | -3.62E+16 |
| 2015-08-01 | 13:30:00 | 1.251089811 | ['Anhur'] | 1.68E+17 | -5.74E+16 |
| 2015-08-01 | 14:00:00 | 1.251060189 | ['Geb'] | 1.85E+17 | -6.31E+16 |
| 2015-08-01 | 14:30:00 | 1.251031698 | ['Geb'] | 1.82E+17 | -6.21E+16 |
| 2015-08-01 | 15:00:00 | 1.251003077 | ['Geb'] | 1.20E+17 | -4.09E+16 |
| 2015-08-01 | 15:30:00 | 1.250973818 | ['Geb'] | 1.11E+17 | -3.80E+16 |
| 2015-08-01 | 16:00:00 | 1.250943774 | ['Geb'] | 1.22E+17 | -4.17E+16 |
| 2015-08-01 | 16:30:00 | 1.250914906 | ['Geb'] | 1.13E+17 | -3.84E+16 |
| 2015-08-01 | 17:00:00 | 1.250886604 | ['Geb'] | 1.23E+17 | -4.18E+16 |
| 2015-08-01 | 17:30:00 | 1.250857692 | ['Geb'] | 1.26E+17 | -4.31E+16 |
| 2015-08-01 | 18:00:00 | 1.250829615 | ['Atum'] | 1.59E+17 | -5.44E+16 |
| 2015-08-01 | 18:30:00 | 1.250800566 | ['Atum'] | 1.27E+17 | -4.35E+16 |
| 2015-08-01 | 19:00:00 | 1.250772642 | ['Khonsu'] | 1.42E+17 | -4.85E+16 |
| 2015-08-01 | 19:30:00 | 1.250744423 | ['Khonsu'] | 1.37E+17 | -4.69E+16 |
| 2015-08-01 | 20:00:00 | 1.250715088 | ['Imhotep1'] | 1.18E+17 | -4.03E+16 |
| 2015-08-01 | 20:30:00 | 1.25068566 | ['Imhotep'] | 1.08E+17 | -3.69E+16 |
| 2015-08-01 | 21:00:00 | 1.250657843 | ['Imhotep'] | 1.10E+17 | -3.77E+16 |
| 2015-08-01 | 21:30:00 | 1.250629623 | ['Imhotep'] | 1.27E+17 | -4.35E+16 |
| 2015-08-01 | 22:00:00 | 1.250601321 | ['Bes'] | 1.36E+17 | -4.65E+16 |
| 2015-08-01 | 22:30:00 | 1.250573529 | ['Bes'] | 1.54E+17 | -5.27E+16 |
| 2015-08-01 | 23:00:00 | 1.250545849 | ['Bes'] | 1.31E+17 | -4.48E+16 |
| 2015-08-01 | 23:30:00 | 1.250517547 | ['Bes'] | 8.92E+16 | -3.05E+16 |
| 2015-08-02 | 0:00:00 | 1.250489811 | ['Bes'] | 7.82E+16 | -2.67E+16 |
| 2015-08-02 | 0:30:00 | 1.250460714 | ['Anhur'] | 7.47E+16 | -2.55E+16 |
| 2015-08-02 | 5:30:00 | 1.250182128 | ['Anhur'] | 1.13E+17 | -3.86E+16 |
| 2015-08-02 | 6:00:00 | 1.250156923 | ['Atum'] | 1.63E+17 | -5.56E+16 |
| 2015-08-02 | 6:30:00 | 1.250135 | ['Atum'] | 1.43E+17 | -4.87E+16 |
| 2015-08-02 | 10:30:00 | 1.249909556 | ['Atum'] | 1.49E+17 | -5.08E+16 |
| 2015-08-02 | 11:00:00 | 1.249896 | ['Bes'] | 1.52E+17 | -5.19E+16 |
| 2015-08-02 | 15:30:00 | 1.249643846 | ['Geb'] | 1.08E+17 | -3.67E+16 |
| 2015-08-02 | 16:00:00 | 1.249618302 | ['Geb'] | 1.24E+17 | -4.24E+16 |
| 2015-08-02 | 16:30:00 | 1.249596364 | ['Geb'] | 1.28E+17 | -4.36E+16 |
| 2015-08-02 | 20:30:00 | 1.24937925 | ['Geb'] | 8.95E+16 | -3.06E+16 |
| 2015-08-02 | 21:00:00 | 1.249366 | ['Bes'] | 9.39E+16 | -3.21E+16 |
| 2015-08-03 | 1:30:00 | 1.249123056 | ['Bes'] | 1.18E+17 | -4.03E+16 |
| 2015-08-03 | 2:30:00 | 1.249077179 | ['Geb'] | 1.64E+17 | -5.61E+16 |

| Date | Time | Value | Region | Col5 | Col6 |
|---|---|---|---|---|---|
| 2015-08-03 | 6:30:00 | 1.24887027 | ['Geb'] | 1.14E+17 | -3.91E+16 |
| 2015-08-03 | 7:00:00 | 1.248856818 | ['Bes'] | 9.39E+16 | -3.21E+16 |
| 2015-08-03 | 11:00:00 | 1.24864 | ['Hatmehit'] | 1.55E+17 | -5.34E+16 |
| 2015-08-05 | 6:00:00 | 1.246747209 | ['Geb'] | 4.73E+16 | -1.62E+16 |
| 2015-08-05 | 7:00:00 | 1.2467056 | ['Hatmehit'] | 1.01E+17 | -3.45E+16 |
| 2015-08-05 | 7:30:00 | 1.2466866 | ['Anubis'] | 1.09E+17 | -3.71E+16 |
| 2015-08-05 | 9:30:00 | 1.246612941 | ['Anubis'] | 3.00E+16 | -1.03E+16 |
| 2015-08-05 | 10:30:00 | 1.24656 | ['Imhotep'] | 1.16E+17 | -3.97E+16 |
| 2015-08-05 | 11:00:00 | 1.246550364 | ['Imhotep'] | 1.14E+17 | -3.89E+16 |
| 2015-08-05 | 11:30:00 | 1.246530377 | ['Imhotep'] | 1.24E+17 | -4.23E+16 |
| 2015-08-05 | 12:00:00 | 1.24651098 | ['Imhotep'] | 1.36E+17 | -4.66E+16 |
| 2015-08-05 | 12:30:00 | 1.246491569 | ['Imhotep'] | 1.04E+17 | -3.55E+16 |
| 2015-08-05 | 13:00:00 | 1.246473269 | ['Khepry'] | 5.84E+16 | -1.99E+16 |
| 2015-08-05 | 15:00:00 | 1.246397115 | ['Anhur'] | 9.27E+16 | -3.16E+16 |
| 2015-08-05 | 15:30:00 | 1.246377719 | ['Anhur'] | 6.63E+16 | -2.26E+16 |
| 2015-08-05 | 16:00:00 | 1.246358163 | ['Wosret'] | 4.97E+16 | -1.70E+16 |
| 2015-08-05 | 16:30:00 | 1.24633898 | ['Wosret'] | 4.08E+16 | -1.39E+16 |
| 2015-08-05 | 17:00:00 | 1.246321321 | ['Wosret'] | 4.52E+16 | -1.54E+16 |
| 2015-08-05 | 17:30:00 | 1.246303019 | ['Wosret'] | 3.87E+16 | -1.32E+16 |
| 2015-08-05 | 18:00:00 | 1.246285 | ['Geb'] | 4.23E+16 | -1.44E+16 |
| 2015-08-05 | 18:30:00 | 1.24626566 | ['Geb'] | 5.06E+16 | -1.73E+16 |
| 2015-08-05 | 19:00:00 | 1.246247059 | ['Hapi'] | 7.97E+16 | -2.72E+16 |
| 2015-08-05 | 19:30:00 | 1.246228846 | ['Anubis'] | 1.16E+17 | -3.96E+16 |
| 2015-08-05 | 20:00:00 | 1.24621 | ['Anubis'] | 1.10E+17 | -3.74E+16 |
| 2015-08-05 | 20:30:00 | 1.2461914 | ['Atum'] | 4.99E+16 | -1.70E+16 |
| 2015-08-05 | 21:30:00 | 1.246154528 | ['Khonsu'] | 4.26E+16 | -1.45E+16 |
| 2015-08-05 | 22:00:00 | 1.246135849 | ['Imhotep1'] | 5.22E+16 | -1.78E+16 |
| 2015-08-05 | 22:30:00 | 1.246118077 | ['Imhotep'] | 7.95E+16 | -2.71E+16 |
| 2015-08-05 | 23:00:00 | 1.246100377 | ['Imhotep'] | 9.99E+16 | -3.41E+16 |
| 2015-08-05 | 23:30:00 | 1.246082453 | ['Imhotep'] | 1.08E+17 | -3.68E+16 |
| 2015-08-06 | 0:00:00 | 1.24606463 | ['Imhotep'] | 1.29E+17 | -4.39E+16 |
| 2015-08-06 | 1:30:00 | 1.24601 | ['Khepry'] | 3.71E+16 | -1.27E+16 |
| 2015-08-06 | 3:00:00 | 1.245956792 | ['Anhur'] | 6.37E+16 | -2.18E+16 |
| 2015-08-06 | 3:30:00 | 1.245939615 | ['Anhur'] | 8.16E+16 | -2.79E+16 |
| 2015-08-06 | 4:00:00 | 1.245922692 | ['Geb'] | 5.42E+16 | -1.85E+16 |
| 2015-08-06 | 4:30:00 | 1.245904464 | ['Wosret'] | 3.92E+16 | -1.34E+16 |
| 2015-08-06 | 5:00:00 | 1.24588566 | ['Wosret'] | 4.44E+16 | -1.51E+16 |
| 2015-08-06 | 5:30:00 | 1.245868491 | ['Wosret'] | 3.82E+16 | -1.31E+16 |
| 2015-08-06 | 6:00:00 | 1.245851321 | ['Wosret'] | 3.56E+16 | -1.22E+16 |
| 2015-08-06 | 6:30:00 | 1.245834231 | ['Geb'] | 4.29E+16 | -1.47E+16 |
| 2015-08-06 | 7:00:00 | 1.245821034 | ['Geb'] | 5.18E+16 | -1.77E+16 |
| 2015-08-06 | 11:30:00 | 1.245664423 | ['Imhotep'] | 7.20E+16 | -2.46E+16 |
| 2015-08-06 | 16:00:00 | 1.2455125 | ['Imhotep'] | 2.90E+16 | -9.91115E+15 |
| 2015-08-06 | 16:30:00 | 1.245498269 | ['Geb'] | 2.59E+16 | -8.84881E+15 |
| 2015-08-08 | 6:00:00 | 1.244426471 | ['Hatmehit'] | 1.69E+16 | -5.75606E+15 |
| 2015-08-08 | 9:00:00 | 1.24435 | ['Hatmehit'] | 6.61E+16 | -2.26E+16 |
| 2015-08-08 | 9:30:00 | 1.244341887 | ['Anubis'] | 5.60E+16 | -1.91E+16 |
| 2015-08-08 | 10:00:00 | 1.24432875 | ['Atum'] | 1.93E+16 | -6.60584E+15 |
| 2015-08-08 | 10:30:00 | 1.244317273 | ['Atum'] | 1.29E+16 | -4.44146E+15 |
| 2015-08-08 | 14:00:00 | 1.24423 | ['Atum'] | 6.81E+16 | -2.33E+16 |
| 2015-08-08 | 14:30:00 | 1.244224906 | ['Khepry'] | 4.75E+16 | -1.62E+16 |
| 2015-08-08 | 20:00:00 | 1.2441025 | ['Anuket'] | 3.05E+16 | -1.04E+16 |
| 2015-08-08 | 20:30:00 | 1.24409375 | ['Geb'] | 4.00E+16 | -1.37E+16 |
| 2015-08-09 | 12:00:00 | 1.24378283 | ['Imhotep1'] | 2.26E+16 | -7.72091E+15 |
| 2015-08-09 | 12:30:00 | 1.243774038 | ['Imhotep1'] | 3.62E+16 | -1.24E+16 |
| 2015-08-12 | 6:00:00 | 1.24304 | ['Hapi'] | 1.56E+16 | -5.35314E+15 |
| 2015-08-12 | 7:00:00 | 1.24304 | ['Maat'] | 6.05E+15 | -2.06892E+15 |
| 2015-08-12 | 8:00:00 | 1.24303 | ['Nut'] | 5.58E+15 | -1.91016E+15 |
| 2015-08-12 | 13:00:00 | 1.24302 | ['Ash'] | 1.55E+16 | -5.28034E+15 |
| 2015-08-12 | 14:00:00 | 1.24301 | ['Imhotep'] | 2.64E+16 | -9.0149E+15 |
| 2015-08-12 | 14:30:00 | 1.24301 | ['Imhotep'] | 2.88E+16 | -9.82157E+15 |
| 2015-08-12 | 15:00:00 | 1.24301 | ['Imhotep'] | 2.80E+16 | -9.5754E+15 |
| 2015-08-12 | 17:00:00 | 1.24301 | ['Babi'] | 3.10E+16 | -1.06E+16 |
| 2015-08-12 | 17:30:00 | 1.24301 | ['Babi'] | 3.47E+16 | -1.18E+16 |
| 2015-08-12 | 20:00:00 | 1.243 | ['Maat'] | 4.21E+15 | -1.45707E+15 |
| 2015-08-12 | 20:30:00 | 1.243 | ['Nut'] | 4.24E+15 | -1.46042E+15 |
| 2015-08-13 | 6:00:00 | 1.243 | ['Nut'] | 1.96E+16 | -6.72581E+15 |
| 2015-08-15 | 9:30:00 | 1.24333 | ['Hathor'] | 2.73E+16 | -9.33063E+15 |
| 2015-08-15 | 10:00:00 | 1.243331154 | ['Anuket'] | 4.99E+16 | -1.70E+16 |
| 2015-08-16 | 15:30:00 | 1.243781053 | ['Anuket'] | 3.66E+16 | -1.25E+16 |
| 2015-08-16 | 16:00:00 | 1.24379 | ['Ash'] | 4.68E+16 | -1.60E+16 |
| 2015-08-16 | 16:30:00 | 1.243799655 | ['Aten'] | 5.86E+16 | -2.00E+16 |
| 2015-08-16 | 17:00:00 | 1.243808333 | ['Aten'] | 4.23E+16 | -1.45E+16 |
| 2015-08-16 | 17:30:00 | 1.24382 | ['Babi'] | 2.57E+16 | -8.80013E+15 |
| 2015-08-16 | 19:30:00 | 1.243856038 | ['Maat'] | 1.25E+16 | -4.29457E+15 |
| 2015-08-16 | 21:30:00 | 1.2439 | ['Serqet'] | 2.11E+16 | -7.21567E+15 |
| 2015-08-16 | 22:00:00 | 1.243903654 | ['Anuket'] | 3.34E+16 | -1.14E+16 |
| 2015-08-17 | 1:00:00 | 1.243964151 | ['Ash'] | 1.12E+16 | -3.86239E+15 |
| 2015-08-17 | 1:30:00 | 1.24397 | ['Ash'] | 2.59E+16 | -8.85628E+15 |
| 2015-08-17 | 2:30:00 | 1.244 | ['Imhotep'] | 4.42E+16 | -1.51E+16 |
| 2015-08-17 | 3:00:00 | 1.244004444 | ['Imhotep'] | 3.84E+16 | -1.31E+16 |
| 2015-08-18 | 3:00:00 | 1.24456 | ['Imhotep'] | 6.47E+16 | -2.23E+16 |
| 2015-08-19 | 6:00:00 | 1.245334091 | ['Babi'] | 5.07E+16 | -1.73E+16 |
| 2015-08-19 | 7:00:00 | 1.245367083 | ['Babi'] | 1.87E+16 | -6.39427E+15 |
| 2015-08-19 | 9:30:00 | 1.24544566 | ['Maat'] | 1.71E+16 | -5.85466E+15 |
| 2015-08-19 | 10:00:00 | 1.245462885 | ['Serqet'] | 2.34E+16 | -8.00398E+15 |
| 2015-08-19 | 20:00:00 | 1.245795893 | ['Bastet'] | 7.91E+15 | -2.71124E+15 |
| 2015-08-20 | 1:00:00 | 1.2459716 | ['Atum'] | 1.80E+16 | -6.13488E+15 |

| Date | Time | Value | Region |  |  |
|---|---|---|---|---|---|
| 2015-08-20 | 6:00:00 | 1.246152979 | ['Hatmehit'] | 5.79E+16 | -1.98E+16 |
| 2015-08-20 | 11:00:00 | 1.246336735 | ['Ash'] | 8.79E+16 | -3.00E+16 |
| 2015-08-20 | 11:30:00 | 1.246356042 | ['Hapi'] | 8.92E+16 | -3.05E+16 |
| 2015-08-20 | 12:00:00 | 1.24637 | ['Anubis'] | 5.32E+16 | -1.82E+16 |
| 2015-08-20 | 16:00:00 | 1.246531667 | ['Hatmehit'] | 1.25E+17 | -4.26E+16 |
| 2015-08-20 | 16:30:00 | 1.246546226 | ['Imhotep'] | 1.10E+17 | -3.74E+16 |
| 2015-08-20 | 17:00:00 | 1.24656566 | ['Khepry'] | 7.82E+16 | -2.67E+16 |
| 2015-08-21 | 2:30:00 | 1.24694434 | ['Imhotep'] | 2.19E+16 | -7.48723E+15 |
| 2015-08-21 | 3:00:00 | 1.246964909 | ['Imhotep'] | 6.94E+16 | -2.37E+16 |
| 2015-08-22 | 5:30:00 | 1.248130625 | ['Hatmehit'] | 7.63E+16 | -2.61E+16 |
| 2015-08-22 | 6:00:00 | 1.248145532 | ['Khepry'] | 7.76E+16 | -2.65E+16 |
| 2015-08-22 | 9:00:00 | 1.2483 | ['Khepry'] | 2.37E+16 | -8.12179E+15 |
| 2015-08-22 | 9:30:00 | 1.248310769 | ['Wosret'] | 2.09E+16 | -7.14042E+15 |
| 2015-08-22 | 10:00:00 | 1.248335263 | ['Wosret'] | 2.59E+16 | -8.85127E+15 |
| 2015-08-22 | 10:30:00 | 1.248355 | ['Wosret'] | 2.38E+16 | -8.12426E+15 |
| 2015-08-22 | 14:00:00 | 1.24854 | ['Wosret'] | 1.98E+16 | -6.79404E+15 |
| 2015-08-22 | 19:30:00 | 1.248796078 | ['Hathor'] | 6.20E+16 | -2.12E+16 |
| 2015-08-22 | 20:00:00 | 1.24882 | ['Anhur'] | 4.49E+16 | -1.53E+16 |
| 2015-08-22 | 20:30:00 | 1.248844528 | ['Anhur'] | 4.00E+16 | -1.36E+16 |
| 2015-08-22 | 21:30:00 | 1.248894423 | ['Wosret'] | 3.12E+16 | -1.06E+16 |
| 2015-08-22 | 22:00:00 | 1.248919074 | ['Wosret'] | 3.46E+16 | -1.18E+16 |
| 2015-08-22 | 22:30:00 | 1.24894463 | ['Wosret'] | 3.86E+16 | -1.32E+16 |
| 2015-08-22 | 23:00:00 | 1.248970377 | ['Geb'] | 6.15E+16 | -2.10E+16 |
| 2015-08-22 | 23:30:00 | 1.248995094 | ['Geb'] | 1.00E+17 | -3.41E+16 |
| 2015-08-23 | 0:30:00 | 1.249045192 | ['Geb'] | 1.04E+17 | -3.56E+16 |
| 2015-08-23 | 2:00:00 | 1.249120377 | ['Atum'] | 4.24E+16 | -1.45E+16 |
| 2015-08-23 | 2:30:00 | 1.249147222 | ['Khonsu'] | 3.12E+16 | -1.06E+16 |
| 2015-08-23 | 3:00:00 | 1.249172308 | ['Khonsu'] | 3.56E+16 | -1.21E+16 |
| 2015-08-23 | 3:30:00 | 1.249198462 | ['Imhotep'] | 8.24E+16 | -2.81E+16 |
| 2015-08-23 | 4:00:00 | 1.24922283 | ['Imhotep'] | 1.32E+17 | -4.50E+16 |
| 2015-08-23 | 4:30:00 | 1.249249057 | ['Imhotep'] | 1.75E+17 | -5.97E+16 |
| 2015-08-23 | 5:30:00 | 1.2493 | ['Bes'] | 1.29E+17 | -4.39E+16 |
| 2015-08-23 | 6:00:00 | 1.249325849 | ['Bes'] | 9.35E+16 | -3.19E+16 |
| 2015-08-23 | 6:30:00 | 1.249350943 | ['Bes'] | 6.11E+16 | -2.09E+16 |
| 2015-08-23 | 7:00:00 | 1.249378214 | ['Bes'] | 6.44E+16 | -2.20E+16 |
| 2015-08-23 | 7:30:00 | 1.24940434 | ['Anhur'] | 6.36E+16 | -2.17E+16 |
| 2015-08-23 | 8:30:00 | 1.249455849 | ['Anhur'] | 6.28E+16 | -2.14E+16 |
| 2015-08-23 | 9:00:00 | 1.249481321 | ['Anhur'] | 4.68E+16 | -1.60E+16 |
| 2015-08-23 | 10:00:00 | 1.249533462 | ['Geb'] | 5.49E+16 | -1.87E+16 |
| 2015-08-23 | 11:00:00 | 1.249577 | ['Geb'] | 4.15E+16 | -1.42E+16 |
| 2015-08-23 | 11:30:00 | 1.249617429 | ['Geb'] | 1.14E+17 | -3.91E+16 |
| 2015-08-23 | 12:00:00 | 1.249639434 | ['Geb'] | 1.33E+17 | -4.55E+16 |
| 2015-08-23 | 12:30:00 | 1.249665577 | ['Geb'] | 1.17E+17 | -3.98E+16 |
| 2015-08-23 | 13:00:00 | 1.249691321 | ['Atum'] | 9.89E+16 | -3.38E+16 |
| 2015-08-23 | 13:30:00 | 1.249718491 | ['Khonsu'] | 7.14E+16 | -2.44E+16 |
| 2015-08-23 | 14:00:00 | 1.249745 | ['Khonsu'] | 5.54E+16 | -1.89E+16 |
| 2015-08-23 | 14:30:00 | 1.249772 | ['Khonsu'] | 4.54E+16 | -1.55E+16 |
| 2015-08-23 | 15:00:00 | 1.249799811 | ['Imhotep1'] | 4.78E+16 | -1.63E+16 |
| 2015-08-23 | 15:30:00 | 1.249825532 | ['Imhotep'] | 7.28E+16 | -2.49E+16 |
| 2015-08-23 | 16:00:00 | 1.249852174 | ['Imhotep'] | 1.07E+17 | -3.66E+16 |
| 2015-08-23 | 16:30:00 | 1.249878667 | ['Imhotep'] | 7.82E+16 | -2.68E+16 |
| 2015-08-24 | 2:00:00 | 1.250399245 | ['Khonsu'] | 6.28E+16 | -2.15E+16 |
| 2015-08-24 | 2:30:00 | 1.250426981 | ['Imhotep1'] | 4.41E+16 | -1.51E+16 |
| 2015-08-24 | 3:00:00 | 1.250454808 | ['Bes'] | 3.43E+16 | -1.17E+16 |
| 2015-08-24 | 4:30:00 | 1.250540577 | ['Bes'] | 7.28E+16 | -2.49E+16 |
| 2015-08-24 | 8:00:00 | 1.250739057 | ['Geb'] | 5.75E+16 | -1.97E+16 |
| 2015-08-24 | 8:30:00 | 1.250767255 | ['Geb'] | 4.13E+16 | -1.41E+16 |
| 2015-08-26 | 5:30:00 | 1.253556 | ['Geb'] | 1.07E+17 | -3.67E+16 |
| 2015-08-26 | 9:30:00 | 1.253807727 | ['Geb'] | 7.65E+16 | -2.61E+16 |
| 2015-08-26 | 14:30:00 | 1.254158438 | ['Geb'] | 1.10E+17 | -3.77E+16 |
| 2015-08-26 | 15:00:00 | 1.254185472 | ['Khonsu'] | 1.21E+17 | -4.12E+16 |
| 2015-08-26 | 15:30:00 | 1.254219623 | ['Imhotep1'] | 1.12E+17 | -3.84E+16 |
| 2015-08-26 | 16:00:00 | 1.254253774 | ['Imhotep'] | 1.26E+17 | -4.32E+16 |
| 2015-08-26 | 16:30:00 | 1.254288 | ['Imhotep'] | 1.41E+17 | -4.81E+16 |
| 2015-08-26 | 17:00:00 | 1.254324231 | ['Imhotep'] | 1.38E+17 | -4.70E+16 |
| 2015-08-26 | 17:30:00 | 1.254357547 | ['Imhotep'] | 1.43E+17 | -4.89E+16 |
| 2015-08-26 | 18:00:00 | 1.254392264 | ['Bes'] | 1.34E+17 | -4.59E+16 |
| 2015-08-26 | 18:30:00 | 1.254425769 | ['Bes'] | 1.24E+17 | -4.23E+16 |
| 2015-08-26 | 19:00:00 | 1.254460192 | ['Bes'] | 1.01E+17 | -3.44E+16 |
| 2015-08-26 | 19:30:00 | 1.254494906 | ['Bes'] | 1.03E+17 | -3.51E+16 |
| 2015-08-26 | 20:00:00 | 1.254529434 | ['Anhur'] | 1.46E+17 | -5.00E+16 |
| 2015-08-26 | 20:30:00 | 1.254564727 | ['Anhur'] | 1.81E+17 | -6.19E+16 |
| 2015-08-26 | 21:00:00 | 1.254600377 | ['Anhur'] | 1.42E+17 | -4.84E+16 |
| 2015-08-26 | 21:30:00 | 1.254635385 | ['Anhur'] | 9.89E+16 | -3.38E+16 |
| 2015-08-26 | 22:00:00 | 1.254669434 | ['Geb'] | 8.02E+16 | -2.74E+16 |
| 2015-08-26 | 22:30:00 | 1.254704528 | ['Geb'] | 9.60E+16 | -3.28E+16 |
| 2015-08-26 | 23:00:00 | 1.254738868 | ['Geb'] | 1.17E+17 | -3.99E+16 |
| 2015-08-26 | 23:30:00 | 1.254773846 | ['Geb'] | 1.16E+17 | -3.95E+16 |
| 2015-08-27 | 0:00:00 | 1.254807692 | ['Geb'] | 1.16E+17 | -3.98E+16 |
| 2015-08-27 | 1:00:00 | 1.254879286 | ['Geb'] | 1.44E+17 | -4.91E+16 |
| 2015-08-27 | 1:30:00 | 1.254915472 | ['Atum'] | 1.41E+17 | -4.82E+16 |
| 2015-08-27 | 2:00:00 | 1.254950566 | ['Khonsu'] | 1.48E+17 | -5.07E+16 |
| 2015-08-27 | 2:30:00 | 1.254985385 | ['Khonsu'] | 1.49E+17 | -5.08E+16 |
| 2015-08-27 | 3:00:00 | 1.255020377 | ['Khonsu'] | 1.38E+17 | -4.70E+16 |
| 2015-08-27 | 3:30:00 | 1.255055472 | ['Imhotep1'] | 1.30E+17 | -4.44E+16 |
| 2015-08-27 | 4:00:00 | 1.255091346 | ['Imhotep'] | 1.36E+17 | -4.64E+16 |
| 2015-08-27 | 4:30:00 | 1.255125577 | ['Imhotep'] | 1.45E+17 | -4.96E+16 |
| 2015-08-27 | 6:00:00 | 1.255234717 | ['Bes'] | 1.31E+17 | -4.49E+16 |
| 2015-08-27 | 6:30:00 | 1.25527 | ['Bes'] | 1.07E+17 | -3.67E+16 |
| 2015-08-27 | 7:00:00 | 1.255295 | ['Bes'] | 8.82E+16 | -3.01E+16 |
| 2015-08-27 | 11:00:00 | 1.255598889 | ['Bes'] | 1.16E+17 | -3.98E+16 |
| 2015-08-27 | 11:30:00 | 1.255617143 | ['Geb'] | 1.25E+17 | -4.27E+16 |
| 2015-08-27 | 16:30:00 | 1.255993654 | ['Imhotep'] | 1.52E+17 | -5.19E+16 |
| 2015-08-27 | 21:30:00 | 1.256365849 | ['Geb'] | 1.38E+17 | -4.70E+16 |
| 2015-08-28 | 0:00:00 | 1.256551154 | ['Seth'] | 1.35E+17 | -4.63E+16 |
| 2015-08-28 | 1:30:00 | 1.256665 | ['Atum'] | 1.37E+17 | -4.69E+16 |
| 2015-08-28 | 2:00:00 | 1.256703208 | ['Khonsu'] | 1.13E+17 | -3.85E+16 |
| 2015-08-28 | 2:30:00 | 1.256740755 | ['Khonsu'] | 1.51E+17 | -5.15E+16 |
| 2015-08-28 | 3:30:00 | 1.256814906 | ['Imhotep'] | 1.28E+17 | -4.38E+16 |
| 2015-08-28 | 4:00:00 | 1.256853208 | ['Ash'] | 1.17E+17 | -3.99E+16 |
| 2015-08-28 | 5:00:00 | 1.256928462 | ['Aten'] | 1.94E+17 | -6.62E+16 |
| 2015-08-28 | 5:30:00 | 1.256967544 | ['Khepry'] | 1.55E+17 | -5.28E+16 |
| 2015-08-28 | 6:00:00 | 1.257006731 | ['Khepry'] | 1.22E+17 | -4.16E+16 |
| 2015-08-28 | 6:30:00 | 1.257044717 | ['Hathor'] | 1.19E+17 | -4.08E+16 |
| 2015-08-28 | 7:30:00 | 1.257120755 | ['Bastet'] | 1.61E+17 | -5.51E+16 |
| 2015-08-28 | 8:00:00 | 1.257158627 | ['Anhur'] | 1.74E+17 | -5.93E+16 |
| 2015-08-28 | 8:30:00 | 1.257196415 | ['Wosret'] | 1.46E+17 | -4.97E+16 |
| 2015-08-28 | 9:00:00 | 1.257234528 | ['Wosret'] | 1.35E+17 | -4.61E+16 |
| 2015-08-28 | 9:30:00 | 1.257272885 | ['Wosret'] | 1.06E+17 | -3.63E+16 |
| 2015-08-28 | 12:00:00 | 1.257466981 | ['Wosret'] | 1.04E+17 | -3.55E+16 |
| 2015-08-28 | 12:30:00 | 1.257505192 | ['Anubis'] | 1.44E+17 | -4.92E+16 |
| 2015-08-28 | 13:30:00 | 1.257580816 | ['Khonsu'] | 1.32E+17 | -4.51E+16 |
| 2015-08-28 | 17:30:00 | 1.2579052 | ['Khonsu'] | 1.26E+17 | -4.31E+16 |
| 2015-08-28 | 18:00:00 | 1.257929211 | ['Bes'] | 1.27E+17 | -4.33E+16 |
| 2015-08-28 | 22:30:00 | 1.258303846 | ['Bes'] | 1.16E+17 | -3.95E+16 |
| 2015-08-28 | 23:00:00 | 1.258331404 | ['Geb'] | 1.29E+17 | -4.42E+16 |
| 2015-08-28 | 23:30:00 | 1.258373077 | ['Geb'] | 1.14E+17 | -3.90E+16 |
| 2015-08-29 | 5:30:00 | 1.258865217 | ['Geb'] | 1.30E+17 | -4.44E+16 |
| 2015-08-29 | 6:00:00 | 1.258893269 | ['Bes'] | 1.19E+17 | -4.05E+16 |
| 2015-08-29 | 6:30:00 | 1.258934151 | ['Anhur'] | 1.22E+17 | -4.17E+16 |
| 2015-08-29 | 7:00:00 | 1.258974528 | ['Anhur'] | 1.33E+17 | -4.54E+16 |
| 2015-08-29 | 7:30:00 | 1.259015094 | ['Anhur'] | 1.67E+17 | -5.70E+16 |
| 2015-08-29 | 8:00:00 | 1.25904 | ['Wosret'] | 2.72E+17 | -9.72E+16 |
| 2015-08-29 | 9:00:00 | 1.259153125 | ['Hatmehit'] | 2.04E+17 | -6.98E+16 |
| 2015-08-29 | 9:30:00 | 1.259179811 | ['Sobek'] | 1.41E+17 | -4.81E+16 |
| 2015-08-29 | 10:00:00 | 1.259220588 | ['Geb'] | 1.13E+17 | -3.88E+16 |
| 2015-08-29 | 10:30:00 | 1.25925 | ['Geb'] | 1.10E+17 | -3.77E+16 |
| 2015-08-29 | 14:30:00 | 1.2595916 | ['Imhotep'] | 1.58E+17 | -5.41E+16 |
| 2015-08-29 | 15:00:00 | 1.259616 | ['Imhotep'] | 1.67E+17 | -5.73E+16 |
| 2015-08-29 | 19:00:00 | 1.259982857 | ['Imhotep'] | 1.91E+17 | -6.53E+16 |
| 2015-08-29 | 19:30:00 | 1.260006981 | ['Neith'] | 1.86E+17 | -6.35E+16 |
| 2015-08-29 | 20:00:00 | 1.260048654 | ['Neith'] | 1.96E+17 | -6.69E+16 |
| 2015-08-29 | 20:30:00 | 1.260081818 | ['Wosret'] | 1.86E+17 | -6.36E+16 |
| 2015-08-30 | 0:00:00 | 1.26041 | ['Wosret'] | 1.69E+17 | -5.82E+16 |
| 2015-08-30 | 0:30:00 | 1.260428302 | ['Khonsu'] | 1.61E+17 | -5.50E+16 |
| 2015-08-30 | 1:00:00 | 1.260453333 | ['Khonsu'] | 1.71E+17 | -5.85E+16 |
| 2015-08-30 | 5:30:00 | 1.2608566 | ['Khonsu'] | 1.35E+17 | -4.61E+16 |
| 2015-08-30 | 6:00:00 | 1.260899245 | ['Anhur'] | 1.33E+17 | -4.55E+16 |
| 2015-08-30 | 6:30:00 | 1.260942692 | ['Sobek'] | 1.34E+17 | -4.56E+16 |
| 2015-08-30 | 7:00:00 | 1.260984902 | ['Neith'] | 1.65E+17 | -5.64E+16 |
| 2015-08-30 | 7:30:00 | 1.261028269 | ['Neith'] | 2.06E+17 | -7.03E+16 |
| 2015-08-30 | 11:00:00 | 1.26131 | ['Seth'] | 9.57E+16 | -3.27E+16 |
| 2015-08-30 | 11:30:00 | 1.261382353 | ['Seth'] | 1.52E+17 | -5.20E+16 |
| 2015-08-30 | 12:00:00 | 1.2614174 | ['Atum'] | 1.68E+17 | -5.74E+16 |
| 2015-08-30 | 12:30:00 | 1.26146 | ['Khonsu'] | 1.67E+17 | -5.69E+16 |
| 2015-08-30 | 13:00:00 | 1.261503585 | ['Khonsu'] | 1.11E+17 | -3.81E+16 |
| 2015-08-30 | 14:30:00 | 1.261637391 | ['Imhotep'] | 1.14E+17 | -3.90E+16 |
| 2015-08-30 | 15:00:00 | 1.261680392 | ['Imhotep'] | 1.31E+17 | -4.46E+16 |
| 2015-08-30 | 15:30:00 | 1.261723462 | ['Imhotep'] | 1.25E+17 | -4.27E+16 |
| 2015-08-30 | 16:00:00 | 1.2617688 | ['Imhotep'] | 1.24E+17 | -4.24E+16 |
| 2015-08-30 | 16:30:00 | 1.261811633 | ['Khepry'] | 1.24E+17 | -4.24E+16 |
| 2015-08-30 | 17:00:00 | 1.261855098 | ['Bes'] | 1.35E+17 | -4.61E+16 |
| 2015-08-30 | 17:30:00 | 1.261898679 | ['Bes'] | 1.39E+17 | -4.74E+16 |
| 2015-08-30 | 18:00:00 | 1.261942692 | ['Hathor'] | 1.36E+17 | -4.65E+16 |
| 2015-08-30 | 18:30:00 | 1.261988302 | ['Sobek'] | 1.40E+17 | -4.77E+16 |
| 2015-08-30 | 19:00:00 | 1.262032941 | ['Neith'] | 1.82E+17 | -6.20E+16 |
| 2015-08-30 | 19:30:00 | 1.262077551 | ['Neith'] | 2.15E+17 | -7.36E+16 |
| 2015-08-30 | 20:00:00 | 1.262120784 | ['Wosret'] | 2.15E+17 | -7.33E+16 |
| 2015-08-30 | 20:30:00 | 1.262165577 | ['Wosret'] | 1.86E+17 | -6.34E+16 |
| 2015-08-30 | 23:00:00 | 1.262387857 | ['Wosret'] | 1.01E+17 | -3.45E+16 |
| 2015-08-30 | 23:30:00 | 1.262433208 | ['Anubis'] | 1.56E+17 | -5.32E+16 |
| 2015-08-31 | 0:00:00 | 1.262478491 | ['Atum'] | 1.84E+17 | -6.30E+16 |
| 2015-08-31 | 0:30:00 | 1.262521569 | ['Atum'] | 2.22E+17 | -7.58E+16 |
| 2015-08-31 | 1:00:00 | 1.262567255 | ['Khonsu'] | 1.42E+17 | -4.89E+16 |
| 2015-08-31 | 6:00:00 | 1.26304 | ['Hatmehit'] | 1.15E+17 | -3.95E+16 |
| 2015-08-31 | 6:30:00 | 1.263070571 | ['Bastet'] | 1.28E+17 | -4.37E+16 |
| 2015-08-31 | 7:00:00 | 1.26309619 | ['Bastet'] | 1.62E+17 | -5.53E+16 |
| 2015-08-31 | 12:00:00 | 1.263566415 | ['Seth'] | 1.50E+17 | -5.13E+16 |
| 2015-08-31 | 12:30:00 | 1.263609348 | ['Atum'] | 1.94E+17 | -6.62E+16 |
| 2015-09-03 | 0:30:00 | 1.269499048 | ['Atum'] | 1.43E+17 | -4.87E+16 |
| 2015-09-03 | 1:00:00 | 1.269535577 | ['Seth'] | 1.69E+17 | -5.76E+16 |
| 2015-09-03 | 2:30:00 | 1.26969 | ['Apis'] | 1.11E+17 | -3.80E+16 |
| 2015-09-03 | 4:00:00 | 1.269842083 | ['Apis'] | 1.11E+17 | -3.79E+16 |
| 2015-09-03 | 6:00:00 | 1.270056863 | ['Apis'] | 1.09E+17 | -3.73E+16 |
| 2015-09-03 | 6:30:00 | 1.270107925 | ['Khepry'] | 7.53E+16 | -2.57E+16 |
| 2015-09-03 | 7:00:00 | 1.270160189 | ['Khepry'] | 6.67E+16 | -2.28E+16 |
| 2015-09-03 | 7:30:00 | 1.27019619 | ['Khepry'] | 6.61E+16 | -2.26E+16 |
| 2015-09-03 | 8:30:00 | 1.270318163 | ['Khepry'] | 1.38E+17 | -4.70E+16 |
| 2015-09-03 | 11:00:00 | 1.270581509 | ['Maftet'] | 9.31E+16 | -3.18E+16 |
| 2015-09-03 | 11:30:00 | 1.270619167 | ['Wosret'] | 9.51E+16 | -3.25E+16 |
| 2015-09-03 | 14:30:00 | 1.270953585 | ['Sobek'] | 7.82E+16 | -2.67E+16 |
| 2015-09-03 | 16:00:00 | 1.271125769 | ['Sobek'] | 1.15E+17 | -3.93E+16 |
| 2015-09-03 | 16:30:00 | 1.271164906 | ['Imhotep'] | 1.25E+17 | -4.27E+16 |
| 2015-09-03 | 17:00:00 | 1.271217736 | ['Imhotep'] | 1.49E+17 | -5.08E+16 |
| 2015-09-03 | 21:00:00 | 1.271666471 | ['Imhotep'] | 3.89E+16 | -1.33E+16 |
| 2015-09-03 | 21:30:00 | 1.271698776 | ['Wosret'] | 5.98E+16 | -2.04E+16 |
| 2015-09-03 | 22:30:00 | 1.271820882 | ['Wosret'] | 6.97E+16 | -2.38E+16 |
| 2015-09-03 | 23:00:00 | 1.27186566 | ['Maftet'] | 6.31E+16 | -2.15E+16 |
| 2015-09-03 | 23:30:00 | 1.271918462 | ['Maftet'] | 5.33E+16 | -1.82E+16 |
| 2015-09-04 | 2:30:00 | 1.272243214 | ['Atum'] | 5.04E+16 | -1.72E+16 |
| 2015-09-04 | 3:00:00 | 1.272298868 | ['Khonsu'] | 4.41E+16 | -1.51E+16 |
| 2015-09-04 | 3:30:00 | 1.27233913 | ['Imhotep1'] | 3.58E+16 | -1.22E+16 |
| 2015-09-04 | 4:30:00 | 1.272461961 | ['Imhotep1'] | 9.58E+16 | -3.27E+16 |
| 2015-09-04 | 5:00:00 | 1.272515094 | ['Imhotep'] | 1.20E+17 | -4.09E+16 |
| 2015-09-04 | 5:30:00 | 1.272561795 | ['Imhotep'] | 1.37E+17 | -4.67E+16 |
| 2015-09-04 | 7:30:00 | 1.272791429 | ['Imhotep'] | 4.54E+16 | -1.55E+16 |
| 2015-09-04 | 8:00:00 | 1.272826875 | ['Hapi'] | 4.18E+16 | -1.43E+16 |
| 2015-09-04 | 9:00:00 | 1.272954694 | ['Hapi'] | 5.20E+16 | -1.78E+16 |
| 2015-09-04 | 9:30:00 | 1.273007059 | ['Bastet'] | 3.80E+16 | -1.30E+16 |
| 2015-09-04 | 12:30:00 | 1.2733475 | ['Bastet'] | 4.33E+16 | -1.48E+16 |
| 2015-09-04 | 13:00:00 | 1.273395283 | ['Seth'] | 7.81E+16 | -2.67E+16 |
| 2015-09-04 | 13:30:00 | 1.273445238 | ['Seth'] | 1.03E+17 | -3.53E+16 |
| 2015-09-04 | 18:00:00 | 1.273941944 | ['Khepry'] | 1.21E+17 | -4.14E+16 |
| 2015-09-04 | 23:00:00 | 1.274510943 | ['Nut'] | 1.16E+16 | -3.96652E+15 |
| 2015-09-05 | 6:00:00 | 1.275304615 | ['Imhotep'] | 1.08E+17 | -3.67E+16 |
| 2015-09-05 | 6:30:00 | 1.2753618 | ['Imhotep'] | 8.77E+16 | -2.99E+16 |
| 2015-09-05 | 9:30:00 | 1.275705185 | ['Bastet'] | 1.99E+16 | -6.78857E+15 |
| 2015-09-05 | 10:00:00 | 1.275763208 | ['Bastet'] | 1.23E+16 | -4.19798E+15 |
| 2015-09-05 | 11:30:00 | 1.275933585 | ['Maftet'] | 1.12E+16 | -3.82799E+15 |
| 2015-09-05 | 12:00:00 | 1.275990943 | ['Anuket'] | 1.98E+16 | -6.76038E+15 |
| 2015-09-05 | 12:30:00 | 1.2760475 | ['Hapi'] | 2.41E+16 | -8.23923E+15 |
| 2015-09-05 | 13:00:00 | 1.276103846 | ['Hapi'] | 4.22E+16 | -1.44E+16 |
| 2015-09-05 | 14:00:00 | 1.276222692 | ['Anubis'] | 7.62E+16 | -2.60E+16 |
| 2015-09-05 | 16:00:00 | 1.276453673 | ['Imhotep1'] | 1.85E+16 | -6.32227E+15 |
| 2015-09-05 | 16:30:00 | 1.276510566 | ['Imhotep'] | 4.05E+16 | -1.38E+16 |
| 2015-09-05 | 17:00:00 | 1.276567308 | ['Imhotep'] | 6.17E+16 | -2.11E+16 |
| 2015-09-05 | 17:30:00 | 1.276625 | ['Imhotep'] | 7.60E+16 | -2.60E+16 |
| 2015-09-05 | 18:00:00 | 1.276685091 | ['Imhotep'] | 8.52E+16 | -2.91E+16 |
| 2015-09-05 | 18:30:00 | 1.276744528 | ['Imhotep'] | 8.46E+16 | -2.89E+16 |
| 2015-09-06 | 0:00:00 | 1.277386038 | ['Anuket'] | 1.23E+16 | -4.18835E+15 |
| 2015-09-06 | 0:30:00 | 1.277444231 | ['Hapi'] | 1.49E+16 | -5.10303E+15 |
| 2015-09-06 | 1:00:00 | 1.277496279 | ['Seth'] | 2.42E+16 | -8.27469E+15 |
| 2015-09-06 | 5:30:00 | 1.278035208 | ['Seth'] | 5.67E+16 | -1.94E+16 |
| 2015-09-06 | 6:00:00 | 1.278091321 | ['Imhotep'] | 5.85E+16 | -2.00E+16 |
| 2015-09-07 | 2:00:00 | 1.280498302 | ['Anubis'] | 3.97E+16 | -1.36E+16 |
| 2015-09-13 | 16:00:00 | 1.30206434 | ['Imhotep'] | 2.54E+16 | -8.68571E+15 |
| 2015-09-21 | 12:30:00 | 1.333387368 | ['Babi'] | 4.44E+15 | -1.5612E+15 |
| 2015-09-21 | 13:30:00 | 1.333536604 | ['Maat'] | 1.36E+16 | -4.66818E+15 |
| 2015-09-21 | 15:30:00 | 1.333902453 | ['Serqet'] | 8.45E+16 | -2.89E+16 |
| 2015-09-21 | 16:00:00 | 1.33399283 | ['Hathor'] | 7.19E+16 | -2.46E+16 |
| 2015-09-21 | 16:30:00 | 1.334083396 | ['Hathor'] | 4.64E+16 | -1.59E+16 |
| 2015-09-21 | 17:30:00 | 1.334262642 | ['Hathor'] | 8.37E+15 | -2.88447E+15 |
| 2015-09-21 | 19:30:00 | 1.334631923 | ['Ash'] | 1.18E+17 | -4.02E+16 |
| 2015-09-21 | 20:30:00 | 1.334811132 | ['Ash'] | 9.25E+16 | -3.16E+16 |
| 2015-09-21 | 21:00:00 | 1.334901698 | ['Ash'] | 7.13E+16 | -2.44E+16 |
| 2015-09-21 | 21:30:00 | 1.334991923 | ['Ash'] | 6.04E+16 | -2.06E+16 |
| 2015-09-21 | 22:00:00 | 1.335082075 | ['Ash'] | 4.35E+16 | -1.49E+16 |
| 2015-09-21 | 22:30:00 | 1.335171373 | ['Ash'] | 4.42E+16 | -1.51E+16 |
| 2015-09-22 | 3:30:00 | 1.336091481 | ['Serqet'] | 9.02E+16 | -3.08E+16 |
| 2015-09-22 | 4:00:00 | 1.336183962 | ['Hathor'] | 7.78E+16 | -2.66E+16 |
| 2015-09-22 | 4:30:00 | 1.336275283 | ['Hathor'] | 6.34E+16 | -2.17E+16 |
| 2015-09-22 | 8:00:00 | 1.33692 | ['Ash'] | 1.31E+17 | -4.48E+16 |
| 2015-09-22 | 8:30:00 | 1.337010962 | ['Ash'] | 1.12E+17 | -3.83E+16 |
| 2015-09-22 | 9:00:00 | 1.3371025 | ['Ash'] | 8.27E+16 | -2.82E+16 |
| 2015-09-22 | 9:30:00 | 1.337193962 | ['Ash'] | 6.52E+16 | -2.23E+16 |
| 2015-09-22 | 10:00:00 | 1.337284717 | ['Ash'] | 5.25E+16 | -1.79E+16 |
| 2015-09-22 | 10:30:00 | 1.337376731 | ['Ash'] | 4.61E+16 | -1.58E+16 |
| 2015-09-22 | 16:00:00 | 1.338390755 | ['Serqet'] | 9.16E+16 | -3.13E+16 |
| 2015-09-22 | 16:30:00 | 1.338485714 | ['Hathor'] | 7.40E+16 | -2.53E+16 |
| 2015-09-22 | 20:30:00 | 1.339226604 | ['Ash'] | 1.51E+17 | -5.17E+16 |
| 2015-09-22 | 21:00:00 | 1.3393225 | ['Ash'] | 1.07E+17 | -3.64E+16 |
| 2015-09-22 | 21:30:00 | 1.339416923 | ['Ash'] | 7.14E+16 | -2.44E+16 |
| 2015-09-22 | 22:00:00 | 1.339509057 | ['Ash'] | 5.54E+16 | -1.89E+16 |
| 2015-09-22 | 22:30:00 | 1.339602308 | ['Ash'] | 4.55E+16 | -1.56E+16 |
| 2015-09-23 | 6:00:00 | 1.340995 | ['Atum'] | 2.39E+16 | -8.15342E+15 |
| 2015-09-23 | 9:30:00 | 1.341663953 | ['Atum'] | 8.61E+16 | -2.94E+16 |
| 2015-09-23 | 10:00:00 | 1.341749623 | ['Khepry'] | 6.76E+16 | -2.31E+16 |
| 2015-09-23 | 10:30:00 | 1.341831053 | ['Aten'] | 6.36E+16 | -2.17E+16 |
| 2015-09-23 | 14:30:00 | 1.342614667 | ['Aten'] | 3.74E+16 | -1.28E+16 |
| 2015-09-23 | 15:00:00 | 1.342670968 | ['Maftet'] | 5.26E+16 | -1.80E+16 |
| 2015-09-23 | 19:30:00 | 1.343571579 | ['Maftet'] | 1.02E+17 | -3.48E+16 |
| 2015-09-23 | 20:00:00 | 1.343635849 | ['Imhotep'] | 1.62E+17 | -5.53E+16 |
| 2015-09-23 | 20:30:00 | 1.343733393 | ['Imhotep'] | 1.79E+17 | -6.10E+16 |
| 2015-09-24 | 7:00:00 | 1.345710606 | ['Imhotep1'] | 5.55E+16 | -1.89E+16 |
| 2015-09-24 | 11:00:00 | 1.346500222 | ['Imhotep1'] | 8.07E+16 | -2.76E+16 |
| 2015-09-24 | 21:30:00 | 1.348517358 | ['Imhotep'] | 1.34E+17 | -4.58E+16 |
| 2015-09-24 | 22:00:00 | 1.348613585 | ['Imhotep'] | 9.96E+16 | -3.40E+16 |
| 2015-09-24 | 22:30:00 | 1.348708462 | ['Khepry'] | 7.65E+16 | -2.61E+16 |
| 2015-09-24 | 23:00:00 | 1.348803962 | ['Khepry'] | 8.26E+16 | -2.82E+16 |
| 2015-09-24 | 23:30:00 | 1.348900377 | ['Khepry'] | 1.16E+17 | -3.98E+16 |
| 2015-09-25 | 0:00:00 | 1.348996792 | ['Aker'] | 1.53E+17 | -5.22E+16 |
| 2015-09-25 | 0:30:00 | 1.349092115 | ['Hapi'] | 1.22E+17 | -4.17E+16 |
| 2015-09-25 | 1:00:00 | 1.349187736 | ['Neith'] | 9.69E+16 | -3.31E+16 |
| 2015-09-25 | 1:30:00 | 1.349287321 | ['Wosret'] | 8.44E+16 | -2.88E+16 |
| 2015-09-25 | 2:00:00 | 1.349386981 | ['Wosret'] | 9.54E+16 | -3.26E+16 |
| 2015-09-25 | 2:30:00 | 1.349483585 | ['Wosret'] | 1.16E+17 | -3.96E+16 |
| 2015-09-25 | 3:30:00 | 1.349675769 | ['Anuket'] | 1.21E+17 | -4.12E+16 |
| 2015-09-25 | 4:00:00 | 1.349771509 | ['Anuket'] | 1.95E+17 | -6.69E+16 |
| 2015-09-25 | 6:00:00 | 1.350162909 | ['Atum'] | 1.41E+17 | -4.81E+16 |
| 2015-09-25 | 6:30:00 | 1.350260577 | ['Atum'] | 8.97E+16 | -3.06E+16 |
| 2015-09-25 | 7:00:00 | 1.350356604 | ['Khonsu'] | 8.34E+16 | -2.85E+16 |
| 2015-09-25 | 7:30:00 | 1.350454038 | ['Imhotep1'] | 9.27E+16 | -3.17E+16 |
| 2015-09-25 | 8:00:00 | 1.350550385 | ['Imhotep'] | 1.25E+17 | -4.26E+16 |
| 2015-09-25 | 8:30:00 | 1.350646346 | ['Imhotep'] | 1.67E+17 | -5.69E+16 |
| 2015-09-25 | 9:00:00 | 1.350742453 | ['Imhotep'] | 1.59E+17 | -5.43E+16 |
| 2015-09-25 | 9:30:00 | 1.350839434 | ['Imhotep'] | 1.41E+17 | -4.82E+16 |
| 2015-09-25 | 10:00:00 | 1.350938545 | ['Imhotep'] | 1.14E+17 | -3.88E+16 |
| 2015-09-25 | 10:30:00 | 1.351038889 | ['Khepry'] | 8.19E+16 | -2.80E+16 |
| 2015-09-25 | 11:00:00 | 1.351136923 | ['Khepry'] | 8.34E+16 | -2.85E+16 |
| 2015-09-25 | 11:30:00 | 1.351233462 | ['Khepry'] | 1.09E+17 | -3.72E+16 |
| 2015-09-25 | 12:00:00 | 1.351329811 | ['Aker'] | 1.88E+17 | -6.45E+16 |
| 2015-09-25 | 12:30:00 | 1.351426981 | ['Hapi'] | 1.35E+17 | -4.62E+16 |
| 2015-09-25 | 13:00:00 | 1.351523269 | ['Neith'] | 1.04E+17 | -3.56E+16 |
| 2015-09-25 | 13:30:00 | 1.351610909 | ['Wosret'] | 9.96E+16 | -3.40E+16 |
| 2015-09-25 | 18:00:00 | 1.352486176 | ['Atum'] | 1.63E+17 | -5.57E+16 |
| 2015-09-25 | 22:30:00 | 1.353428571 | ['Hatmehit'] | 1.01E+17 | -3.47E+16 |
| 2015-09-25 | 23:00:00 | 1.353482222 | ['Bes'] | 9.03E+16 | -3.09E+16 |
| 2015-09-25 | 23:30:00 | 1.353583962 | ['Anhur'] | 1.08E+17 | -3.69E+16 |
| 2015-09-26 | 0:30:00 | 1.353781346 | ['Anhur'] | 7.67E+16 | -2.62E+16 |
| 2015-09-26 | 1:00:00 | 1.353879811 | ['Hapi'] | 1.06E+17 | -3.60E+16 |
| 2015-09-26 | 1:30:00 | 1.353977115 | ['Wosret'] | 1.18E+17 | -4.02E+16 |
| 2015-09-26 | 2:00:00 | 1.354073654 | ['Wosret'] | 1.41E+17 | -4.81E+16 |
| 2015-09-26 | 2:30:00 | 1.354173077 | ['Wosret'] | 2.03E+17 | -6.97E+16 |
| 2015-09-26 | 6:00:00 | 1.354858 | ['Atum'] | 2.25E+17 | -7.72E+16 |
| 2015-09-26 | 9:00:00 | 1.355502857 | ['Atum'] | 1.58E+17 | -5.41E+16 |
| 2015-09-26 | 9:30:00 | 1.35556 | ['Imhotep'] | 1.64E+17 | -5.60E+16 |
| 2015-09-26 | 10:00:00 | 1.355657451 | ['Khepry'] | 1.60E+17 | -5.48E+16 |
| 2015-09-26 | 10:30:00 | 1.355721765 | ['Bes'] | 1.45E+17 | -5.01E+16 |

| Date | Time | Value | Region | E+ | E- |
|---|---|---|---|---|---|
| 2015-09-26 | 14:00:00 | 1.3565 | ['Bes'] | 1.00E+17 | -3.43E+16 |
| 2015-09-26 | 14:30:00 | 1.356552075 | ['Wosret'] | 1.52E+17 | -5.21E+16 |
| 2015-09-26 | 19:30:00 | 1.357546346 | ['Wosret'] | 1.50E+17 | -5.14E+16 |
| 2015-09-26 | 20:00:00 | 1.357644906 | ['Imhotep'] | 1.38E+17 | -4.72E+16 |
| 2015-09-28 | 22:30:00 | 1.36796 | ['Khepry'] | 9.36E+16 | -3.21E+16 |
| 2015-09-29 | 2:30:00 | 1.3687425 | ['Khepry'] | 1.11E+17 | -3.86E+16 |
| 2015-09-30 | 9:30:00 | 1.375236905 | ['Hatmehit'] | 1.09E+17 | -3.73E+16 |
| 2015-09-30 | 10:00:00 | 1.375331321 | ['Hatmehit'] | 1.15E+17 | -3.94E+16 |
| 2015-09-30 | 10:30:00 | 1.375417353 | ['Hatmehit'] | 1.08E+17 | -3.68E+16 |
| 2015-09-30 | 14:30:00 | 1.376304242 | ['Hatmehit'] | 1.22E+17 | -4.17E+16 |
| 2015-09-30 | 15:00:00 | 1.376366923 | ['Wosret'] | 1.02E+17 | -3.50E+16 |
| 2015-09-30 | 19:30:00 | 1.37738 | ['Wosret'] | 1.22E+17 | -4.19E+16 |
| 2015-09-30 | 20:00:00 | 1.377447925 | ['Imhotep'] | 1.19E+17 | -4.06E+16 |
| 2015-09-30 | 20:30:00 | 1.377552885 | ['Imhotep'] | 1.20E+17 | -4.10E+16 |
| 2015-10-01 | 11:00:00 | 1.380669429 | ['Imhotep'] | 1.24E+17 | -4.25E+16 |
| 2015-10-01 | 11:30:00 | 1.380727 | ['Imhotep'] | 1.14E+17 | -3.90E+16 |
| 2015-10-01 | 16:00:00 | 1.381753913 | ['Imhotep'] | 1.35E+17 | -4.64E+16 |
| 2015-10-01 | 16:30:00 | 1.381836429 | ['Imhotep'] | 1.14E+17 | -3.90E+16 |
| 2015-10-01 | 21:00:00 | 1.382848 | ['Hatmehit'] | 9.65E+16 | -3.31E+16 |
| 2015-10-01 | 21:30:00 | 1.382903023 | ['Hatmehit'] | 1.25E+17 | -4.29E+16 |
| 2015-10-02 | 2:00:00 | 1.38394 | ['Hatmehit'] | 2.21E+17 | -7.66E+16 |
| 2015-10-02 | 2:30:00 | 1.383999623 | ['Bastet'] | 2.01E+17 | -6.87E+16 |
| 2015-10-02 | 3:00:00 | 1.384106538 | ['Wosret'] | 1.34E+17 | -4.58E+16 |
| 2015-10-02 | 3:30:00 | 1.384175789 | ['Wosret'] | 1.30E+17 | -4.45E+16 |
| 2015-10-02 | 12:30:00 | 1.38619129 | ['Wosret'] | 1.53E+17 | -5.27E+16 |
| 2015-10-02 | 13:00:00 | 1.386276415 | ['Wosret'] | 1.61E+17 | -5.53E+16 |
| 2015-10-02 | 13:30:00 | 1.386373721 | ['Wosret'] | 1.91E+17 | -6.57E+16 |
| 2015-10-02 | 22:30:00 | 1.388398182 | ['Hatmehit'] | 9.78E+16 | -3.39E+16 |
| 2015-10-02 | 23:00:00 | 1.388464231 | ['Hatmehit'] | 1.41E+17 | -4.84E+16 |
| 2015-10-03 | 0:00:00 | 1.38865 | ['Hatmehit'] | 7.51E+16 | -2.63E+16 |
| 2015-10-03 | 5:30:00 | 1.389931111 | ['Hatmehit'] | 8.04E+16 | -2.75E+16 |
| 2015-10-03 | 6:00:00 | 1.389993542 | ['Hatmehit'] | 8.42E+16 | -2.88E+16 |
| 2015-10-03 | 9:30:00 | 1.390768868 | ['Hatmehit'] | 7.00E+16 | -2.39E+16 |
| 2015-10-03 | 10:00:00 | 1.390878679 | ['Hatmehit'] | 5.43E+16 | -1.86E+16 |
| 2015-10-03 | 10:30:00 | 1.390953125 | ['Hatmehit'] | 4.33E+16 | -1.49E+16 |
| 2015-10-03 | 14:00:00 | 1.39182 | ['Hatmehit'] | 1.11E+17 | -3.85E+16 |
| 2015-10-03 | 14:30:00 | 1.391874 | ['Wosret'] | 1.31E+17 | -4.49E+16 |
| 2015-10-03 | 19:30:00 | 1.392986154 | ['Khonsu'] | 9.95E+16 | -3.40E+16 |
| 2015-10-03 | 20:00:00 | 1.393095094 | ['Khonsu'] | 8.10E+16 | -2.77E+16 |
| 2015-10-03 | 20:30:00 | 1.393174783 | ['Khonsu'] | 8.71E+16 | -2.98E+16 |
| 2015-10-04 | 0:30:00 | 1.394094906 | ['Wosret'] | 1.23E+17 | -4.21E+16 |
| 2015-10-04 | 1:00:00 | 1.394159167 | ['Hathor'] | 1.33E+17 | -4.56E+16 |
| 2015-10-04 | 5:30:00 | 1.3952096 | ['Hathor'] | 7.83E+16 | -2.68E+16 |
| 2015-10-04 | 6:00:00 | 1.39531717 | ['Hathor'] | 8.78E+16 | -3.00E+16 |
| 2015-10-04 | 6:30:00 | 1.395401071 | ['Atum'] | 7.65E+16 | -2.62E+16 |
| 2015-10-05 | 1:30:00 | 1.399707647 | ['Atum'] | 9.62E+16 | -3.29E+16 |
| 2015-10-05 | 2:00:00 | 1.399799804 | ['Wosret'] | 9.00E+16 | -3.08E+16 |
| 2015-10-05 | 2:30:00 | 1.399899048 | ['Wosret'] | 6.55E+16 | -2.24E+16 |
| 2015-10-10 | 0:30:00 | 1.427327679 | ['Bes'] | 4.13E+16 | -1.41E+16 |
| 2015-10-10 | 1:00:00 | 1.427450755 | ['Imhotep'] | 6.30E+16 | -2.15E+16 |
| 2015-10-10 | 1:30:00 | 1.427569811 | ['Imhotep'] | 1.12E+17 | -3.81E+16 |
| 2015-10-10 | 2:00:00 | 1.427687059 | ['Imhotep'] | 1.78E+17 | -6.07E+16 |
| 2015-10-10 | 2:30:00 | 1.42780566 | ['Imhotep'] | 1.66E+17 | -5.68E+16 |
| 2015-10-10 | 3:00:00 | 1.427924808 | ['Khepry'] | 1.38E+17 | -4.72E+16 |
| 2015-10-10 | 3:30:00 | 1.42804434 | ['Khepry'] | 9.86E+16 | -3.37E+16 |
| 2015-10-10 | 4:00:00 | 1.428162308 | ['Khepry'] | 7.74E+16 | -2.64E+16 |
| 2015-10-10 | 4:30:00 | 1.428280566 | ['Hathor'] | 8.39E+16 | -2.86E+16 |
| 2015-10-10 | 5:00:00 | 1.428403929 | ['Bastet'] | 9.53E+16 | -3.25E+16 |
| 2015-10-10 | 5:30:00 | 1.428526792 | ['Sobek'] | 8.71E+16 | -2.98E+16 |
| 2015-10-10 | 6:00:00 | 1.428637333 | ['Neith'] | 7.14E+16 | -2.44E+16 |
| 2015-10-10 | 9:00:00 | 1.429415 | ['Hatmehit'] | 9.88E+16 | -3.39E+16 |
| 2015-10-10 | 9:30:00 | 1.429483091 | ['Seth'] | 1.25E+17 | -4.27E+16 |
| 2015-10-10 | 10:00:00 | 1.429604717 | ['Anubis'] | 1.31E+17 | -4.48E+16 |
| 2015-10-10 | 14:00:00 | 1.43062 | ['Anubis'] | 1.48E+17 | -5.06E+16 |
| 2015-10-10 | 14:30:00 | 1.430684423 | ['Imhotep'] | 1.58E+17 | -5.39E+16 |
| 2015-10-10 | 19:30:00 | 1.431889245 | ['Imhotep'] | 6.19E+16 | -2.11E+16 |
| 2015-10-10 | 20:00:00 | 1.432008269 | ['Wosret'] | 7.50E+16 | -2.56E+16 |
| 2015-10-10 | 20:30:00 | 1.432095185 | ['Hapi'] | 6.97E+16 | -2.38E+16 |
| 2015-10-11 | 0:30:00 | 1.433095686 | ['Hapi'] | 4.72E+16 | -1.61E+16 |
| 2015-10-11 | 1:00:00 | 1.433212885 | ['Imhotep'] | 5.39E+16 | -1.84E+16 |
| 2015-10-11 | 1:30:00 | 1.433331698 | ['Imhotep'] | 7.07E+16 | -2.42E+16 |
| 2015-10-11 | 2:00:00 | 1.433452075 | ['Imhotep'] | 1.27E+17 | -4.33E+16 |
| 2015-10-11 | 2:30:00 | 1.43357625 | ['Imhotep'] | 1.59E+17 | -5.44E+16 |
| 2015-10-11 | 3:00:00 | 1.433700755 | ['Imhotep'] | 1.42E+17 | -4.86E+16 |
| 2015-10-11 | 3:30:00 | 1.433821321 | ['Khepry'] | 1.22E+17 | -4.15E+16 |
| 2015-10-11 | 4:00:00 | 1.433941176 | ['Khepry'] | 9.71E+16 | -3.32E+16 |
| 2015-10-11 | 4:30:00 | 1.434059811 | ['Khepry'] | 7.46E+16 | -2.55E+16 |
| 2015-10-11 | 5:00:00 | 1.434179423 | ['Sobek'] | 6.40E+16 | -2.18E+16 |
| 2015-10-11 | 7:30:00 | 1.434789245 | ['Wosret'] | 4.54E+16 | -1.55E+16 |
| 2015-10-11 | 8:00:00 | 1.434909231 | ['Wosret'] | 6.06E+16 | -2.07E+16 |
| 2015-10-11 | 8:30:00 | 1.435029423 | ['Seth'] | 6.85E+16 | -2.34E+16 |
| 2015-10-11 | 9:00:00 | 1.435149057 | ['Seth'] | 6.63E+16 | -2.26E+16 |
| 2015-10-11 | 9:30:00 | 1.435269811 | ['Anubis'] | 8.65E+16 | -2.95E+16 |
| 2015-10-11 | 10:00:00 | 1.435390377 | ['Anubis'] | 1.24E+17 | -4.22E+16 |
| 2015-10-11 | 10:30:00 | 1.435510192 | ['Atum'] | 1.18E+17 | -4.03E+16 |
| 2015-10-11 | 11:00:00 | 1.435583 | ['Atum'] | 9.50E+16 | -3.25E+16 |
| 2015-10-11 | 11:30:00 | 1.435778529 | ['Hatmehit'] | 7.21E+16 | -2.46E+16 |
| 2015-10-11 | 12:00:00 | 1.435876538 | ['Khonsu'] | 7.60E+16 | -2.60E+16 |
| 2015-10-11 | 12:30:00 | 1.435997358 | ['Khonsu'] | 5.31E+16 | -1.81E+16 |
| 2015-10-11 | 13:00:00 | 1.436118302 | ['Imhotep'] | 4.62E+16 | -1.58E+16 |
| 2015-10-11 | 13:30:00 | 1.436238077 | ['Imhotep'] | 5.31E+16 | -1.81E+16 |
| 2015-10-11 | 14:00:00 | 1.436360545 | ['Imhotep'] | 9.60E+16 | -3.28E+16 |
| 2015-10-11 | 14:30:00 | 1.436485556 | ['Imhotep'] | 1.46E+17 | -4.98E+16 |
| 2015-10-11 | 15:00:00 | 1.436608302 | ['Imhotep'] | 1.36E+17 | -4.66E+16 |
| 2015-10-11 | 15:30:00 | 1.436729245 | ['Bes'] | 1.08E+17 | -3.70E+16 |
| 2015-10-11 | 16:00:00 | 1.436850377 | ['Khepry'] | 8.85E+16 | -3.02E+16 |
| 2015-10-11 | 16:30:00 | 1.436970385 | ['Khepry'] | 6.55E+16 | -2.24E+16 |
| 2015-10-11 | 20:30:00 | 1.437944038 | ['Seth'] | 4.99E+16 | -1.70E+16 |
| 2015-10-11 | 21:00:00 | 1.43806434 | ['Anubis'] | 5.02E+16 | -1.71E+16 |
| 2015-10-11 | 21:30:00 | 1.43818566 | ['Anubis'] | 5.83E+16 | -1.99E+16 |
| 2015-10-11 | 22:00:00 | 1.43830717 | ['Anubis'] | 9.60E+16 | -3.28E+16 |
| 2015-10-11 | 22:30:00 | 1.4384274 | ['Atum'] | 1.09E+17 | -3.71E+16 |
| 2015-10-12 | 1:00:00 | 1.439041887 | ['Imhotep'] | 3.04E+16 | -1.04E+16 |
| 2015-10-12 | 1:30:00 | 1.439162353 | ['Imhotep'] | 2.88E+16 | -9.82288E+15 |
| 2015-10-12 | 2:00:00 | 1.43928283 | ['Imhotep'] | 5.51E+16 | -1.88E+16 |
| 2015-10-12 | 2:30:00 | 1.439390488 | ['Imhotep'] | 9.03E+16 | -3.08E+16 |
| 2015-10-14 | 5:30:00 | 1.452033 | ['Hatmehit'] | 5.69E+16 | -1.94E+16 |
| 2015-10-14 | 6:00:00 | 1.452109286 | ['Aten'] | 5.53E+16 | -1.89E+16 |
| 2015-10-14 | 10:00:00 | 1.453116415 | ['Hatmehit'] | 2.70E+16 | -9.24065E+15 |
| 2015-10-14 | 10:30:00 | 1.453241961 | ['Maat'] | 3.77E+16 | -1.29E+16 |
| 2015-10-14 | 11:00:00 | 1.453364528 | ['Seth'] | 4.58E+16 | -1.56E+16 |
| 2015-10-14 | 12:00:00 | 1.453616667 | ['Seth'] | 1.53E+17 | -5.25E+16 |
| 2015-10-14 | 14:00:00 | 1.454117547 | ['Ash'] | 2.79E+16 | -9.55899E+15 |
| 2015-10-14 | 14:30:00 | 1.454240962 | ['Ash'] | 1.65E+16 | -5.65204E+15 |
| 2015-10-14 | 15:00:00 | 1.454366604 | ['Ash'] | 3.18E+16 | -1.09E+16 |
| 2015-10-14 | 15:30:00 | 1.454491321 | ['Ash'] | 9.65E+16 | -3.29E+16 |
| 2015-10-14 | 16:30:00 | 1.45474 | ['Ash'] | 1.22E+17 | -4.16E+16 |
| 2015-10-14 | 17:00:00 | 1.454867857 | ['Ash'] | 1.04E+17 | -3.54E+16 |
| 2015-10-14 | 17:30:00 | 1.454996038 | ['Aten'] | 7.90E+16 | -2.70E+16 |
| 2015-10-14 | 18:00:00 | 1.455120962 | ['Aten'] | 6.58E+16 | -2.25E+16 |
| 2015-10-14 | 20:30:00 | 1.455743208 | ['Hapi'] | 2.78E+16 | -9.48356E+15 |
| 2015-10-14 | 21:00:00 | 1.455874615 | ['Bastet'] | 2.36E+16 | -8.09058E+15 |
| 2015-10-14 | 21:30:00 | 1.455999057 | ['Maat'] | 3.13E+16 | -1.07E+16 |
| 2015-10-14 | 22:00:00 | 1.4561225 | ['Maat'] | 2.27E+16 | -7.76777E+15 |
| 2015-10-14 | 23:30:00 | 1.456496792 | ['Seth'] | 4.85E+16 | -1.66E+16 |
| 2015-10-15 | 0:00:00 | 1.456619216 | ['Seth'] | 1.11E+17 | -3.78E+16 |
| 2015-10-15 | 6:30:00 | 1.458259808 | ['Ash'] | 5.58E+16 | -1.91E+16 |
| 2015-10-15 | 7:00:00 | 1.4583475 | ['Aten'] | 5.73E+16 | -1.96E+16 |
| 2015-10-15 | 11:00:00 | 1.459413889 | ['Aten'] | 2.90E+16 | -9.91666E+15 |
| 2015-10-15 | 11:30:00 | 1.459487143 | ['Nut'] | 3.75E+16 | -1.28E+16 |
| 2015-10-15 | 16:00:00 | 1.460677353 | ['Nut'] | 1.08E+17 | -3.70E+16 |
| 2015-10-15 | 16:30:00 | 1.46078 | ['Imhotep'] | 1.12E+17 | -3.82E+16 |

| Date | Time | Value | Name | Col5 | Col6 |
|---|---|---|---|---|---|
| 2015-10-15 | 17:00:00 | 1.460906038 | ['Imhotep'] | 9.89E+16 | -3.38E+16 |
| 2015-10-16 | 3:00:00 | 1.463440714 | ['Hathor'] | 1.43E+16 | -4.89926E+15 |
| 2015-10-16 | 7:30:00 | 1.464585556 | ['Hathor'] | 4.95E+16 | -1.69E+16 |
| 2015-10-16 | 13:30:00 | 1.46609619 | ['Hathor'] | 7.95E+16 | -2.72E+16 |
| 2015-10-16 | 17:30:00 | 1.467157143 | ['Hatmehit'] | 7.07E+16 | -2.41E+16 |
| 2015-10-16 | 18:00:00 | 1.467237436 | ['Ash'] | 7.24E+16 | -2.47E+16 |
| 2015-10-17 | 5:30:00 | 1.47024375 | ['Ash'] | 6.83E+16 | -2.33E+16 |
| 2015-10-17 | 6:00:00 | 1.470318043 | ['Imhotep'] | 7.23E+16 | -2.47E+16 |
| 2015-10-17 | 9:00:00 | 1.471152857 | ['Imhotep'] | 2.93E+16 | -1.01E+16 |
| 2015-10-17 | 9:30:00 | 1.471219423 | ['Hapi'] | 2.20E+16 | -7.50229E+15 |
| 2015-10-17 | 10:00:00 | 1.471352679 | ['Hapi'] | 1.56E+16 | -5.31316E+15 |
| 2015-10-17 | 10:30:00 | 1.471440714 | ['Maat'] | 1.62E+16 | -5.55673E+15 |
| 2015-10-17 | 19:30:00 | 1.473795577 | ['Maat'] | 5.59E+16 | -1.91E+16 |
| 2015-10-17 | 20:00:00 | 1.473922453 | ['Aten'] | 4.25E+16 | -1.45E+16 |
| 2015-10-18 | 6:00:00 | 1.476500385 | ['Imhotep'] | 8.07E+16 | -2.75E+16 |
| 2015-10-18 | 6:30:00 | 1.476601379 | ['Imhotep'] | 7.14E+16 | -2.44E+16 |
| 2015-10-19 | 1:30:00 | 1.481578286 | ['Imhotep'] | 4.68E+16 | -1.60E+16 |
| 2015-10-19 | 2:00:00 | 1.481687736 | ['Geb'] | 8.85E+16 | -3.02E+16 |
| 2015-10-19 | 2:30:00 | 1.481805122 | ['Seth'] | 6.65E+16 | -2.27E+16 |
| 2015-10-24 | 9:30:00 | 1.515551538 | ['Imhotep1'] | 4.38E+16 | -1.50E+16 |
| 2015-10-24 | 10:00:00 | 1.515690877 | ['Imhotep'] | 5.65E+16 | -1.93E+16 |
| 2015-10-24 | 10:30:00 | 1.515783077 | ['Imhotep'] | 5.95E+16 | -2.03E+16 |
| 2015-10-24 | 14:30:00 | 1.516910377 | ['Hapi'] | 2.67E+16 | -9.10604E+15 |
| 2015-10-24 | 19:30:00 | 1.518292 | ['Hapi'] | 3.14E+16 | -1.07E+16 |
| 2015-10-24 | 20:00:00 | 1.518402642 | ['Anubis'] | 1.81E+16 | -6.17844E+15 |
| 2015-10-24 | 20:30:00 | 1.518499583 | ['Atum'] | 1.54E+16 | -5.25114E+15 |
| 2015-10-25 | 0:30:00 | 1.5196274 | ['Atum'] | 4.55E+16 | -1.55E+16 |
| 2015-10-25 | 5:30:00 | 1.520988627 | ['Atum'] | 2.66E+16 | -9.07069E+15 |
| 2015-10-25 | 10:30:00 | 1.522347885 | ['Imhotep1'] | 7.27E+16 | -2.48E+16 |
| 2015-10-25 | 11:00:00 | 1.522429091 | ['Imhotep'] | 8.01E+16 | -2.73E+16 |
| 2015-10-25 | 12:00:00 | 1.522749792 | ['Imhotep'] | 7.70E+16 | -2.63E+16 |
| 2015-10-25 | 15:30:00 | 1.523715957 | ['Bastet'] | 3.46E+16 | -1.18E+16 |
| 2015-10-26 | 1:30:00 | 1.52644566 | ['Bastet'] | 3.83E+16 | -1.31E+16 |
| 2015-10-26 | 2:00:00 | 1.526581346 | ['Khepry'] | 5.12E+16 | -1.75E+16 |
| 2015-10-26 | 2:30:00 | 1.5267015 | ['Khepry'] | 7.22E+16 | -2.47E+16 |
| 2015-10-26 | 6:00:00 | 1.52775 | ['Hatmehit'] | 4.60E+16 | -1.57E+16 |
| 2015-10-26 | 6:30:00 | 1.52781463 | ['Neith'] | 5.58E+16 | -1.91E+16 |
| 2015-10-26 | 11:30:00 | 1.529207895 | ['Neith'] | 1.07E+17 | -3.66E+16 |
| 2015-10-26 | 12:00:00 | 1.529329423 | ['Imhotep'] | 9.90E+16 | -3.38E+16 |
| 2015-10-26 | 12:30:00 | 1.529464906 | ['Imhotep'] | 8.63E+16 | -2.95E+16 |
| 2015-10-26 | 16:00:00 | 1.530429811 | ['Geb'] | 7.71E+16 | -2.63E+16 |
| 2015-10-26 | 16:30:00 | 1.530565769 | ['Anhur'] | 6.58E+16 | -2.25E+16 |
| 2015-10-26 | 17:00:00 | 1.5307014 | ['Anhur'] | 7.47E+16 | -2.55E+16 |
| 2015-10-26 | 17:30:00 | 1.53084 | ['Anhur'] | 8.25E+16 | -2.82E+16 |

| Date | Time | Value | Label | Col5 | Col6 |
|---|---|---|---|---|---|
| 2015-10-26 | 18:00:00 | 1.5309754 | ['Anhur'] | 7.27E+16 | -2.48E+16 |
| 2015-10-26 | 18:30:00 | 1.531108824 | ['Anhur'] | 6.60E+16 | -2.25E+16 |
| 2015-10-26 | 19:00:00 | 1.531247551 | ['Anhur'] | 7.61E+16 | -2.60E+16 |
| 2015-10-26 | 19:30:00 | 1.531383585 | ['Anhur'] | 8.44E+16 | -2.88E+16 |
| 2015-10-26 | 21:30:00 | 1.531938491 | ['Atum'] | 5.50E+16 | -1.88E+16 |
| 2015-10-26 | 22:00:00 | 1.532073725 | ['Khonsu'] | 6.70E+16 | -2.29E+16 |
| 2015-10-26 | 22:30:00 | 1.532215106 | ['Khonsu'] | 8.19E+16 | -2.80E+16 |
| 2015-10-26 | 23:00:00 | 1.532349167 | ['Imhotep1'] | 8.96E+16 | -3.06E+16 |
| 2015-10-26 | 23:30:00 | 1.532484717 | ['Imhotep1'] | 1.08E+17 | -3.69E+16 |
| 2015-10-27 | 0:00:00 | 1.532624727 | ['Imhotep'] | 1.09E+17 | -3.72E+16 |
| 2015-10-27 | 0:30:00 | 1.532766667 | ['Imhotep'] | 1.05E+17 | -3.58E+16 |
| 2015-10-27 | 1:00:00 | 1.532905472 | ['Imhotep'] | 9.13E+16 | -3.12E+16 |
| 2015-10-27 | 1:30:00 | 1.533039574 | ['Bes'] | 7.08E+16 | -2.42E+16 |
| 2015-10-27 | 2:00:00 | 1.533180222 | ['Bes'] | 6.09E+16 | -2.08E+16 |
| 2015-10-27 | 2:30:00 | 1.533318511 | ['Bes'] | 6.64E+16 | -2.27E+16 |
| 2015-10-28 | 5:30:00 | 1.540825294 | ['Bes'] | 6.37E+16 | -2.18E+16 |
| 2015-10-28 | 6:00:00 | 1.540911282 | ['Sobek'] | 8.12E+16 | -2.77E+16 |
| 2015-10-28 | 9:30:00 | 1.541898913 | ['Sobek'] | 7.64E+16 | -2.61E+16 |
| 2015-10-28 | 10:00:00 | 1.542028868 | ['Atum'] | 7.61E+16 | -2.60E+16 |
| 2015-10-28 | 10:30:00 | 1.542153659 | ['Khonsu'] | 7.48E+16 | -2.55E+16 |
| 2015-10-28 | 11:30:00 | 1.542454231 | ['Khonsu'] | 6.82E+16 | -2.33E+16 |
| 2015-10-28 | 12:00:00 | 1.542590417 | ['Khonsu'] | 9.94E+16 | -3.39E+16 |
| 2015-10-28 | 13:30:00 | 1.543006364 | ['Imhotep'] | 9.73E+16 | -3.32E+16 |
| 2015-10-28 | 14:00:00 | 1.543142791 | ['Imhotep'] | 5.86E+16 | -2.00E+16 |
| 2015-10-28 | 14:30:00 | 1.543285 | ['Imhotep'] | 5.74E+16 | -1.96E+16 |
| 2015-10-28 | 15:00:00 | 1.543418372 | ['Bes'] | 5.80E+16 | -1.98E+16 |
| 2015-10-28 | 16:00:00 | 1.54370125 | ['Khepry'] | 8.07E+16 | -2.76E+16 |
| 2015-10-28 | 17:00:00 | 1.543976415 | ['Khepry'] | 1.11E+17 | -3.78E+16 |
| 2015-10-28 | 17:30:00 | 1.544117407 | ['Babi'] | 7.86E+16 | -2.68E+16 |
| 2015-10-28 | 18:30:00 | 1.5443925 | ['Babi'] | 8.52E+16 | -2.91E+16 |
| 2015-10-28 | 19:00:00 | 1.5445375 | ['Wosret'] | 9.15E+16 | -3.13E+16 |
| 2015-10-28 | 19:30:00 | 1.544679423 | ['Wosret'] | 9.81E+16 | -3.35E+16 |
| 2015-10-28 | 20:00:00 | 1.54481717 | ['Geb'] | 8.06E+16 | -2.75E+16 |
| 2015-10-28 | 20:30:00 | 1.544955849 | ['Geb'] | 1.06E+17 | -3.61E+16 |
| 2015-10-28 | 21:30:00 | 1.545234146 | ['Geb'] | 9.51E+16 | -3.25E+16 |
| 2015-10-28 | 23:30:00 | 1.545792075 | ['Bes'] | 8.15E+16 | -2.78E+16 |
| 2015-10-29 | 0:00:00 | 1.545929808 | ['Imhotep'] | 8.83E+16 | -3.02E+16 |
| 2015-10-29 | 0:30:00 | 1.54607 | ['Imhotep'] | 1.19E+17 | -4.05E+16 |
| 2015-10-29 | 1:00:00 | 1.546205192 | ['Imhotep'] | 1.08E+17 | -3.69E+16 |
| 2015-10-29 | 1:30:00 | 1.546285714 | ['Bes'] | 1.05E+17 | -3.59E+16 |
| 2015-10-29 | 6:00:00 | 1.547603725 | ['Bes'] | 1.26E+17 | -4.29E+16 |
| 2015-10-29 | 6:30:00 | 1.547743455 | ['Anhur'] | 1.08E+17 | -3.70E+16 |
| 2015-10-29 | 7:00:00 | 1.547855 | ['Anhur'] | 1.02E+17 | -3.50E+16 |
| 2015-10-29 | 11:00:00 | 1.549017674 | ['Anhur'] | 1.00E+17 | -3.42E+16 |

| Date | Time | Value | Region | Col5 | Col6 |
|---|---|---|---|---|---|
| 2015-10-29 | 11:30:00 | 1.549111923 | ['Khonsu'] | 9.60E+16 | -3.28E+16 |
| 2015-10-29 | 16:00:00 | 1.550432667 | ['Khonsu'] | 1.02E+17 | -3.48E+16 |
| 2015-10-29 | 16:30:00 | 1.550542115 | ['Bes'] | 1.24E+17 | -4.22E+16 |
| 2015-10-29 | 17:00:00 | 1.550680816 | ['Anhur'] | 1.29E+17 | -4.42E+16 |
| 2015-10-29 | 21:00:00 | 1.55184381 | ['Anhur'] | 1.14E+17 | -3.90E+16 |
| 2015-10-29 | 21:30:00 | 1.551941346 | ['Geb'] | 1.10E+17 | -3.75E+16 |
| 2015-10-29 | 22:00:00 | 1.55208 | ['Sobek'] | 1.06E+17 | -3.61E+16 |
| 2015-10-29 | 22:30:00 | 1.552219038 | ['Sobek'] | 1.07E+17 | -3.65E+16 |
| 2015-10-29 | 23:00:00 | 1.552357885 | ['Sobek'] | 9.64E+16 | -3.29E+16 |
| 2015-10-29 | 23:30:00 | 1.55249902 | ['Hapi'] | 9.10E+16 | -3.11E+16 |
| 2015-10-30 | 0:00:00 | 1.55264 | ['Atum'] | 9.00E+16 | -3.07E+16 |
| 2015-10-30 | 0:30:00 | 1.552783774 | ['Khonsu'] | 1.05E+17 | -3.58E+16 |
| 2015-10-30 | 1:00:00 | 1.552924906 | ['Khonsu'] | 1.16E+17 | -3.97E+16 |
| 2015-10-30 | 1:30:00 | 1.55306434 | ['Imhotep1'] | 1.14E+17 | -3.91E+16 |
| 2015-10-30 | 2:00:00 | 1.553203077 | ['Imhotep1'] | 1.20E+17 | -4.10E+16 |
| 2015-10-30 | 2:30:00 | 1.553340755 | ['Imhotep'] | 1.10E+17 | -3.76E+16 |
| 2015-10-30 | 3:00:00 | 1.553481132 | ['Imhotep'] | 9.14E+16 | -3.12E+16 |
| 2015-10-30 | 3:30:00 | 1.553619808 | ['Bes'] | 8.29E+16 | -2.83E+16 |
| 2015-10-30 | 4:30:00 | 1.5539025 | ['Bes'] | 9.28E+16 | -3.17E+16 |
| 2015-10-30 | 5:00:00 | 1.554048302 | ['Bes'] | 1.09E+17 | -3.71E+16 |
| 2015-10-30 | 5:30:00 | 1.5541875 | ['Anhur'] | 1.23E+17 | -4.21E+16 |
| 2015-10-30 | 6:00:00 | 1.554326154 | ['Anhur'] | 1.48E+17 | -5.04E+16 |
| 2015-10-30 | 6:30:00 | 1.554465577 | ['Anhur'] | 1.36E+17 | -4.65E+16 |
| 2015-10-30 | 7:00:00 | 1.554604423 | ['Anhur'] | 1.26E+17 | -4.31E+16 |
| 2015-10-30 | 7:30:00 | 1.554745192 | ['Anhur'] | 1.23E+17 | -4.20E+16 |
| 2015-10-30 | 8:00:00 | 1.554838824 | ['Anhur'] | 1.15E+17 | -3.93E+16 |
| 2015-10-30 | 12:30:00 | 1.556172571 | ['Anhur'] | 1.00E+17 | -3.41E+16 |
| 2015-10-30 | 13:00:00 | 1.556290943 | ['Khonsu'] | 1.11E+17 | -3.80E+16 |
| 2015-10-30 | 17:30:00 | 1.557601304 | ['Hatmehit'] | 1.13E+17 | -3.87E+16 |
| 2015-10-30 | 18:00:00 | 1.557678788 | ['Anhur'] | 1.24E+17 | -4.25E+16 |
| 2015-10-30 | 22:30:00 | 1.559024 | ['Anhur'] | 1.00E+17 | -3.42E+16 |
| 2015-10-30 | 23:00:00 | 1.55911098 | ['Geb'] | 8.62E+16 | -2.94E+16 |
| 2015-10-31 | 0:00:00 | 1.559350455 | ['Wosret'] | 1.31E+17 | -4.49E+16 |
| 2015-10-31 | 5:30:00 | 1.560987368 | ['Hatmehit'] | 1.18E+17 | -4.02E+16 |
| 2015-10-31 | 6:00:00 | 1.561075208 | ['Geb'] | 1.14E+17 | -3.90E+16 |
| 2015-10-31 | 9:00:00 | 1.5619875 | ['Hatmehit'] | 1.30E+17 | -4.43E+16 |
| 2015-10-31 | 9:30:00 | 1.562071346 | ['Geb'] | 1.35E+17 | -4.59E+16 |
| 2015-10-31 | 10:00:00 | 1.562209811 | ['Geb'] | 1.04E+17 | -3.55E+16 |
| 2015-10-31 | 10:30:00 | 1.562352593 | ['Geb'] | 8.71E+16 | -2.97E+16 |
| 2015-10-31 | 11:00:00 | 1.562498727 | ['Atum'] | 8.33E+16 | -2.84E+16 |
| 2015-10-31 | 11:30:00 | 1.56264283 | ['Atum'] | 8.08E+16 | -2.76E+16 |
| 2015-10-31 | 12:00:00 | 1.562724545 | ['Khonsu'] | 8.26E+16 | -2.82E+16 |
| 2015-10-31 | 13:00:00 | 1.5631 | ['Imhotep1'] | 1.13E+17 | -3.85E+16 |
| 2015-10-31 | 13:30:00 | 1.563202453 | ['Imhotep'] | 1.13E+17 | -3.87E+16 |

| Date | Time | Value | Label | Col5 | Col6 |
|---|---|---|---|---|---|
| 2015-10-31 | 14:00:00 | 1.563343585 | ['Imhotep'] | 1.15E+17 | -3.91E+16 |
| 2015-10-31 | 14:30:00 | 1.563483269 | ['Bes'] | 9.31E+16 | -3.18E+16 |
| 2015-10-31 | 15:00:00 | 1.5636275 | ['Bes'] | 1.02E+17 | -3.47E+16 |
| 2015-10-31 | 15:30:00 | 1.563773774 | ['Bes'] | 9.33E+16 | -3.19E+16 |
| 2015-10-31 | 17:00:00 | 1.564195962 | ['Bes'] | 8.46E+16 | -2.89E+16 |
| 2015-10-31 | 17:30:00 | 1.564301379 | ['Anhur'] | 1.00E+17 | -3.41E+16 |
| 2015-10-31 | 19:00:00 | 1.564756863 | ['Anhur'] | 1.43E+17 | -4.88E+16 |
| 2015-10-31 | 19:30:00 | 1.564903684 | ['Geb'] | 1.48E+17 | -5.04E+16 |
| 2015-10-31 | 20:30:00 | 1.565189623 | ['Geb'] | 1.74E+17 | -5.95E+16 |
| 2015-10-31 | 21:00:00 | 1.565330755 | ['Sobek'] | 1.72E+17 | -5.87E+16 |
| 2015-10-31 | 21:30:00 | 1.565470577 | ['Sobek'] | 1.48E+17 | -5.06E+16 |
| 2015-10-31 | 22:00:00 | 1.565610377 | ['Geb'] | 1.26E+17 | -4.29E+16 |
| 2015-10-31 | 22:30:00 | 1.565751509 | ['Geb'] | 1.00E+17 | -3.42E+16 |
| 2015-10-31 | 23:00:00 | 1.565894902 | ['Atum'] | 9.11E+16 | -3.11E+16 |
| 2015-10-31 | 23:30:00 | 1.5660325 | ['Atum'] | 8.84E+16 | -3.02E+16 |
| 2015-11-11 | 14:30:00 | 1.640238667 | ['Atum'] | 2.95E+16 | -1.01E+16 |
| 2015-11-11 | 15:00:00 | 1.640325806 | ['Aten'] | 2.84E+16 | -9.69788E+15 |
| 2015-11-11 | 20:30:00 | 1.641995714 | ['Hathor'] | 3.25E+16 | -1.11E+16 |
| 2015-11-12 | 1:00:00 | 1.643333654 | ['Imhotep'] | 3.33E+16 | -1.14E+16 |
| 2015-11-12 | 6:00:00 | 1.644830769 | ['Hapi'] | 1.07E+16 | -3.65004E+15 |
| 2015-11-12 | 6:30:00 | 1.64498 | ['Maat'] | 1.22E+16 | -4.16475E+15 |
| 2015-11-12 | 11:00:00 | 1.646341892 | ['Hatmehit'] | 1.22E+16 | -4.16546E+15 |
| 2015-11-12 | 16:00:00 | 1.647848077 | ['Hatmehit'] | 2.29E+16 | -7.83448E+15 |
| 2015-11-12 | 16:30:00 | 1.647961458 | ['Aten'] | 2.40E+16 | -8.19078E+15 |
| 2015-11-12 | 17:00:00 | 1.648107358 | ['Babi'] | 2.46E+16 | -8.39144E+15 |
| 2015-11-12 | 21:00:00 | 1.649358125 | ['Babi'] | 1.39E+16 | -4.77373E+15 |
| 2015-11-12 | 21:30:00 | 1.649451509 | ['Serqet'] | 2.70E+16 | -9.23935E+15 |
| 2015-11-12 | 22:00:00 | 1.649599038 | ['Anuket'] | 3.82E+16 | -1.31E+16 |
| 2015-11-12 | 22:30:00 | 1.64975125 | ['Anuket'] | 3.98E+16 | -1.36E+16 |
| 2015-11-12 | 23:00:00 | 1.649905283 | ['Hapi'] | 2.26E+16 | -7.73975E+15 |
| 2015-11-13 | 1:00:00 | 1.650500784 | ['Ash'] | 1.60E+16 | -5.4621E+15 |
| 2015-11-13 | 1:30:00 | 1.650647547 | ['Ash'] | 2.30E+16 | -7.84914E+15 |
| 2015-11-13 | 2:00:00 | 1.650796415 | ['Imhotep'] | 3.10E+16 | -1.06E+16 |
| 2015-11-13 | 2:30:00 | 1.650945472 | ['Imhotep'] | 3.94E+16 | -1.35E+16 |
| 2015-11-13 | 3:00:00 | 1.651099286 | ['Imhotep'] | 4.12E+16 | -1.41E+16 |
| 2015-11-13 | 3:30:00 | 1.651253208 | ['Imhotep'] | 3.56E+16 | -1.22E+16 |
| 2015-11-13 | 4:00:00 | 1.651401346 | ['Imhotep'] | 2.93E+16 | -9.99829E+15 |
| 2015-11-13 | 4:30:00 | 1.651548679 | ['Ash'] | 2.41E+16 | -8.2362E+15 |
| 2015-11-13 | 5:00:00 | 1.651697736 | ['Aten'] | 2.20E+16 | -7.52761E+15 |
| 2015-11-13 | 6:00:00 | 1.651993208 | ['Babi'] | 2.71E+16 | -9.27315E+15 |
| 2015-11-13 | 6:30:00 | 1.652142264 | ['Babi'] | 1.80E+16 | -6.15998E+15 |
| 2015-11-13 | 7:00:00 | 1.652291509 | ['Hapi'] | 1.02E+16 | -3.49163E+15 |
| 2015-11-13 | 7:30:00 | 1.652445273 | ['Maat'] | 9.63E+15 | -3.2925E+15 |
| 2015-11-13 | 8:00:00 | 1.652599434 | ['Hatmehit'] | 1.04E+16 | -3.5414E+15 |

| Date | Time | Value | Region | Col5 | Col6 |
|---|---|---|---|---|---|
| 2015-11-13 | 9:00:00 | 1.652894717 | ['Maat'] | 8.03E+15 | -2.74786E+15 |
| 2015-11-13 | 9:30:00 | 1.653043962 | ['Nut'] | 1.29E+16 | -4.39277E+15 |
| 2015-11-13 | 10:00:00 | 1.653193019 | ['Anuket'] | 2.44E+16 | -8.33967E+15 |
| 2015-11-13 | 10:30:00 | 1.653340962 | ['Anuket'] | 3.55E+16 | -1.21E+16 |
| 2015-11-13 | 11:00:00 | 1.653488679 | ['Anuket'] | 3.61E+16 | -1.23E+16 |
| 2015-11-14 | 9:30:00 | 1.6602425 | ['Nut'] | 1.04E+16 | -3.56166E+15 |
| 2015-11-14 | 10:00:00 | 1.660390755 | ['Anuket'] | 1.38E+16 | -4.7165E+15 |
| 2015-11-14 | 14:30:00 | 1.661745472 | ['Imhotep'] | 3.05E+16 | -1.04E+16 |
| 2015-11-14 | 20:30:00 | 1.663517586 | ['Hatmehit'] | 1.42E+16 | -4.84593E+15 |
| 2015-11-15 | 0:30:00 | 1.664750962 | ['Hatmehit'] | 3.18E+16 | -1.09E+16 |
| 2015-11-15 | 5:30:00 | 1.666263774 | ['Hatmehit'] | 3.66E+16 | -1.25E+16 |
| 2015-11-15 | 6:00:00 | 1.666412308 | ['Babi'] | 3.11E+16 | -1.06E+16 |
| 2015-11-15 | 6:30:00 | 1.666523571 | ['Aker'] | 2.79E+16 | -9.51415E+15 |
| 2015-11-15 | 10:30:00 | 1.667772955 | ['Aker'] | 1.86E+16 | -6.3427E+15 |
| 2015-11-15 | 11:00:00 | 1.66786 | ['Anuket'] | 2.44E+16 | -8.34596E+15 |
| 2015-11-15 | 15:30:00 | 1.669289767 | ['Hatmehit'] | 5.34E+16 | -1.82E+16 |
| 2015-11-15 | 16:00:00 | 1.669424906 | ['Imhotep'] | 5.88E+16 | -2.01E+16 |
| 2015-11-15 | 16:30:00 | 1.669551351 | ['Imhotep'] | 6.59E+16 | -2.25E+16 |
| 2015-11-15 | 20:30:00 | 1.67079775 | ['Imhotep'] | 2.09E+16 | -7.13543E+15 |
| 2015-11-16 | 1:30:00 | 1.672307561 | ['Imhotep'] | 1.91E+16 | -6.50737E+15 |
| 2015-11-16 | 2:00:00 | 1.6724475 | ['Khonsu'] | 1.52E+16 | -5.20331E+15 |
| 2015-11-16 | 2:30:00 | 1.672580857 | ['Apis'] | 1.95E+16 | -6.64768E+15 |
| 2015-11-16 | 11:30:00 | 1.675348788 | ['Hatmehit'] | 4.47E+16 | -1.53E+16 |
| 2015-11-16 | 12:00:00 | 1.675472075 | ['Hapi'] | 6.20E+16 | -2.12E+16 |
| 2015-11-16 | 12:30:00 | 1.675616087 | ['Seth'] | 6.65E+16 | -2.27E+16 |
| 2015-11-16 | 16:30:00 | 1.676865161 | ['Seth'] | 7.63E+16 | -2.61E+16 |
| 2015-11-16 | 17:00:00 | 1.676947667 | ['Imhotep'] | 7.53E+16 | -2.57E+16 |
| 2015-11-16 | 21:30:00 | 1.678381379 | ['Imhotep'] | 3.14E+16 | -1.07E+16 |
| 2015-11-16 | 22:00:00 | 1.678499811 | ['Wosret'] | 3.82E+16 | -1.30E+16 |
| 2015-11-16 | 22:30:00 | 1.678650755 | ['Wosret'] | 3.48E+16 | -1.19E+16 |
| 2015-11-17 | 2:30:00 | 1.679898077 | ['Wosret'] | 3.62E+16 | -1.24E+16 |
| 2015-11-17 | 3:00:00 | 1.679993171 | ['Khonsu'] | 3.73E+16 | -1.27E+16 |
| 2015-11-17 | 7:30:00 | 1.681406452 | ['Khonsu'] | 5.20E+16 | -1.78E+16 |
| 2015-11-17 | 8:00:00 | 1.681526038 | ['Anhur'] | 7.21E+16 | -2.46E+16 |
| 2015-11-17 | 8:30:00 | 1.681675577 | ['Anhur'] | 6.96E+16 | -2.38E+16 |
| 2015-11-17 | 9:00:00 | 1.681769 | ['Sobek'] | 5.57E+16 | -1.90E+16 |
| 2015-11-17 | 12:00:00 | 1.6827575 | ['Sobek'] | 5.97E+16 | -2.04E+16 |
| 2015-11-17 | 12:30:00 | 1.682891346 | ['Seth'] | 7.46E+16 | -2.55E+16 |
| 2015-11-17 | 13:00:00 | 1.683029545 | ['Anubis'] | 7.78E+16 | -2.66E+16 |
| 2015-11-17 | 17:30:00 | 1.684465714 | ['Hatmehit'] | 8.48E+16 | -2.90E+16 |
| 2015-11-17 | 18:00:00 | 1.684567222 | ['Imhotep'] | 6.97E+16 | -2.38E+16 |
| 2015-11-17 | 18:30:00 | 1.684719811 | ['Bes'] | 4.84E+16 | -1.65E+16 |
| 2015-11-17 | 19:00:00 | 1.6848025 | ['Bes'] | 3.55E+16 | -1.21E+16 |
| 2015-11-17 | 22:30:00 | 1.685990833 | ['Hatmehit'] | 6.03E+16 | -2.06E+16 |
| 2015-11-17 | 23:00:00 | 1.686086415 | ['Geb'] | 6.34E+16 | -2.16E+16 |
| 2015-11-17 | 23:30:00 | 1.686236538 | ['Geb'] | 5.90E+16 | -2.02E+16 |
| 2015-11-18 | 0:00:00 | 1.686347407 | ['Geb'] | 4.13E+16 | -1.41E+16 |
| 2015-11-18 | 5:30:00 | 1.688086875 | ['Geb'] | 6.89E+16 | -2.35E+16 |
| 2015-11-18 | 6:00:00 | 1.688198043 | ['Imhotep'] | 7.87E+16 | -2.69E+16 |
| 2015-11-18 | 7:00:00 | 1.688523529 | ['Imhotep'] | 4.27E+16 | -1.46E+16 |
| 2015-11-18 | 7:30:00 | 1.688670577 | ['Bes'] | 3.87E+16 | -1.32E+16 |
| 2015-11-18 | 8:00:00 | 1.688820566 | ['Bes'] | 5.00E+16 | -1.71E+16 |
| 2015-11-18 | 8:30:00 | 1.688971154 | ['Sobek'] | 7.46E+16 | -2.55E+16 |
| 2015-11-18 | 9:00:00 | 1.689121961 | ['Anhur'] | 7.63E+16 | -2.61E+16 |
| 2015-11-18 | 9:30:00 | 1.689272453 | ['Wosret'] | 8.69E+16 | -2.97E+16 |
| 2015-11-18 | 10:00:00 | 1.689424151 | ['Wosret'] | 5.95E+16 | -2.03E+16 |
| 2015-11-18 | 10:30:00 | 1.689567872 | ['Wosret'] | 4.76E+16 | -1.63E+16 |
| 2015-11-18 | 14:30:00 | 1.69081375 | ['Wosret'] | 5.20E+16 | -1.78E+16 |
| 2015-11-18 | 15:00:00 | 1.690929 | ['Khonsu'] | 5.03E+16 | -1.72E+16 |
| 2015-11-18 | 19:30:00 | 1.692354444 | ['Khonsu'] | 5.27E+16 | -1.80E+16 |
| 2015-11-18 | 20:00:00 | 1.692469245 | ['Anhur'] | 6.30E+16 | -2.15E+16 |
| 2015-11-18 | 20:30:00 | 1.692625536 | ['Anhur'] | 8.05E+16 | -2.75E+16 |
| 2015-11-19 | 0:30:00 | 1.693899333 | ['Anhur'] | 4.30E+16 | -1.47E+16 |
| 2015-11-19 | 1:00:00 | 1.6940008 | ['Geb'] | 6.05E+16 | -2.07E+16 |
| 2015-11-19 | 7:00:00 | 1.695795667 | ['Bes'] | 8.96E+16 | -3.06E+16 |
| 2015-11-19 | 11:00:00 | 1.697068 | ['Bes'] | 9.86E+16 | -3.37E+16 |
| 2015-11-19 | 11:30:00 | 1.697155833 | ['Geb'] | 7.88E+16 | -2.69E+16 |
| 2015-11-19 | 16:00:00 | 1.698612963 | ['Hatmehit'] | 8.24E+16 | -2.81E+16 |
| 2015-11-19 | 16:30:00 | 1.698725283 | ['Imhotep'] | 7.76E+16 | -2.65E+16 |
| 2015-11-19 | 17:30:00 | 1.699023878 | ['Bes'] | 8.02E+16 | -2.74E+16 |
| 2015-11-19 | 19:30:00 | 1.699642745 | ['Neith'] | 8.12E+16 | -2.77E+16 |
| 2015-11-19 | 20:30:00 | 1.699947255 | ['Wosret'] | 6.86E+16 | -2.34E+16 |
| 2015-11-20 | 2:30:00 | 1.70178537 | ['Bes'] | 6.41E+16 | -2.19E+16 |
| 2015-11-20 | 3:00:00 | 1.701939434 | ['Bes'] | 5.72E+16 | -1.95E+16 |
| 2015-11-20 | 3:30:00 | 1.702091509 | ['Bes'] | 6.21E+16 | -2.12E+16 |
| 2015-11-20 | 7:30:00 | 1.70332375 | ['Bes'] | 7.91E+16 | -2.70E+16 |
| 2015-11-20 | 8:30:00 | 1.70361902 | ['Bes'] | 6.21E+16 | -2.12E+16 |
| 2015-11-20 | 12:30:00 | 1.704843654 | ['Bes'] | 8.80E+16 | -3.01E+16 |
| 2015-11-20 | 13:00:00 | 1.704994808 | ['Geb'] | 9.20E+16 | -3.14E+16 |
| 2015-11-20 | 13:30:00 | 1.705132727 | ['Bes'] | 8.13E+16 | -2.78E+16 |
| 2015-11-20 | 17:30:00 | 1.7064175 | ['Bes'] | 6.41E+16 | -2.19E+16 |
| 2015-11-20 | 18:00:00 | 1.706501892 | ['Bes'] | 6.85E+16 | -2.34E+16 |
| 2015-11-20 | 23:30:00 | 1.708211429 | ['Bes'] | 9.59E+16 | -3.27E+16 |
| 2015-11-21 | 0:00:00 | 1.708333077 | ['Imhotep'] | 1.04E+17 | -3.55E+16 |
| 2015-11-21 | 5:30:00 | 1.710092963 | ['Imhotep'] | 6.74E+16 | -2.30E+16 |
| 2015-11-21 | 6:00:00 | 1.710207358 | ['Bes'] | 6.94E+16 | -2.37E+16 |
| 2015-11-21 | 9:30:00 | 1.711279623 | ['Anhur'] | 3.98E+16 | -1.36E+16 |
| 2015-11-21 | 10:00:00 | 1.71143283 | ['Anhur'] | 3.90E+16 | -1.33E+16 |

| Date | Time | Value | Label | Col5 | Col6 |
|---|---|---|---|---|---|
| 2015-11-21 | 14:30:00 | 1.712814615 | ['Geb'] | 9.10E+16 | -3.11E+16 |
| 2015-11-21 | 19:30:00 | 1.714350943 | ['Bes'] | 7.70E+16 | -2.63E+16 |
| 2015-11-21 | 20:00:00 | 1.714504528 | ['Bes'] | 7.84E+16 | -2.68E+16 |
| 2015-11-21 | 20:30:00 | 1.714615769 | ['Bes'] | 8.01E+16 | -2.74E+16 |
| 2015-11-22 | 0:30:00 | 1.715887255 | ['Bes'] | 1.02E+17 | -3.48E+16 |
| 2015-11-22 | 1:00:00 | 1.715977273 | ['Geb'] | 1.22E+17 | -4.18E+16 |
| 2015-11-22 | 5:30:00 | 1.717430408 | ['Hatmehit'] | 5.24E+16 | -1.79E+16 |
| 2015-11-22 | 6:00:00 | 1.717576346 | ['Bes'] | 4.88E+16 | -1.66E+16 |
| 2015-11-22 | 6:30:00 | 1.717698182 | ['Bes'] | 5.96E+16 | -2.03E+16 |
| 2015-11-22 | 10:30:00 | 1.71897 | ['Bes'] | 4.55E+16 | -1.55E+16 |
| 2015-11-22 | 11:00:00 | 1.719053 | ['Geb'] | 4.12E+16 | -1.41E+16 |
| 2015-11-22 | 11:30:00 | 1.719299143 | ['Geb'] | 5.35E+16 | -1.83E+16 |
| 2015-11-22 | 12:00:00 | 1.719422885 | ['Geb'] | 7.11E+16 | -2.43E+16 |
| 2015-11-22 | 12:30:00 | 1.719574717 | ['Geb'] | 8.69E+16 | -2.97E+16 |
| 2015-11-22 | 13:00:00 | 1.719727925 | ['Geb'] | 1.02E+17 | -3.48E+16 |
| 2015-11-22 | 13:30:00 | 1.719881321 | ['Geb'] | 1.03E+17 | -3.51E+16 |
| 2015-11-22 | 14:00:00 | 1.720036111 | ['Geb'] | 1.10E+17 | -3.77E+16 |
| 2015-11-22 | 14:30:00 | 1.720195741 | ['Geb'] | 1.06E+17 | -3.62E+16 |
| 2015-11-22 | 15:00:00 | 1.72035 | ['Geb'] | 9.37E+16 | -3.20E+16 |
| 2015-11-22 | 15:30:00 | 1.720502075 | ['Geb'] | 8.11E+16 | -2.77E+16 |
| 2015-11-22 | 16:00:00 | 1.720655577 | ['Geb'] | 7.84E+16 | -2.68E+16 |
| 2015-11-22 | 16:30:00 | 1.720808491 | ['Geb'] | 7.85E+16 | -2.68E+16 |
| 2015-11-22 | 17:00:00 | 1.720960769 | ['Bes'] | 9.49E+16 | -3.24E+16 |
| 2015-11-22 | 17:30:00 | 1.721113019 | ['Bes'] | 9.00E+16 | -3.07E+16 |
| 2015-11-22 | 18:00:00 | 1.721266226 | ['Bes'] | 5.68E+16 | -1.94E+16 |
| 2015-11-22 | 18:30:00 | 1.721424107 | ['Bes'] | 4.54E+16 | -1.55E+16 |
| 2015-11-22 | 19:00:00 | 1.721582642 | ['Bes'] | 5.96E+16 | -2.03E+16 |
| 2015-11-22 | 19:30:00 | 1.721735192 | ['Bes'] | 6.29E+16 | -2.15E+16 |
| 2015-11-22 | 20:00:00 | 1.72188717 | ['Bes'] | 6.60E+16 | -2.25E+16 |
| 2015-11-22 | 20:30:00 | 1.722040377 | ['Bes'] | 6.91E+16 | -2.36E+16 |
| 2015-11-22 | 21:30:00 | 1.722346154 | ['Bes'] | 7.12E+16 | -2.43E+16 |
| 2015-11-22 | 22:00:00 | 1.722498113 | ['Bes'] | 6.45E+16 | -2.20E+16 |
| 2015-11-22 | 22:30:00 | 1.722651321 | ['Bes'] | 5.10E+16 | -1.74E+16 |
| 2015-11-22 | 23:00:00 | 1.722809821 | ['Geb'] | 4.44E+16 | -1.52E+16 |
| 2015-11-22 | 23:30:00 | 1.722968491 | ['Anhur'] | 4.02E+16 | -1.37E+16 |
| 2015-11-23 | 0:00:00 | 1.723121887 | ['Anhur'] | 5.10E+16 | -1.74E+16 |
| 2015-11-23 | 0:30:00 | 1.723274038 | ['Geb'] | 7.04E+16 | -2.40E+16 |
| 2015-11-23 | 1:00:00 | 1.723426415 | ['Geb'] | 8.69E+16 | -2.97E+16 |
| 2015-11-23 | 1:30:00 | 1.723579623 | ['Geb'] | 9.60E+16 | -3.28E+16 |
| 2015-11-23 | 2:00:00 | 1.723733019 | ['Geb'] | 9.47E+16 | -3.23E+16 |
| 2015-11-23 | 2:30:00 | 1.72386878 | ['Geb'] | 9.82E+16 | -3.35E+16 |
| 2015-11-23 | 11:30:00 | 1.726678182 | ['Geb'] | 4.87E+16 | -1.66E+16 |
| 2015-11-23 | 12:00:00 | 1.726820943 | ['Sobek'] | 5.06E+16 | -1.73E+16 |
| 2015-11-23 | 12:30:00 | 1.726975094 | ['Sobek'] | 6.04E+16 | -2.06E+16 |
| 2015-11-23 | 13:00:00 | 1.727103056 | ['Sobek'] | 7.45E+16 | -2.54E+16 |
| 2015-11-23 | 16:30:00 | 1.728242333 | ['Sobek'] | 6.85E+16 | -2.34E+16 |
| 2015-11-23 | 17:00:00 | 1.728328621 | ['Atum'] | 6.44E+16 | -2.20E+16 |
| 2015-11-23 | 21:30:00 | 1.729791111 | ['Atum'] | 6.85E+16 | -2.34E+16 |
| 2015-11-23 | 22:00:00 | 1.729904038 | ['Bes'] | 6.86E+16 | -2.34E+16 |
| 2015-11-23 | 22:30:00 | 1.730055769 | ['Bes'] | 6.73E+16 | -2.30E+16 |
| 2015-11-23 | 23:00:00 | 1.730210189 | ['Anhur'] | 7.33E+16 | -2.50E+16 |
| 2015-11-23 | 23:30:00 | 1.730364151 | ['Sobek'] | 6.83E+16 | -2.33E+16 |
| 2015-11-24 | 0:00:00 | 1.730522364 | ['Neith'] | 6.30E+16 | -2.15E+16 |
| 2015-11-24 | 0:30:00 | 1.730681887 | ['Neith'] | 6.12E+16 | -2.09E+16 |
| 2015-11-24 | 1:00:00 | 1.730835098 | ['Wosret'] | 7.50E+16 | -2.56E+16 |
| 2015-11-24 | 1:30:00 | 1.7309825 | ['Neith'] | 8.99E+16 | -3.07E+16 |
| 2015-11-24 | 4:00:00 | 1.731753585 | ['Anubis'] | 8.21E+16 | -2.80E+16 |
| 2015-11-24 | 4:30:00 | 1.731914 | ['Anubis'] | 6.37E+16 | -2.18E+16 |
| 2015-11-24 | 5:00:00 | 1.732071509 | ['Atum'] | 6.13E+16 | -2.09E+16 |
| 2015-11-24 | 5:30:00 | 1.732224231 | ['Khonsu'] | 6.32E+16 | -2.16E+16 |
| 2015-11-24 | 6:00:00 | 1.732376604 | ['Khonsu'] | 6.80E+16 | -2.32E+16 |
| 2015-11-24 | 9:00:00 | 1.733304727 | ['Khepry'] | 6.51E+16 | -2.22E+16 |
| 2015-11-24 | 9:30:00 | 1.733461698 | ['Bes'] | 6.96E+16 | -2.38E+16 |
| 2015-11-24 | 10:00:00 | 1.733616038 | ['Bes'] | 6.59E+16 | -2.25E+16 |
| 2015-11-24 | 11:00:00 | 1.733920943 | ['Bes'] | 5.67E+16 | -1.94E+16 |
| 2015-11-24 | 11:30:00 | 1.734072 | ['Neith'] | 7.34E+16 | -2.51E+16 |
| 2015-11-24 | 12:00:00 | 1.734229245 | ['Neith'] | 7.75E+16 | -2.65E+16 |
| 2015-11-24 | 12:30:00 | 1.734381346 | ['Wosret'] | 7.03E+16 | -2.40E+16 |
| 2015-11-24 | 13:00:00 | 1.734534717 | ['Wosret'] | 7.09E+16 | -2.42E+16 |
| 2015-11-24 | 23:30:00 | 1.737789623 | ['Wosret'] | 5.54E+16 | -1.89E+16 |
| 2015-11-25 | 0:00:00 | 1.73793 | ['Neith'] | 6.90E+16 | -2.36E+16 |
| 2015-11-25 | 5:30:00 | 1.739669444 | ['Neith'] | 4.83E+16 | -1.65E+16 |
| 2015-11-25 | 6:00:00 | 1.73978913 | ['Anhur'] | 5.20E+16 | -1.77E+16 |
| 2015-11-25 | 9:30:00 | 1.740885577 | ['Anhur'] | 7.10E+16 | -2.42E+16 |
| 2015-11-25 | 10:00:00 | 1.74104 | ['Geb'] | 6.65E+16 | -2.27E+16 |
| 2015-11-25 | 10:30:00 | 1.74117425 | ['Atum'] | 6.65E+16 | -2.27E+16 |
| 2015-11-25 | 13:30:00 | 1.742178571 | ['Atum'] | 6.18E+16 | -2.12E+16 |
| 2015-11-25 | 14:00:00 | 1.742280962 | ['Anhur'] | 8.96E+16 | -3.06E+16 |
| 2015-11-25 | 14:30:00 | 1.742433962 | ['Imhotep'] | 7.68E+16 | -2.62E+16 |
| 2015-11-25 | 15:00:00 | 1.7425685 | ['Bes'] | 5.89E+16 | -2.01E+16 |
| 2015-11-25 | 19:30:00 | 1.744018966 | ['Bes'] | 6.58E+16 | -2.25E+16 |
| 2015-11-25 | 20:30:00 | 1.744296842 | ['Anhur'] | 8.40E+16 | -2.87E+16 |
| 2015-11-26 | 0:30:00 | 1.745584091 | ['Anhur'] | 6.66E+16 | -2.28E+16 |
| 2015-11-26 | 1:00:00 | 1.745693019 | ['Imhotep'] | 8.54E+16 | -2.92E+16 |
| 2015-11-26 | 6:00:00 | 1.747252041 | ['Imhotep'] | 4.37E+16 | -1.49E+16 |
| 2015-11-26 | 6:30:00 | 1.747400755 | ['Anhur'] | 5.62E+16 | -1.92E+16 |
| 2015-11-26 | 11:00:00 | 1.748817568 | ['Anhur'] | 5.94E+16 | -2.03E+16 |
| 2015-11-26 | 11:30:00 | 1.748907273 | ['Khonsu'] | 5.85E+16 | -2.00E+16 |
| 2015-11-26 | 16:00:00 | 1.750389231 | ['Khonsu'] | 3.46E+16 | -1.18E+16 |
| 2015-11-26 | 16:30:00 | 1.750503846 | ['Anhur'] | 4.35E+16 | -1.49E+16 |
| 2015-11-26 | 17:00:00 | 1.75065549 | ['Anhur'] | 3.82E+16 | -1.30E+16 |
| 2015-11-26 | 21:00:00 | 1.751959375 | ['Anhur'] | 8.07E+16 | -2.76E+16 |
| 2015-11-26 | 21:30:00 | 1.752050909 | ['Geb'] | 7.81E+16 | -2.67E+16 |
| 2015-11-27 | 2:30:00 | 1.753615 | ['Imhotep'] | 9.49E+16 | -3.24E+16 |
| 2015-11-27 | 3:00:00 | 1.753776275 | ['Imhotep'] | 6.44E+16 | -2.20E+16 |
| 2015-11-27 | 7:30:00 | 1.75517283 | ['Imhotep'] | 2.88E+16 | -9.82847E+15 |
| 2015-11-27 | 13:30:00 | 1.757070909 | ['Imhotep'] | 5.29E+16 | -1.81E+16 |
| 2015-11-27 | 18:00:00 | 1.758414359 | ['Anhur'] | 2.17E+16 | -7.42626E+15 |
| 2015-11-27 | 23:30:00 | 1.760152182 | ['Atum'] | 1.61E+16 | -5.48269E+15 |
| 2015-11-28 | 6:00:00 | 1.762180784 | ['Aker'] | 1.89E+16 | -6.45907E+15 |
| 2015-11-28 | 6:30:00 | 1.76233434 | ['Hapi'] | 1.51E+16 | -5.14929E+15 |
| 2015-11-28 | 7:00:00 | 1.762488269 | ['Bastet'] | 9.96E+15 | -3.40399E+15 |
| 2015-11-28 | 9:30:00 | 1.7632714 | ['Anuket'] | 4.34E+16 | -1.48E+16 |
| 2015-11-28 | 10:00:00 | 1.763429231 | ['Anuket'] | 5.07E+16 | -1.73E+16 |
| 2015-11-28 | 12:00:00 | 1.764045577 | ['Atum'] | 1.35E+16 | -4.61822E+15 |
| 2015-11-28 | 13:30:00 | 1.764523725 | ['Imhotep1'] | 1.76E+16 | -6.03022E+15 |
| 2015-11-28 | 14:00:00 | 1.764677843 | ['Imhotep'] | 4.53E+16 | -1.55E+16 |
| 2015-11-28 | 14:30:00 | 1.764830192 | ['Imhotep'] | 8.25E+16 | -2.82E+16 |
| 2015-11-28 | 15:00:00 | 1.764985 | ['Imhotep'] | 9.68E+16 | -3.31E+16 |
| 2015-11-28 | 15:30:00 | 1.765144314 | ['Imhotep'] | 6.06E+16 | -2.07E+16 |
| 2015-11-28 | 16:00:00 | 1.765295577 | ['Imhotep'] | 4.02E+16 | -1.37E+16 |
| 2015-11-28 | 16:30:00 | 1.765449811 | ['Khepry'] | 1.89E+16 | -6.46016E+15 |
| 2015-11-28 | 17:00:00 | 1.765604423 | ['Khepry'] | 2.21E+16 | -7.54331E+15 |
| 2015-11-28 | 17:30:00 | 1.765765536 | ['Khepry'] | 1.94E+16 | -6.62765E+15 |
| 2015-11-28 | 18:00:00 | 1.765926346 | ['Aker'] | 1.42E+16 | -4.86326E+15 |
| 2015-11-28 | 18:30:00 | 1.766081923 | ['Aker'] | 1.19E+16 | -4.06969E+15 |
| 2015-11-28 | 19:00:00 | 1.766235769 | ['Bastet'] | 7.67E+15 | -2.63054E+15 |
| 2015-11-28 | 20:30:00 | 1.766699423 | ['Maftet'] | 1.10E+16 | -3.77667E+15 |
| 2015-11-28 | 21:00:00 | 1.766853396 | ['Nut'] | 1.95E+16 | -6.66196E+15 |
| 2015-11-28 | 21:30:00 | 1.767009245 | ['Anuket'] | 2.65E+16 | -9.06513E+15 |
| 2015-11-29 | 0:00:00 | 1.767794151 | ['Anubis'] | 1.08E+16 | -3.6786E+15 |
| 2015-11-29 | 2:00:00 | 1.768416364 | ['Imhotep'] | 2.88E+16 | -9.85338E+15 |
| 2015-11-29 | 3:00:00 | 1.76873434 | ['Imhotep'] | 8.26E+16 | -2.82E+16 |
| 2015-11-29 | 3:30:00 | 1.768889231 | ['Imhotep'] | 6.20E+16 | -2.12E+16 |
| 2015-11-29 | 4:00:00 | 1.769045472 | ['Imhotep'] | 4.51E+16 | -1.54E+16 |
| 2015-11-29 | 4:30:00 | 1.769199615 | ['Imhotep'] | 2.32E+16 | -7.92358E+15 |
| 2015-11-29 | 5:00:00 | 1.769353585 | ['Aten'] | 2.02E+16 | -6.89953E+15 |
| 2015-11-29 | 5:30:00 | 1.769509245 | ['Khepry'] | 1.65E+16 | -5.66632E+15 |
| 2015-11-29 | 6:00:00 | 1.769664906 | ['Aker'] | 1.51E+16 | -5.18196E+15 |
| 2015-11-29 | 6:30:00 | 1.769822653 | ['Aker'] | 1.02E+16 | -3.52174E+15 |
| 2015-11-29 | 8:30:00 | 1.770450213 | ['Seth'] | 5.83E+15 | -2.02751E+15 |
| 2015-11-29 | 10:00:00 | 1.770913462 | ['Seth'] | 3.94E+16 | -1.35E+16 |
| 2015-11-29 | 13:30:00 | 1.772012308 | ['Khepry'] | 4.63E+15 | -1.63462E+15 |
| 2015-11-29 | 14:00:00 | 1.7721632 | ['Imhotep'] | 1.28E+16 | -4.3674E+15 |
| 2015-11-29 | 14:30:00 | 1.772323 | ['Ash'] | 3.73E+16 | -1.28E+16 |
| 2015-11-29 | 15:00:00 | 1.77247549 | ['Imhotep'] | 6.00E+16 | -2.05E+16 |
| 2015-11-29 | 15:30:00 | 1.772632778 | ['Bastet'] | 5.93E+16 | -2.02E+16 |
| 2015-11-29 | 16:00:00 | 1.772794 | ['Imhotep'] | 4.08E+16 | -1.39E+16 |
| 2015-11-29 | 16:30:00 | 1.772952642 | ['Imhotep'] | 2.72E+16 | -9.30528E+15 |
| 2015-11-29 | 21:30:00 | 1.774513137 | ['Serqet'] | 9.94E+15 | -3.43221E+15 |
| 2015-11-29 | 22:00:00 | 1.774670566 | ['Anuket'] | 3.85E+16 | -1.32E+16 |
| 2015-11-29 | 22:30:00 | 1.774826038 | ['Geb'] | 4.39E+16 | -1.50E+16 |
| 2015-11-30 | 2:30:00 | 1.776077547 | ['Imhotep'] | 2.59E+16 | -8.85558E+15 |
| 2015-11-30 | 3:00:00 | 1.776197333 | ['Imhotep'] | 3.99E+16 | -1.36E+16 |
| 2015-11-30 | 12:00:00 | 1.779046792 | ['Anubis'] | 1.62E+16 | -5.55159E+15 |
| 2015-11-30 | 16:30:00 | 1.780491667 | ['Anubis'] | 2.11E+16 | -7.19901E+15 |
| 2015-11-30 | 17:00:00 | 1.780581563 | ['Aten'] | 1.63E+16 | -5.57237E+15 |
| 2015-12-02 | 0:00:00 | 1.790339038 | ['Hapi'] | 1.83E+16 | -6.24008E+15 |
| 2015-12-03 | 1:00:00 | 1.798193846 | ['Hapi'] | 1.91E+16 | -6.52722E+15 |
| 2015-12-03 | 6:30:00 | 1.799923077 | ['Aten'] | 5.42E+15 | -1.88675E+15 |
| 2015-12-03 | 21:00:00 | 1.80454125 | ['Hatmehit'] | 9.06E+15 | -3.18822E+15 |
| 2015-12-04 | 2:30:00 | 1.806222075 | ['Seth'] | 1.14E+16 | -3.89893E+15 |
| 2015-12-05 | 6:30:00 | 1.815044906 | ['Ash'] | 4.37E+15 | -1.50203E+15 |
| 2015-12-05 | 7:00:00 | 1.815198824 | ['Imhotep'] | 6.43E+15 | -2.20213E+15 |
| 2015-12-05 | 9:30:00 | 1.815995769 | ['Aten'] | 9.29E+15 | -3.18501E+15 |
| 2015-12-05 | 14:30:00 | 1.817571538 | ['Serqet'] | 3.12E+15 | -1.09212E+15 |
| 2015-12-05 | 19:30:00 | 1.819146731 | ['Imhotep'] | 1.09E+16 | -3.72382E+15 |
| 2015-12-05 | 20:00:00 | 1.819303846 | ['Imhotep'] | 1.37E+16 | -4.66802E+15 |
| 2015-12-06 | 6:30:00 | 1.822630189 | ['Ash'] | 3.89E+15 | -1.34363E+15 |
| 2015-12-06 | 16:00:00 | 1.82563 | ['Seth'] | 1.13E+16 | -3.85746E+15 |
| 2015-12-06 | 16:30:00 | 1.825785849 | ['Seth'] | 2.31E+16 | -7.88966E+15 |
| 2015-12-06 | 21:00:00 | 1.827211154 | ['Imhotep'] | 3.63E+16 | -1.24E+16 |
| 2015-12-06 | 21:30:00 | 1.827368113 | ['Khepry'] | 3.64E+16 | -1.24E+16 |
| 2015-12-06 | 22:00:00 | 1.827524038 | ['Aten'] | 2.46E+16 | -8.41059E+15 |
| 2015-12-06 | 22:30:00 | 1.827680189 | ['Babi'] | 1.62E+16 | -5.52599E+15 |
| 2015-12-06 | 23:00:00 | 1.827842857 | ['Aker'] | 1.17E+16 | -3.99055E+15 |
| 2015-12-06 | 23:30:00 | 1.828005849 | ['Bastet'] | 1.57E+16 | -5.38144E+15 |
| 2015-12-07 | 0:00:00 | 1.828163396 | ['Bastet'] | 1.95E+16 | -6.64979E+15 |
| 2015-12-07 | 0:30:00 | 1.828319615 | ['Bastet'] | 1.79E+16 | -6.11026E+15 |
| 2015-12-07 | 1:30:00 | 1.828633396 | ['Hatmehit'] | 8.09E+15 | -2.76731E+15 |
| 2015-12-07 | 2:00:00 | 1.828788431 | ['Hatmehit'] | 9.31E+15 | -3.19565E+15 |
| 2015-12-07 | 2:30:00 | 1.828946863 | ['Maat'] | 6.98E+15 | -2.3915E+15 |
| 2015-12-07 | 3:00:00 | 1.829103585 | ['Maftet'] | 9.27E+15 | -3.16979E+15 |
| 2015-12-07 | 3:30:00 | 1.829266429 | ['Hapi'] | 9.58E+15 | -3.27994E+15 |
| 2015-12-07 | 4:00:00 | 1.829427692 | ['Seth'] | 1.21E+16 | -4.12555E+15 |
| 2015-12-07 | 4:30:00 | 1.829586604 | ['Seth'] | 2.18E+16 | -7.44392E+15 |
| 2015-12-07 | 5:00:00 | 1.82974434 | ['Anubis'] | 3.21E+16 | -1.10E+16 |
| 2015-12-07 | 5:30:00 | 1.829900392 | ['Atum'] | 2.46E+16 | -8.39781E+15 |
| 2015-12-07 | 11:30:00 | 1.831828 | ['Hatmehit'] | 2.04E+16 | -6.97529E+15 |
| 2015-12-07 | 12:00:00 | 1.831952264 | ['Bastet'] | 2.03E+16 | -6.9461E+15 |
| 2015-12-07 | 12:30:00 | 1.832115179 | ['Bastet'] | 2.48E+16 | -8.46396E+15 |
| 2015-12-07 | 13:00:00 | 1.832279038 | ['Bastet'] | 1.64E+16 | -5.5942E+15 |
| 2015-12-07 | 14:00:00 | 1.832591923 | ['Hatmehit'] | 1.04E+16 | -3.5699E+15 |
| 2015-12-07 | 14:30:00 | 1.832748302 | ['Maftet'] | 9.33E+15 | -3.19089E+15 |
| 2015-12-07 | 15:00:00 | 1.832906038 | ['Maftet'] | 9.84E+15 | -3.36257E+15 |
| 2015-12-07 | 15:30:00 | 1.833063585 | ['Seth'] | 1.23E+16 | -4.19221E+15 |
| 2015-12-07 | 16:00:00 | 1.833218627 | ['Seth'] | 1.25E+16 | -4.27598E+15 |
| 2015-12-07 | 16:30:00 | 1.833377778 | ['Anubis'] | 2.06E+16 | -7.03234E+15 |
| 2015-12-07 | 17:00:00 | 1.833511471 | ['Anubis'] | 3.28E+16 | -1.12E+16 |
| 2015-12-07 | 21:00:00 | 1.834871111 | ['Hatmehit'] | 5.02E+16 | -1.72E+16 |
| 2015-12-07 | 21:30:00 | 1.834998519 | ['Hatmehit'] | 5.42E+16 | -1.85E+16 |
| 2015-12-07 | 22:00:00 | 1.835119792 | ['Khepry'] | 5.22E+16 | -1.78E+16 |
| 2015-12-07 | 22:30:00 | 1.835286346 | ['Khepry'] | 2.85E+16 | -9.73881E+15 |
| 2015-12-08 | 2:30:00 | 1.836591538 | ['Khepry'] | 1.05E+16 | -3.59467E+15 |
| 2015-12-08 | 3:00:00 | 1.836707347 | ['Maftet'] | 1.15E+16 | -3.94152E+15 |
| 2015-12-08 | 4:00:00 | 1.837030213 | ['Nut'] | 1.32E+16 | -4.52915E+15 |
| 2015-12-08 | 5:00:00 | 1.837341887 | ['Khonsu'] | 3.24E+16 | -1.11E+16 |
| 2015-12-08 | 7:30:00 | 1.83817087 | ['Khonsu'] | 1.20E+16 | -4.11229E+15 |
| 2015-12-08 | 8:00:00 | 1.838293273 | ['Imhotep'] | 2.46E+16 | -8.39596E+15 |
| 2015-12-08 | 8:30:00 | 1.83845434 | ['Imhotep'] | 3.32E+16 | -1.13E+16 |
| 2015-12-08 | 9:30:00 | 1.838768627 | ['Imhotep'] | 4.63E+16 | -1.58E+16 |
| 2015-12-08 | 11:30:00 | 1.839396296 | ['Bastet'] | 2.05E+16 | -7.00114E+15 |
| 2015-12-08 | 12:30:00 | 1.839765909 | ['Hatmehit'] | 3.36E+16 | -1.15E+16 |
| 2015-12-08 | 13:00:00 | 1.839868889 | ['Bastet'] | 3.96E+16 | -1.35E+16 |
| 2015-12-08 | 17:30:00 | 1.841350909 | ['Hatmehit'] | 4.29E+16 | -1.47E+16 |
| 2015-12-08 | 18:00:00 | 1.841461346 | ['Atum'] | 2.90E+16 | -9.91272E+15 |
| 2015-12-08 | 18:30:00 | 1.841622941 | ['Atum'] | 1.96E+16 | -6.69609E+15 |
| 2015-12-08 | 22:30:00 | 1.8429324 | ['Atum'] | 3.37E+16 | -1.15E+16 |
| 2015-12-08 | 23:00:00 | 1.843045769 | ['Khepry'] | 2.41E+16 | -8.24733E+15 |
| 2015-12-08 | 23:30:00 | 1.843203269 | ['Bastet'] | 1.99E+16 | -6.80247E+15 |
| 2015-12-09 | 6:00:00 | 1.845255476 | ['Atum'] | 3.81E+16 | -1.30E+16 |
| 2015-12-09 | 9:30:00 | 1.846376809 | ['Atum'] | 6.25E+16 | -2.14E+16 |
| 2015-12-09 | 10:00:00 | 1.846533774 | ['Imhotep'] | 6.30E+16 | -2.15E+16 |
| 2015-12-09 | 20:00:00 | 1.849706981 | ['Imhotep1'] | 3.23E+16 | -1.10E+16 |
| 2015-12-09 | 20:30:00 | 1.849865094 | ['Imhotep'] | 3.81E+16 | -1.30E+16 |
| 2015-12-09 | 21:00:00 | 1.850025 | ['Imhotep'] | 5.04E+16 | -1.72E+16 |
| 2015-12-09 | 21:30:00 | 1.850188909 | ['Imhotep'] | 6.17E+16 | -2.11E+16 |
| 2015-12-09 | 22:00:00 | 1.85035 | ['Imhotep'] | 6.52E+16 | -2.23E+16 |
| 2015-12-09 | 22:30:00 | 1.850506471 | ['Bes'] | 5.56E+16 | -1.90E+16 |
| 2015-12-09 | 23:00:00 | 1.850663077 | ['Bes'] | 4.38E+16 | -1.50E+16 |

| Date | Time | Value | Label | Col5 | Col6 |
|---|---|---|---|---|---|
| 2015-12-09 | 23:30:00 | 1.850821765 | ['Bes'] | 2.94E+16 | -1.01E+16 |
| 2015-12-10 | 0:00:00 | 1.85098 | ['Anhur'] | 2.60E+16 | -8.87082E+15 |
| 2015-12-10 | 0:30:00 | 1.851136346 | ['Sobek'] | 2.92E+16 | -9.96969E+15 |
| 2015-12-10 | 1:00:00 | 1.851292692 | ['Sobek'] | 4.35E+16 | -1.48E+16 |
| 2015-12-10 | 1:30:00 | 1.85145625 | ['Neith'] | 5.84E+16 | -1.99E+16 |
| 2015-12-10 | 2:00:00 | 1.85162 | ['Wosret'] | 6.23E+16 | -2.13E+16 |
| 2015-12-10 | 2:30:00 | 1.851778302 | ['Wosret'] | 5.07E+16 | -1.73E+16 |
| 2015-12-10 | 3:00:00 | 1.851936078 | ['Sobek'] | 3.29E+16 | -1.12E+16 |
| 2015-12-10 | 3:30:00 | 1.852091887 | ['Sobek'] | 3.59E+16 | -1.22E+16 |
| 2015-12-10 | 4:00:00 | 1.852249811 | ['Hapi'] | 4.09E+16 | -1.40E+16 |
| 2015-12-10 | 4:30:00 | 1.852408269 | ['Seth'] | 3.46E+16 | -1.18E+16 |
| 2015-12-10 | 5:00:00 | 1.852564808 | ['Anubis'] | 3.18E+16 | -1.09E+16 |
| 2015-12-10 | 5:30:00 | 1.852721509 | ['Anubis'] | 4.20E+16 | -1.44E+16 |
| 2015-12-10 | 6:00:00 | 1.852886111 | ['Atum'] | 5.49E+16 | -1.87E+16 |
| 2015-12-10 | 6:30:00 | 1.853049808 | ['Atum'] | 4.17E+16 | -1.42E+16 |
| 2015-12-10 | 7:00:00 | 1.853206792 | ['Khonsu'] | 4.23E+16 | -1.44E+16 |
| 2015-12-10 | 7:30:00 | 1.853364906 | ['Khonsu'] | 4.09E+16 | -1.40E+16 |
| 2015-12-10 | 8:00:00 | 1.853523137 | ['Imhotep1'] | 3.88E+16 | -1.32E+16 |
| 2015-12-10 | 8:30:00 | 1.853678679 | ['Imhotep'] | 4.04E+16 | -1.38E+16 |
| 2015-12-10 | 9:00:00 | 1.853836792 | ['Imhotep'] | 4.19E+16 | -1.43E+16 |
| 2015-12-10 | 9:30:00 | 1.853991961 | ['Imhotep'] | 5.00E+16 | -1.71E+16 |
| 2015-12-10 | 10:00:00 | 1.854150566 | ['Imhotep'] | 5.86E+16 | -2.00E+16 |
| 2015-12-10 | 10:30:00 | 1.854314464 | ['Bes'] | 5.71E+16 | -1.95E+16 |
| 2015-12-10 | 11:00:00 | 1.85447717 | ['Bes'] | 5.45E+16 | -1.86E+16 |
| 2015-12-10 | 11:30:00 | 1.854586 | ['Bes'] | 4.77E+16 | -1.63E+16 |
| 2015-12-10 | 16:00:00 | 1.856103704 | ['Bes'] | 3.54E+16 | -1.21E+16 |
| 2015-12-10 | 16:30:00 | 1.856222642 | ['Seth'] | 3.50E+16 | -1.19E+16 |
| 2015-12-10 | 17:00:00 | 1.85637383 | ['Seth'] | 3.08E+16 | -1.05E+16 |
| 2015-12-10 | 21:00:00 | 1.85771 | ['Seth'] | 4.22E+16 | -1.44E+16 |
| 2015-12-10 | 21:30:00 | 1.857810192 | ['Imhotep'] | 4.30E+16 | -1.47E+16 |
| 2015-12-11 | 2:30:00 | 1.859404808 | ['Sobek'] | 7.45E+16 | -2.54E+16 |
| 2015-12-11 | 3:00:00 | 1.859561923 | ['Sobek'] | 8.41E+16 | -2.87E+16 |
| 2015-12-11 | 3:30:00 | 1.859673913 | ['Sobek'] | 9.59E+16 | -3.28E+16 |
| 2015-12-11 | 7:30:00 | 1.860991765 | ['Sobek'] | 4.60E+16 | -1.57E+16 |
| 2015-12-11 | 12:30:00 | 1.862602051 | ['Sobek'] | 3.61E+16 | -1.23E+16 |
| 2015-12-11 | 13:00:00 | 1.862738462 | ['Sobek'] | 3.38E+16 | -1.15E+16 |
| 2015-12-11 | 13:30:00 | 1.862895962 | ['Sobek'] | 4.29E+16 | -1.46E+16 |
| 2015-12-11 | 14:00:00 | 1.863029189 | ['Sobek'] | 5.76E+16 | -1.97E+16 |
| 2015-12-11 | 17:30:00 | 1.86422 | ['Sobek'] | 5.86E+16 | -2.00E+16 |
| 2015-12-11 | 18:00:00 | 1.86430641 | ['Atum'] | 6.11E+16 | -2.09E+16 |
| 2015-12-11 | 22:30:00 | 1.86582875 | ['Atum'] | 6.08E+16 | -2.08E+16 |
| 2015-12-11 | 23:00:00 | 1.865915385 | ['Bes'] | 5.52E+16 | -1.89E+16 |
| 2015-12-11 | 23:30:00 | 1.866079107 | ['Bes'] | 4.38E+16 | -1.49E+16 |
| 2015-12-12 | 5:30:00 | 1.868017027 | ['Bes'] | 6.79E+16 | -2.32E+16 |

| Date | Time | Value | Label | Col5 | Col6 |
|---|---|---|---|---|---|
| 2015-12-12 | 6:00:00 | 1.868148462 | ['Bes'] | 5.99E+16 | -2.04E+16 |
| 2015-12-12 | 6:30:00 | 1.868305962 | ['Bes'] | 4.73E+16 | -1.61E+16 |
| 2015-12-12 | 7:00:00 | 1.868464902 | ['Bes'] | 3.63E+16 | -1.24E+16 |
| 2015-12-12 | 7:30:00 | 1.868622642 | ['Bes'] | 3.09E+16 | -1.06E+16 |
| 2015-12-12 | 9:30:00 | 1.869264151 | ['Bes'] | 4.51E+16 | -1.54E+16 |
| 2015-12-12 | 10:00:00 | 1.869423019 | ['Bes'] | 4.89E+16 | -1.67E+16 |
| 2015-12-12 | 10:30:00 | 1.869532727 | ['Bes'] | 4.66E+16 | -1.59E+16 |
| 2015-12-12 | 14:30:00 | 1.870855094 | ['Geb'] | 3.37E+16 | -1.15E+16 |
| 2015-12-12 | 15:00:00 | 1.871015 | ['Geb'] | 4.46E+16 | -1.52E+16 |
| 2015-12-12 | 15:30:00 | 1.871173 | ['Geb'] | 5.15E+16 | -1.76E+16 |
| 2015-12-12 | 16:00:00 | 1.871327925 | ['Bes'] | 6.08E+16 | -2.08E+16 |
| 2015-12-12 | 16:30:00 | 1.871486792 | ['Bes'] | 6.95E+16 | -2.37E+16 |
| 2015-12-12 | 17:00:00 | 1.871650714 | ['Bes'] | 8.12E+16 | -2.77E+16 |
| 2015-12-12 | 17:30:00 | 1.87181434 | ['Bes'] | 7.06E+16 | -2.41E+16 |
| 2015-12-12 | 18:00:00 | 1.871973019 | ['Bes'] | 6.15E+16 | -2.10E+16 |
| 2015-12-12 | 18:30:00 | 1.872130192 | ['Bes'] | 4.94E+16 | -1.69E+16 |
| 2015-12-12 | 19:00:00 | 1.87228717 | ['Bes'] | 3.80E+16 | -1.30E+16 |
| 2015-12-12 | 19:30:00 | 1.872445577 | ['Bes'] | 3.49E+16 | -1.19E+16 |
| 2015-12-12 | 20:00:00 | 1.872604717 | ['Bes'] | 3.65E+16 | -1.25E+16 |
| 2015-12-12 | 20:30:00 | 1.872761731 | ['Bes'] | 4.08E+16 | -1.39E+16 |
| 2015-12-12 | 21:00:00 | 1.872920926 | ['Bes'] | 3.78E+16 | -1.29E+16 |
| 2015-12-12 | 21:30:00 | 1.873084909 | ['Bes'] | 3.67E+16 | -1.25E+16 |
| 2015-12-12 | 22:00:00 | 1.873245577 | ['Geb'] | 4.13E+16 | -1.41E+16 |
| 2015-12-12 | 22:30:00 | 1.873405472 | ['Anhur'] | 4.41E+16 | -1.50E+16 |
| 2015-12-12 | 23:00:00 | 1.873564151 | ['Anhur'] | 4.30E+16 | -1.47E+16 |
| 2015-12-12 | 23:30:00 | 1.873721538 | ['Sobek'] | 3.73E+16 | -1.27E+16 |
| 2015-12-13 | 0:00:00 | 1.8738798 | ['Sobek'] | 3.54E+16 | -1.21E+16 |
| 2015-12-13 | 0:30:00 | 1.874037547 | ['Sobek'] | 3.69E+16 | -1.26E+16 |
| 2015-12-13 | 1:00:00 | 1.874193922 | ['Sobek'] | 3.58E+16 | -1.22E+16 |
| 2015-12-13 | 1:30:00 | 1.874356607 | ['Sobek'] | 3.31E+16 | -1.13E+16 |
| 2015-12-13 | 2:00:00 | 1.874520943 | ['Geb'] | 3.00E+16 | -1.02E+16 |
| 2015-12-13 | 2:30:00 | 1.874679623 | ['Geb'] | 2.83E+16 | -9.66136E+15 |
| 2015-12-13 | 3:00:00 | 1.874838491 | ['Bes'] | 3.55E+16 | -1.21E+16 |
| 2015-12-13 | 3:30:00 | 1.874997358 | ['Bes'] | 4.19E+16 | -1.43E+16 |
| 2015-12-13 | 4:30:00 | 1.875311698 | ['Bes'] | 5.19E+16 | -1.77E+16 |
| 2015-12-13 | 5:00:00 | 1.875470377 | ['Bes'] | 5.53E+16 | -1.89E+16 |
| 2015-12-13 | 5:30:00 | 1.875629057 | ['Bes'] | 5.62E+16 | -1.92E+16 |
| 2015-12-13 | 6:00:00 | 1.875792857 | ['Bes'] | 5.59E+16 | -1.91E+16 |
| 2015-12-13 | 10:30:00 | 1.877234314 | ['Bes'] | 3.61E+16 | -1.23E+16 |
| 2015-12-13 | 11:00:00 | 1.877322857 | ['Anhur'] | 3.73E+16 | -1.28E+16 |
| 2015-12-13 | 16:00:00 | 1.878980727 | ['Bes'] | 3.19E+16 | -1.09E+16 |
| 2015-12-13 | 16:30:00 | 1.879117143 | ['Bes'] | 3.50E+16 | -1.20E+16 |
| 2015-12-13 | 20:30:00 | 1.88043 | ['Bes'] | 5.70E+16 | -1.95E+16 |
| 2015-12-13 | 21:00:00 | 1.880506667 | ['Anhur'] | 6.84E+16 | -2.34E+16 |

| Date | Time | Value | Name | Col5 | Col6 |
|---|---|---|---|---|---|
| 2015-12-14 | 1:30:00 | 1.882027895 | ['Anhur'] | 3.24E+16 | -1.11E+16 |
| 2015-12-14 | 11:30:00 | 1.885228529 | ['Anhur'] | 3.17E+16 | -1.08E+16 |
| 2015-12-14 | 12:00:00 | 1.885360943 | ['Neith'] | 2.33E+16 | -7.95688E+15 |
| 2015-12-14 | 16:30:00 | 1.886818684 | ['Neith'] | 3.68E+16 | -1.26E+16 |
| 2015-12-14 | 17:00:00 | 1.88692 | ['Imhotep'] | 4.30E+16 | -1.47E+16 |
| 2015-12-14 | 21:30:00 | 1.888425667 | ['Hatmehit'] | 5.25E+16 | -1.79E+16 |
| 2015-12-14 | 22:00:00 | 1.888547115 | ['Anhur'] | 6.45E+16 | -2.20E+16 |
| 2015-12-14 | 22:30:00 | 1.888704717 | ['Sobek'] | 9.45E+16 | -3.23E+16 |
| 2015-12-15 | 8:00:00 | 1.891738947 | ['Bes'] | 4.13E+16 | -1.41E+16 |
| 2015-12-15 | 8:30:00 | 1.891903462 | ['Anhur'] | 4.46E+16 | -1.52E+16 |
| 2015-12-15 | 12:30:00 | 1.893223889 | ['Hatmehit'] | 3.87E+16 | -1.32E+16 |
| 2015-12-15 | 13:00:00 | 1.893322128 | ['Anuket'] | 4.69E+16 | -1.60E+16 |
| 2015-12-15 | 17:30:00 | 1.894816364 | ['Anuket'] | 6.18E+16 | -2.11E+16 |
| 2015-12-15 | 18:00:00 | 1.894934259 | ['Imhotep'] | 5.66E+16 | -1.93E+16 |
| 2015-12-15 | 18:30:00 | 1.895095962 | ['Imhotep'] | 4.47E+16 | -1.53E+16 |
| 2015-12-15 | 19:00:00 | 1.89519 | ['Imhotep'] | 4.45E+16 | -1.52E+16 |
| 2015-12-15 | 22:30:00 | 1.89642 | ['Imhotep'] | 4.69E+16 | -1.60E+16 |
| 2015-12-15 | 23:00:00 | 1.896532115 | ['Wosret'] | 5.81E+16 | -1.98E+16 |
| 2015-12-16 | 5:30:00 | 1.89862 | ['Wosret'] | 5.49E+16 | -1.88E+16 |
| 2015-12-16 | 6:00:00 | 1.898752143 | ['Imhotep'] | 5.97E+16 | -2.04E+16 |
| 2015-12-16 | 10:00:00 | 1.900039434 | ['Anhur'] | 3.11E+16 | -1.06E+16 |
| 2015-12-16 | 20:00:00 | 1.9032975 | ['Bes'] | 4.47E+16 | -1.53E+16 |
| 2015-12-16 | 20:30:00 | 1.903394038 | ['Anhur'] | 4.30E+16 | -1.47E+16 |
| 2015-12-17 | 1:00:00 | 1.9048372 | ['Sobek'] | 4.24E+16 | -1.45E+16 |
| 2015-12-17 | 6:00:00 | 1.906444898 | ['Sobek'] | 5.72E+16 | -1.95E+16 |
| 2015-12-17 | 6:30:00 | 1.906596923 | ['Imhotep'] | 4.66E+16 | -1.59E+16 |
| 2015-12-17 | 7:00:00 | 1.906717667 | ['Imhotep'] | 3.64E+16 | -1.24E+16 |
| 2015-12-17 | 11:00:00 | 1.908058857 | ['Imhotep'] | 3.78E+16 | -1.29E+16 |
| 2015-12-17 | 16:00:00 | 1.909673846 | ['Imhotep'] | 3.32E+16 | -1.14E+16 |
| 2015-12-17 | 16:30:00 | 1.909789038 | ['Khonsu'] | 2.77E+16 | -9.46263E+15 |
| 2015-12-17 | 17:00:00 | 1.909949245 | ['Imhotep1'] | 3.34E+16 | -1.14E+16 |
| 2015-12-17 | 21:30:00 | 1.911387358 | ['Aker'] | 1.73E+16 | -5.91753E+15 |
| 2015-12-17 | 22:00:00 | 1.911547547 | ['Aker'] | 1.73E+16 | -5.92485E+15 |
| 2015-12-17 | 22:30:00 | 1.911711429 | ['Anhur'] | 2.21E+16 | -7.56148E+15 |
| 2015-12-17 | 23:00:00 | 1.911876038 | ['Wosret'] | 2.74E+16 | -9.36863E+15 |
| 2015-12-17 | 23:30:00 | 1.912035577 | ['Wosret'] | 2.79E+16 | -9.52955E+15 |
| 2015-12-18 | 0:30:00 | 1.912351132 | ['Wosret'] | 3.36E+16 | -1.15E+16 |
| 2015-12-18 | 1:00:00 | 1.912511731 | ['Neith'] | 4.30E+16 | -1.47E+16 |
| 2015-12-18 | 1:30:00 | 1.912668462 | ['Anuket'] | 4.35E+16 | -1.49E+16 |
| 2015-12-18 | 2:00:00 | 1.912826604 | ['Anuket'] | 3.69E+16 | -1.26E+16 |
| 2015-12-18 | 2:30:00 | 1.91298566 | ['Anuket'] | 2.74E+16 | -9.36939E+15 |
| 2015-12-18 | 3:00:00 | 1.913150893 | ['Anubis'] | 2.36E+16 | -8.0691E+15 |
| 2015-12-18 | 3:30:00 | 1.913314906 | ['Atum'] | 2.00E+16 | -6.82697E+15 |
| 2015-12-18 | 4:00:00 | 1.913474038 | ['Khonsu'] | 1.93E+16 | -6.60999E+15 |

| Date | Time | Value | Region | Col5 | Col6 |
|---|---|---|---|---|---|
| 2015-12-18 | 4:30:00 | 1.913632308 | ['Imhotep1'] | 1.83E+16 | -6.24551E+15 |
| 2015-12-18 | 5:00:00 | 1.913790377 | ['Imhotep1'] | 2.35E+16 | -8.03301E+15 |
| 2015-12-18 | 5:30:00 | 1.913949434 | ['Imhotep'] | 3.78E+16 | -1.29E+16 |
| 2015-12-18 | 6:00:00 | 1.914105714 | ['Imhotep'] | 6.16E+16 | -2.10E+16 |
| 2015-12-18 | 6:30:00 | 1.914266731 | ['Imhotep'] | 5.93E+16 | -2.02E+16 |
| 2015-12-18 | 7:00:00 | 1.914428 | ['Imhotep'] | 4.17E+16 | -1.42E+16 |
| 2015-12-18 | 7:30:00 | 1.914592963 | ['Imhotep'] | 2.05E+16 | -6.99325E+15 |
| 2015-12-18 | 8:00:00 | 1.914753774 | ['Khepry'] | 1.38E+16 | -4.70856E+15 |
| 2015-12-18 | 8:30:00 | 1.914913846 | ['Khepry'] | 1.78E+16 | -6.08042E+15 |
| 2015-12-18 | 9:00:00 | 1.91507283 | ['Khepry'] | 1.58E+16 | -5.39333E+15 |
| 2015-12-18 | 9:30:00 | 1.915231569 | ['Aker'] | 1.37E+16 | -4.67776E+15 |
| 2015-12-18 | 10:00:00 | 1.915388491 | ['Aker'] | 1.49E+16 | -5.08502E+15 |
| 2015-12-18 | 10:30:00 | 1.915547736 | ['Wosret'] | 1.13E+16 | -3.87038E+15 |
| 2015-12-18 | 11:00:00 | 1.91570717 | ['Wosret'] | 1.23E+16 | -4.2124E+15 |
| 2015-12-18 | 11:30:00 | 1.915871429 | ['Wosret'] | 1.42E+16 | -4.84961E+15 |
| 2015-12-18 | 12:30:00 | 1.916194423 | ['Maftet'] | 2.41E+16 | -8.21897E+15 |
| 2015-12-18 | 13:00:00 | 1.916352308 | ['Anuket'] | 3.12E+16 | -1.07E+16 |
| 2015-12-18 | 13:30:00 | 1.91650087 | ['Anuket'] | 3.69E+16 | -1.26E+16 |
| 2015-12-18 | 18:00:00 | 1.917930233 | ['Imhotep'] | 5.19E+16 | -1.77E+16 |
| 2015-12-18 | 18:30:00 | 1.918118235 | ['Imhotep'] | 7.33E+16 | -2.50E+16 |
| 2015-12-18 | 23:30:00 | 1.919714906 | ['Hatmehit'] | 8.21E+15 | -2.81858E+15 |
| 2015-12-19 | 6:00:00 | 1.92179 | ['Imhotep'] | 4.90E+16 | -1.67E+16 |
| 2015-12-19 | 14:30:00 | 1.924514231 | ['Anuket'] | 1.94E+16 | -6.64631E+15 |
| 2015-12-19 | 19:30:00 | 1.92611717 | ['Imhotep'] | 3.27E+16 | -1.12E+16 |
| 2015-12-19 | 20:00:00 | 1.926275769 | ['Khepry'] | 1.63E+16 | -5.56437E+15 |
| 2015-12-19 | 20:30:00 | 1.926433396 | ['Aten'] | 1.02E+16 | -3.4903E+15 |
| 2015-12-19 | 21:00:00 | 1.926593019 | ['Babi'] | 1.18E+16 | -4.05444E+15 |
| 2015-12-19 | 21:30:00 | 1.926753019 | ['Babi'] | 8.08E+15 | -2.80002E+15 |
| 2015-12-20 | 0:00:00 | 1.927558302 | ['Hatmehit'] | 6.27E+15 | -2.20962E+15 |
| 2015-12-20 | 0:30:00 | 1.927716981 | ['Nut'] | 7.50E+15 | -2.58798E+15 |
| 2015-12-20 | 2:00:00 | 1.92819283 | ['Anuket'] | 2.67E+16 | -9.1094E+15 |
| 2015-12-20 | 2:30:00 | 1.92835283 | ['Anuket'] | 2.16E+16 | -7.39105E+15 |
| 2015-12-20 | 3:00:00 | 1.928518393 | ['Anuket'] | 1.12E+16 | -3.84366E+15 |
| 2015-12-20 | 3:30:00 | 1.928682075 | ['Anuket'] | 6.63E+15 | -2.27301E+15 |
| 2015-12-20 | 4:00:00 | 1.928841569 | ['Seth'] | 4.00E+15 | -1.37864E+15 |
| 2015-12-20 | 6:00:00 | 1.929473846 | ['Imhotep'] | 2.12E+16 | -7.23431E+15 |
| 2015-12-20 | 6:30:00 | 1.929633774 | ['Imhotep'] | 3.96E+16 | -1.35E+16 |
| 2015-12-20 | 7:00:00 | 1.929797091 | ['Imhotep'] | 4.36E+16 | -1.49E+16 |
| 2015-12-20 | 7:30:00 | 1.929962222 | ['Imhotep'] | 2.97E+16 | -1.01E+16 |
| 2015-12-20 | 8:00:00 | 1.930124231 | ['Imhotep'] | 1.59E+16 | -5.4475E+15 |
| 2015-12-20 | 8:30:00 | 1.930281731 | ['Ash'] | 8.66E+15 | -2.95874E+15 |
| 2015-12-20 | 9:00:00 | 1.930439057 | ['Aten'] | 9.72E+15 | -3.33353E+15 |
| 2015-12-20 | 9:30:00 | 1.930599057 | ['Babi'] | 6.51E+15 | -2.2388E+15 |
| 2015-12-20 | 11:00:00 | 1.931014545 | ['Maat'] | 5.00E+15 | -1.89641E+15 |

| Date | Time | Value | Name | E1 | E2 |
|---|---|---|---|---|---|
| 2015-12-21 | 2:00:00 | 1.935885962 | ['Hathor'] | 1.95E+16 | -6.67426E+15 |
| 2015-12-21 | 6:30:00 | 1.9373575 | ['Hathor'] | 1.44E+16 | -4.9184E+15 |
| 2015-12-21 | 13:30:00 | 1.939578654 | ['Seth'] | 6.05E+15 | -2.1028E+15 |
| 2015-12-22 | 8:30:00 | 1.945669434 | ['Ash'] | 7.15E+15 | -2.45413E+15 |
| 2015-12-22 | 14:00:00 | 1.947431071 | ['Seth'] | 6.82E+15 | -2.38443E+15 |
| 2015-12-22 | 23:00:00 | 1.950318462 | ['Babi'] | 3.27E+15 | -1.16292E+15 |
| 2015-12-24 | 1:00:00 | 1.958661455 | ['Hathor'] | 4.81E+15 | -1.65978E+15 |
| 2015-12-24 | 6:00:00 | 1.96026549 | ['Hathor'] | 5.72E+15 | -2.01705E+15 |
| 2015-12-25 | 13:00:00 | 1.970216731 | ['Babi'] | 5.21E+15 | -1.82117E+15 |
| 2015-12-25 | 13:30:00 | 1.970377358 | ['Babi'] | 4.97E+15 | -1.70643E+15 |
| 2015-12-25 | 17:30:00 | 1.971710909 | ['Babi'] | 3.07E+15 | -1.08368E+15 |
| 2015-12-25 | 18:00:00 | 1.971801951 | ['Hapi'] | 2.65E+15 | -9.23971E+14 |
| 2015-12-25 | 23:00:00 | 1.973426604 | ['Imhotep'] | 3.73E+15 | -1.2869E+15 |
| 2015-12-26 | 7:30:00 | 1.976156078 | ['Atum'] | 9.55E+15 | -3.26877E+15 |
| 2015-12-26 | 8:00:00 | 1.976315849 | ['Atum'] | 1.21E+16 | -4.14708E+15 |
| 2015-12-26 | 10:00:00 | 1.976965714 | ['Ash'] | 2.43E+15 | -8.7297E+14 |
| 2015-12-26 | 10:30:00 | 1.977123774 | ['Ash'] | 3.66E+15 | -1.25658E+15 |
| 2015-12-26 | 11:00:00 | 1.977283962 | ['Ash'] | 5.04E+15 | -1.72734E+15 |
| 2015-12-26 | 11:30:00 | 1.977444151 | ['Imhotep'] | 6.89E+15 | -2.35703E+15 |
| 2015-12-26 | 12:00:00 | 1.9776025 | ['Ash'] | 7.96E+15 | -2.7186E+15 |
| 2015-12-26 | 12:30:00 | 1.977762963 | ['Aten'] | 8.12E+15 | -2.77608E+15 |
| 2015-12-26 | 14:30:00 | 1.978410962 | ['Hapi'] | 6.26E+15 | -2.16124E+15 |
| 2015-12-26 | 15:00:00 | 1.97857 | ['Hathor'] | 6.46E+15 | -2.2441E+15 |
| 2015-12-26 | 16:00:00 | 1.978888491 | ['Maat'] | 3.38E+15 | -1.18127E+15 |
| 2015-12-26 | 17:00:00 | 1.979212407 | ['Maat'] | 4.27E+15 | -1.47615E+15 |
| 2015-12-26 | 18:00:00 | 1.9795375 | ['Hathor'] | 2.33E+15 | -8.1517E+14 |
| 2015-12-26 | 19:00:00 | 1.979856038 | ['Seth'] | 2.53E+15 | -9.16167E+14 |
| 2015-12-26 | 19:30:00 | 1.980014314 | ['Seth'] | 6.00E+15 | -2.09567E+15 |
| 2015-12-26 | 20:30:00 | 1.980333396 | ['Seth'] | 1.35E+16 | -4.64463E+15 |
| 2015-12-26 | 22:30:00 | 1.980982692 | ['Imhotep'] | 4.26E+15 | -1.4586E+15 |
| 2015-12-26 | 23:00:00 | 1.981141321 | ['Imhotep'] | 7.49E+15 | -2.55925E+15 |
| 2015-12-26 | 23:30:00 | 1.981301961 | ['Imhotep'] | 1.04E+16 | -3.58013E+15 |
| 2015-12-27 | 0:00:00 | 1.9814625 | ['Khepry'] | 1.20E+16 | -4.08661E+15 |
| 2015-12-27 | 6:30:00 | 1.983552407 | ['Hapi'] | 1.83E+15 | -6.4893E+14 |
| 2015-12-27 | 10:30:00 | 1.984845625 | ['Hatmehit'] | 3.94E+15 | -1.35027E+15 |
| 2015-12-27 | 11:00:00 | 1.98493625 | ['Imhotep'] | 7.48E+15 | -2.57488E+15 |
| 2015-12-27 | 14:00:00 | 1.985964231 | ['Bastet'] | 6.95E+15 | -2.39505E+15 |
| 2015-12-27 | 16:00:00 | 1.986603019 | ['Maat'] | 7.26E+15 | -2.50911E+15 |
| 2015-12-27 | 23:30:00 | 1.989015849 | ['Khepry'] | 1.21E+16 | -4.13033E+15 |
| 2015-12-28 | 2:00:00 | 1.989813396 | ['Babi'] | 1.05E+16 | -3.57481E+15 |
| 2015-12-28 | 6:30:00 | 1.991279211 | ['Babi'] | 5.41E+15 | -1.85933E+15 |
| 2015-12-28 | 11:30:00 | 1.9929 | ['Babi'] | 1.60E+16 | -5.46693E+15 |
| 2015-12-28 | 12:00:00 | 1.99302875 | ['Imhotep'] | 2.41E+16 | -8.24737E+15 |
| 2015-12-28 | 12:30:00 | 1.993187917 | ['Imhotep'] | 3.01E+16 | -1.03E+16 |

| Date | Time | Value | Region | Col5 | Col6 |
|---|---|---|---|---|---|
| 2015-12-28 | 17:00:00 | 1.99463566 | ['Hatmehit'] | 7.15E+15 | -2.44611E+15 |
| 2015-12-28 | 17:30:00 | 1.994795 | ['Hatmehit'] | 7.84E+15 | -2.68292E+15 |
| 2015-12-28 | 18:00:00 | 1.994953585 | ['Maftet'] | 6.76E+15 | -2.31138E+15 |
| 2015-12-28 | 18:30:00 | 1.99511875 | ['Maftet'] | 6.29E+15 | -2.15192E+15 |
| 2015-12-28 | 19:00:00 | 1.995285283 | ['Seth'] | 6.66E+15 | -2.28607E+15 |
| 2015-12-28 | 19:30:00 | 1.995444906 | ['Seth'] | 8.46E+15 | -2.89253E+15 |
| 2015-12-28 | 20:00:00 | 1.99560566 | ['Seth'] | 1.33E+16 | -4.5408E+15 |
| 2015-12-28 | 20:30:00 | 1.995764423 | ['Anubis'] | 2.11E+16 | -7.20118E+15 |
| 2015-12-28 | 21:00:00 | 1.995923019 | ['Atum'] | 1.98E+16 | -6.7636E+15 |
| 2015-12-28 | 21:30:00 | 1.996083208 | ['Atum'] | 1.01E+16 | -3.47316E+15 |
| 2015-12-28 | 22:00:00 | 1.996243396 | ['Atum'] | 7.22E+15 | -2.48365E+15 |
| 2015-12-28 | 22:30:00 | 1.996401731 | ['Apis'] | 5.81E+15 | -1.99235E+15 |
| 2015-12-29 | 8:00:00 | 1.999458545 | ['Seth'] | 1.47E+16 | -5.0103E+15 |
| 2015-12-29 | 8:30:00 | 1.999624259 | ['Anubis'] | 2.37E+16 | -8.09075E+15 |
| 2015-12-29 | 13:00:00 | 2.001061556 | ['Imhotep'] | 3.35E+16 | -1.14E+16 |
| 2015-12-29 | 18:00:00 | 2.002675714 | ['Maftet'] | 9.68E+15 | -3.31481E+15 |
| 2015-12-29 | 18:30:00 | 2.002840192 | ['Maftet'] | 9.97E+15 | -3.40809E+15 |
| 2015-12-29 | 19:00:00 | 2.002933333 | ['Seth'] | 8.28E+15 | -2.87399E+15 |
| 2015-12-29 | 23:00:00 | 2.004285962 | ['Imhotep'] | 1.19E+16 | -4.07504E+15 |
| 2015-12-29 | 23:30:00 | 2.004444528 | ['Imhotep'] | 2.09E+16 | -7.12977E+15 |
| 2015-12-30 | 6:00:00 | 2.006517391 | ['Wosret'] | 1.34E+16 | -4.56809E+15 |
| 2015-12-30 | 10:00:00 | 2.007816538 | ['Atum'] | 1.53E+16 | -5.23259E+15 |
| 2015-12-30 | 20:00:00 | 2.011030577 | ['Seth'] | 1.82E+16 | -6.22417E+15 |
| 2015-12-30 | 20:30:00 | 2.011195789 | ['Anubis'] | 2.31E+16 | -7.89049E+15 |
| 2015-12-31 | 1:00:00 | 2.012643725 | ['Imhotep'] | 3.83E+16 | -1.31E+16 |
| 2015-12-31 | 6:00:00 | 2.0142568 | ['Imhotep'] | 3.88E+16 | -1.32E+16 |
| 2015-12-31 | 6:30:00 | 2.014412642 | ['Sobek'] | 2.46E+16 | -8.3849E+15 |
| 2015-12-31 | 7:00:00 | 2.014531481 | ['Hapi'] | 2.13E+16 | -7.29665E+15 |
| 2015-12-31 | 9:00:00 | 2.015236 | ['Hapi'] | 2.61E+16 | -8.9168E+15 |
| 2015-12-31 | 11:00:00 | 2.015878421 | ['Hapi'] | 2.35E+16 | -8.01327E+15 |
| 2015-12-31 | 11:30:00 | 2.015972917 | ['Imhotep'] | 2.74E+16 | -9.35426E+15 |
| 2015-12-31 | 14:00:00 | 2.016830357 | ['Imhotep'] | 3.47E+16 | -1.19E+16 |
| 2015-12-31 | 16:00:00 | 2.017505556 | ['Imhotep'] | 2.81E+16 | -9.58609E+15 |
| 2015-12-31 | 16:30:00 | 2.017624231 | ['Sobek'] | 3.79E+16 | -1.30E+16 |
| 2015-12-31 | 17:00:00 | 2.017783019 | ['Sobek'] | 5.08E+16 | -1.74E+16 |
| 2015-12-31 | 21:30:00 | 2.019231509 | ['Khonsu'] | 2.48E+16 | -8.46462E+15 |
| 2015-12-31 | 22:00:00 | 2.019397857 | ['Khonsu'] | 2.58E+16 | -8.81069E+15 |
| 2015-12-31 | 22:30:00 | 2.019562941 | ['Imhotep1'] | 2.80E+16 | -9.55975E+15 |
| 2015-12-31 | 23:00:00 | 2.019720566 | ['Imhotep'] | 2.70E+16 | -9.23354E+15 |
| 2015-12-31 | 23:30:00 | 2.019879231 | ['Imhotep'] | 2.82E+16 | -9.62918E+15 |
| 2016-01-01 | 0:00:00 | 2.020040577 | ['Imhotep'] | 2.80E+16 | -9.56974E+15 |
| 2016-01-01 | 0:30:00 | 2.020199423 | ['Bes'] | 3.15E+16 | -1.08E+16 |
| 2016-01-01 | 1:00:00 | 2.020357925 | ['Bes'] | 3.65E+16 | -1.25E+16 |
| 2016-01-01 | 1:30:00 | 2.020518113 | ['Bes'] | 3.52E+16 | -1.20E+16 |

| Date | Time | Value | Label | Col5 | Col6 |
|---|---|---|---|---|---|
| 2016-01-01 | 2:00:00 | 2.020681455 | ['Bes'] | 2.99E+16 | -1.02E+16 |
| 2016-01-01 | 4:00:00 | 2.021326038 | ['Sobek'] | 1.54E+16 | -5.26653E+15 |
| 2016-01-01 | 6:30:00 | 2.02213 | ['Geb'] | 5.20E+16 | -1.78E+16 |
| 2016-01-01 | 7:00:00 | 2.022295472 | ['Geb'] | 5.14E+16 | -1.75E+16 |
| 2016-01-01 | 7:30:00 | 2.02245566 | ['Geb'] | 4.71E+16 | -1.61E+16 |
| 2016-01-01 | 8:00:00 | 2.02261566 | ['Geb'] | 3.64E+16 | -1.24E+16 |
| 2016-01-01 | 9:00:00 | 2.02293283 | ['Bes'] | 2.90E+16 | -9.91212E+15 |
| 2016-01-01 | 9:30:00 | 2.023093019 | ['Bes'] | 2.24E+16 | -7.64817E+15 |
| 2016-01-01 | 10:00:00 | 2.023253208 | ['Imhotep1'] | 1.81E+16 | -6.17809E+15 |
| 2016-01-01 | 10:30:00 | 2.023411538 | ['Imhotep'] | 1.88E+16 | -6.42105E+15 |
| 2016-01-01 | 11:00:00 | 2.023577193 | ['Imhotep'] | 2.15E+16 | -7.35866E+15 |
| 2016-01-01 | 11:30:00 | 2.023743725 | ['Bes'] | 2.65E+16 | -9.03328E+15 |
| 2016-01-01 | 12:00:00 | 2.023900943 | ['Bes'] | 2.90E+16 | -9.89085E+15 |
| 2016-01-01 | 12:30:00 | 2.024061132 | ['Bes'] | 2.76E+16 | -9.43274E+15 |
| 2016-01-01 | 13:00:00 | 2.024221887 | ['Bes'] | 2.99E+16 | -1.02E+16 |
| 2016-01-01 | 13:30:00 | 2.024380962 | ['Bes'] | 2.95E+16 | -1.01E+16 |
| 2016-01-01 | 14:00:00 | 2.024517949 | ['Anhur'] | 2.76E+16 | -9.42797E+15 |
| 2016-01-01 | 23:00:00 | 2.027432157 | ['Bes'] | 1.64E+16 | -5.61493E+15 |
| 2016-01-01 | 23:30:00 | 2.027592075 | ['Bes'] | 1.71E+16 | -5.85975E+15 |
| 2016-01-02 | 5:30:00 | 2.029547209 | ['Hatmehit'] | 3.11E+16 | -1.06E+16 |
| 2016-01-02 | 6:00:00 | 2.029691132 | ['Sobek'] | 3.09E+16 | -1.06E+16 |
| 2016-01-02 | 7:00:00 | 2.030008302 | ['Geb'] | 3.91E+16 | -1.33E+16 |
| 2016-01-02 | 7:30:00 | 2.030168302 | ['Geb'] | 4.40E+16 | -1.50E+16 |
| 2016-01-02 | 11:00:00 | 2.0312998 | ['Imhotep'] | 1.45E+16 | -4.95025E+15 |
| 2016-01-02 | 11:30:00 | 2.031456538 | ['Imhotep'] | 1.55E+16 | -5.28402E+15 |
| 2016-01-02 | 12:00:00 | 2.031615094 | ['Imhotep'] | 1.95E+16 | -6.64484E+15 |
| 2016-01-02 | 12:30:00 | 2.031775283 | ['Bes'] | 2.27E+16 | -7.76687E+15 |
| 2016-01-02 | 13:00:00 | 2.031938545 | ['Bes'] | 2.44E+16 | -8.32854E+15 |
| 2016-01-02 | 13:30:00 | 2.032104259 | ['Bes'] | 2.34E+16 | -8.00238E+15 |
| 2016-01-02 | 14:00:00 | 2.032266038 | ['Bes'] | 2.27E+16 | -7.74599E+15 |
| 2016-01-02 | 18:00:00 | 2.033552642 | ['Sobek'] | 2.39E+16 | -8.15721E+15 |
| 2016-01-02 | 19:00:00 | 2.03387283 | ['Geb'] | 3.29E+16 | -1.12E+16 |
| 2016-01-02 | 19:30:00 | 2.034031538 | ['Geb'] | 3.58E+16 | -1.22E+16 |
| 2016-01-02 | 20:30:00 | 2.034312759 | ['Geb'] | 3.71E+16 | -1.27E+16 |
| 2016-01-03 | 0:30:00 | 2.0356362 | ['Geb'] | 1.97E+16 | -6.71494E+15 |
| 2016-01-03 | 6:00:00 | 2.037402453 | ['Sobek'] | 1.55E+16 | -5.30164E+15 |
| 2016-01-03 | 6:30:00 | 2.037538611 | ['Geb'] | 1.81E+16 | -6.18646E+15 |
| 2016-01-03 | 11:00:00 | 2.03895 | ['Bes'] | 1.58E+16 | -5.42928E+15 |
| 2016-01-03 | 16:00:00 | 2.040625636 | ['Sobek'] | 1.28E+16 | -4.3857E+15 |
| 2016-01-03 | 16:30:00 | 2.040768421 | ['Sobek'] | 1.48E+16 | -5.07394E+15 |
| 2016-01-04 | 1:30:00 | 2.043682909 | ['Bes'] | 2.41E+16 | -8.25325E+15 |
| 2016-01-04 | 2:00:00 | 2.043844808 | ['Geb'] | 2.14E+16 | -7.31095E+15 |
| 2016-01-04 | 12:00:00 | 2.047059811 | ['Bes'] | 3.01E+16 | -1.03E+16 |
| 2016-01-04 | 15:30:00 | 2.048184286 | ['Bastet'] | 2.38E+16 | -8.12346E+15 |
| 2016-01-04 | 16:00:00 | 2.048348077 | ['Wosret'] | 1.82E+16 | -6.20889E+15 |
| 2016-01-04 | 16:30:00 | 2.048506415 | ['Wosret'] | 1.59E+16 | -5.43658E+15 |
| 2016-01-04 | 21:00:00 | 2.049954808 | ['Khonsu'] | 2.92E+16 | -9.97765E+15 |
| 2016-01-04 | 21:30:00 | 2.050113396 | ['Khonsu'] | 2.53E+16 | -8.64372E+15 |
| 2016-01-04 | 22:00:00 | 2.050273396 | ['Imhotep'] | 1.59E+16 | -5.44377E+15 |
| 2016-01-04 | 23:30:00 | 2.050758113 | ['Wosret'] | 2.48E+16 | -8.48279E+15 |
| 2016-01-05 | 1:30:00 | 2.051395472 | ['Geb'] | 3.90E+16 | -1.33E+16 |
| 2016-01-05 | 3:00:00 | 2.051865897 | ['Anhur'] | 2.43E+16 | -8.31171E+15 |
| 2016-01-05 | 4:00:00 | 2.052208113 | ['Imhotep1'] | 1.36E+16 | -4.66556E+15 |
| 2016-01-05 | 4:30:00 | 2.052366923 | ['Khonsu'] | 1.98E+16 | -6.75662E+15 |
| 2016-01-05 | 6:00:00 | 2.052846226 | ['Anuket'] | 1.90E+16 | -6.48635E+15 |
| 2016-01-05 | 6:30:00 | 2.053005686 | ['Bes'] | 1.72E+16 | -5.89017E+15 |
| 2016-01-05 | 8:00:00 | 2.05349 | ['Imhotep1'] | 2.37E+16 | -8.09309E+15 |
| 2016-01-05 | 8:30:00 | 2.053654906 | ['Imhotep1'] | 2.43E+16 | -8.29332E+15 |
| 2016-01-05 | 9:30:00 | 2.053973962 | ['Bastet'] | 2.20E+16 | -7.51556E+15 |
| 2016-01-05 | 10:00:00 | 2.054132692 | ['Anhur'] | 2.27E+16 | -7.75798E+15 |
| 2016-01-05 | 11:30:00 | 2.054611509 | ['Bes'] | 1.72E+16 | -5.89986E+15 |
| 2016-01-05 | 13:00:00 | 2.055099608 | ['Anhur'] | 3.11E+16 | -1.06E+16 |
| 2016-01-05 | 18:00:00 | 2.0567075 | ['Atum'] | 2.51E+16 | -8.56081E+15 |
| 2016-01-05 | 18:30:00 | 2.056866226 | ['Khonsu'] | 2.00E+16 | -6.83088E+15 |
| 2016-01-05 | 19:00:00 | 2.056971176 | ['Khonsu'] | 1.78E+16 | -6.12722E+15 |
| 2016-01-05 | 23:00:00 | 2.058314423 | ['Bes'] | 1.78E+16 | -6.07088E+15 |
| 2016-01-05 | 23:30:00 | 2.058473019 | ['Bes'] | 2.34E+16 | -8.00531E+15 |
| 2016-01-06 | 6:00:00 | 2.060556047 | ['Atum'] | 2.91E+16 | -9.94935E+15 |
| 2016-01-06 | 6:30:00 | 2.0608 | ['Atum'] | 3.79E+16 | -1.34E+16 |
| 2016-01-06 | 8:30:00 | 2.061370566 | ['Atum'] | 3.46E+16 | -1.18E+16 |
| 2016-01-06 | 9:00:00 | 2.061530566 | ['Khonsu'] | 3.42E+16 | -1.17E+16 |
| 2016-01-06 | 9:30:00 | 2.061690962 | ['Imhotep1'] | 3.16E+16 | -1.08E+16 |
| 2016-01-06 | 10:30:00 | 2.061982778 | ['Bes'] | 2.53E+16 | -8.62967E+15 |
| 2016-01-06 | 14:30:00 | 2.063333571 | ['Bes'] | 3.67E+16 | -1.25E+16 |
| 2016-01-06 | 20:00:00 | 2.065063396 | ['Imhotep1'] | 2.00E+16 | -6.86442E+15 |
| 2016-01-06 | 20:30:00 | 2.065228571 | ['Imhotep'] | 2.65E+16 | -9.07263E+15 |
| 2016-01-06 | 21:00:00 | 2.065394151 | ['Imhotep'] | 3.49E+16 | -1.19E+16 |
| 2016-01-06 | 21:30:00 | 2.065554151 | ['Imhotep1'] | 3.70E+16 | -1.27E+16 |
| 2016-01-06 | 22:00:00 | 2.06571434 | ['Imhotep1'] | 2.75E+16 | -9.39202E+15 |
| 2016-01-07 | 1:00:00 | 2.066675789 | ['Anhur'] | 1.30E+16 | -4.43585E+15 |
| 2016-01-07 | 3:30:00 | 2.067479423 | ['Neith'] | 3.71E+16 | -1.27E+16 |
| 2016-01-07 | 4:00:00 | 2.067638302 | ['Wosret'] | 3.11E+16 | -1.06E+16 |
| 2016-01-07 | 4:30:00 | 2.067798491 | ['Wosret'] | 2.49E+16 | -8.50249E+15 |
| 2016-01-07 | 5:00:00 | 2.067958491 | ['Wosret'] | 2.71E+16 | -9.2683E+15 |
| 2016-01-07 | 5:30:00 | 2.068124286 | ['Wosret'] | 2.58E+16 | -8.82553E+15 |
| 2016-01-07 | 6:00:00 | 2.068289245 | ['Anuket'] | 2.20E+16 | -7.49711E+15 |
| 2016-01-07 | 6:30:00 | 2.068447885 | ['Atum'] | 2.03E+16 | -6.93529E+15 |
| 2016-01-07 | 8:30:00 | 2.069085192 | ['Atum'] | 2.91E+16 | -9.9495E+15 |
| 2016-01-07 | 9:00:00 | 2.069243774 | ['Khonsu'] | 3.80E+16 | -1.30E+16 |
| 2016-01-07 | 9:30:00 | 2.069407091 | ['Imhotep1'] | 4.21E+16 | -1.44E+16 |
| 2016-01-07 | 10:00:00 | 2.069572963 | ['Imhotep1'] | 2.84E+16 | -9.68869E+15 |
| 2016-01-07 | 16:00:00 | 2.071533043 | ['Imhotep1'] | 2.07E+16 | -7.07832E+15 |
| 2016-01-07 | 16:30:00 | 2.07166881 | ['Anuket'] | 2.58E+16 | -8.79778E+15 |
| 2016-01-07 | 17:00:00 | 2.071818679 | ['Anuket'] | 2.25E+16 | -7.68537E+15 |
| 2016-01-07 | 21:30:00 | 2.0732666 | ['Imhotep'] | 3.74E+16 | -1.28E+16 |
| 2016-01-08 | 23:00:00 | 2.081461346 | ['Imhotep'] | 9.95E+15 | -3.41384E+15 |
| 2016-01-08 | 23:30:00 | 2.081627193 | ['Aten'] | 5.02E+15 | -1.75658E+15 |
| 2016-01-09 | 6:00:00 | 2.083717736 | ['Anuket'] | 8.74E+15 | -2.99758E+15 |
| 2016-01-09 | 6:30:00 | 2.083877925 | ['Anuket'] | 5.34E+15 | -1.85706E+15 |
| 2016-01-09 | 7:00:00 | 2.084036346 | ['Anubis'] | 3.02E+15 | -1.0449E+15 |
| 2016-01-09 | 7:30:00 | 2.084195094 | ['Atum'] | 4.32E+15 | -1.51919E+15 |
| 2016-01-09 | 8:00:00 | 2.08436 | ['Ash'] | 5.79E+15 | -1.99484E+15 |
| 2016-01-09 | 9:00:00 | 2.084685849 | ['Imhotep'] | 1.72E+16 | -5.89654E+15 |
| 2016-01-09 | 9:30:00 | 2.084845849 | ['Imhotep'] | 2.91E+16 | -9.93446E+15 |
| 2016-01-09 | 10:00:00 | 2.085004615 | ['Imhotep'] | 2.95E+16 | -1.01E+16 |
| 2016-01-09 | 10:30:00 | 2.085163019 | ['Imhotep'] | 2.13E+16 | -7.27459E+15 |
| 2016-01-09 | 11:00:00 | 2.085323208 | ['Imhotep'] | 1.25E+16 | -4.26815E+15 |
| 2016-01-09 | 11:30:00 | 2.085481731 | ['Aten'] | 4.63E+15 | -1.59793E+15 |
| 2016-01-09 | 12:00:00 | 2.085642353 | ['Babi'] | 7.80E+15 | -2.68506E+15 |
| 2016-01-09 | 17:00:00 | 2.087254286 | ['Anuket'] | 1.26E+16 | -4.31338E+15 |
| 2016-01-09 | 17:30:00 | 2.087419434 | ['Anuket'] | 1.34E+16 | -4.60178E+15 |
| 2016-01-09 | 18:00:00 | 2.087578077 | ['Anuket'] | 8.08E+15 | -2.77042E+15 |
| 2016-01-09 | 18:30:00 | 2.087736538 | ['Anuket'] | 5.86E+15 | -2.00723E+15 |
| 2016-01-09 | 19:00:00 | 2.087896792 | ['Seth'] | 3.52E+15 | -1.26049E+15 |
| 2016-01-10 | 0:30:00 | 2.089661509 | ['Seth'] | 4.13E+15 | -1.44637E+15 |
| 2016-01-10 | 5:30:00 | 2.091272308 | ['Seth'] | 9.82E+15 | -3.38471E+15 |
| 2016-01-10 | 6:00:00 | 2.091430755 | ['Anuket'] | 7.57E+15 | -2.60209E+15 |
| 2016-01-10 | 10:30:00 | 2.092874615 | ['Bes'] | 1.88E+16 | -6.42791E+15 |
| 2016-01-10 | 13:00:00 | 2.093678113 | ['Anhur'] | 4.54E+15 | -1.6098E+15 |
| 2016-01-10 | 23:00:00 | 2.096895789 | ['Imhotep'] | 1.17E+16 | -4.02099E+15 |
| 2016-01-10 | 23:30:00 | 2.097061154 | ['Ash'] | 4.14E+15 | -1.42917E+15 |
| 2016-01-11 | 11:00:00 | 2.100825 | ['Hatmehit'] | 2.63E+16 | -1.15E+16 |
| 2016-01-15 | 20:30:00 | 2.134631731 | ['Seth'] | 9.44E+15 | -3.23911E+15 |
| 2016-01-15 | 21:30:00 | 2.134948491 | ['Seth'] | 6.31E+15 | -2.17554E+15 |
| 2016-01-16 | 8:30:00 | 2.138483962 | ['Anuket'] | 1.22E+16 | -4.19798E+15 |
| 2016-01-16 | 20:00:00 | 2.142166346 | ['Anuket'] | 1.29E+16 | -4.41845E+15 |
| 2016-01-16 | 20:30:00 | 2.142325094 | ['Hapi'] | 1.05E+16 | -3.60424E+15 |
| 2016-01-17 | 0:30:00 | 2.143610377 | ['Ash'] | 5.85E+15 | -2.0143E+15 |
| 2016-01-17 | 6:30:00 | 2.145542264 | ['Serqet'] | 7.79E+15 | -2.71832E+15 |
| 2016-01-17 | 11:00:00 | 2.146912857 | ['Ash'] | 3.46E+15 | -1.40109E+15 |
| 2016-01-18 | 11:30:00 | 2.154864667 | ['Ash'] | 1.39E+16 | -4.77458E+15 |
| 2016-01-18 | 12:00:00 | 2.154995636 | ['Imhotep'] | 1.50E+16 | -5.13696E+15 |
| 2016-01-18 | 12:30:00 | 2.155159615 | ['Imhotep'] | 1.08E+16 | -3.70727E+15 |
| 2016-01-18 | 21:30:00 | 2.158044615 | ['Ash'] | 4.41E+15 | -1.55773E+15 |
| 2016-01-18 | 22:30:00 | 2.158360566 | ['Imhotep1'] | 9.20E+15 | -3.16564E+15 |
| 2016-01-18 | 23:00:00 | 2.15852 | ['Imhotep'] | 1.63E+16 | -5.5825E+15 |
| 2016-01-18 | 23:30:00 | 2.158680189 | ['Imhotep'] | 1.85E+16 | -6.31918E+15 |
| 2016-01-19 | 0:00:00 | 2.158837885 | ['Imhotep'] | 1.86E+16 | -6.35894E+15 |
| 2016-01-19 | 0:30:00 | 2.158995849 | ['Imhotep'] | 1.57E+16 | -5.36468E+15 |
| 2016-01-19 | 1:00:00 | 2.159161429 | ['Khepry'] | 1.40E+16 | -4.78377E+15 |
| 2016-01-19 | 3:00:00 | 2.159802885 | ['Aker'] | 6.60E+15 | -2.25947E+15 |
| 2016-01-19 | 3:30:00 | 2.159961132 | ['Bastet'] | 3.72E+15 | -1.29525E+15 |
| 2016-01-19 | 4:00:00 | 2.160120943 | ['Maat'] | 4.01E+15 | -1.41055E+15 |
| 2016-01-19 | 4:30:00 | 2.160280189 | ['Hatmehit'] | 5.20E+15 | -1.7894E+15 |
| 2016-01-19 | 5:00:00 | 2.16044 | ['Hatmehit'] | 4.38E+15 | -1.51963E+15 |
| 2016-01-19 | 5:30:00 | 2.160605536 | ['Nut'] | 6.12E+15 | -2.10701E+15 |
| 2016-01-19 | 7:00:00 | 2.161086226 | ['Anuket'] | 2.04E+16 | -6.98001E+15 |
| 2016-01-19 | 7:30:00 | 2.161245094 | ['Anuket'] | 1.61E+16 | -5.48221E+15 |
| 2016-01-19 | 8:00:00 | 2.161403654 | ['Seth'] | 1.30E+16 | -4.44255E+15 |
| 2016-01-19 | 8:30:00 | 2.161561887 | ['Anubis'] | 7.39E+15 | -2.53658E+15 |
| 2016-01-19 | 9:00:00 | 2.161720943 | ['Atum'] | 4.85E+15 | -1.6699E+15 |
| 2016-01-19 | 10:30:00 | 2.162210377 | ['Imhotep'] | 1.11E+16 | -3.81589E+15 |
| 2016-01-19 | 11:00:00 | 2.162369412 | ['Imhotep'] | 2.02E+16 | -6.90439E+15 |
| 2016-01-19 | 11:30:00 | 2.162525686 | ['Imhotep'] | 2.48E+16 | -8.48481E+15 |
| 2016-01-19 | 12:00:00 | 2.162686226 | ['Imhotep'] | 2.60E+16 | -8.88802E+15 |
| 2016-01-19 | 12:30:00 | 2.162846226 | ['Imhotep'] | 1.96E+16 | -6.69849E+15 |
| 2016-01-19 | 13:00:00 | 2.162987143 | ['Imhotep'] | 1.47E+16 | -5.01676E+15 |
| 2016-01-19 | 18:00:00 | 2.164605192 | ['Serqet'] | 6.46E+15 | -2.22885E+15 |
| 2016-01-19 | 19:00:00 | 2.1648625 | ['Anuket'] | 2.24E+16 | -7.83238E+15 |
| 2016-01-19 | 22:30:00 | 2.166084444 | ['Anuket'] | 1.57E+16 | -5.43404E+15 |
| 2016-01-19 | 23:00:00 | 2.166211346 | ['Imhotep'] | 2.41E+16 | -8.23773E+15 |
| 2016-01-19 | 23:30:00 | 2.166369245 | ['Imhotep'] | 3.39E+16 | -1.16E+16 |
| 2016-01-20 | 6:30:00 | 2.168617547 | ['Seth'] | 1.69E+16 | -5.76616E+15 |
| 2016-01-20 | 10:00:00 | 2.169732857 | ['Ash'] | 6.38E+15 | -2.20623E+15 |
| 2016-01-20 | 10:30:00 | 2.169898302 | ['Imhotep1'] | 1.15E+16 | -3.95578E+15 |
| 2016-01-20 | 11:00:00 | 2.17005717 | ['Ash'] | 2.12E+16 | -7.25457E+15 |
| 2016-01-20 | 11:30:00 | 2.170217692 | ['Ash'] | 2.80E+16 | -9.57493E+15 |
| 2016-01-20 | 14:30:00 | 2.171174386 | ['Hathor'] | 8.92E+15 | -3.06045E+15 |
| 2016-01-20 | 20:00:00 | 2.172937091 | ['Anuket'] | 1.52E+16 | -5.20864E+15 |
| 2016-01-20 | 20:30:00 | 2.173100189 | ['Hapi'] | 8.02E+15 | -2.74508E+15 |
| 2016-01-21 | 1:00:00 | 2.174536538 | ['Khepry'] | 1.67E+16 | -5.71661E+15 |
| 2016-01-21 | 6:00:00 | 2.176137885 | ['Serqet'] | 9.65E+15 | -3.31295E+15 |
| 2016-01-21 | 11:00:00 | 2.177735636 | ['Serqet'] | 3.64E+16 | -1.24E+16 |
| 2016-01-21 | 11:30:00 | 2.177856522 | ['Imhotep'] | 3.72E+16 | -1.27E+16 |
| 2016-01-21 | 16:00:00 | 2.179371471 | ['Hatmehit'] | 6.72E+15 | -2.3709E+15 |
| 2016-01-21 | 21:30:00 | 2.181091915 | ['Atum'] | 5.83E+15 | -2.02488E+15 |

| Date | Time | Value | Region | A | B |
|---|---|---|---|---|---|
| 2016-01-22 | 2:30:00 | 2.182701455 | ['Khepry'] | 1.49E+16 | -5.12861E+15 |
| 2016-01-22 | 3:00:00 | 2.182863077 | ['Aker'] | 1.09E+16 | -3.71803E+15 |
| 2016-01-22 | 7:30:00 | 2.184301538 | ['Hatmehit'] | 2.97E+16 | -1.01E+16 |
| 2016-01-22 | 8:00:00 | 2.184402941 | ['Anuket'] | 2.75E+16 | -9.42485E+15 |
| 2016-01-22 | 12:30:00 | 2.185919231 | ['Anuket'] | 2.47E+16 | -8.44757E+15 |
| 2016-01-22 | 13:00:00 | 2.18605717 | ['Imhotep'] | 1.80E+16 | -6.15849E+15 |
| 2016-01-22 | 13:30:00 | 2.186215849 | ['Imhotep'] | 1.07E+16 | -3.66182E+15 |
| 2016-01-22 | 18:00:00 | 2.187632308 | ['Wosret'] | 1.46E+16 | -4.99191E+15 |
| 2016-01-22 | 22:30:00 | 2.189159167 | ['Wosret'] | 1.66E+16 | -5.72873E+15 |
| 2016-01-22 | 23:00:00 | 2.189252453 | ['Imhotep1'] | 2.73E+16 | -9.32628E+15 |
| 2016-01-22 | 23:30:00 | 2.189415273 | ['Imhotep'] | 4.09E+16 | -1.40E+16 |
| 2016-01-23 | 0:00:00 | 2.189545357 | ['Imhotep'] | 4.99E+16 | -1.70E+16 |
| 2016-01-23 | 9:00:00 | 2.192525 | ['Imhotep'] | 1.43E+16 | -4.99446E+15 |
| 2016-01-23 | 9:30:00 | 2.192616604 | ['Atum'] | 1.25E+16 | -4.28942E+15 |
| 2016-01-23 | 10:00:00 | 2.192776226 | ['Khonsu'] | 1.71E+16 | -5.83218E+15 |
| 2016-01-23 | 14:00:00 | 2.194115833 | ['Hatmehit'] | 1.84E+16 | -6.35008E+15 |
| 2016-01-23 | 15:00:00 | 2.194373208 | ['Aker'] | 1.65E+16 | -5.6332E+15 |
| 2016-01-23 | 16:00:00 | 2.194689216 | ['Hapi'] | 1.11E+16 | -3.79033E+15 |
| 2016-01-23 | 16:30:00 | 2.194848113 | ['Sobek'] | 1.14E+16 | -3.9074E+15 |
| 2016-01-23 | 17:00:00 | 2.195010182 | ['Wosret'] | 1.83E+16 | -6.2404E+15 |
| 2016-01-23 | 17:30:00 | 2.195176604 | ['Wosret'] | 2.23E+16 | -7.63243E+15 |
| 2016-01-23 | 18:00:00 | 2.195336038 | ['Wosret'] | 1.86E+16 | -6.35597E+15 |
| 2016-01-23 | 19:00:00 | 2.195653269 | ['Neith'] | 2.84E+16 | -9.68869E+15 |
| 2016-01-23 | 19:30:00 | 2.195810377 | ['Anuket'] | 2.69E+16 | -9.17555E+15 |
| 2016-01-23 | 20:00:00 | 2.195969423 | ['Anuket'] | 2.18E+16 | -7.45541E+15 |
| 2016-01-23 | 20:30:00 | 2.196128077 | ['Hapi'] | 2.03E+16 | -6.94033E+15 |
| 2016-01-23 | 21:00:00 | 2.196286731 | ['Anubis'] | 1.86E+16 | -6.3495E+15 |
| 2016-01-23 | 21:30:00 | 2.196451053 | ['Atum'] | 1.60E+16 | -5.46756E+15 |
| 2016-01-23 | 22:00:00 | 2.196615769 | ['Khonsu'] | 1.92E+16 | -6.57558E+15 |
| 2016-01-23 | 23:30:00 | 2.197091698 | ['Imhotep'] | 3.52E+16 | -1.20E+16 |
| 2016-01-24 | 0:00:00 | 2.197249412 | ['Imhotep'] | 4.19E+16 | -1.43E+16 |
| 2016-01-24 | 0:30:00 | 2.197407358 | ['Imhotep'] | 3.07E+16 | -1.05E+16 |
| 2016-01-24 | 1:00:00 | 2.197566415 | ['Imhotep'] | 2.06E+16 | -7.05292E+15 |
| 2016-01-24 | 1:30:00 | 2.197726038 | ['Imhotep'] | 1.51E+16 | -5.17121E+15 |
| 2016-01-24 | 2:00:00 | 2.19789 | ['Khepry'] | 1.56E+16 | -5.31998E+15 |
| 2016-01-24 | 2:30:00 | 2.198054528 | ['Bes'] | 1.59E+16 | -5.4426E+15 |
| 2016-01-24 | 3:00:00 | 2.198213725 | ['Khepry'] | 1.76E+16 | -6.03438E+15 |
| 2016-01-24 | 3:30:00 | 2.198369811 | ['Aker'] | 1.72E+16 | -5.86152E+15 |
| 2016-01-24 | 4:00:00 | 2.198529245 | ['Aker'] | 1.35E+16 | -4.62822E+15 |
| 2016-01-24 | 4:30:00 | 2.198688491 | ['Sobek'] | 1.38E+16 | -4.73163E+15 |
| 2016-01-24 | 5:30:00 | 2.199004151 | ['Wosret'] | 2.05E+16 | -7.00478E+15 |
| 2016-01-24 | 6:00:00 | 2.19916283 | ['Wosret'] | 2.10E+16 | -7.17004E+15 |
| 2016-01-24 | 6:30:00 | 2.199327679 | ['Wosret'] | 2.08E+16 | -7.11617E+15 |
| 2016-01-24 | 9:00:00 | 2.200131481 | ['Sobek'] | 2.19E+16 | -7.48169E+15 |
| 2016-01-24 | 9:30:00 | 2.200290577 | ['Atum'] | 2.07E+16 | -7.08368E+15 |
| 2016-01-24 | 10:00:00 | 2.200448302 | ['Khonsu'] | 2.07E+16 | -7.05552E+15 |
| 2016-01-24 | 11:00:00 | 2.200701 | ['Khonsu'] | 2.28E+16 | -7.80896E+15 |
| 2016-01-24 | 12:00:00 | 2.201085094 | ['Ash'] | 4.07E+16 | -1.39E+16 |
| 2016-01-24 | 12:30:00 | 2.201244528 | ['Ash'] | 3.37E+16 | -1.15E+16 |
| 2016-01-24 | 13:00:00 | 2.201401731 | ['Ash'] | 2.45E+16 | -8.37751E+15 |
| 2016-01-24 | 13:30:00 | 2.201559811 | ['Khepry'] | 1.63E+16 | -5.56706E+15 |
| 2016-01-24 | 14:00:00 | 2.201724286 | ['Khepry'] | 1.65E+16 | -5.6296E+15 |
| 2016-01-24 | 15:30:00 | 2.202234167 | ['Anhur'] | 2.10E+16 | -7.19683E+15 |
| 2016-01-24 | 16:00:00 | 2.202364038 | ['Anhur'] | 1.91E+16 | -6.5345E+15 |
| 2016-01-24 | 18:30:00 | 2.203169423 | ['Anhur'] | 2.72E+16 | -9.28755E+15 |
| 2016-01-24 | 19:00:00 | 2.203325577 | ['Anhur'] | 2.79E+16 | -9.54205E+15 |
| 2016-01-24 | 20:00:00 | 2.203642642 | ['Anuket'] | 2.06E+16 | -7.05077E+15 |
| 2016-01-24 | 20:30:00 | 2.203801321 | ['Anuket'] | 2.57E+16 | -8.79762E+15 |
| 2016-01-24 | 23:30:00 | 2.204763962 | ['Imhotep'] | 2.65E+16 | -9.05848E+15 |
| 2016-01-25 | 0:00:00 | 2.204922642 | ['Imhotep'] | 3.28E+16 | -1.12E+16 |
| 2016-01-25 | 0:30:00 | 2.205082308 | ['Imhotep'] | 2.92E+16 | -9.98519E+15 |
| 2016-01-25 | 1:00:00 | 2.205239615 | ['Imhotep'] | 2.96E+16 | -1.01E+16 |
| 2016-01-25 | 1:30:00 | 2.20539717 | ['Imhotep'] | 2.64E+16 | -9.00312E+15 |
| 2016-01-25 | 2:00:00 | 2.205556415 | ['Imhotep'] | 1.89E+16 | -6.44049E+15 |
| 2016-01-25 | 12:00:00 | 2.208749245 | ['Imhotep'] | 2.93E+16 | -1.00E+16 |
| 2016-01-25 | 12:30:00 | 2.2089044 | ['Imhotep'] | 2.89E+16 | -9.88847E+15 |
| 2016-01-25 | 16:30:00 | 2.210222813 | ['Hatmehit'] | 5.11E+16 | -1.75E+16 |
| 2016-01-25 | 17:00:00 | 2.21032697 | ['Anhur'] | 5.30E+16 | -1.81E+16 |
| 2016-01-25 | 22:00:00 | 2.211949808 | ['Anubis'] | 1.73E+16 | -5.90549E+15 |
| 2016-01-25 | 22:30:00 | 2.212106981 | ['Atum'] | 1.53E+16 | -5.2207E+15 |
| 2016-01-26 | 3:00:00 | 2.213542885 | ['Bes'] | 1.71E+16 | -5.83965E+15 |
| 2016-01-26 | 8:00:00 | 2.215140377 | ['Neith'] | 1.87E+16 | -6.41109E+15 |
| 2016-01-26 | 13:00:00 | 2.216726444 | ['Imhotep'] | 2.70E+16 | -9.2063E+15 |
| 2016-01-26 | 17:30:00 | 2.218170244 | ['Wosret'] | 4.96E+16 | -1.70E+16 |
| 2016-01-26 | 18:00:00 | 2.218334808 | ['Anhur'] | 3.78E+16 | -1.29E+16 |
| 2016-01-26 | 18:30:00 | 2.218492264 | ['Sobek'] | 2.98E+16 | -1.02E+16 |
| 2016-01-26 | 23:30:00 | 2.220086038 | ['Khonsu'] | 1.51E+16 | -5.16894E+15 |
| 2016-01-27 | 6:00:00 | 2.222143043 | ['Anhur'] | 3.00E+16 | -1.02E+16 |
| 2016-01-27 | 7:00:00 | 2.22247875 | ['Hatmehit'] | 2.61E+16 | -8.91967E+15 |
| 2016-01-27 | 7:30:00 | 2.222642642 | ['Wosret'] | 2.15E+16 | -7.33352E+15 |
| 2016-01-27 | 8:00:00 | 2.222799808 | ['Wosret'] | 2.20E+16 | -7.51436E+15 |
| 2016-01-27 | 8:30:00 | 2.222957358 | ['Wosret'] | 1.93E+16 | -6.59074E+15 |
| 2016-01-27 | 9:00:00 | 2.223116226 | ['Maftet'] | 1.97E+16 | -6.7293E+15 |
| 2016-01-27 | 9:30:00 | 2.223275283 | ['Anuket'] | 1.19E+16 | -4.08634E+15 |
| 2016-01-27 | 10:00:00 | 2.223432308 | ['Geb'] | 9.42E+15 | -3.22163E+15 |
| 2016-01-27 | 10:30:00 | 2.223589811 | ['Geb'] | 8.51E+15 | -2.91112E+15 |
| 2016-01-27 | 11:00:00 | 2.223752 | ['Imhotep1'] | 1.12E+16 | -3.83847E+15 |
| 2016-01-27 | 12:30:00 | 2.224235849 | ['Atum'] | 2.32E+16 | -7.9121E+15 |
| 2016-01-27 | 13:30:00 | 2.224550566 | ['Khonsu'] | 1.93E+16 | -6.60314E+15 |
| 2016-01-27 | 14:00:00 | 2.224709434 | ['Bes'] | 1.51E+16 | -5.16889E+15 |
| 2016-01-27 | 14:30:00 | 2.224868302 | ['Imhotep1'] | 1.57E+16 | -5.36857E+15 |
| 2016-01-27 | 15:00:00 | 2.225025577 | ['Imhotep'] | 8.95E+15 | -3.0593E+15 |
| 2016-01-27 | 15:30:00 | 2.225189474 | ['Bes'] | 1.11E+16 | -3.77949E+15 |
| 2016-01-27 | 16:00:00 | 2.225354038 | ['Bes'] | 1.34E+16 | -4.58737E+15 |
| 2016-01-27 | 16:30:00 | 2.225511321 | ['Bes'] | 1.36E+16 | -4.66339E+15 |
| 2016-01-27 | 17:00:00 | 2.22567 | ['Bes'] | 2.50E+16 | -8.52918E+15 |
| 2016-01-27 | 18:00:00 | 2.225986154 | ['Anhur'] | 2.65E+16 | -9.06717E+15 |
| 2016-01-27 | 18:30:00 | 2.226143585 | ['Anhur'] | 2.26E+16 | -7.72171E+15 |
| 2016-01-27 | 19:00:00 | 2.226302453 | ['Anhur'] | 1.99E+16 | -6.81485E+15 |
| 2016-01-27 | 19:30:00 | 2.226461321 | ['Anhur'] | 2.18E+16 | -7.46415E+15 |
| 2016-01-27 | 21:00:00 | 2.226946731 | ['Sobek'] | 1.47E+16 | -5.02372E+15 |
| 2016-01-27 | 21:30:00 | 2.227103962 | ['Wosret'] | 1.45E+16 | -4.95162E+15 |
| 2016-01-27 | 22:00:00 | 2.22726283 | ['Atum'] | 1.28E+16 | -4.39373E+15 |
| 2016-01-27 | 22:30:00 | 2.227421509 | ['Wosret'] | 1.20E+16 | -4.11827E+15 |
| 2016-01-27 | 23:00:00 | 2.227578846 | ['Khonsu'] | 9.90E+15 | -3.3937E+15 |
| 2016-01-27 | 23:30:00 | 2.227736226 | ['Anuket'] | 1.19E+16 | -4.08251E+15 |
| 2016-01-28 | 1:30:00 | 2.228383077 | ['Imhotep'] | 1.69E+16 | -5.78757E+15 |
| 2016-01-28 | 2:00:00 | 2.228539423 | ['Atum'] | 1.62E+16 | -5.53644E+15 |
| 2016-01-28 | 2:30:00 | 2.228696604 | ['Bes'] | 1.43E+16 | -4.87343E+15 |
| 2016-01-28 | 3:00:00 | 2.228855283 | ['Khonsu'] | 1.39E+16 | -4.75278E+15 |
| 2016-01-28 | 3:30:00 | 2.229014151 | ['Bes'] | 1.33E+16 | -4.54997E+15 |
| 2016-01-28 | 4:00:00 | 2.229171346 | ['Imhotep'] | 1.35E+16 | -4.61362E+15 |
| 2016-01-28 | 4:30:00 | 2.229332182 | ['Anhur'] | 1.27E+16 | -4.32726E+15 |
| 2016-01-28 | 5:00:00 | 2.229496296 | ['Ash'] | 1.39E+16 | -4.76063E+15 |
| 2016-01-28 | 6:00:00 | 2.22981566 | ['Bes'] | 2.03E+16 | -6.94559E+15 |
| 2016-01-28 | 6:30:00 | 2.229974528 | ['Anhur'] | 2.28E+16 | -7.77655E+15 |
| 2016-01-28 | 7:00:00 | 2.230130208 | ['Anhur'] | 2.57E+16 | -8.77686E+15 |
| 2016-01-28 | 7:30:00 | 2.230287736 | ['Bastet'] | 2.44E+16 | -8.33326E+15 |
| 2016-01-28 | 8:00:00 | 2.230445962 | ['Wosret'] | 2.40E+16 | -8.19303E+15 |
| 2016-01-28 | 13:00:00 | 2.232041887 | ['Khonsu'] | 1.92E+16 | -6.5577E+15 |
| 2016-01-28 | 16:30:00 | 2.233159231 | ['Imhotep'] | 1.18E+16 | -4.03399E+15 |
| 2016-01-28 | 17:00:00 | 2.233316604 | ['Imhotep'] | 1.00E+16 | -3.42087E+15 |
| 2016-01-28 | 18:30:00 | 2.233802778 | ['Hatmehit'] | 2.05E+16 | -7.03965E+15 |
| 2016-01-28 | 21:00:00 | 2.234648333 | ['Hatmehit'] | 2.17E+16 | -7.43384E+15 |
| 2016-01-28 | 21:30:00 | 2.234745 | ['Wosret'] | 2.30E+16 | -7.86838E+15 |
| 2016-01-29 | 2:30:00 | 2.236350769 | ['Khonsu'] | 1.35E+16 | -4.62533E+15 |
| 2016-01-29 | 3:00:00 | 2.236507925 | ['Imhotep1'] | 1.41E+16 | -4.81727E+15 |
| 2016-01-29 | 3:30:00 | 2.236643793 | ['Imhotep'] | 1.51E+16 | -5.15336E+15 |
| 2016-01-29 | 7:30:00 | 2.237949574 | ['Hatmehit'] | 2.44E+16 | -8.34332E+15 |
| 2016-01-29 | 13:30:00 | 2.239849057 | ['Hatmehit'] | 1.45E+16 | -4.94666E+15 |
| 2016-01-29 | 14:00:00 | 2.240006154 | ['Atum'] | 1.59E+16 | -5.43692E+15 |
| 2016-01-29 | 14:30:00 | 2.240169825 | ['Khonsu'] | 1.74E+16 | -5.94592E+15 |
| 2016-01-29 | 15:00:00 | 2.240333846 | ['Khonsu'] | 1.27E+16 | -4.34298E+15 |
| 2016-01-29 | 15:30:00 | 2.240490943 | ['Imhotep1'] | 1.54E+16 | -5.26617E+15 |
| 2016-01-29 | 16:00:00 | 2.240649623 | ['Imhotep'] | 1.94E+16 | -6.64244E+15 |
| 2016-01-29 | 16:30:00 | 2.240808302 | ['Imhotep'] | 1.74E+16 | -5.95854E+15 |
| 2016-01-29 | 17:00:00 | 2.240965385 | ['Imhotep'] | 1.31E+16 | -4.50099E+15 |
| 2016-01-29 | 17:30:00 | 2.241122642 | ['Imhotep'] | 1.37E+16 | -4.69266E+15 |
| 2016-01-29 | 18:00:00 | 2.241282115 | ['Khepry'] | 1.65E+16 | -5.65482E+15 |
| 2016-01-29 | 18:30:00 | 2.241439811 | ['Aten'] | 1.14E+16 | -3.88947E+15 |
| 2016-01-29 | 19:00:00 | 2.241603929 | ['Khepry'] | 1.29E+16 | -4.39447E+15 |
| 2016-01-29 | 21:00:00 | 2.242240377 | ['Bastet'] | 2.87E+16 | -9.8139E+15 |
| 2016-01-29 | 21:30:00 | 2.242398077 | ['Wosret'] | 2.47E+16 | -8.42395E+15 |
| 2016-01-29 | 22:00:00 | 2.242556154 | ['Wosret'] | 2.07E+16 | -7.0809E+15 |
| 2016-01-29 | 22:30:00 | 2.242713208 | ['Wosret'] | 1.88E+16 | -6.43244E+15 |
| 2016-01-29 | 23:00:00 | 2.242873333 | ['Maftet'] | 1.50E+16 | -5.12966E+15 |
| 2016-01-29 | 23:30:00 | 2.243037455 | ['Maftet'] | 1.92E+16 | -6.56014E+15 |
| 2016-01-30 | 5:30:00 | 2.244982667 | ['Maftet'] | 2.05E+16 | -7.01839E+15 |
| 2016-01-30 | 6:00:00 | 2.245104717 | ['Bes'] | 1.96E+16 | -6.69254E+15 |
| 2016-01-30 | 6:30:00 | 2.245263396 | ['Anhur'] | 1.35E+16 | -4.63339E+15 |
| 2016-01-30 | 7:00:00 | 2.245420577 | ['Anhur'] | 1.64E+16 | -5.6085E+15 |
| 2016-01-30 | 7:30:00 | 2.245579608 | ['Anhur'] | 2.29E+16 | -7.82712E+15 |
| 2016-01-30 | 8:00:00 | 2.245743929 | ['Sobek'] | 3.24E+16 | -1.11E+16 |
| 2016-01-30 | 8:30:00 | 2.245906731 | ['Neith'] | 3.63E+16 | -1.24E+16 |
| 2016-01-30 | 9:00:00 | 2.246063774 | ['Wosret'] | 3.28E+16 | -1.12E+16 |
| 2016-01-30 | 9:30:00 | 2.246222453 | ['Wosret'] | 2.48E+16 | -8.46457E+15 |
| 2016-01-30 | 10:00:00 | 2.246379615 | ['Neith'] | 2.22E+16 | -7.59426E+15 |
| 2016-01-30 | 10:30:00 | 2.246535769 | ['Sobek'] | 1.73E+16 | -5.91361E+15 |
| 2016-01-30 | 11:00:00 | 2.246695094 | ['Anuket'] | 1.58E+16 | -5.38362E+15 |
| 2016-01-30 | 11:30:00 | 2.246853774 | ['Hapi'] | 1.75E+16 | -5.99622E+15 |
| 2016-01-30 | 12:00:00 | 2.247010577 | ['Seth'] | 1.62E+16 | -5.54673E+15 |
| 2016-01-30 | 12:30:00 | 2.247174912 | ['Anubis'] | 1.29E+16 | -4.40312E+15 |
| 2016-01-30 | 13:00:00 | 2.247338462 | ['Atum'] | 1.57E+16 | -5.37254E+15 |
| 2016-01-30 | 13:30:00 | 2.247495283 | ['Khonsu'] | 1.63E+16 | -5.57252E+15 |
| 2016-01-30 | 14:00:00 | 2.247653269 | ['Khonsu'] | 1.54E+16 | -5.2673E+15 |
| 2016-01-30 | 14:30:00 | 2.247812642 | ['Khonsu'] | 1.41E+16 | -4.80491E+15 |
| 2016-01-30 | 15:00:00 | 2.247969423 | ['Imhotep'] | 1.21E+16 | -4.12206E+15 |
| 2016-01-30 | 15:30:00 | 2.248126604 | ['Imhotep'] | 1.41E+16 | -4.80145E+15 |
| 2016-01-30 | 16:00:00 | 2.248284906 | ['Imhotep'] | 1.51E+16 | -5.1531E+15 |
| 2016-01-30 | 16:30:00 | 2.248445185 | ['Imhotep'] | 1.61E+16 | -5.49205E+15 |
| 2016-01-30 | 17:00:00 | 2.248609455 | ['Imhotep'] | 1.61E+16 | -5.49645E+15 |
| 2016-01-30 | 17:30:00 | 2.248770943 | ['Bes'] | 2.15E+16 | -7.33187E+15 |
| 2016-01-30 | 18:00:00 | 2.248928077 | ['Bes'] | 2.40E+16 | -8.18769E+15 |
| 2016-01-30 | 18:30:00 | 2.249085283 | ['Bes'] | 1.33E+16 | -4.55011E+15 |
| 2016-01-30 | 19:00:00 | 2.249243269 | ['Anhur'] | 1.61E+16 | -5.48533E+15 |
| 2016-01-30 | 19:30:00 | 2.249402264 | ['Anhur'] | 2.26E+16 | -7.72178E+15 |
| 2016-01-30 | 23:00:00 | 2.250519434 | ['Anhur'] | 1.44E+16 | -4.93814E+15 |
| 2016-01-30 | 23:30:00 | 2.250678113 | ['Geb'] | 1.60E+16 | -5.46262E+15 |
| 2016-01-31 | 0:30:00 | 2.250993654 | ['Khonsu'] | 1.47E+16 | -5.01668E+15 |
| 2016-01-31 | 3:00:00 | 2.251786667 | ['Ash'] | 1.44E+16 | -4.93463E+15 |
| 2016-01-31 | 3:30:00 | 2.251948727 | ['Ash'] | 1.32E+16 | -4.52721E+15 |
| 2016-01-31 | 4:00:00 | 2.252113208 | ['Imhotep'] | 1.76E+16 | -6.02104E+15 |
| 2016-01-31 | 6:30:00 | 2.252901154 | ['Bes'] | 1.25E+16 | -4.29472E+15 |
| 2016-01-31 | 9:00:00 | 2.253702453 | ['Bastet'] | 1.75E+16 | -5.9835E+15 |
| 2016-01-31 | 10:30:00 | 2.254176538 | ['Wosret'] | 1.14E+16 | -3.89096E+15 |
| 2016-01-31 | 11:00:00 | 2.25427 | ['Neith'] | 1.02E+16 | -3.52107E+15 |
| 2016-01-31 | 12:00:00 | 2.254651923 | ['Bes'] | 1.49E+16 | -5.09302E+15 |
| 2016-01-31 | 12:30:00 | 2.254808868 | ['Bes'] | 1.50E+16 | -5.1402E+15 |
| 2016-01-31 | 13:00:00 | 2.254967358 | ['Khonsu'] | 1.36E+16 | -4.62933E+15 |
| 2016-01-31 | 13:30:00 | 2.255125849 | ['Khonsu'] | 1.24E+16 | -4.23708E+15 |
| 2016-01-31 | 14:00:00 | 2.255290727 | ['Khonsu'] | 1.08E+16 | -3.7086E+15 |
| 2016-01-31 | 14:30:00 | 2.255453019 | ['Khonsu'] | 1.45E+16 | -4.94398E+15 |
| 2016-01-31 | 15:00:00 | 2.25561 | ['Khonsu'] | 1.64E+16 | -5.59416E+15 |
| 2016-01-31 | 15:30:00 | 2.255766981 | ['Imhotep1'] | 1.51E+16 | -5.15185E+15 |
| 2016-01-31 | 16:00:00 | 2.255925283 | ['Imhotep'] | 2.00E+16 | -6.81428E+15 |
| 2016-01-31 | 16:30:00 | 2.256083774 | ['Imhotep'] | 2.06E+16 | -7.03303E+15 |
| 2016-01-31 | 17:00:00 | 2.256238163 | ['Imhotep'] | 2.06E+16 | -7.05738E+15 |
| 2016-01-31 | 19:00:00 | 2.256881887 | ['Babi'] | 7.68E+15 | -2.63071E+15 |
| 2016-01-31 | 19:30:00 | 2.257040189 | ['Aker'] | 1.05E+16 | -3.60009E+15 |
| 2016-01-31 | 20:00:00 | 2.257198868 | ['Hapi'] | 1.69E+16 | -5.76043E+15 |
| 2016-01-31 | 20:30:00 | 2.257355769 | ['Sobek'] | 1.51E+16 | -5.15651E+15 |
| 2016-01-31 | 21:00:00 | 2.257514906 | ['Neith'] | 1.86E+16 | -6.33929E+15 |
| 2016-01-31 | 21:30:00 | 2.257672692 | ['Geb'] | 1.96E+16 | -6.68796E+15 |
| 2016-01-31 | 22:00:00 | 2.257829434 | ['Geb'] | 1.46E+16 | -4.98248E+15 |
| 2016-01-31 | 22:30:00 | 2.257987925 | ['Geb'] | 1.07E+16 | -3.65447E+15 |
| 2016-01-31 | 23:00:00 | 2.25815125 | ['Geb'] | 9.26E+15 | -3.16462E+15 |
| 2016-01-31 | 23:30:00 | 2.258313462 | ['Anuket'] | 9.27E+15 | -3.16891E+15 |
| 2016-02-01 | 0:00:00 | 2.258475098 | ['Anuket'] | 1.41E+16 | -4.8251E+15 |
| 2016-02-01 | 2:00:00 | 2.259103962 | ['Atum'] | 9.89E+15 | -3.38777E+15 |
| 2016-02-01 | 3:00:00 | 2.259417547 | ['Atum'] | 1.16E+16 | -3.94699E+15 |
| 2016-02-01 | 3:30:00 | 2.259582545 | ['Khonsu'] | 1.61E+16 | -5.48646E+15 |
| 2016-02-01 | 4:00:00 | 2.259744717 | ['Imhotep1'] | 1.68E+16 | -5.74795E+15 |
| 2016-02-01 | 4:30:00 | 2.259903019 | ['Imhotep1'] | 2.19E+16 | -7.4757E+15 |
| 2016-02-01 | 5:00:00 | 2.260061509 | ['Imhotep'] | 2.73E+16 | -9.33894E+15 |
| 2016-02-01 | 5:30:00 | 2.260218462 | ['Imhotep'] | 1.77E+16 | -6.0406E+15 |
| 2016-02-01 | 6:00:00 | 2.260374423 | ['Imhotep'] | 1.28E+16 | -4.37881E+15 |
| 2016-02-01 | 7:30:00 | 2.260850377 | ['Khepry'] | 8.41E+15 | -2.88577E+15 |
| 2016-02-01 | 8:00:00 | 2.261014107 | ['Khepry'] | 1.19E+16 | -4.06742E+15 |
| 2016-02-01 | 8:30:00 | 2.261175962 | ['Khepry'] | 1.33E+16 | -4.53392E+15 |
| 2016-02-01 | 9:00:00 | 2.261332642 | ['Aker'] | 1.01E+16 | -3.45624E+15 |
| 2016-02-01 | 9:30:00 | 2.261491132 | ['Hapi'] | 7.58E+15 | -2.60123E+15 |
| 2016-02-01 | 10:30:00 | 2.261805962 | ['Wosret'] | 6.63E+15 | -2.28314E+15 |
| 2016-02-01 | 11:00:00 | 2.261963019 | ['Wosret'] | 7.10E+15 | -2.4308E+15 |
| 2016-02-01 | 11:30:00 | 2.262122115 | ['Wosret'] | 7.89E+15 | -2.702E+15 |
| 2016-02-01 | 12:00:00 | 2.262282909 | ['Anuket'] | 1.10E+16 | -3.76633E+15 |
| 2016-02-01 | 12:30:00 | 2.262446667 | ['Seth'] | 1.61E+16 | -5.49482E+15 |
| 2016-02-01 | 13:00:00 | 2.262534286 | ['Atum'] | 1.46E+16 | -5.01509E+15 |
| 2016-02-01 | 16:30:00 | 2.263740541 | ['Atum'] | 2.47E+16 | -8.4425E+15 |
| 2016-02-01 | 22:00:00 | 2.265464717 | ['Wosret'] | 6.20E+15 | -2.12437E+15 |
| 2016-02-01 | 22:30:00 | 2.265623208 | ['Maftet'] | 5.94E+15 | -2.03721E+15 |
| 2016-02-02 | 0:30:00 | 2.266255 | ['Anuket'] | 1.46E+16 | -5.00441E+15 |
| 2016-02-02 | 3:00:00 | 2.267032432 | ['Apis'] | 5.10E+15 | -1.76585E+15 |
| 2016-02-02 | 17:30:00 | 2.271698519 | ['Apis'] | 1.68E+16 | -5.7641E+15 |
| 2016-02-02 | 18:00:00 | 2.271818113 | ['Ash'] | 1.17E+16 | -3.99534E+15 |
| 2016-02-02 | 18:30:00 | 2.271974423 | ['Aten'] | 8.17E+15 | -2.79291E+15 |
| 2016-02-02 | 23:00:00 | 2.273403077 | ['Nut'] | 3.48E+15 | -1.22951E+15 |
| 2016-02-02 | 23:30:00 | 2.273560392 | ['Serqet'] | 3.15E+15 | -1.12414E+15 |
| 2016-02-03 | 5:30:00 | 2.275493514 | ['Serqet'] | 1.68E+16 | -5.73225E+15 |
| 2016-02-03 | 10:00:00 | 2.276893396 | ['Serqet'] | 3.86E+15 | -1.34446E+15 |
| 2016-02-03 | 20:30:00 | 2.280229107 | ['Bastet'] | 9.27E+15 | -3.17123E+15 |
| 2016-02-04 | 6:00:00 | 2.283244423 | ['Khepry'] | 1.32E+16 | -4.51705E+15 |
| 2016-02-04 | 6:30:00 | 2.283400943 | ['Khepry'] | 9.81E+15 | -3.35769E+15 |
| 2016-02-04 | 13:00:00 | 2.285463393 | ['Seth'] | 1.31E+16 | -4.50167E+15 |
| 2016-02-04 | 13:30:00 | 2.285624423 | ['Hatmehit'] | 1.23E+16 | -4.21099E+15 |
| 2016-02-04 | 14:00:00 | 2.285781132 | ['Anuket'] | 8.77E+15 | -3.00382E+15 |
| 2016-02-04 | 17:00:00 | 2.286729091 | ['Imhotep'] | 2.32E+16 | -7.93779E+15 |
| 2016-02-04 | 18:30:00 | 2.2872062 | ['Ash'] | 1.09E+16 | -3.72161E+15 |
| 2016-02-04 | 19:00:00 | 2.287362264 | ['Imhotep'] | 7.46E+15 | -2.55792E+15 |
| 2016-02-04 | 19:30:00 | 2.28752 | ['Khepry'] | 6.73E+15 | -2.31038E+15 |
| 2016-02-04 | 20:00:00 | 2.287677925 | ['Neith'] | 9.05E+15 | -3.1128E+15 |
| 2016-02-05 | 0:30:00 | 2.289107143 | ['Wosret'] | 1.16E+16 | -3.97281E+15 |
| 2016-02-05 | 1:30:00 | 2.289421481 | ['Atum'] | 1.34E+16 | -4.59401E+15 |
| 2016-02-05 | 3:00:00 | 2.289903962 | ['Khonsu'] | 1.17E+16 | -4.00532E+15 |
| 2016-02-05 | 4:00:00 | 2.290213774 | ['Anuket'] | 1.28E+16 | -4.36601E+15 |
| 2016-02-05 | 5:00:00 | 2.290532545 | ['Imhotep'] | 1.68E+16 | -5.74377E+15 |
| 2016-02-05 | 5:30:00 | 2.290697547 | ['Khonsu'] | 1.62E+16 | -5.53595E+15 |
| 2016-02-05 | 6:30:00 | 2.291012115 | ['Khonsu'] | 1.33E+16 | -4.53667E+15 |
| 2016-02-05 | 7:00:00 | 2.291168302 | ['Imhotep'] | 1.09E+16 | -3.72407E+15 |
| 2016-02-05 | 9:30:00 | 2.29195717 | ['Hathor'] | 1.67E+16 | -5.69683E+15 |
| 2016-02-05 | 10:30:00 | 2.292283137 | ['Khonsu'] | 1.33E+16 | -4.57099E+15 |
| 2016-02-05 | 11:30:00 | 2.292595849 | ['Maat'] | 1.04E+16 | -3.54681E+15 |
| 2016-02-05 | 23:30:00 | 2.296392308 | ['Hapi'] | 1.39E+16 | -4.76326E+15 |
| 2016-02-06 | 0:00:00 | 2.296520606 | ['Atum'] | 1.35E+16 | -4.63116E+15 |
| 2016-02-06 | 2:30:00 | 2.297408182 | ['Hatmehit'] | 1.21E+16 | -4.23108E+15 |
| 2016-02-06 | 3:00:00 | 2.29750283 | ['Khonsu'] | 1.09E+16 | -3.7177E+15 |
| 2016-02-06 | 3:30:00 | 2.297660377 | ['Imhotep'] | 1.02E+16 | -3.49216E+15 |
| 2016-02-06 | 4:00:00 | 2.297818113 | ['Imhotep'] | 1.02E+16 | -3.49527E+15 |
| 2016-02-06 | 4:30:00 | 2.297976226 | ['Imhotep'] | 1.23E+16 | -4.2041E+15 |
| 2016-02-06 | 5:00:00 | 2.298139107 | ['Imhotep'] | 1.41E+16 | -4.8298E+15 |
| 2016-02-06 | 5:30:00 | 2.298300192 | ['Bes'] | 1.29E+16 | -4.42357E+15 |
| 2016-02-06 | 19:30:00 | 2.302730577 | ['Hatmehit'] | 1.37E+16 | -4.69357E+15 |
| 2016-02-06 | 20:00:00 | 2.302886981 | ['Anhur'] | 1.63E+16 | -5.5675E+15 |
| 2016-02-07 | 2:00:00 | 2.304783519 | ['Khonsu'] | 9.10E+15 | -3.12854E+15 |
| 2016-02-07 | 2:30:00 | 2.304946545 | ['Imhotep1'] | 1.14E+16 | -3.90369E+15 |
| 2016-02-07 | 3:00:00 | 2.305107547 | ['Imhotep'] | 1.08E+16 | -3.69584E+15 |
| 2016-02-07 | 3:30:00 | 2.305263654 | ['Imhotep'] | 1.29E+16 | -4.40514E+15 |
| 2016-02-07 | 4:00:00 | 2.305419811 | ['Imhotep'] | 1.43E+16 | -4.87784E+15 |
| 2016-02-07 | 4:30:00 | 2.305577692 | ['Bes'] | 1.44E+16 | -4.91238E+15 |
| 2016-02-07 | 5:00:00 | 2.305734906 | ['Bes'] | 1.11E+16 | -3.80304E+15 |
| 2016-02-07 | 6:00:00 | 2.30604717 | ['Bes'] | 9.53E+15 | -3.27841E+15 |
| 2016-02-07 | 6:30:00 | 2.306209643 | ['Anhur'] | 1.00E+16 | -3.42457E+15 |
| 2016-02-07 | 7:00:00 | 2.30637283 | ['Anhur'] | 1.38E+16 | -4.72667E+15 |
| 2016-02-07 | 7:30:00 | 2.306530189 | ['Sobek'] | 1.70E+16 | -5.82176E+15 |
| 2016-02-07 | 8:00:00 | 2.306688113 | ['Neith'] | 1.80E+16 | -6.14988E+15 |
| 2016-02-07 | 9:00:00 | 2.307 | ['Wosret'] | 1.78E+16 | -6.10197E+15 |
| 2016-02-07 | 10:00:00 | 2.307315283 | ['Anuket'] | 1.49E+16 | -5.10992E+15 |
| 2016-02-07 | 13:00:00 | 2.308264528 | ['Khonsu'] | 7.54E+15 | -2.59922E+15 |
| 2016-02-07 | 13:30:00 | 2.308420577 | ['Khonsu'] | 9.93E+15 | -3.40272E+15 |
| 2016-02-07 | 14:00:00 | 2.30858 | ['Imhotep1'] | 1.17E+16 | -4.00978E+15 |
| 2016-02-07 | 14:30:00 | 2.308742963 | ['Imhotep'] | 1.38E+16 | -4.7037E+15 |
| 2016-02-07 | 15:00:00 | 2.308902264 | ['Imhotep'] | 1.39E+16 | -4.75637E+15 |
| 2016-02-07 | 16:00:00 | 2.309216981 | ['Khonsu'] | 1.14E+16 | -3.91878E+15 |
| 2016-02-07 | 16:30:00 | 2.309347429 | ['Bes'] | 1.13E+16 | -3.85391E+15 |
| 2016-02-08 | 2:30:00 | 2.312521739 | ['Imhotep'] | 1.53E+16 | -5.24071E+15 |
| 2016-02-08 | 12:00:00 | 2.315544906 | ['Imhotep'] | 1.24E+16 | -4.23414E+15 |
| 2016-02-09 | 1:00:00 | 2.319616765 | ['Atum'] | 1.42E+16 | -4.87411E+15 |
| 2016-02-10 | 7:00:00 | 2.329122 | ['Atum'] | 1.62E+16 | -5.55138E+15 |
| 2016-02-10 | 7:30:00 | 2.329282264 | ['Neith'] | 1.68E+16 | -5.74871E+15 |
| 2016-02-10 | 8:00:00 | 2.329439245 | ['Wosret'] | 1.65E+16 | -5.65844E+15 |
| 2016-02-10 | 8:30:00 | 2.329595 | ['Wosret'] | 1.54E+16 | -5.27763E+15 |
| 2016-02-10 | 11:00:00 | 2.330381429 | ['Hapi'] | 1.29E+16 | -4.42237E+15 |
| 2016-02-10 | 11:30:00 | 2.330543774 | ['Anubis'] | 1.31E+16 | -4.48034E+15 |
| 2016-02-10 | 12:00:00 | 2.330701132 | ['Atum'] | 1.34E+16 | -4.60744E+15 |
| 2016-02-10 | 12:30:00 | 2.330858302 | ['Khonsu'] | 1.21E+16 | -4.15756E+15 |
| 2016-02-10 | 14:00:00 | 2.331326792 | ['Imhotep'] | 1.90E+16 | -6.49166E+15 |
| 2016-02-10 | 14:30:00 | 2.331483774 | ['Imhotep'] | 2.14E+16 | -7.33E+15 |
| 2016-02-10 | 15:30:00 | 2.331803393 | ['Imhotep'] | 1.35E+16 | -4.60013E+15 |
| 2016-02-10 | 16:00:00 | 2.33196566 | ['Imhotep'] | 1.01E+16 | -3.45483E+15 |
| 2016-02-10 | 16:30:00 | 2.332121346 | ['Bes'] | 9.77E+15 | -3.34534E+15 |
| 2016-02-10 | 17:30:00 | 2.332434151 | ['Bes'] | 1.01E+16 | -3.47157E+15 |
| 2016-02-10 | 18:30:00 | 2.332746731 | ['Anhur'] | 9.43E+15 | -3.23053E+15 |
| 2016-02-10 | 22:30:00 | 2.334006038 | ['Anuket'] | 1.10E+16 | -3.77318E+15 |
| 2016-02-10 | 23:00:00 | 2.334167679 | ['Atum'] | 1.21E+16 | -4.14578E+15 |
| 2016-02-10 | 23:30:00 | 2.334330377 | ['Imhotep1'] | 1.02E+16 | -3.47628E+15 |
| 2016-02-11 | 0:00:00 | 2.334487547 | ['Anuket'] | 9.93E+15 | -3.39947E+15 |
| 2016-02-11 | 0:30:00 | 2.334643077 | ['Sobek'] | 1.11E+16 | -3.80818E+15 |
| 2016-02-11 | 2:00:00 | 2.335114038 | ['Ash'] | 2.11E+16 | -7.20442E+15 |
| 2016-02-11 | 2:30:00 | 2.335269623 | ['Imhotep'] | 2.22E+16 | -7.59454E+15 |
| 2016-02-11 | 3:00:00 | 2.335425577 | ['Imhotep'] | 1.57E+16 | -5.37069E+15 |
| 2016-02-11 | 3:30:00 | 2.335583774 | ['Imhotep'] | 1.01E+16 | -3.46108E+15 |
| 2016-02-11 | 15:00:00 | 2.339211538 | ['Imhotep'] | 1.71E+16 | -5.83711E+15 |
| 2016-02-11 | 17:00:00 | 2.339844364 | ['Khepry'] | 8.13E+15 | -2.7861E+15 |
| 2016-02-11 | 22:00:00 | 2.341416667 | ['Maftet'] | 1.86E+16 | -6.34894E+15 |
| 2016-02-11 | 22:30:00 | 2.341579091 | ['Maftet'] | 1.79E+16 | -6.11175E+15 |
| 2016-02-11 | 23:00:00 | 2.341736667 | ['Anuket'] | 1.26E+16 | -4.33493E+15 |
| 2016-02-11 | 23:30:00 | 2.341893208 | ['Anuket'] | 9.98E+15 | -3.4149E+15 |
| 2016-02-12 | 0:00:00 | 2.342050189 | ['Anuket'] | 7.25E+15 | -2.49246E+15 |
| 2016-02-12 | 1:30:00 | 2.342518113 | ['Atum'] | 1.06E+16 | -3.63363E+15 |
| 2016-02-12 | 2:00:00 | 2.342675094 | ['Khonsu'] | 1.91E+16 | -6.51278E+15 |
| 2016-02-12 | 3:00:00 | 2.342999245 | ['Imhotep'] | 1.89E+16 | -6.46923E+15 |
| 2016-02-12 | 3:30:00 | 2.343157347 | ['Imhotep'] | 1.18E+16 | -4.01807E+15 |
| 2016-02-12 | 4:00:00 | 2.343310189 | ['Imhotep'] | 8.19E+15 | -2.81249E+15 |
| 2016-02-12 | 4:30:00 | 2.343465192 | ['Hatmehit'] | 6.97E+15 | -2.38361E+15 |
| 2016-02-12 | 7:00:00 | 2.344258679 | ['Babi'] | 4.83E+15 | -1.66854E+15 |
| 2016-02-12 | 8:00:00 | 2.344571731 | ['Atum'] | 4.21E+15 | -1.45487E+15 |
| 2016-02-12 | 9:00:00 | 2.344883208 | ['Wosret'] | 6.84E+15 | -2.34581E+15 |
| 2016-02-12 | 9:30:00 | 2.345039811 | ['Maftet'] | 1.06E+16 | -3.62342E+15 |
| 2016-02-12 | 10:00:00 | 2.345196981 | ['Maftet'] | 1.59E+16 | -5.4235E+15 |
| 2016-02-12 | 10:30:00 | 2.345350392 | ['Maftet'] | 1.55E+16 | -5.30496E+15 |
| 2016-02-12 | 11:00:00 | 2.345514386 | ['Anuket'] | 1.29E+16 | -4.40995E+15 |
| 2016-02-12 | 11:30:00 | 2.345676346 | ['Anuket'] | 8.84E+15 | -3.02117E+15 |
| 2016-02-12 | 12:00:00 | 2.345831698 | ['Anuket'] | 7.14E+15 | -2.44918E+15 |
| 2016-02-12 | 14:30:00 | 2.346579333 | ['Apis'] | 2.18E+16 | -7.47971E+15 |
| 2016-02-12 | 15:30:00 | 2.346965938 | ['Apis'] | 1.53E+16 | -5.24108E+15 |
| 2016-02-12 | 21:00:00 | 2.348618095 | ['Wosret'] | 6.87E+15 | -2.39514E+15 |
| 2016-02-12 | 22:00:00 | 2.349006471 | ['Hatmehit'] | 1.29E+16 | -4.42764E+15 |
| 2016-02-12 | 22:30:00 | 2.349133962 | ['Maftet'] | 1.15E+16 | -3.95666E+15 |
| 2016-02-12 | 23:30:00 | 2.349445962 | ['Hapi'] | 4.21E+15 | -1.46728E+15 |
| 2016-02-13 | 8:30:00 | 2.352278491 | ['Bastet'] | 4.35E+15 | -1.49812E+15 |
| 2016-02-13 | 10:00:00 | 2.352758868 | ['Anuket'] | 1.47E+16 | -5.05363E+15 |
| 2016-02-13 | 10:30:00 | 2.352914038 | ['Anuket'] | 1.14E+16 | -3.9108E+15 |
| 2016-02-13 | 11:00:00 | 2.353069245 | ['Anuket'] | 5.34E+15 | -1.85308E+15 |
| 2016-02-13 | 11:30:00 | 2.353226038 | ['Atum'] | 5.90E+15 | -2.03594E+15 |
| 2016-02-13 | 12:30:00 | 2.353538077 | ['Anuket'] | 3.39E+15 | -1.21082E+15 |
| 2016-02-13 | 14:00:00 | 2.354016981 | ['Imhotep'] | 1.12E+16 | -3.84691E+15 |
| 2016-02-13 | 15:30:00 | 2.354484314 | ['Khepry'] | 1.29E+16 | -4.40803E+15 |
| 2016-02-13 | 17:00:00 | 2.35495434 | ['Hapi'] | 8.72E+15 | -2.99516E+15 |
| 2016-02-13 | 18:00:00 | 2.355278113 | ['Bastet'] | 3.62E+15 | -1.27824E+15 |
| 2016-02-13 | 19:00:00 | 2.355588302 | ['Wosret'] | 3.56E+15 | -1.28984E+15 |
| 2016-02-13 | 20:00:00 | 2.355901698 | ['Wosret'] | 6.34E+15 | -2.18165E+15 |
| 2016-02-13 | 22:00:00 | 2.356530536 | ['Anuket'] | 1.68E+16 | -5.75696E+15 |
| 2016-02-13 | 22:30:00 | 2.356692692 | ['Anuket'] | 1.40E+16 | -4.78262E+15 |
| 2016-02-13 | 23:00:00 | 2.356849615 | ['Anuket'] | 1.08E+16 | -3.70228E+15 |
| 2016-02-13 | 23:30:00 | 2.357004423 | ['Seth'] | 6.68E+15 | -2.28408E+15 |
| 2016-02-14 | 2:00:00 | 2.357783019 | ['Imhotep'] | 1.31E+16 | -4.48517E+15 |
| 2016-02-14 | 2:30:00 | 2.357945 | ['Imhotep'] | 1.96E+16 | -6.70758E+15 |
| 2016-02-14 | 3:00:00 | 2.358106604 | ['Imhotep'] | 2.03E+16 | -6.94644E+15 |
| 2016-02-14 | 3:30:00 | 2.358263396 | ['Imhotep'] | 1.38E+16 | -4.71399E+15 |
| 2016-02-14 | 4:00:00 | 2.358419811 | ['Imhotep'] | 8.06E+15 | -2.75797E+15 |
| 2016-02-14 | 6:00:00 | 2.359043208 | ['Anhur'] | 6.86E+15 | -2.37612E+15 |
| 2016-02-14 | 6:30:00 | 2.359201667 | ['Hapi'] | 5.34E+15 | -1.85127E+15 |
| 2016-02-14 | 7:00:00 | 2.359363818 | ['Sobek'] | 4.13E+15 | -1.43787E+15 |
| 2016-02-14 | 9:30:00 | 2.360145192 | ['Neith'] | 9.53E+15 | -3.26784E+15 |
| 2016-02-14 | 10:00:00 | 2.360300566 | ['Anuket'] | 1.51E+16 | -5.15758E+15 |
| 2016-02-14 | 11:00:00 | 2.360560769 | ['Anuket'] | 1.30E+16 | -4.53383E+15 |
| 2016-02-14 | 12:00:00 | 2.360933962 | ['Atum'] | 8.75E+15 | -3.00852E+15 |
| 2016-02-14 | 13:00:00 | 2.361243962 | ['Khonsu'] | 8.69E+15 | -2.98278E+15 |
| 2016-02-14 | 13:30:00 | 2.361400755 | ['Imhotep1'] | 5.73E+15 | -1.99167E+15 |
| 2016-02-14 | 14:00:00 | 2.361558519 | ['Imhotep'] | 1.12E+16 | -3.85719E+15 |
| 2016-02-14 | 14:30:00 | 2.361720909 | ['Imhotep'] | 1.98E+16 | -6.75919E+15 |
| 2016-02-14 | 15:00:00 | 2.361880377 | ['Imhotep'] | 2.10E+16 | -7.17075E+15 |
| 2016-02-14 | 15:30:00 | 2.362035769 | ['Imhotep'] | 1.95E+16 | -6.67544E+15 |
| 2016-02-14 | 16:00:00 | 2.362190755 | ['Imhotep'] | 9.25E+15 | -3.16719E+15 |
| 2016-02-14 | 16:30:00 | 2.362347358 | ['Khepry'] | 7.05E+15 | -2.41889E+15 |
| 2016-02-14 | 17:00:00 | 2.362500816 | ['Khepry'] | 7.44E+15 | -2.54781E+15 |
| 2016-02-14 | 17:30:00 | 2.362658269 | ['Bes'] | 7.50E+15 | -2.57758E+15 |
| 2016-02-14 | 18:00:00 | 2.36281566 | ['Bes'] | 1.01E+16 | -3.47017E+15 |
| 2016-02-14 | 18:30:00 | 2.362977818 | ['Anhur'] | 7.03E+15 | -2.4364E+15 |
| 2016-02-14 | 20:00:00 | 2.363449057 | ['Wosret'] | 7.34E+15 | -2.53019E+15 |
| 2016-02-15 | 0:00:00 | 2.364707736 | ['Atum'] | 9.12E+15 | -3.12255E+15 |
| 2016-02-15 | 1:00:00 | 2.365018846 | ['Khonsu'] | 1.11E+16 | -3.78879E+15 |
| 2016-02-15 | 2:30:00 | 2.36548717 | ['Imhotep'] | 1.61E+16 | -5.52453E+15 |
| 2016-02-15 | 3:00:00 | 2.365643462 | ['Imhotep'] | 1.91E+16 | -6.52311E+15 |
| 2016-02-15 | 3:30:00 | 2.365799623 | ['Bes'] | 1.49E+16 | -5.10236E+15 |
| 2016-02-15 | 4:00:00 | 2.365961429 | ['Bes'] | 1.44E+16 | -4.92088E+15 |
| 2016-02-15 | 4:30:00 | 2.366122453 | ['Bes'] | 1.19E+16 | -4.06653E+15 |
| 2016-02-15 | 5:00:00 | 2.366279057 | ['Bes'] | 9.99E+15 | -3.4207E+15 |
| 2016-02-15 | 5:30:00 | 2.366434038 | ['Bes'] | 6.20E+15 | -2.18234E+15 |
| 2016-02-15 | 9:30:00 | 2.367692115 | ['Anuket'] | 1.01E+16 | -3.47512E+15 |
| 2016-02-15 | 10:00:00 | 2.367846981 | ['Anuket'] | 1.28E+16 | -4.39401E+15 |
| 2016-02-15 | 10:30:00 | 2.368003396 | ['Anuket'] | 1.17E+16 | -4.02709E+15 |
| 2016-02-15 | 11:00:00 | 2.36816 | ['Hapi'] | 1.11E+16 | -3.7911E+15 |
| 2016-02-15 | 12:00:00 | 2.36847 | ['Hathor'] | 1.08E+16 | -3.69592E+15 |
| 2016-02-15 | 12:30:00 | 2.368623137 | ['Khonsu'] | 1.05E+16 | -3.58708E+15 |
| 2016-02-15 | 15:00:00 | 2.369410323 | ['Khonsu'] | 1.66E+16 | -5.82729E+15 |
| 2016-02-15 | 16:30:00 | 2.369909714 | ['Khonsu'] | 1.40E+16 | -4.81764E+15 |
| 2016-02-15 | 20:00:00 | 2.370990667 | ['Anubis'] | 2.37E+16 | -8.09489E+15 |
| 2016-02-15 | 22:00:00 | 2.371612364 | ['Anubis'] | 1.41E+16 | -4.8434E+15 |
| 2016-02-16 | 3:00:00 | 2.373180851 | ['Ash'] | 1.52E+16 | -5.20623E+15 |
| 2016-02-16 | 6:00:00 | 2.37412434 | ['Hatmehit'] | 7.76E+15 | -2.69292E+15 |
| 2016-02-16 | 14:00:00 | 2.376638868 | ['Hatmehit'] | 7.77E+15 | -2.70193E+15 |
| 2016-02-16 | 14:30:00 | 2.376795283 | ['Khonsu'] | 9.72E+15 | -3.35713E+15 |
| 2016-02-16 | 15:30:00 | 2.377105849 | ['Anuket'] | 7.98E+15 | -2.74477E+15 |
| 2016-02-16 | 16:30:00 | 2.377415849 | ['Imhotep'] | 8.55E+15 | -2.92707E+15 |
| 2016-02-16 | 22:30:00 | 2.379363333 | ['Hatmehit'] | 7.61E+15 | -2.72663E+15 |
| 2016-02-16 | 23:00:00 | 2.379456939 | ['Anuket'] | 6.10E+15 | -2.14121E+15 |
| 2016-02-17 | 0:00:00 | 2.3797344 | ['Geb'] | 5.42E+15 | -1.92366E+15 |
| 2016-02-17 | 5:30:00 | 2.381522105 | ['Geb'] | 6.48E+15 | -2.24449E+15 |
| 2016-02-17 | 7:00:00 | 2.381967857 | ['Wosret'] | 8.95E+15 | -3.06821E+15 |
| 2016-02-17 | 7:30:00 | 2.3821275 | ['Anhur'] | 1.07E+16 | -3.66039E+15 |
| 2016-02-17 | 8:00:00 | 2.382281887 | ['Anhur'] | 1.30E+16 | -4.47308E+15 |
| 2016-02-17 | 8:30:00 | 2.382438113 | ['Anhur'] | 1.58E+16 | -5.41057E+15 |
| 2016-02-17 | 9:00:00 | 2.38259434 | ['Anhur'] | 1.55E+16 | -5.30731E+15 |
| 2016-02-17 | 11:00:00 | 2.383218909 | ['Sobek'] | 5.60E+15 | -1.95759E+15 |
| 2016-02-17 | 11:30:00 | 2.383380741 | ['Sobek'] | 5.98E+15 | -2.06142E+15 |
| 2016-02-17 | 13:00:00 | 2.383847736 | ['Atum'] | 4.78E+15 | -1.64657E+15 |
| 2016-02-17 | 13:30:00 | 2.384003962 | ['Atum'] | 4.50E+15 | -1.56292E+15 |
| 2016-02-17 | 14:00:00 | 2.384159811 | ['Khonsu'] | 5.44E+15 | -1.88381E+15 |
| 2016-02-17 | 14:30:00 | 2.384314423 | ['Khonsu'] | 7.99E+15 | -2.80343E+15 |
| 2016-02-17 | 15:30:00 | 2.38463 | ['Imhotep'] | 1.04E+16 | -3.55227E+15 |
| 2016-02-17 | 16:00:00 | 2.384791509 | ['Imhotep'] | 9.68E+15 | -3.30748E+15 |
| 2016-02-17 | 16:30:00 | 2.384947736 | ['Imhotep'] | 8.45E+15 | -2.88895E+15 |
| 2016-02-17 | 17:30:00 | 2.385258462 | ['Bes'] | 6.18E+15 | -2.11925E+15 |
| 2016-02-17 | 19:00:00 | 2.385725283 | ['Anhur'] | 1.01E+16 | -3.46295E+15 |
| 2016-02-17 | 19:30:00 | 2.385879615 | ['Anhur'] | 1.73E+16 | -5.93322E+15 |
| 2016-02-17 | 20:30:00 | 2.386202308 | ['Anhur'] | 2.02E+16 | -6.91149E+15 |
| 2016-02-17 | 22:00:00 | 2.386669057 | ['Anuket'] | 1.02E+16 | -3.52016E+15 |
| 2016-02-17 | 23:00:00 | 2.386977115 | ['Atum'] | 5.92E+15 | -2.04895E+15 |
| 2016-02-18 | 0:30:00 | 2.387453091 | ['Neith'] | 6.01E+15 | -2.07441E+15 |
| 2016-02-18 | 2:00:00 | 2.387921923 | ['Imhotep'] | 6.57E+15 | -2.27788E+15 |
| 2016-02-18 | 4:30:00 | 2.388702 | ['Imhotep'] | 8.50E+15 | -2.92249E+15 |
| 2016-02-18 | 5:00:00 | 2.388863333 | ['Bes'] | 6.79E+15 | -2.36242E+15 |
| 2016-02-18 | 7:00:00 | 2.389486038 | ['Anhur'] | 1.54E+16 | -5.25923E+15 |
| 2016-02-18 | 10:30:00 | 2.390521667 | ['Wosret'] | 1.22E+16 | -4.25662E+15 |
| 2016-02-18 | 11:30:00 | 2.390859655 | ['Wosret'] | 1.07E+16 | -3.68767E+15 |
| 2016-02-18 | 16:30:00 | 2.392459057 | ['Wosret'] | 1.44E+16 | -4.92195E+15 |
| 2016-02-18 | 17:00:00 | 2.392614906 | ['Imhotep'] | 1.01E+16 | -3.44884E+15 |
| 2016-02-18 | 17:30:00 | 2.392773137 | ['Khepry'] | 8.00E+15 | -2.74392E+15 |
| 2016-02-18 | 22:00:00 | 2.394181455 | ['Hatmehit'] | 9.68E+15 | -3.31266E+15 |
| 2016-02-19 | 3:00:00 | 2.39575451 | ['Imhotep'] | 1.29E+16 | -4.42207E+15 |
| 2016-02-19 | 3:30:00 | 2.395907736 | ['Imhotep'] | 1.83E+16 | -6.23745E+15 |
| 2016-02-19 | 10:00:00 | 2.397935283 | ['Hatmehit'] | 6.97E+15 | -2.40418E+15 |
| 2016-02-19 | 13:00:00 | 2.398877358 | ['Anuket'] | 6.59E+15 | -2.26222E+15 |
| 2016-02-20 | 16:00:00 | 2.407318077 | ['Apis'] | 2.12E+16 | -7.22781E+15 |
| 2016-02-21 | 6:00:00 | 2.411688627 | ['Babi'] | 5.71E+15 | -1.97934E+15 |
| 2016-02-21 | 16:00:00 | 2.414813019 | ['Aten'] | 6.07E+15 | -2.09419E+15 |
| 2016-02-27 | 8:30:00 | 2.457286429 | ['Imhotep'] | 1.97E+15 | -6.96958E+14 |
| 2016-02-27 | 9:00:00 | 2.45744566 | ['Ash'] | 2.43E+15 | -8.62893E+14 |
| 2016-02-27 | 20:00:00 | 2.460849434 | ['Imhotep'] | 2.13E+15 | -7.51621E+14 |
| 2016-02-27 | 20:30:00 | 2.460968966 | ['Imhotep'] | 5.42E+15 | -1.88764E+15 |
| 2016-02-28 | 21:30:00 | 2.468751509 | ['Imhotep1'] | 5.77E+15 | -1.98623E+15 |
| 2016-02-28 | 22:00:00 | 2.468905283 | ['Imhotep'] | 6.31E+15 | -2.16711E+15 |
| 2016-02-29 | 5:30:00 | 2.471229811 | ['Neith'] | 4.16E+15 | -1.44572E+15 |
| 2016-02-29 | 22:00:00 | 2.476339057 | ['Neith'] | 4.57E+15 | -1.59869E+15 |
| 2016-03-01 | 1:30:00 | 2.477415094 | ['Wosret'] | 6.68E+15 | -2.31182E+15 |
| 2016-03-01 | 2:00:00 | 2.477570556 | ['Khepry'] | 4.23E+15 | -1.50227E+15 |
| 2016-03-03 | 11:30:00 | 2.495332182 | ['Hatmehit'] | 4.37E+15 | -1.51953E+15 |
| 2016-03-03 | 12:00:00 | 2.495487692 | ['Wosret'] | 4.97E+15 | -1.74127E+15 |
| 2016-03-03 | 12:30:00 | 2.495639245 | ['Wosret'] | 4.50E+15 | -1.58138E+15 |
| 2016-03-03 | 13:00:00 | 2.49579283 | ['Maftet'] | 4.56E+15 | -1.58521E+15 |
| 2016-03-03 | 16:00:00 | 2.496721509 | ['Atum'] | 4.12E+15 | -1.49307E+15 |
| 2016-03-03 | 23:00:00 | 2.498871923 | ['Wosret'] | 4.68E+15 | -1.64475E+15 |
| 2016-03-03 | 23:30:00 | 2.499026038 | ['Wosret'] | 4.74E+15 | -1.63862E+15 |
| 2016-03-04 | 0:30:00 | 2.499336842 | ['Maftet'] | 3.53E+15 | -1.22283E+15 |
| 2016-03-08 | 22:30:00 | 2.535595385 | ['Aten'] | 4.73E+15 | -1.62803E+15 |
| 2016-03-09 | 7:30:00 | 2.538359245 | ['Khonsu'] | 2.94E+15 | -1.0371E+15 |
| 2016-03-09 | 9:00:00 | 2.538813774 | ['Imhotep'] | 2.43E+15 | -8.66955E+14 |
| 2016-03-09 | 10:00:00 | 2.539117115 | ['Imhotep'] | 3.96E+15 | -1.41529E+15 |
| 2016-03-09 | 10:30:00 | 2.539268302 | ['Bes'] | 2.83E+15 | -1.08931E+15 |
| 2016-03-09 | 13:30:00 | 2.540190755 | ['Neith'] | 4.83E+15 | -1.67906E+15 |
| 2016-03-10 | 23:00:00 | 2.550438868 | ['Anhur'] | 6.04E+15 | -2.10149E+15 |
| 2016-03-15 | 11:00:00 | 2.583349811 | ['Imhotep1'] | 2.58E+15 | -9.12038E+14 |
| 2016-03-15 | 15:30:00 | 2.5847148 | ['Anhur'] | 5.46E+15 | -1.90227E+15 |
| 2016-03-19 | 11:00:00 | 2.612447925 | ['Imhotep'] | 2.95E+15 | -1.08596E+15 |
| 2016-03-19 | 11:30:00 | 2.612602 | ['Imhotep'] | 3.02E+15 | -1.12203E+15 |
| 2016-03-23 | 7:00:00 | 2.640197091 | ['Atum'] | 2.51E+15 | -9.0817E+14 |
| 2016-03-23 | 8:00:00 | 2.640434444 | ['Khonsu'] | 6.46E+15 | -2.26849E+15 |
| 2016-04-09 | 12:30:00 | 2.763127037 | ['Imhotep'] | 2.24E+15 | -7.92784E+14 |
| 2016-04-09 | 19:00:00 | 2.765031961 | ['Anuket'] | 2.53E+15 | -8.75282E+14 |
| 2016-04-09 | 19:30:00 | 2.765177736 | ['Hapi'] | 2.56E+15 | -8.9161E+14 |
| 2016-04-09 | 20:00:00 | 2.765323774 | ['Seth'] | 2.15E+15 | -7.51646E+14 |
| 2016-04-09 | 22:30:00 | 2.766061698 | ['Imhotep'] | 2.07E+15 | -7.33521E+14 |
| 2016-04-09 | 23:00:00 | 2.766207736 | ['Imhotep'] | 4.93E+15 | -1.69501E+15 |
| 2016-04-09 | 23:30:00 | 2.766353269 | ['Imhotep'] | 4.06E+15 | -1.39348E+15 |
| 2016-04-10 | 0:00:00 | 2.766497917 | ['Khepry'] | 3.32E+15 | -1.14395E+15 |
| 2016-04-10 | 6:00:00 | 2.768264231 | ['Atum'] | 4.52E+15 | -1.56065E+15 |
| 2016-04-10 | 6:30:00 | 2.76840875 | ['Anubis'] | 3.52E+15 | -1.24955E+15 |
| 2016-04-10 | 18:30:00 | 2.77193125 | ['Anubis'] | 3.26E+15 | -1.14602E+15 |
| 2016-04-10 | 21:30:00 | 2.772804286 | ['Imhotep'] | 5.24E+15 | -1.80904E+15 |
| 2016-04-10 | 22:30:00 | 2.773100714 | ['Imhotep'] | 2.52E+15 | -8.88958E+14 |
| 2016-04-11 | 22:00:00 | 2.78001375 | ['Imhotep'] | 3.90E+15 | -1.39856E+15 |